\documentclass[twocolumn, twocolappendix]{aastex631}

\newcommand{\angstrom}{\mbox{\normalfont\AA}}  

\revised{\today.}
\usepackage{natbib}
\usepackage{amsmath}
\usepackage{subfigure}
\usepackage{graphicx}
\usepackage{threeparttable}
\usepackage{textcomp} 
\usepackage{subfiles}       

\begin{document}

\title{Infrared-Selected Active Galactic Nuclei in the Kepler Fields}

\author[0000-0002-1667-2544]{Tran Tsan}
\affiliation{Physics Division, Lawrence Berkeley National Laboratory, Berkeley, CA 94720, USA}

\author[0000-0001-6919-1237]{Matthew Malkan}
\affiliation{Department of Physics and Astronomy, University of California Los Angeles, Los Angeles, CA 90095-1547}

\author{Rick Edelson}
\affiliation{Eureka Scientific Inc., 2453 Delmer St., Suite 100, Oakland CA 94602}

\author{Krista Smith}
\affiliation{Department of Physics and Astronomy, Texas A \& M University, College Station, TX 77843-4242}

\author{Daniel Stern}
\affiliation{Jet Propulsion Laboratory, California Institute of Technology, Pasadena, CA 91109-8001}

\author{Matthew Graham}
\affiliation{Center for Advanced Computing Research, California Institute of Technology, Pasadena, CA 91125-0001}

\begin{abstract}

We utilized the \cite{EM12} and \cite{Stern} selection techniques and other methods to identify AGN candidates that were monitored during the Kepler prime and K2 missions.
Subsequent to those observations, we obtained 125 long-slit optical spectra with the Lick 3-m telescope, 58 spectra with the Palomar 5-m telescope, and three with the Keck 10-m telescope to test these identifications.
Of these 186 AGN candidates, 105 were confirmed as Type 1 AGN and 35 as Type 2 AGN, while the remaining 46 were found to have other identifications (e.g., stars and normal galaxies).
This indicated an overall reliability of $\sim$75\%, while the two main methods had much higher reliability, 87\%-96\%.
The spectra indicated redshifts out to $ z = 3.4 $.
Then, we examined the AGN sample properties through the Baldwin, Phillips \& Terlevich diagram and compared the AGN's spectral energy distributions (SEDs) with those from the literature. 
We found that our sample yielded the same AGN population as those identified through other methods, such as optical spectroscopy.

\end{abstract}

\section{Introduction} \label{sec:intro}
Time variability is a key characteristic of many Active Galactic Nuclei (AGN), observed across timescales ranging from less than a day to millions or even billions of years. Quantitatively characterizing this variability with typical ground-based observations poses challenges, even with a CCD detector on a dedicated telescope. These challenges include measurement uncertainties of a few percent, sampling once per night, and frequent interruptions from bad weather and bright moonlight. Despite these challenges, AGN variability, paired with spectroscopic data, provides a powerful means of calculating black hole masses and accretion rates. For instance, techniques such as reverberation mapping of broad emission line widths (e.g., H$\beta$), AGN light curves, and luminosity measurements provide robust methods for calculating black hole masses and determining Eddington ratios \citep{Peterson}.

In 2009, NASA launched the Kepler mission to detect exoplanets. However, after a second reaction wheel failure in 2013, NASA repurposed the mission as K2, redirecting its focus to fields along the ecliptic plane. While addressing these limitations was not part of Kepler's original mission, its 42 CCDs demonstrated exceptional sensitivity, producing highly accurate and continuous light curves for compact objects as faint as $V \sim $ 17 mag. Because Kepler's 116-square-degree field of view (F.O.V.) allowed only a small fraction of `live' pixels to be read out and saved \citep{KeplerHandbook}, identifying AGN targets \textit{in advance} of the observations became essential. To address this challenge, we selected the candidates using the methods discussed in Section \ref{sec:sample_sel}. We then monitored them with Kepler to obtain their light curves and confirmed them as AGN using ground-based spectroscopy from the Lick, Palomar, and Keck observatories.

Significant progress has been made in analyzing Kepler AGN light curves. Studies by \cite{Mushotzky2011, Edelson2013, Edelson_2014, Smith18a} and \citet{Aranzana2018} reveal that Kepler Seyferts exhibit steeper power spectral slopes than those observed in ground-based AGN variability studies. These findings suggest that the damped random walk model cannot fully capture the optical variability of radio-quiet Seyferts. By overcoming the limitations of ground-based observations, Kepler enabled more detailed studies of AGN variability. When combined with ground-based spectroscopic data, the studies become a powerful tool for estimating black hole masses. For example, \citet{Pei2014} uses reverberation mapping of the Kepler AGN, KA1858+4850, to measure its broad line region and combines this with spectroscopic data from the Lick 3 m telescope to estimate the black hole mass. Additionally, \citet{Smith18a} examines a small subset of spectroscopically confirmed AGN from this study and finds significant characteristic variability timescales ranging from 9 to 53 days, suggesting a potential correlation between variability timescale and black hole mass.

In this paper, we focus on identifying new AGN in the Kepler fields from the selected AGN candidates through their spectroscopic features. This identification serves as a crucial first step toward an eventual analysis of their variability properties measured in Kepler light curves. That analysis is deferred to future work. For a comprehensive overview of Kepler AGN studies, including previously mentioned results, see \cite{KeplerOverview_Ch6}. 

\begin{figure}
\begin{center}
    \includegraphics[width=0.48\textwidth]{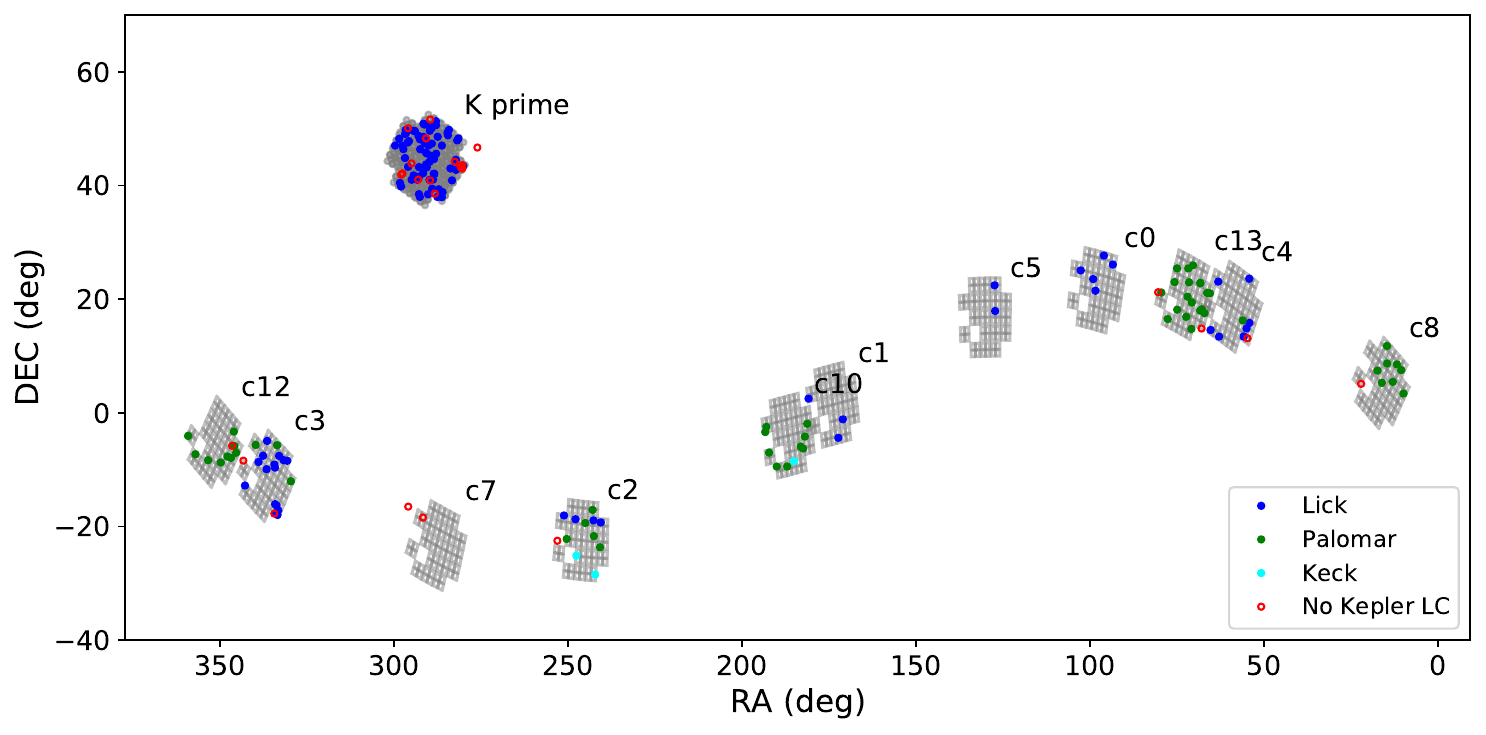}
\end{center}
\caption{Sky map of Lick, Palomar, and Keck candidates and the corresponding Kepler campaigns. The grey panels mark the positions of the CCDs during the campaigns: K2\textsuperscript{1} (c\#) and the Kepler Prime field\textsuperscript{2}. The solid circles indicate objects observed by Kepler and have light curve (LC) data, while empty circles mean they do not have any LC data. These objects were either slightly outside the Kepler campaign fields, between CCD slots, or part of bad CCD slots. Thus, they did not have any data.} 
\small\textsuperscript{1} \small{K2 footprints: https://github.com/KeplerGO/K2FootprintFiles.}
\small\textsuperscript{2} \small{K prime footprint: https://archive.stsci.edu/missions-and-data/kepler/field}.
\label{fig:KeplerCMap}
\end{figure}

The observations cover Kepler prime field and the K2 Fields \# 0-5, 7, 8, 10, 12, and 13 (Figure \ref{fig:KeplerCMap}). We surveyed the AGN candidates at the aforementioned ground-based observatories and supplemented the analysis with broadband photometry from Galaxy Evolution Explorer (GALEX,\dataset[10.17909/25nx-tp05]{http://dx.doi.org/10.17909/25nx-tp05}), SDSS, Two Micron All-Sky Survey (2MASS, \citet{2MASS}), and the Wide-field Infrared Survey Explorer (WISE, \citet{WISE}). 

The paper is structured as follows. Section \ref{sec:sample_sel} details the Kepler AGN candidate selection. Sections \ref{sec:obs} and \ref{sec:RM} describe the ground-based observations and data reduction processes. For the measurements in Section \ref{sec:RM}, we assumed the following cosmological parameters: $H_0 = 73.8$ km s$^{-1}$ Mpc$^{-1}$, $\Omega_M = 0.27$, and $\Omega_{vac} = 1 - \Omega_M$. Section \ref{sec:sample_reliability} examines the reliability of the Kepler AGN candidate sample. Section \ref{sec:Ana_properties} checks the sample AGN properties against established trends through the Baldwin, Phillips \& Terlevich (BPT) diagram \citep{baldwin} and Spectral Energy Distributions (SEDs) for Type 1 AGN. Finally, Section \ref{sec:conclusion} summarizes the results.

\section{Kepler AGN Candidate Selection} \label{sec:sample_sel}

As previously mentioned, the design of Kepler required that all targets be identified beforehand so their pixels could be masked and downloaded during the campaign. We used a four-step process to observe AGN and highly likely AGN candidates in each Kepler field.

First, we searched for known AGN within 7 degrees of the expected Kepler boresight center for the original (Kepler prime) field and subsequent K2 fields, as their positions were announced.
This was done using the \cite{Veron} AGN catalog on the BROWSE interface on the NASA/HEASARC website\footnote{\url{https://heasarc.gsfc.nasa.gov/cgi-bin/W3Browse/w3browse.pl}}, and the SDSS survey \citep{Shen,Schneider}.
This yielded an insufficient number of known AGN because of a dearth of all-sky AGN surveys down to a depth comparable to Kepler's sensitivity limit.
Because these objects are already known AGN, they will not be discussed further in this paper.

Second, we identified highly-likely AGN candidates using several approaches, most notably \cite[][EM12 hereafter]{EM12} and \cite{Stern}.
We combined these two lists to form a preliminary target list for each field. 

Third, we performed miscellaneous checks of the observability of these targets, then proposed them for Kepler observations.
Slight differences between the predicted and actual Kepler boresight position caused some of these potential targets to actually come close but not actually fall on active detectors, but rather in the spaces between detectors or just outside the edges.
Those targets that were accepted and fell on active detectors were then observed by Kepler. 

Fourth, and finally, we performed follow-up spectroscopy at the Lick, Palomar, and Keck observatories to test whether the AGN candidates were actual AGN. This paper reports the results of this last step. Table \ref{tab:logsheet} outlines details of the observations of the AGN candidates and their selection methods. 
 In the end, this approach led to Kepler observing a total of 1155 AGN and AGN candidates proposed by our group. 

\section{Observations} \label{sec:obs}
This section outlines the ground-based observations done at Lick, Palomar, and Keck. We confirmed spectroscopically 186 AGN candidates using the Kast spectrograph on the Lick 3.0-meter reflector, the Double Spectrograph (DBSP) on the Palomar 5-meter reflector \citep{Oke}, and the Low-Resolution Imaging Spectrograph (LRIS) on Keck 10-m reflector. 

\begin{figure}
\plotone{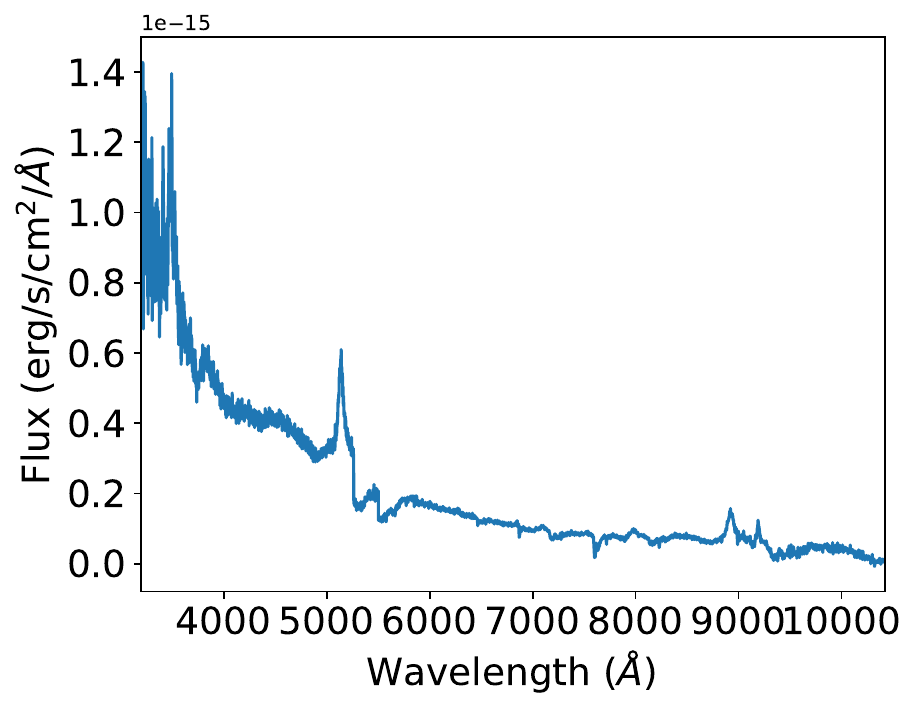}
\caption{Flux for the Lick, Palomar, and Keck Spectra. This figure is a set of spectra from 186 Lick, Palomar, and Keck objects. We show one of the objects here: 0039+0322. The full set is shown in the HTML edition. See Table \ref{tab:linetable} for the observed spectral features and line measurements.}
\label{fig:flux_spectra}
\end{figure}

The spectroscopic observations spanned from 2011 to 2016. We observed 125 objects at Lick, 58 at Palomar, and 3 at Keck, mostly in clear weather. The total integration times ranged from 300 to 1800 seconds. The Kast spectrograph operated at a spectral resolution of $R \sim$ 2000 in the red and 1500 in the blue, with slit widths of mostly 2 arcseconds. The wavelength coverage extended from 300-800 nm. DBSP on Palomar performed at $R \sim$ 800 with a 1-arcsecond slit and covered wavelengths from 300-1000 nm. Keck LRIS provided resolution ranging from $R = $ 300 to 5000, with slit widths of 0.7, 1.0, 1.5, and 8.7 arcseconds. The wavelength coverage followed similarly to the previous two spectrographs: 300 to 900 nm in the optical. Figure \ref{fig:flux_spectra} presents 186 spectra from Lick, Palomar, and Keck. 

\figsetstart
\figsetnum{2}
\figsettitle{186 spectra of AGN candidates from Lick, Palomar, and Keck}

\figsetgrpstart
\figsetgrpnum{2.1}
\figsetgrptitle{Flux of 0039+0322}
\figsetplot{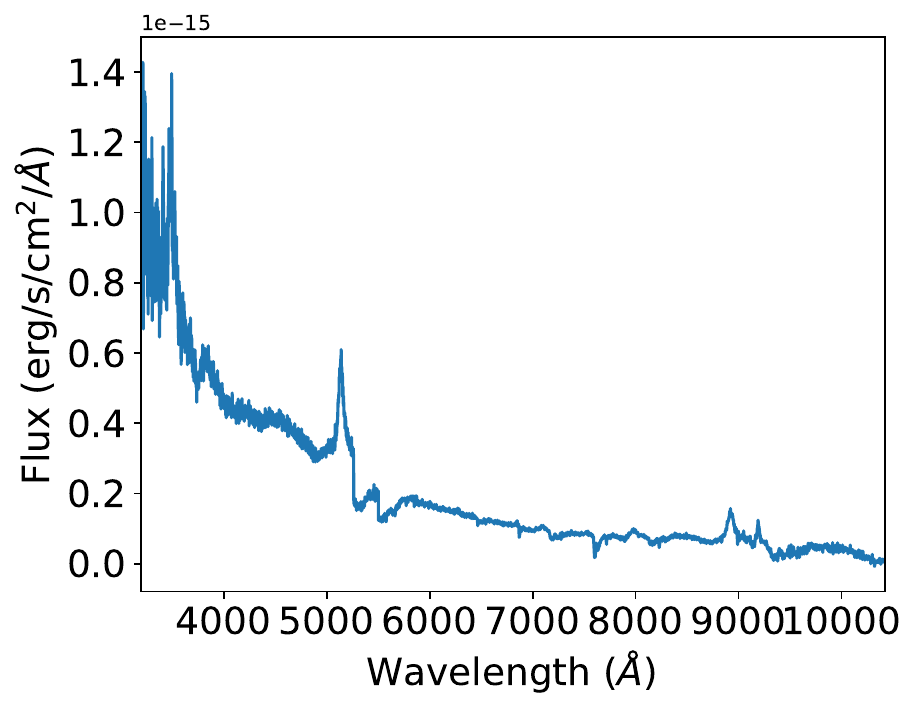}
\figsetgrpnote{Flux for the Lick, Palomar, and Keck spectra. See Table 3 for the observed spectral features and line measurements.}
\figsetgrpend

\figsetgrpstart
\figsetgrpnum{2.2}
\figsetgrptitle{Flux of 0042+0730}
\figsetplot{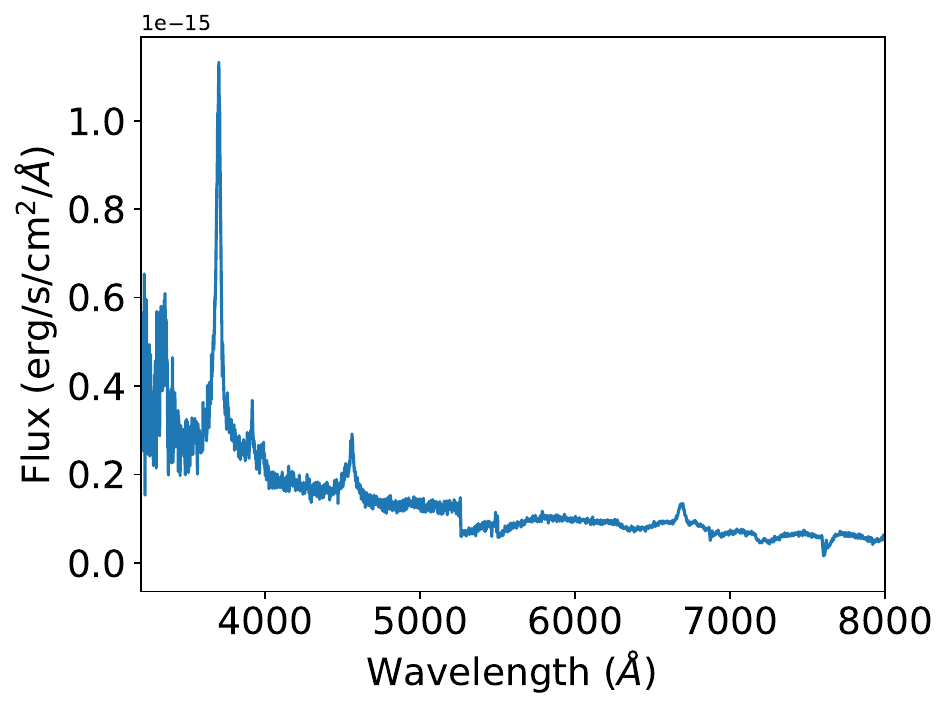}
\figsetgrpnote{Flux for the Lick, Palomar, and Keck spectra. See Table 3 for the observed spectral features and line measurements.}
\figsetgrpend

\figsetgrpstart
\figsetgrpnum{2.3}
\figsetgrptitle{Flux of 0047+0831}
\figsetplot{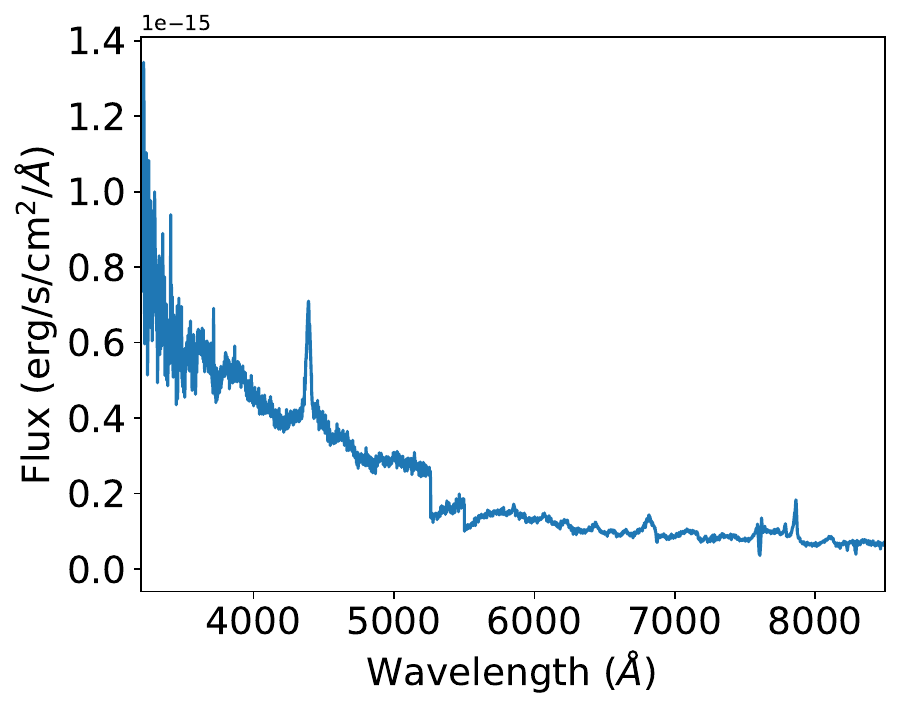}
\figsetgrpnote{Flux for the Lick, Palomar, and Keck spectra. See Table 3 for the observed spectral features and line measurements.}
\figsetgrpend

\figsetgrpstart
\figsetgrpnum{2.4}
\figsetgrptitle{Flux of 0052+0526}
\figsetplot{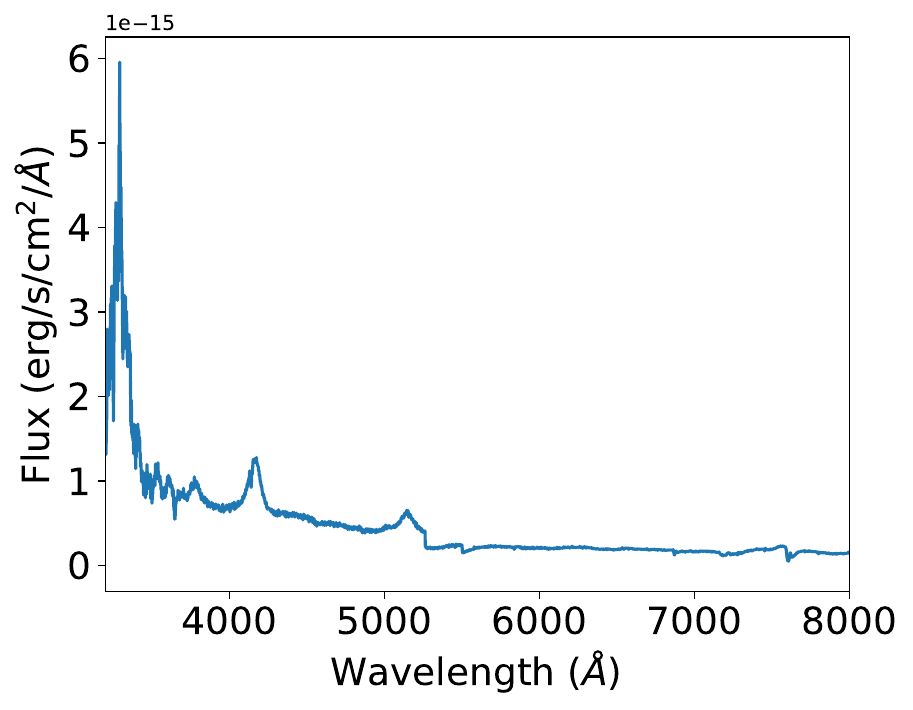}
\figsetgrpnote{Flux for the Lick, Palomar, and Keck spectra. See Table 3 for the observed spectral features and line measurements.}
\figsetgrpend

\figsetgrpstart
\figsetgrpnum{2.5}
\figsetgrptitle{Flux of 0058+0841}
\figsetplot{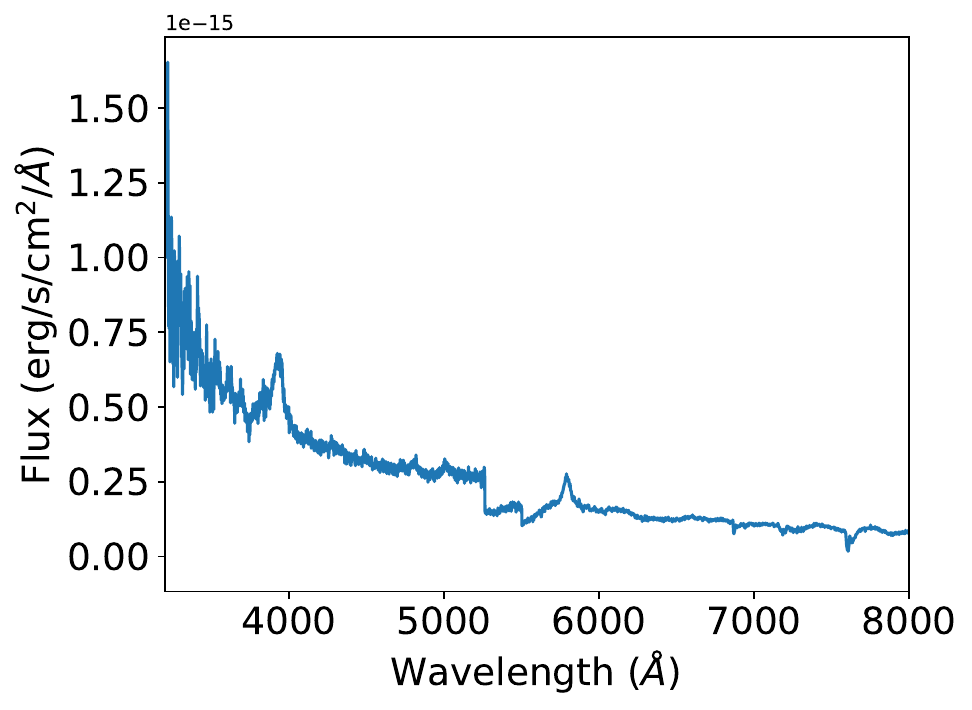}
\figsetgrpnote{Flux for the Lick, Palomar, and Keck spectra. See Table 3 for the observed spectral features and line measurements.}
\figsetgrpend

\figsetgrpstart
\figsetgrpnum{2.6}
\figsetgrptitle{Flux of 0058+1145}
\figsetplot{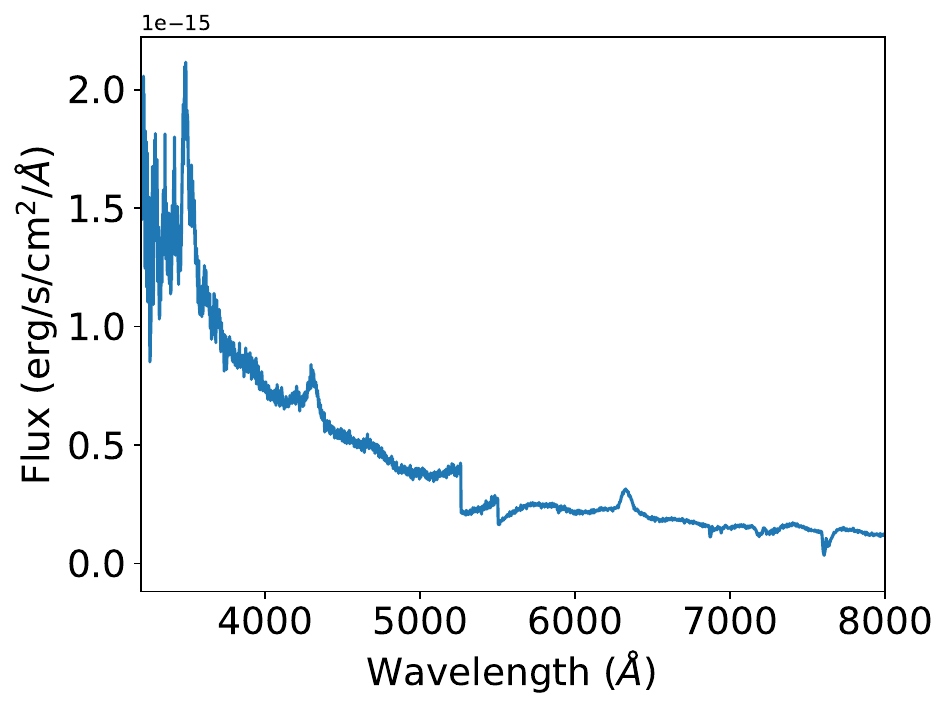}
\figsetgrpnote{Flux for the Lick, Palomar, and Keck spectra. See Table 3 for the observed spectral features and line measurements.}
\figsetgrpend

\figsetgrpstart
\figsetgrpnum{2.7}
\figsetgrptitle{Flux of 0104+0516}
\figsetplot{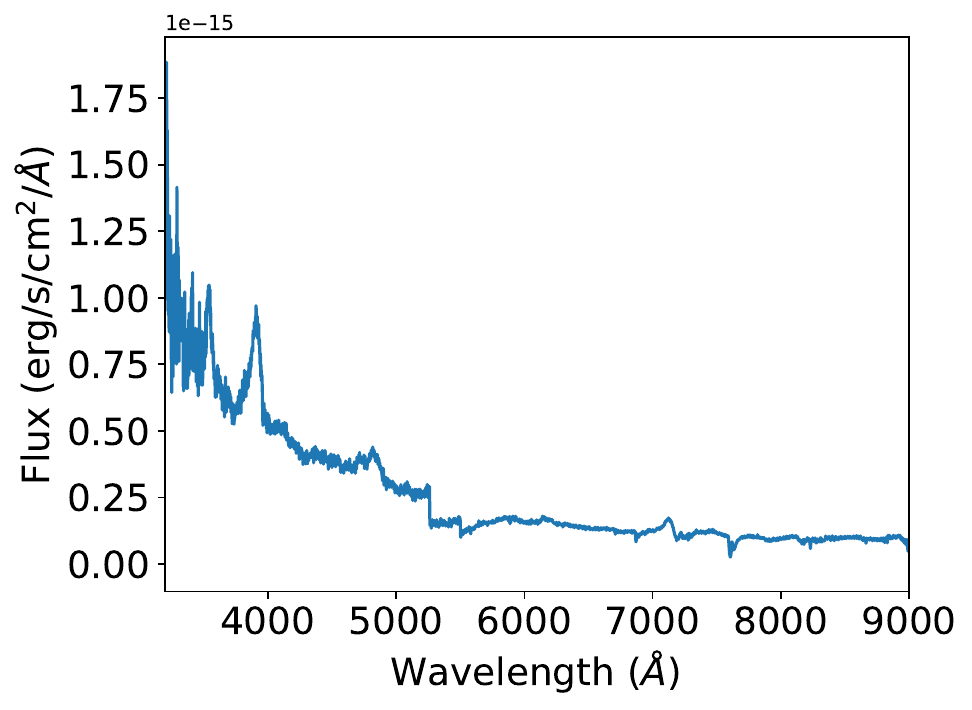}
\figsetgrpnote{Flux for the Lick, Palomar, and Keck spectra. See Table 3 for the observed spectral features and line measurements.}
\figsetgrpend

\figsetgrpstart
\figsetgrpnum{2.8}
\figsetgrptitle{Flux of 0109+0724}
\figsetplot{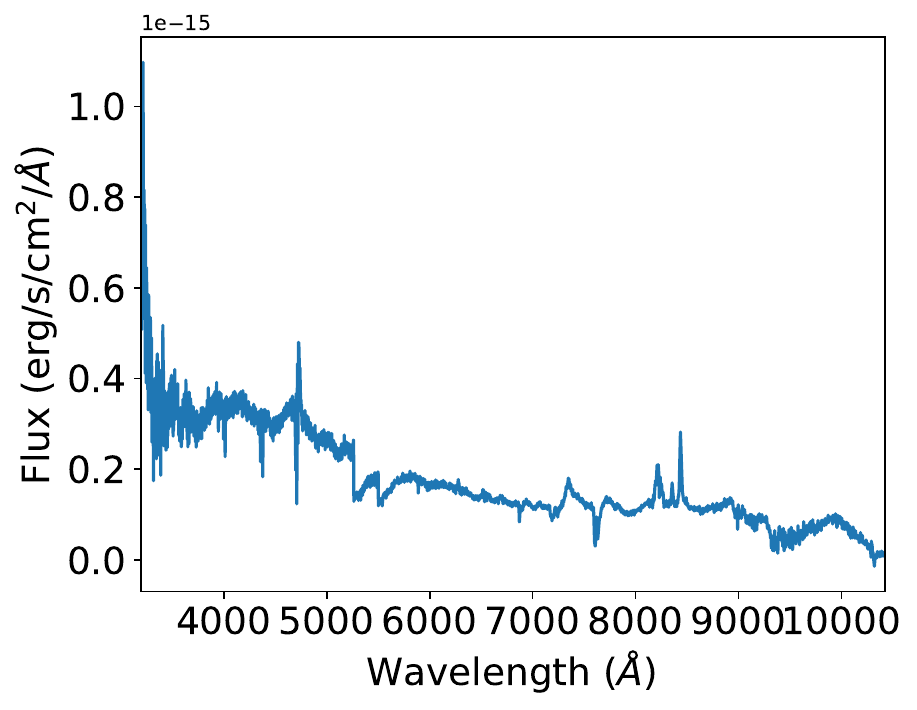}
\figsetgrpnote{Flux for the Lick, Palomar, and Keck spectra. See Table 3 for the observed spectral features and line measurements.}
\figsetgrpend

\figsetgrpstart
\figsetgrpnum{2.9}
\figsetgrptitle{Flux of 0128+0505}
\figsetplot{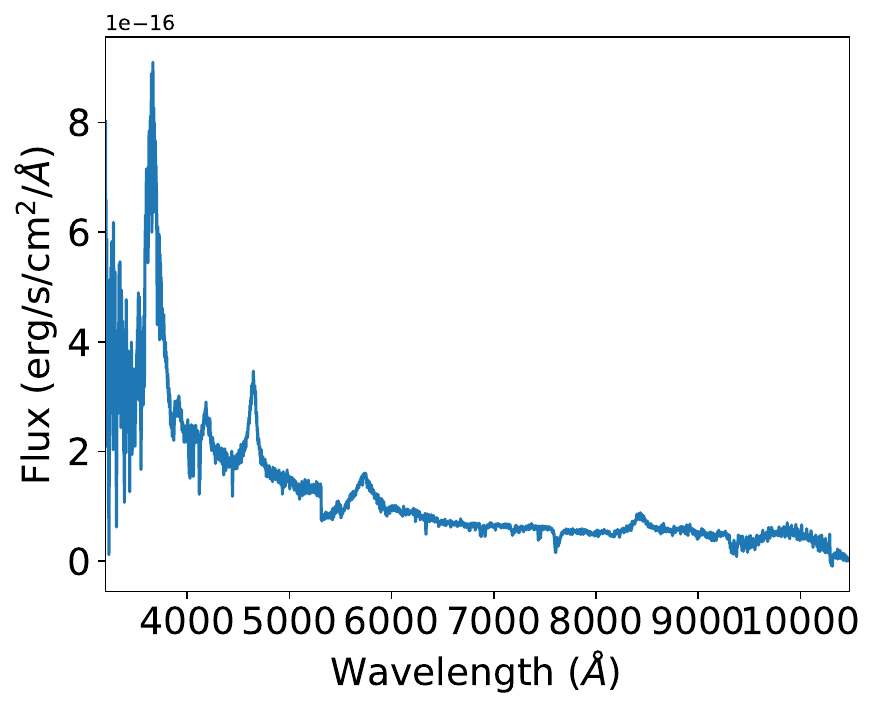}
\figsetgrpnote{Flux for the Lick, Palomar, and Keck spectra. See Table 3 for the observed spectral features and line measurements.}
\figsetgrpend

\figsetgrpstart
\figsetgrpnum{2.10}
\figsetgrptitle{Flux of 0336+1547}
\figsetplot{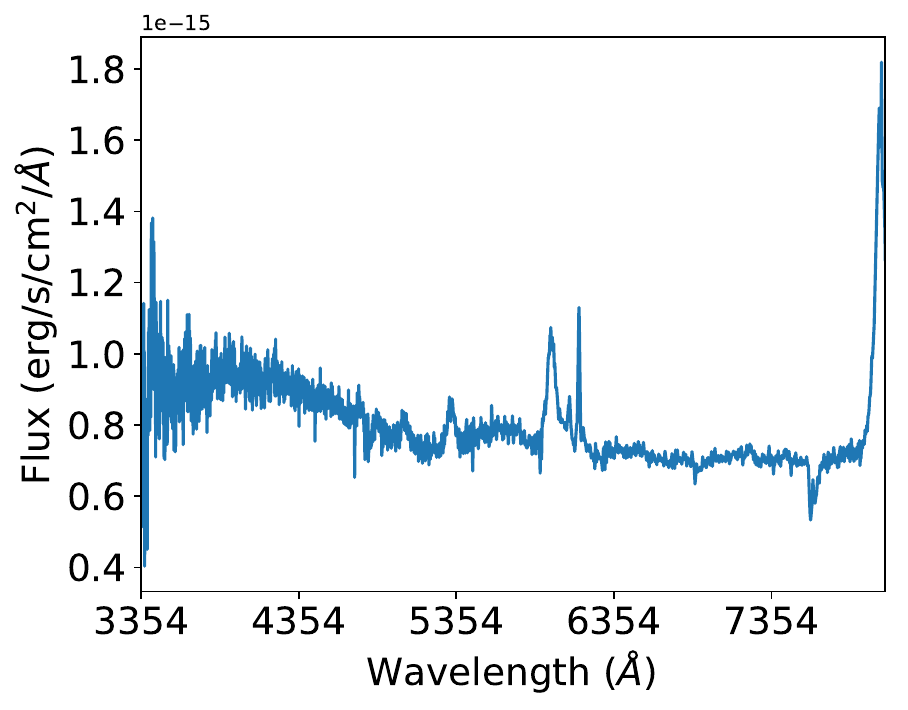}
\figsetgrpnote{Flux for the Lick, Palomar, and Keck spectra. See Table 3 for the observed spectral features and line measurements.}
\figsetgrpend

\figsetgrpstart
\figsetgrpnum{2.11}
\figsetgrptitle{Flux of 0337+2335}
\figsetplot{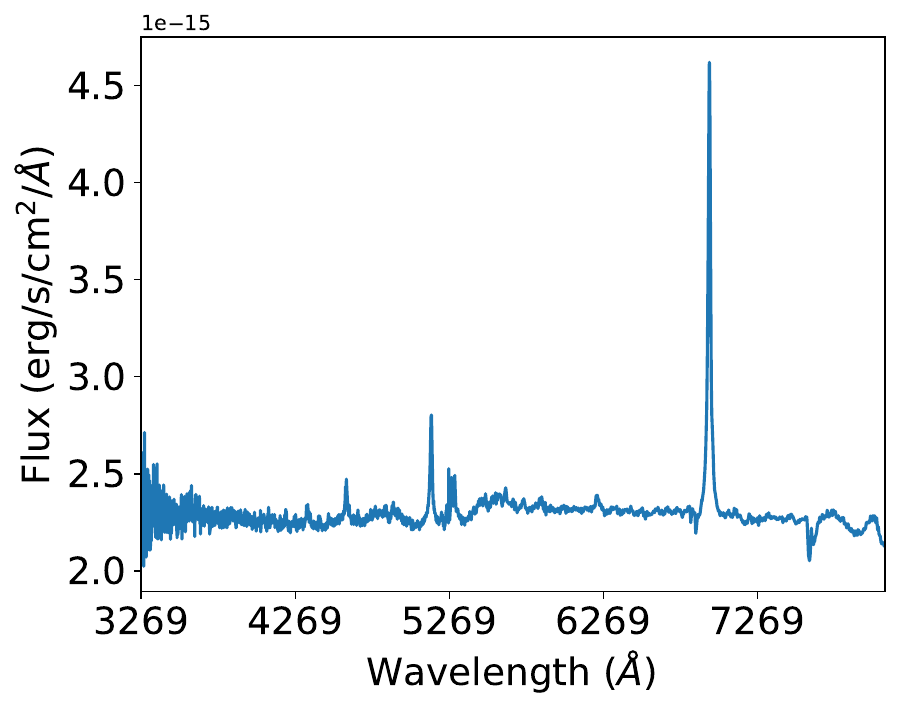}
\figsetgrpnote{Flux for the Lick, Palomar, and Keck spectra. See Table 3 for the observed spectral features and line measurements.}
\figsetgrpend

\figsetgrpstart
\figsetgrpnum{2.12}
\figsetgrptitle{Flux of 0339+1306}
\figsetplot{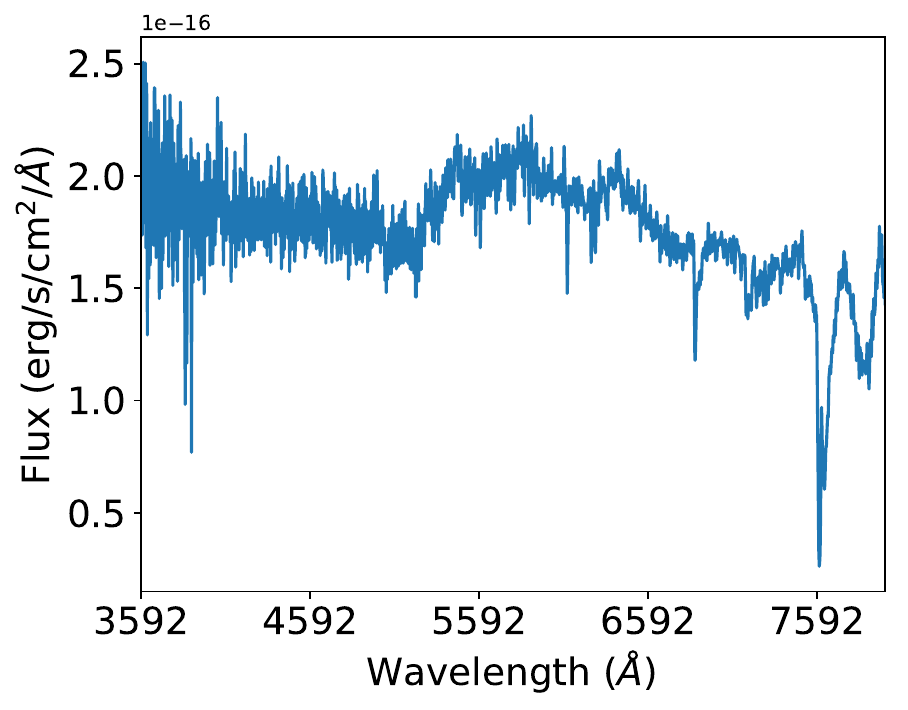}
\figsetgrpnote{Flux for the Lick, Palomar, and Keck spectra. See Table 3 for the observed spectral features and line measurements.}
\figsetgrpend

\figsetgrpstart
\figsetgrpnum{2.13}
\figsetgrptitle{Flux of 0340+1450}
\figsetplot{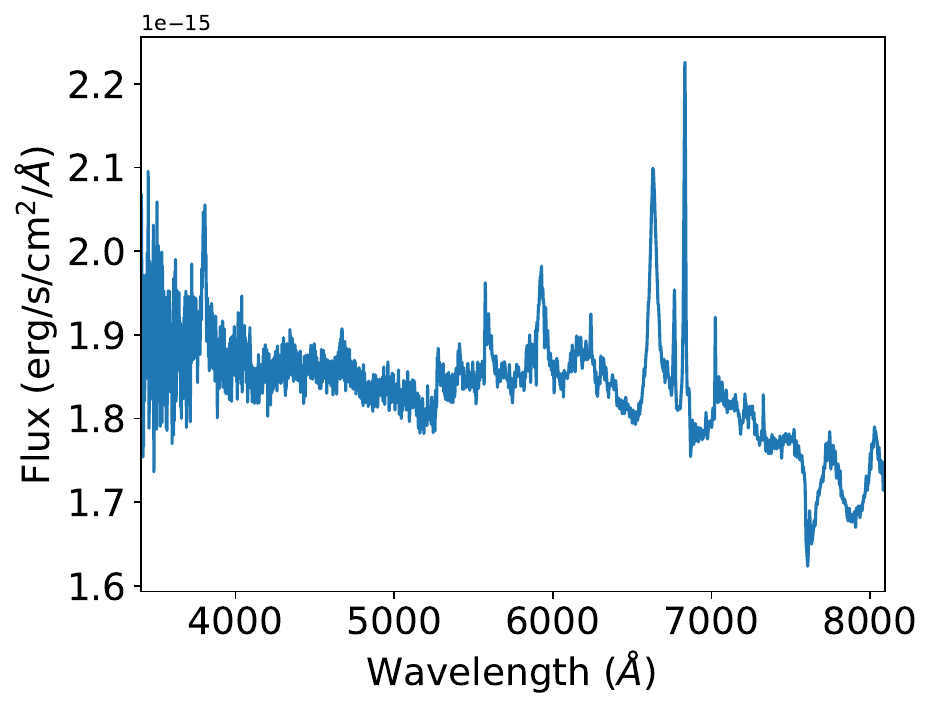}
\figsetgrpnote{Flux for the Lick, Palomar, and Keck spectra. See Table 3 for the observed spectral features and line measurements.}
\figsetgrpend

\figsetgrpstart
\figsetgrpnum{2.14}
\figsetgrptitle{Flux of 0343+1325}
\figsetplot{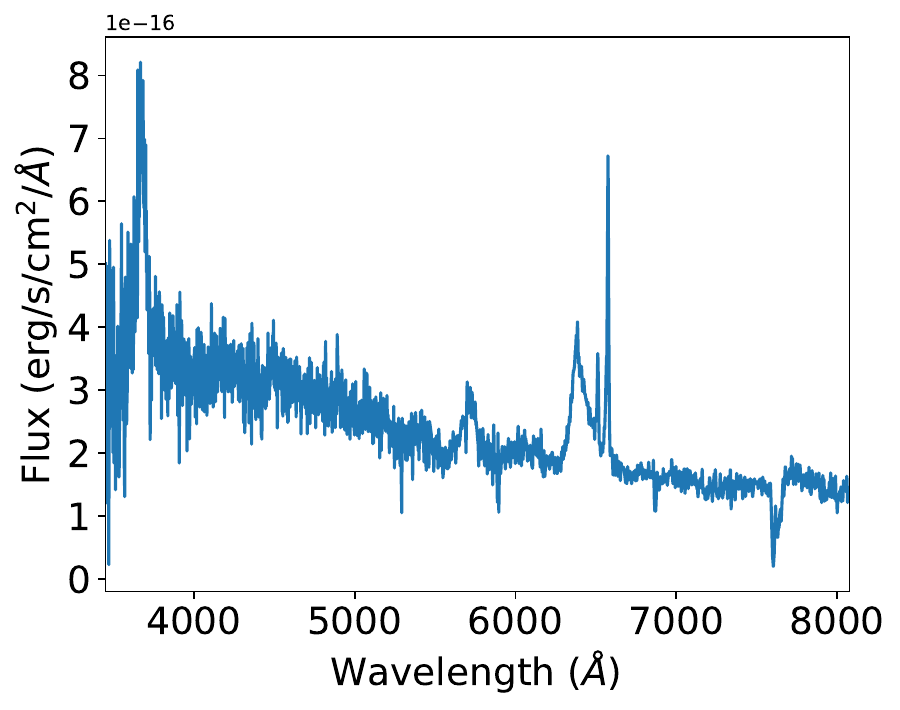}
\figsetgrpnote{Flux for the Lick, Palomar, and Keck spectra. See Table 3 for the observed spectral features and line measurements.}
\figsetgrpend

\figsetgrpstart
\figsetgrpnum{2.15}
\figsetgrptitle{Flux of 0344+1615}
\figsetplot{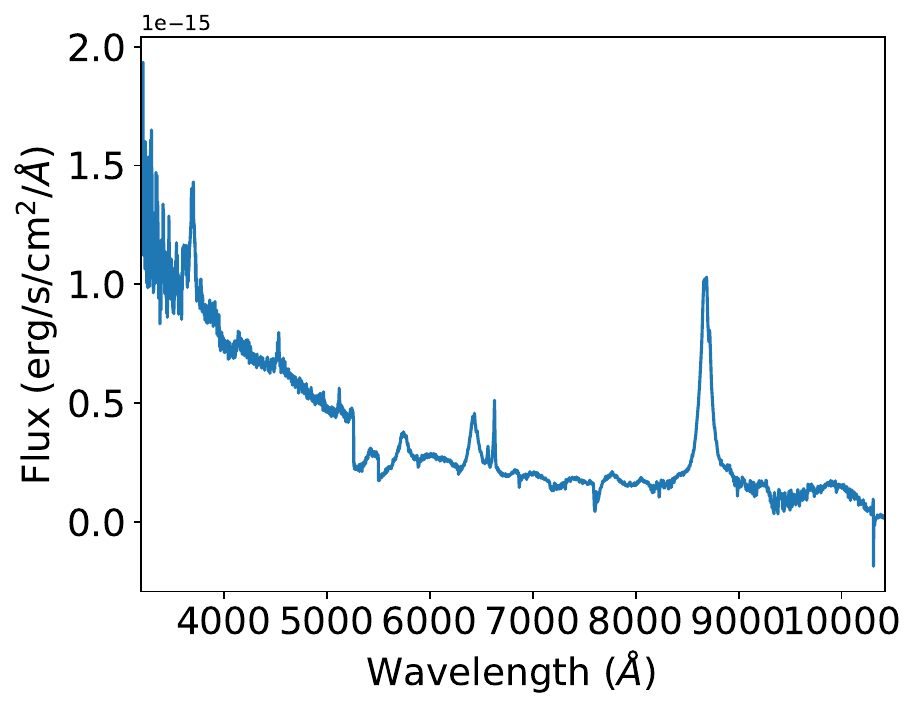}
\figsetgrpnote{Flux for the Lick, Palomar, and Keck spectra. See Table 3 for the observed spectral features and line measurements.}
\figsetgrpend

\figsetgrpstart
\figsetgrpnum{2.16}
\figsetgrptitle{Flux of 0411+1324}
\figsetplot{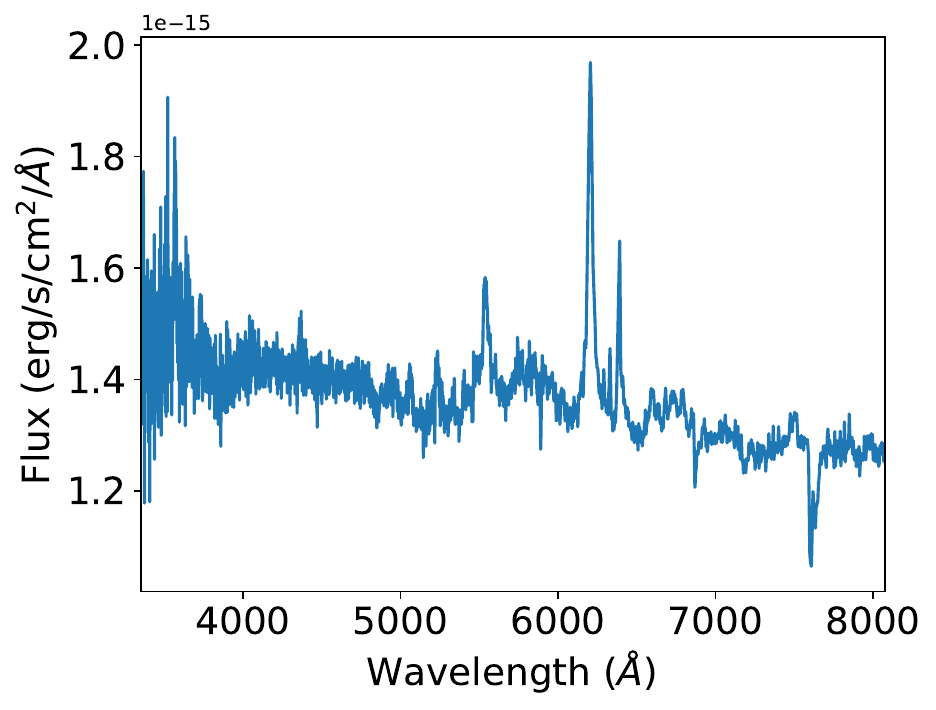}
\figsetgrpnote{Flux for the Lick, Palomar, and Keck spectra. See Table 3 for the observed spectral features and line measurements.}
\figsetgrpend

\figsetgrpstart
\figsetgrpnum{2.17}
\figsetgrptitle{Flux of 0412+2305}
\figsetplot{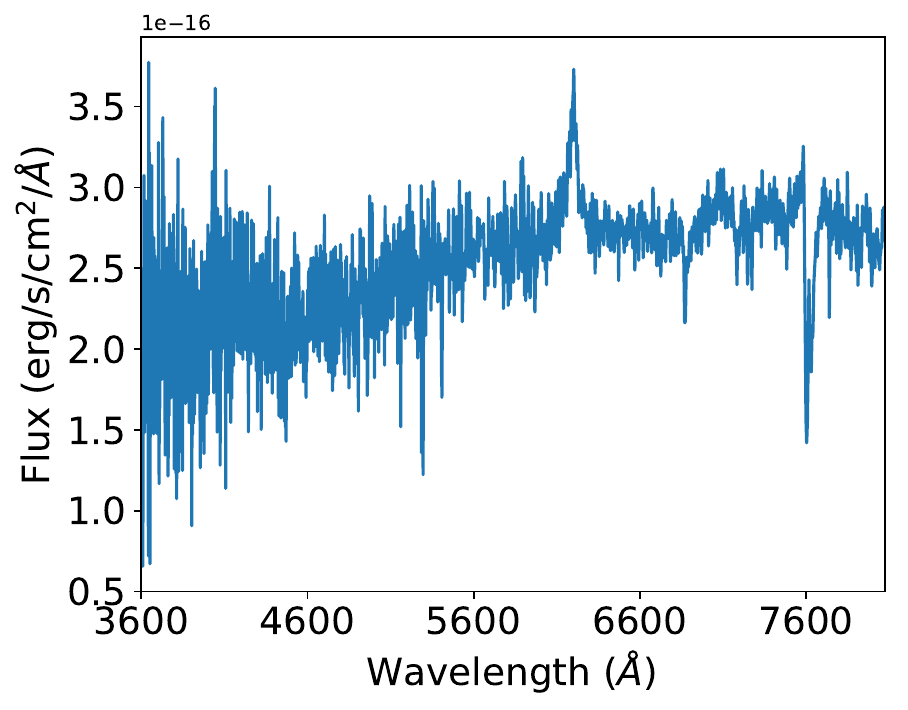}
\figsetgrpnote{Flux for the Lick, Palomar, and Keck spectra. See Table 3 for the observed spectral features and line measurements.}
\figsetgrpend

\figsetgrpstart
\figsetgrpnum{2.18}
\figsetgrptitle{Flux of 0421+1433}
\figsetplot{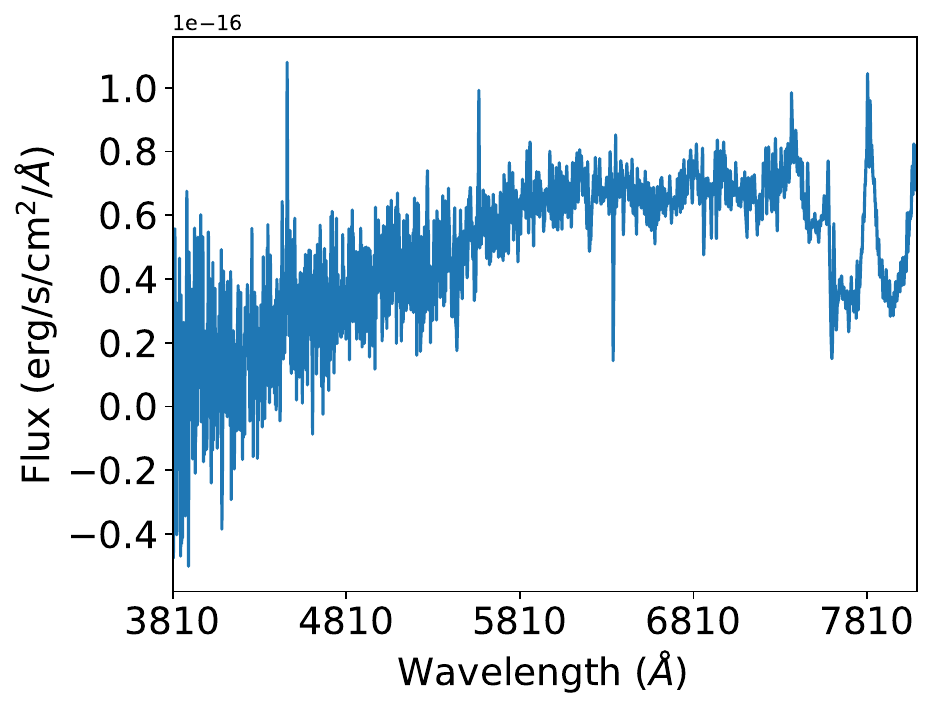}
\figsetgrpnote{Flux for the Lick, Palomar, and Keck spectra. See Table 3 for the observed spectral features and line measurements.}
\figsetgrpend

\figsetgrpstart
\figsetgrpnum{2.19}
\figsetgrptitle{Flux of 0422+2102}
\figsetplot{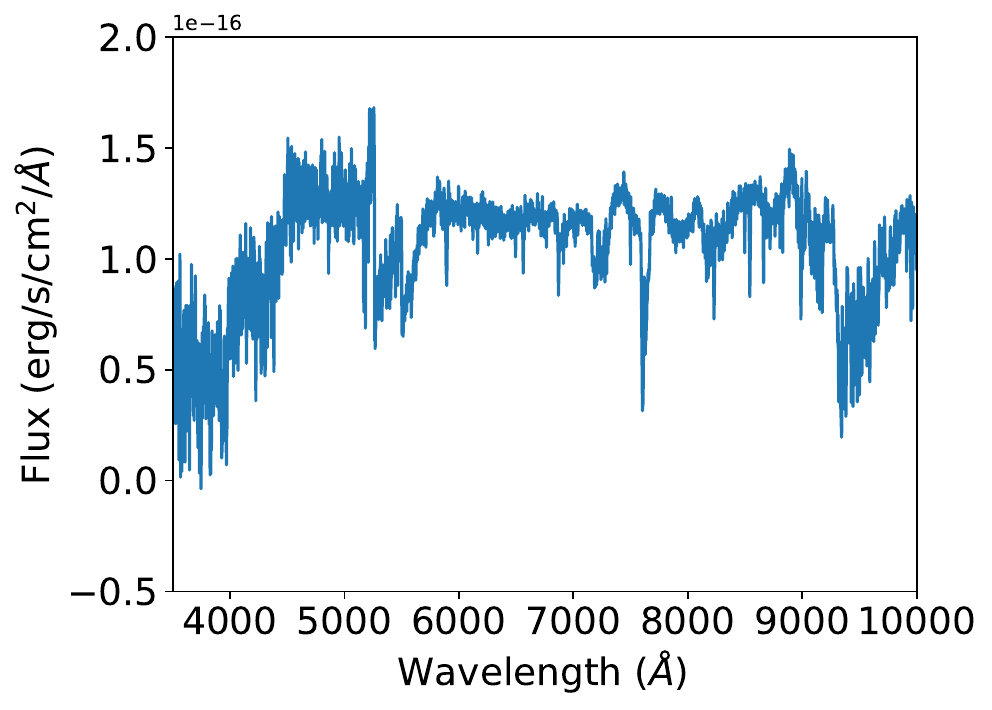}
\figsetgrpnote{Flux for the Lick, Palomar, and Keck spectra. See Table 3 for the observed spectral features and line measurements.}
\figsetgrpend

\figsetgrpstart
\figsetgrpnum{2.20}
\figsetgrptitle{Flux of 0425+2104}
\figsetplot{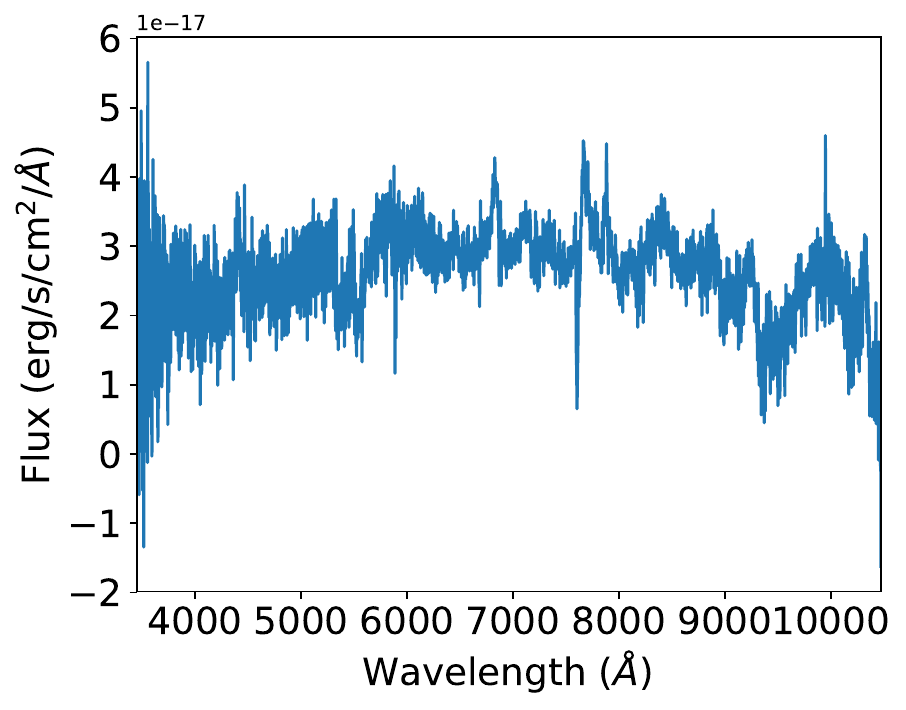}
\figsetgrpnote{Flux for the Lick, Palomar, and Keck spectra. See Table 3 for the observed spectral features and line measurements.}
\figsetgrpend

\figsetgrpstart
\figsetgrpnum{2.21}
\figsetgrptitle{Flux of 0428+2104}
\figsetplot{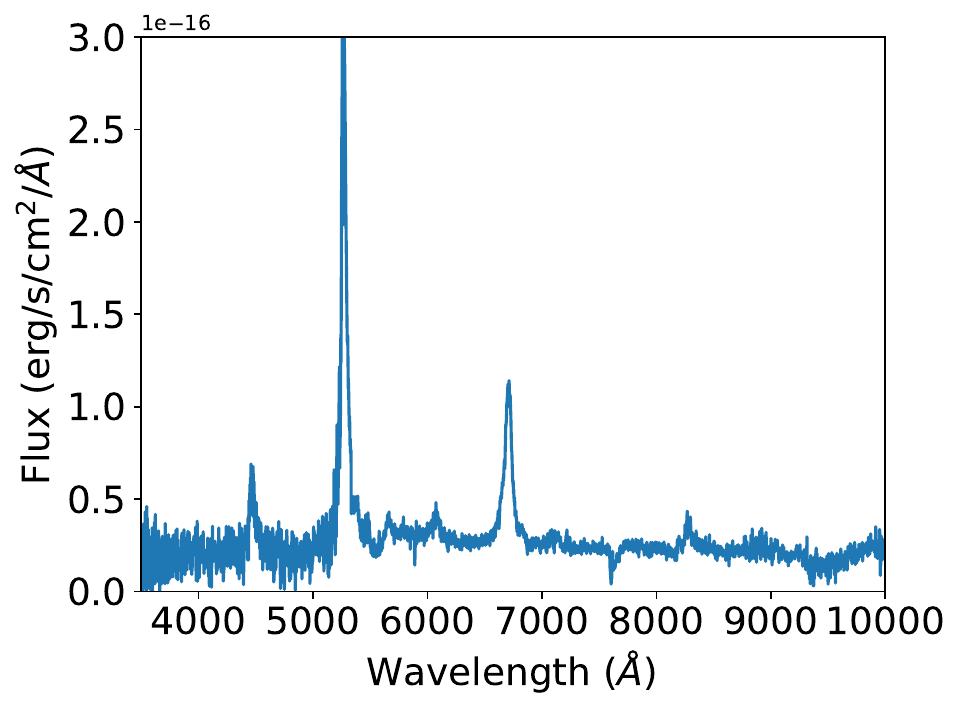}
\figsetgrpnote{Flux for the Lick, Palomar, and Keck spectra. See Table 3 for the observed spectral features and line measurements.}
\figsetgrpend

\figsetgrpstart
\figsetgrpnum{2.22}
\figsetgrptitle{Flux of 0431+1812}
\figsetplot{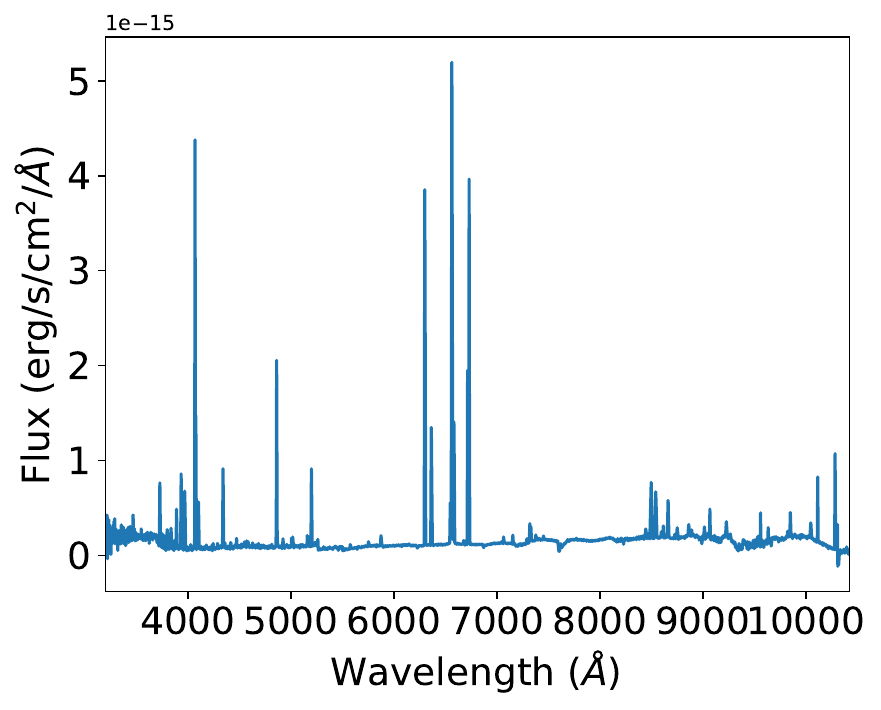}
\figsetgrpnote{Flux for the Lick, Palomar, and Keck spectra. See Table 3 for the observed spectral features and line measurements.}
\figsetgrpend

\figsetgrpstart
\figsetgrpnum{2.23}
\figsetgrptitle{Flux of 0431+1459}
\figsetplot{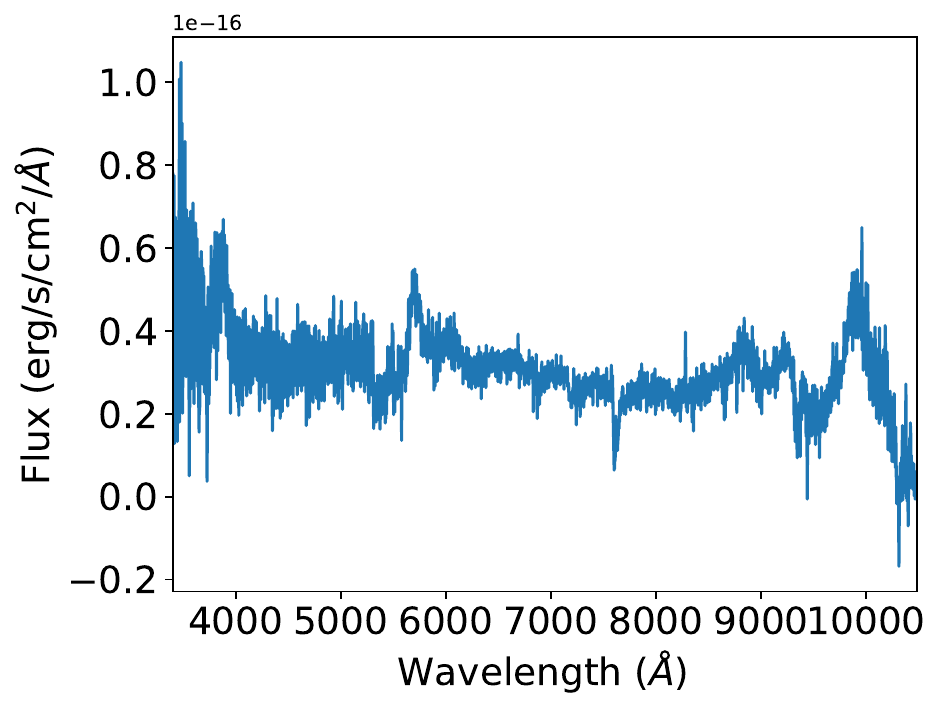}
\figsetgrpnote{Flux for the Lick, Palomar, and Keck spectra. See Table 3 for the observed spectral features and line measurements.}
\figsetgrpend

\figsetgrpstart
\figsetgrpnum{2.24}
\figsetgrptitle{Flux of 0433+2253}
\figsetplot{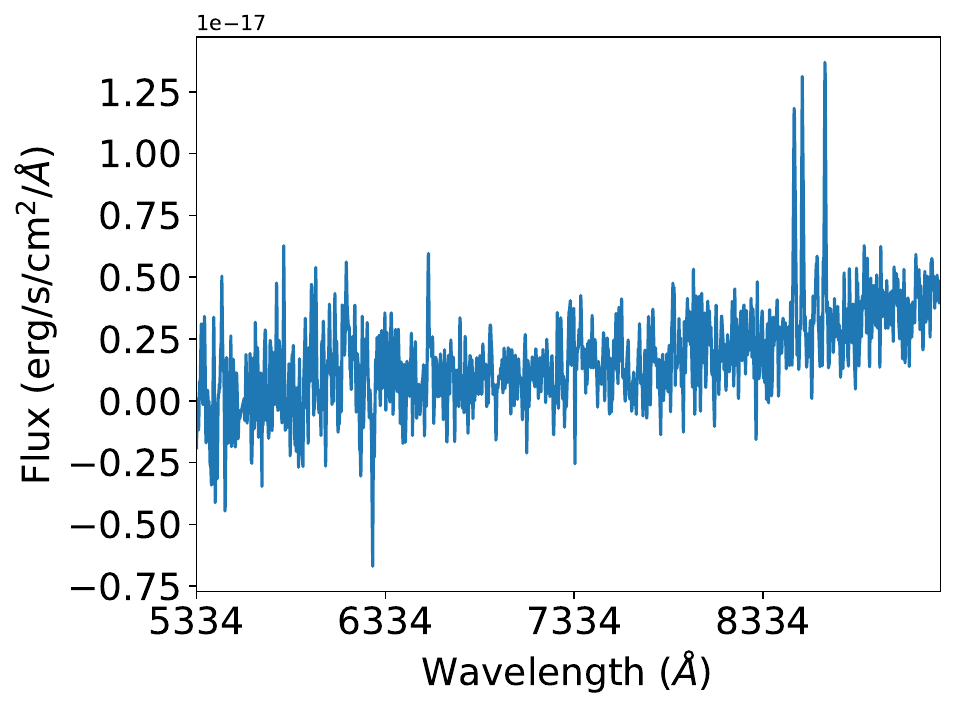}
\figsetgrpnote{Flux for the Lick, Palomar, and Keck spectra. See Table 3 for the observed spectral features and line measurements.}
\figsetgrpend

\figsetgrpstart
\figsetgrpnum{2.25}
\figsetgrptitle{Flux of 0433+2246}
\figsetplot{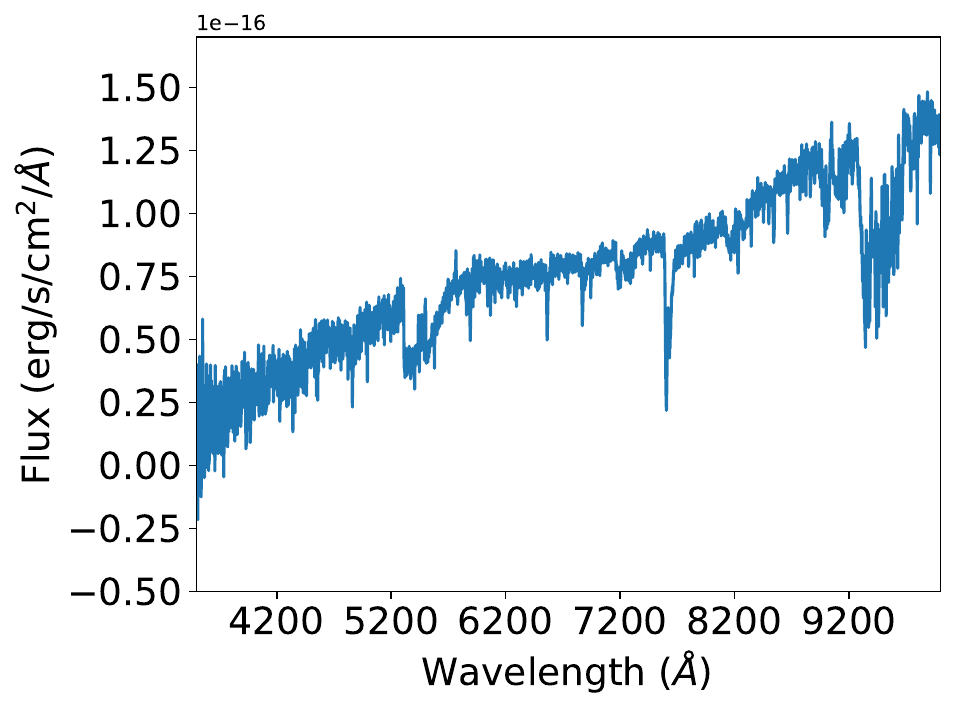}
\figsetgrpnote{Flux for the Lick, Palomar, and Keck spectra. See Table 3 for the observed spectral features and line measurements.}
\figsetgrpend

\figsetgrpstart
\figsetgrpnum{2.26}
\figsetgrptitle{Flux of 0433+1758}
\figsetplot{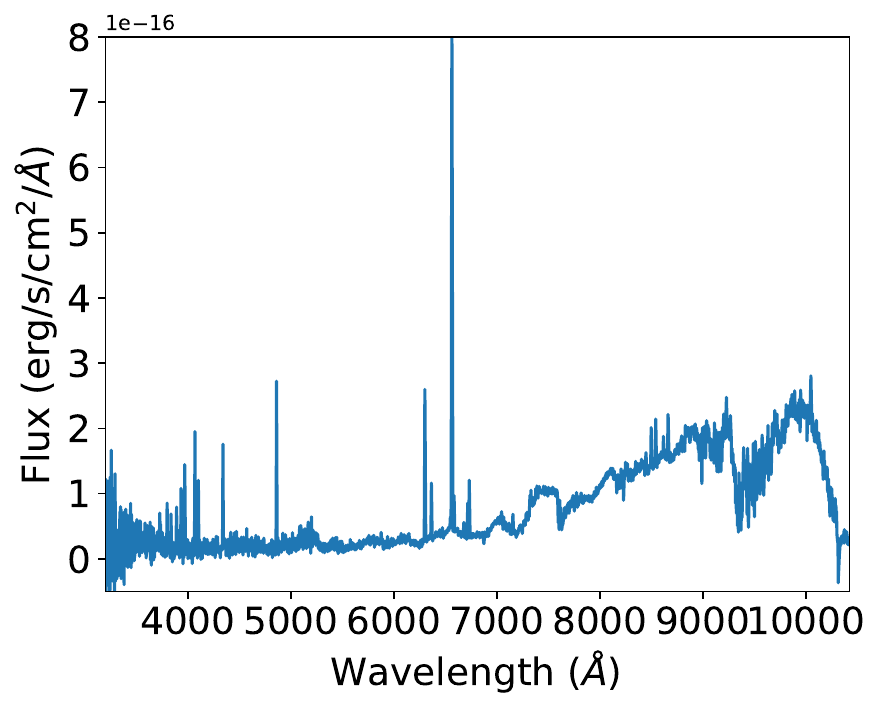}
\figsetgrpnote{Flux for the Lick, Palomar, and Keck spectra. See Table 3 for the observed spectral features and line measurements.}
\figsetgrpend

\figsetgrpstart
\figsetgrpnum{2.27}
\figsetgrptitle{Flux of 0441+2556}
\figsetplot{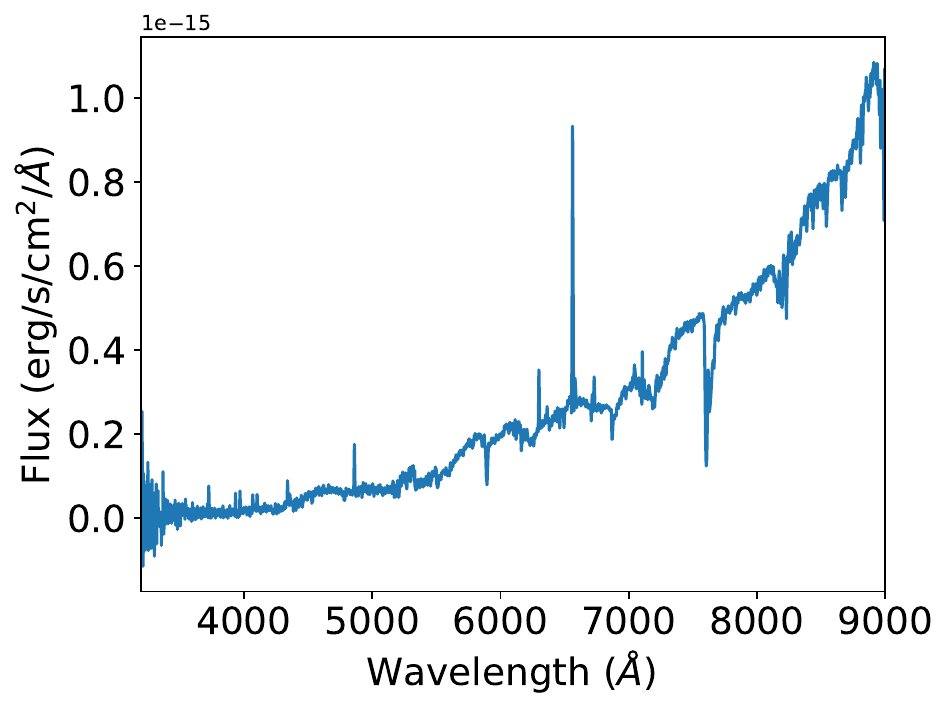}
\figsetgrpnote{Flux for the Lick, Palomar, and Keck spectra. See Table 3 for the observed spectral features and line measurements.}
\figsetgrpend

\figsetgrpstart
\figsetgrpnum{2.28}
\figsetgrptitle{Flux of 0443+1925}
\figsetplot{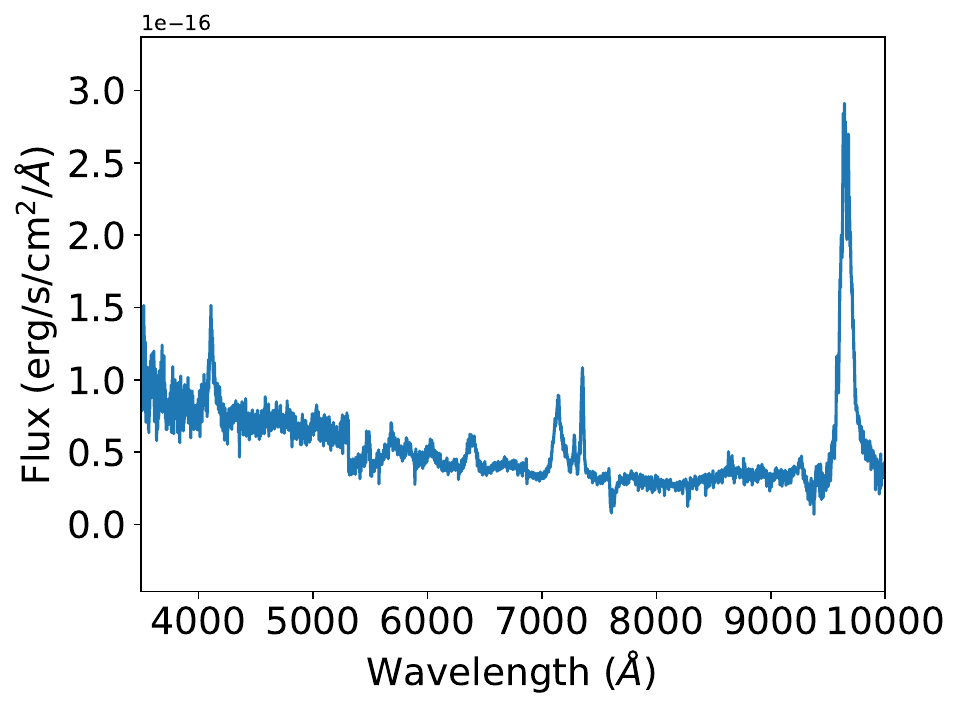}
\figsetgrpnote{Flux for the Lick, Palomar, and Keck spectra. See Table 3 for the observed spectral features and line measurements.}
\figsetgrpend

\figsetgrpstart
\figsetgrpnum{2.29}
\figsetgrptitle{Flux of 0443+1445}
\figsetplot{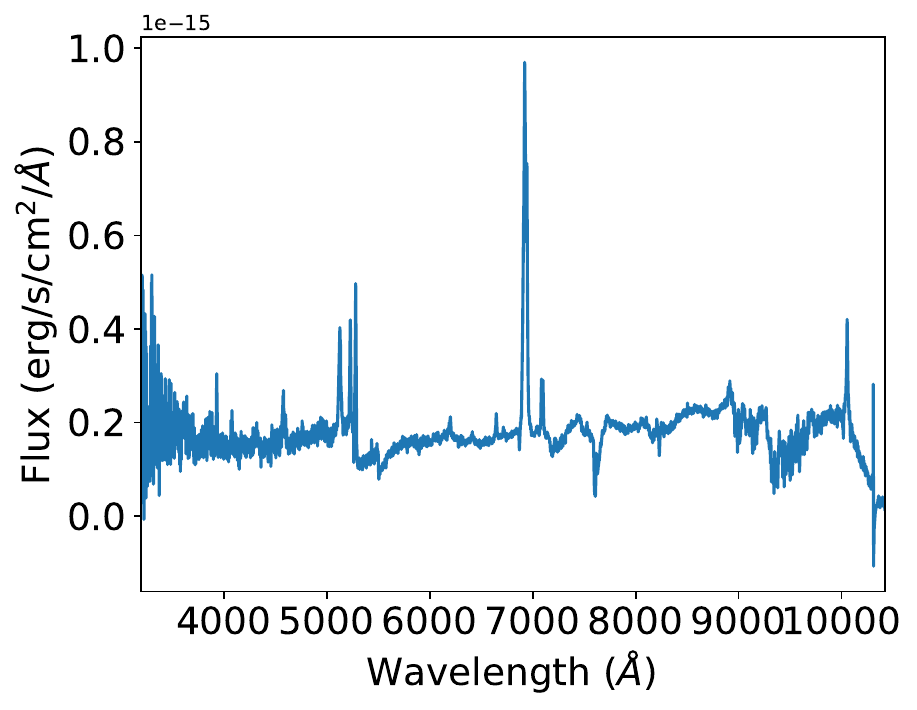}
\figsetgrpnote{Flux for the Lick, Palomar, and Keck spectra. See Table 3 for the observed spectral features and line measurements.}
\figsetgrpend

\figsetgrpstart
\figsetgrpnum{2.30}
\figsetgrptitle{Flux of 0446+2257}
\figsetplot{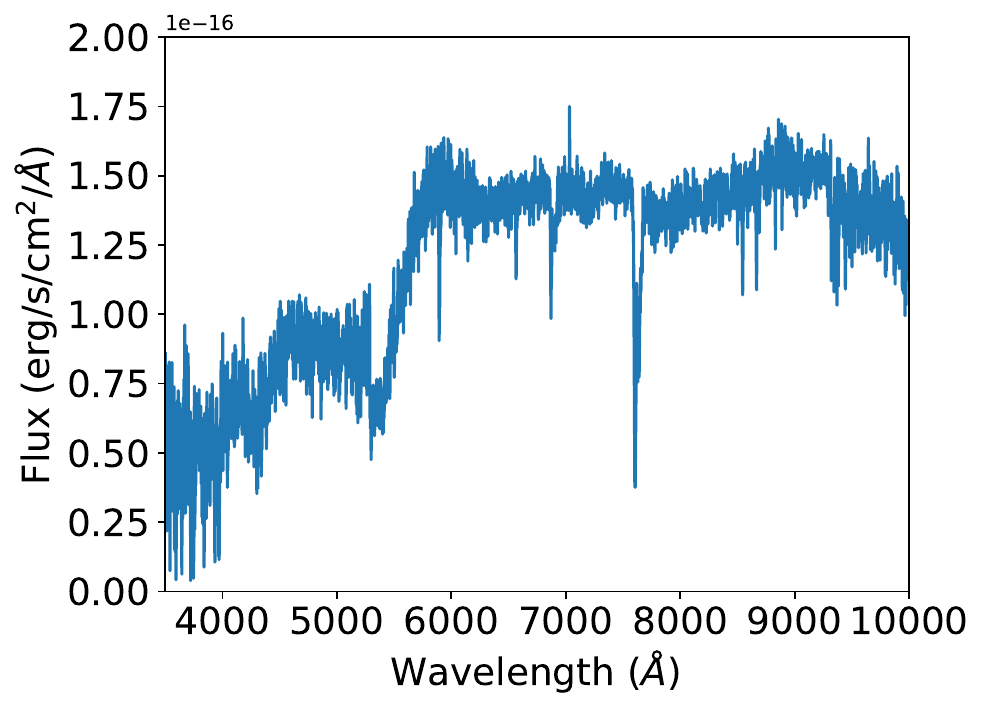}
\figsetgrpnote{Flux for the Lick, Palomar, and Keck spectra. See Table 3 for the observed spectral features and line measurements.}
\figsetgrpend

\figsetgrpstart
\figsetgrpnum{2.31}
\figsetgrptitle{Flux of 0447+2524}
\figsetplot{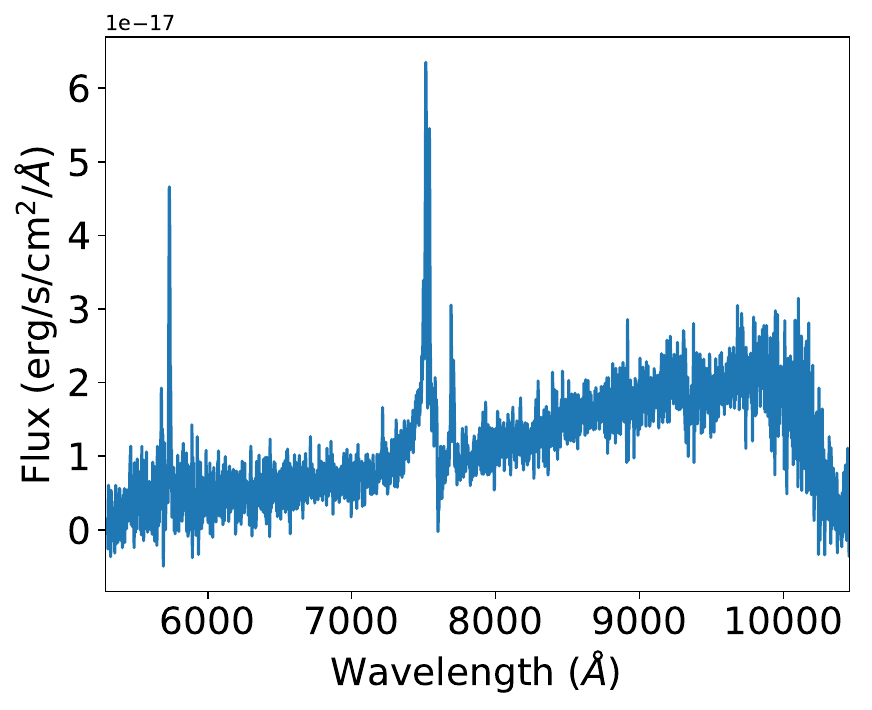}
\figsetgrpnote{Flux for the Lick, Palomar, and Keck spectra. See Table 3 for the observed spectral features and line measurements.}
\figsetgrpend

\figsetgrpstart
\figsetgrpnum{2.32}
\figsetgrptitle{Flux of 0447+2025}
\figsetplot{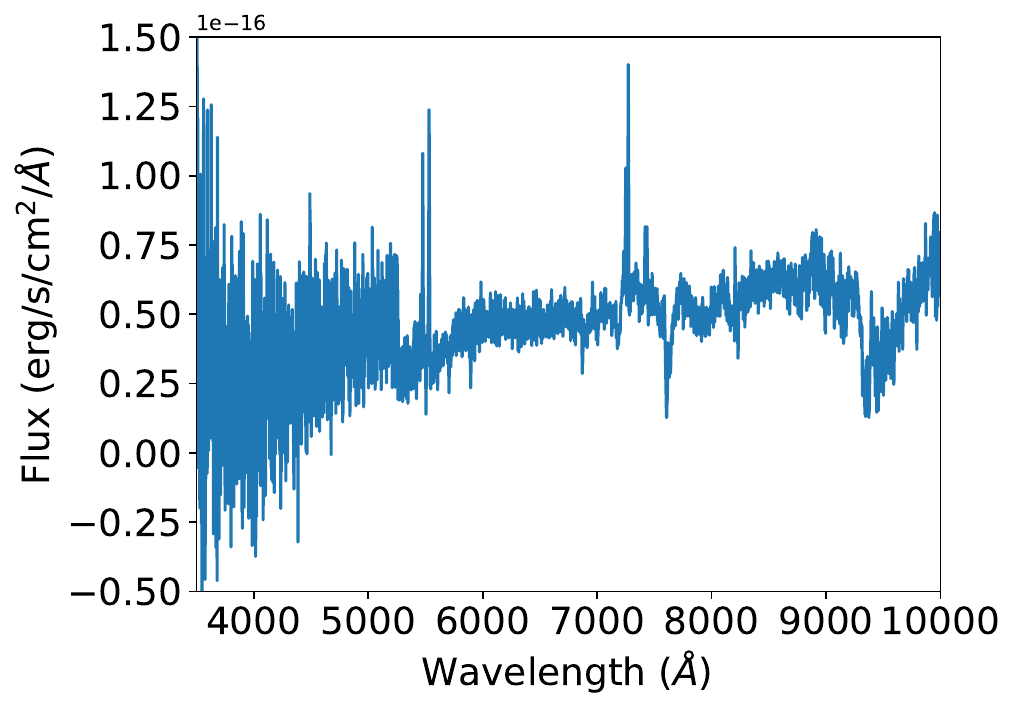}
\figsetgrpnote{Flux for the Lick, Palomar, and Keck spectra. See Table 3 for the observed spectral features and line measurements.}
\figsetgrpend

\figsetgrpstart
\figsetgrpnum{2.33}
\figsetgrptitle{Flux of 0449+1653}
\figsetplot{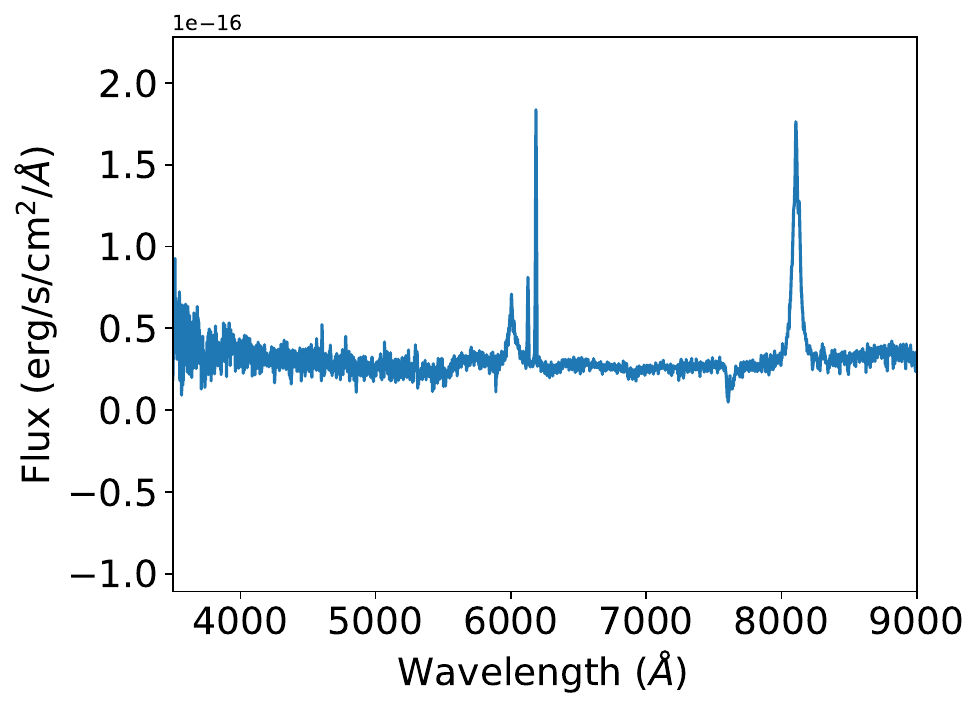}
\figsetgrpnote{Flux for the Lick, Palomar, and Keck spectra. See Table 3 for the observed spectral features and line measurements.}
\figsetgrpend

\figsetgrpstart
\figsetgrpnum{2.34}
\figsetgrptitle{Flux of 0459+1808}
\figsetplot{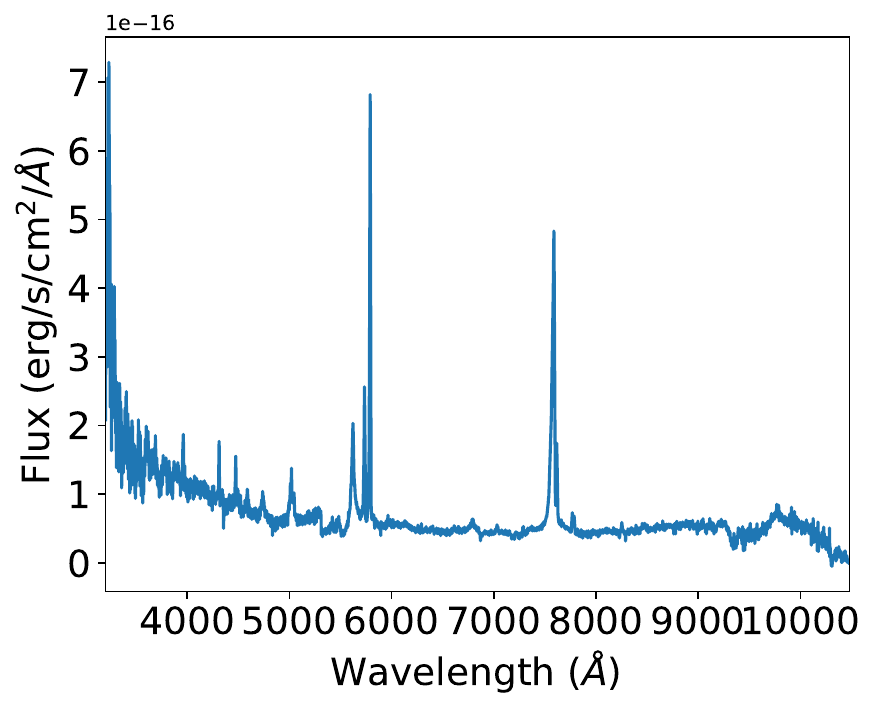}
\figsetgrpnote{Flux for the Lick, Palomar, and Keck spectra. See Table 3 for the observed spectral features and line measurements.}
\figsetgrpend

\figsetgrpstart
\figsetgrpnum{2.35}
\figsetgrptitle{Flux of 0500+2525}
\figsetplot{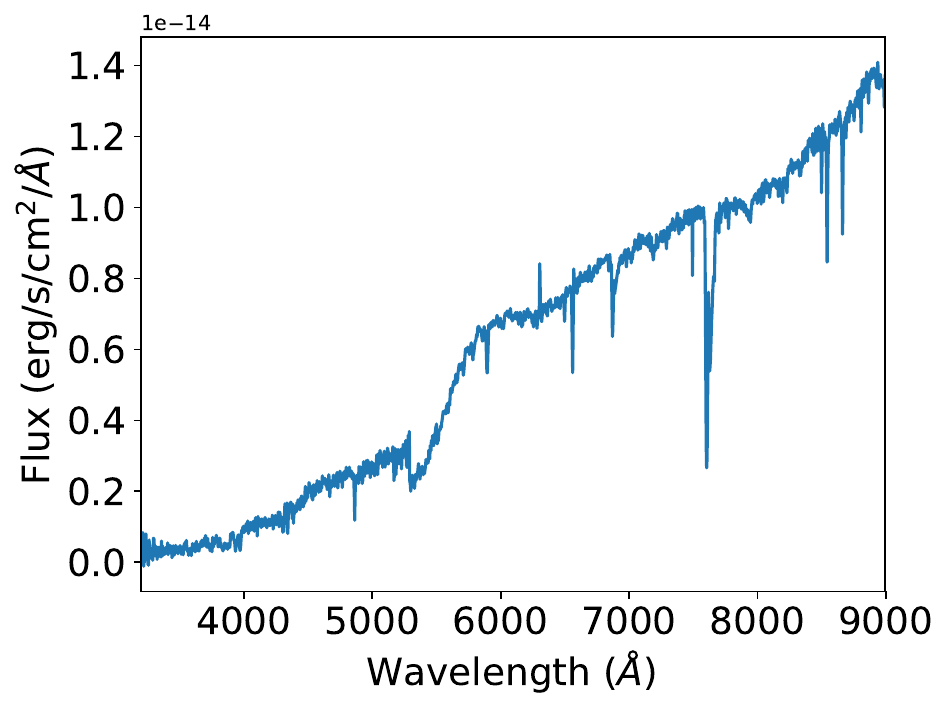}
\figsetgrpnote{Flux for the Lick, Palomar, and Keck spectra. See Table 3 for the observed spectral features and line measurements.}
\figsetgrpend

\figsetgrpstart
\figsetgrpnum{2.36}
\figsetgrptitle{Flux of 0502+2259}
\figsetplot{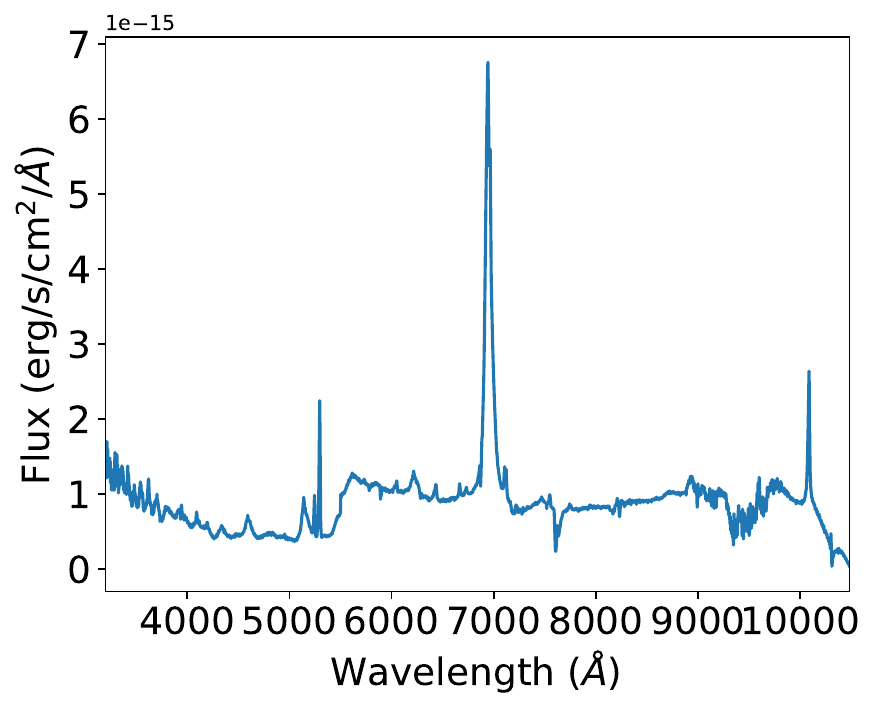}
\figsetgrpnote{Flux for the Lick, Palomar, and Keck spectra. See Table 3 for the observed spectral features and line measurements.}
\figsetgrpend

\figsetgrpstart
\figsetgrpnum{2.37}
\figsetgrptitle{Flux of 0510+1630}
\figsetplot{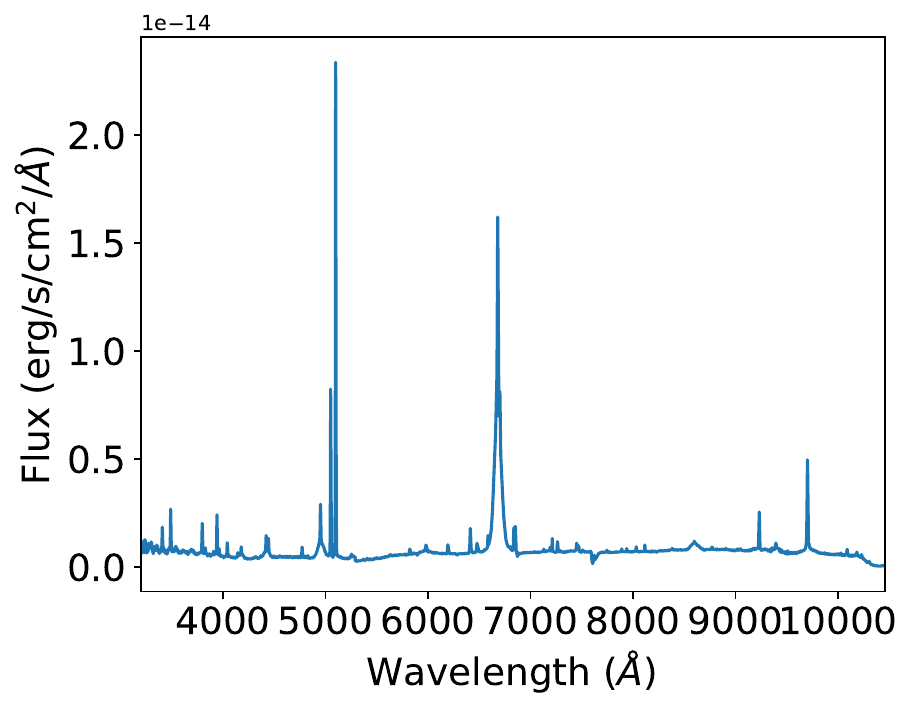}
\figsetgrpnote{Flux for the Lick, Palomar, and Keck spectra. See Table 3 for the observed spectral features and line measurements.}
\figsetgrpend

\figsetgrpstart
\figsetgrpnum{2.38}
\figsetgrptitle{Flux of 0518+2110}
\figsetplot{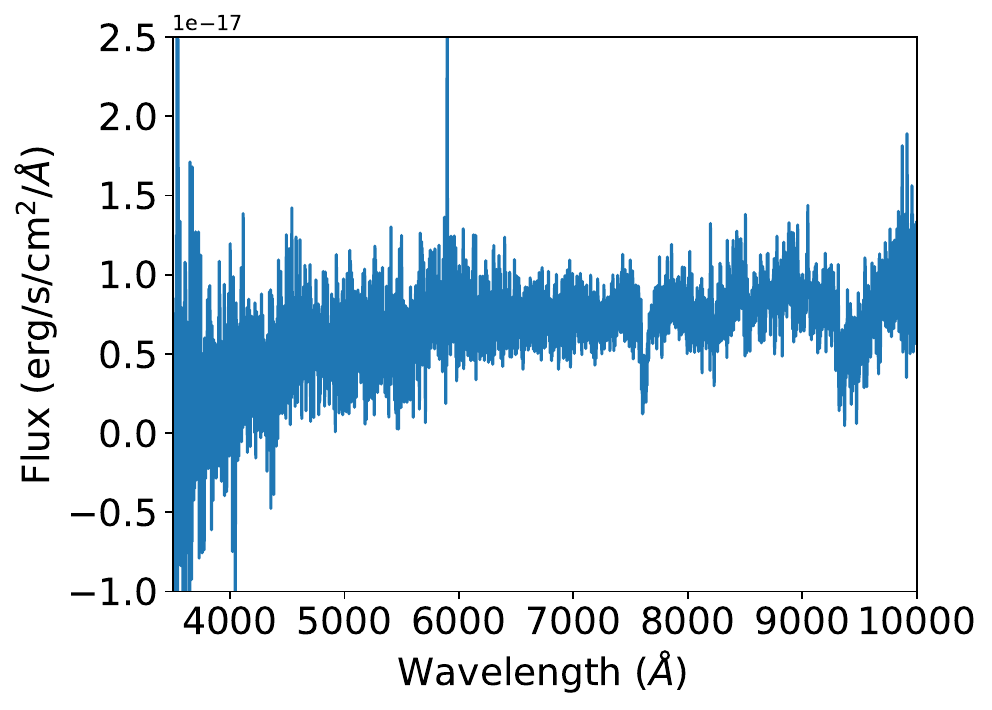}
\figsetgrpnote{Flux for the Lick, Palomar, and Keck spectra. See Table 3 for the observed spectral features and line measurements.}
\figsetgrpend

\figsetgrpstart
\figsetgrpnum{2.39}
\figsetgrptitle{Flux of 0521+2112}
\figsetplot{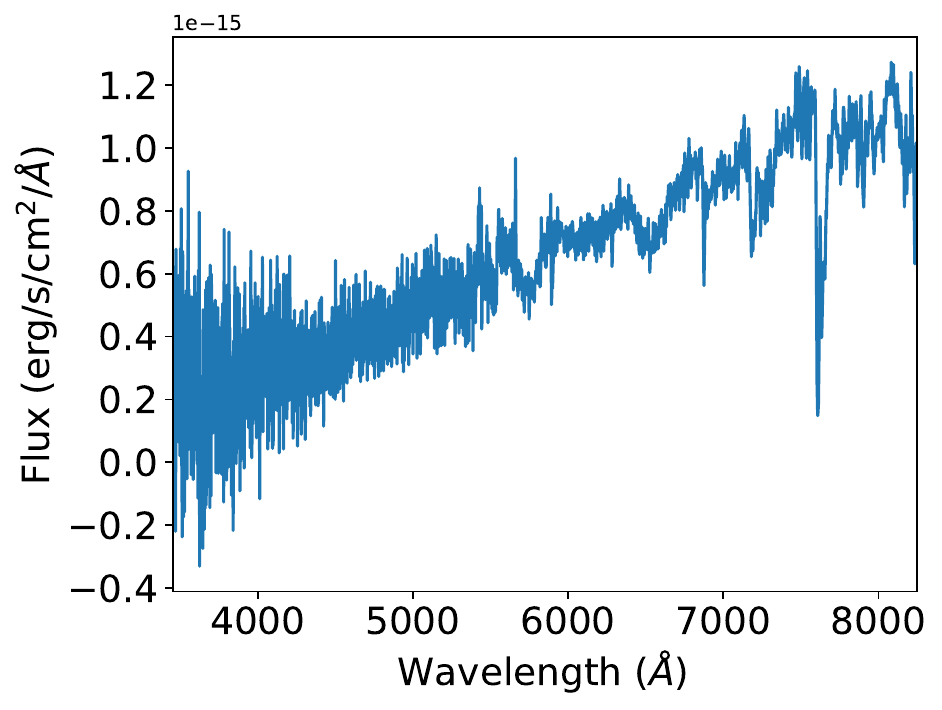}
\figsetgrpnote{Flux for the Lick, Palomar, and Keck spectra. See Table 3 for the observed spectral features and line measurements.}
\figsetgrpend

\figsetgrpstart
\figsetgrpnum{2.40}
\figsetgrptitle{Flux of 0613+2604}
\figsetplot{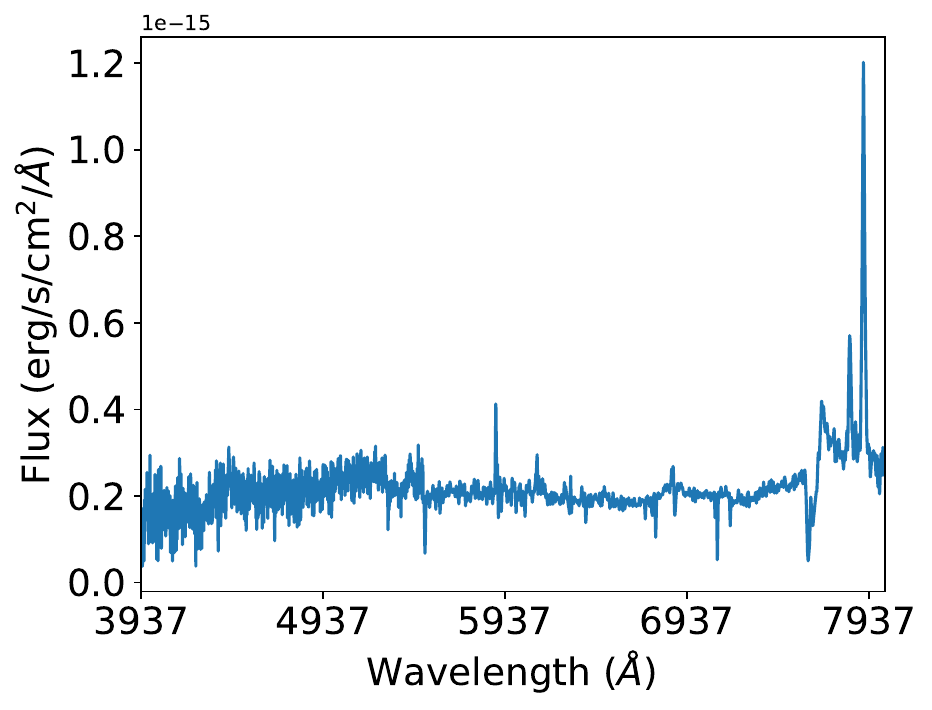}
\figsetgrpnote{Flux for the Lick, Palomar, and Keck spectra. See Table 3 for the observed spectral features and line measurements.}
\figsetgrpend

\figsetgrpstart
\figsetgrpnum{2.41}
\figsetgrptitle{Flux of 0624+2739}
\figsetplot{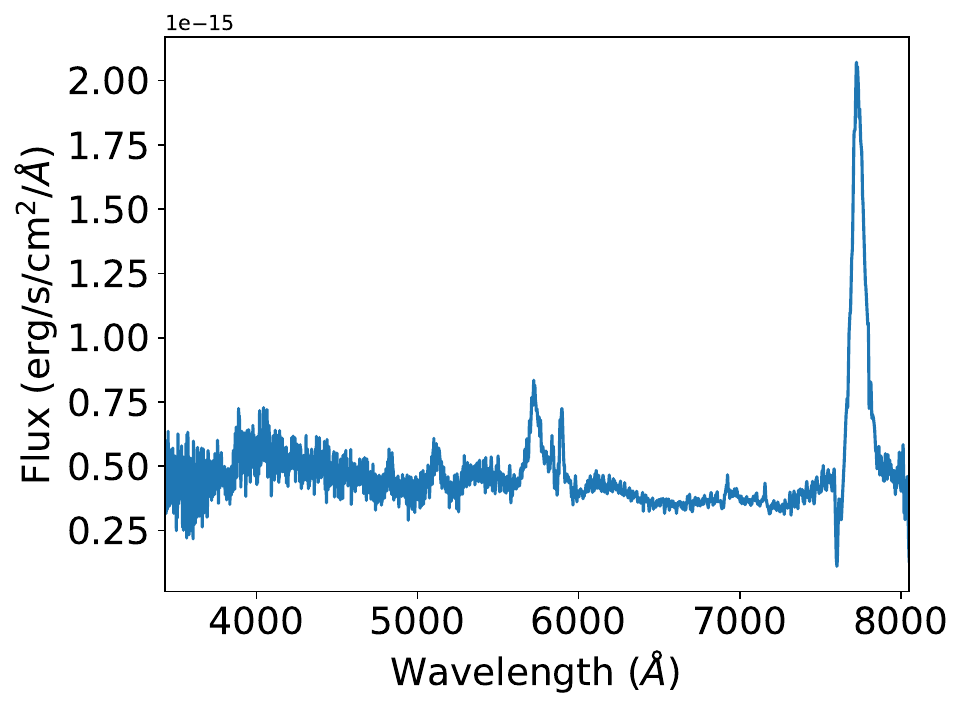}
\figsetgrpnote{Flux for the Lick, Palomar, and Keck spectra. See Table 3 for the observed spectral features and line measurements.}
\figsetgrpend

\figsetgrpstart
\figsetgrpnum{2.42}
\figsetgrptitle{Flux of 0633+2128}
\figsetplot{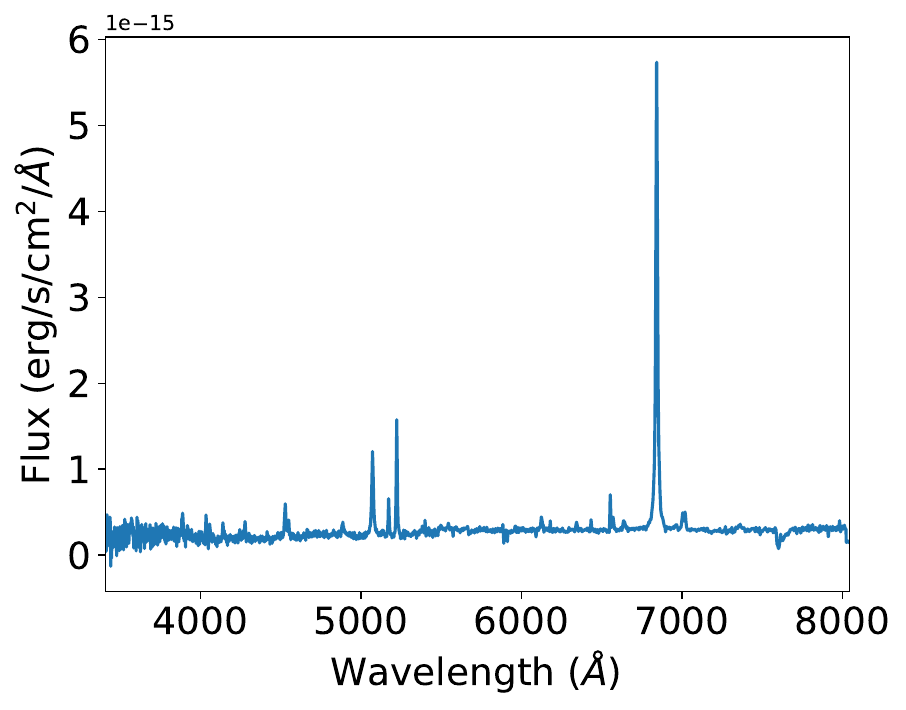}
\figsetgrpnote{Flux for the Lick, Palomar, and Keck spectra. See Table 3 for the observed spectral features and line measurements.}
\figsetgrpend

\figsetgrpstart
\figsetgrpnum{2.43}
\figsetgrptitle{Flux of 0636+2331}
\figsetplot{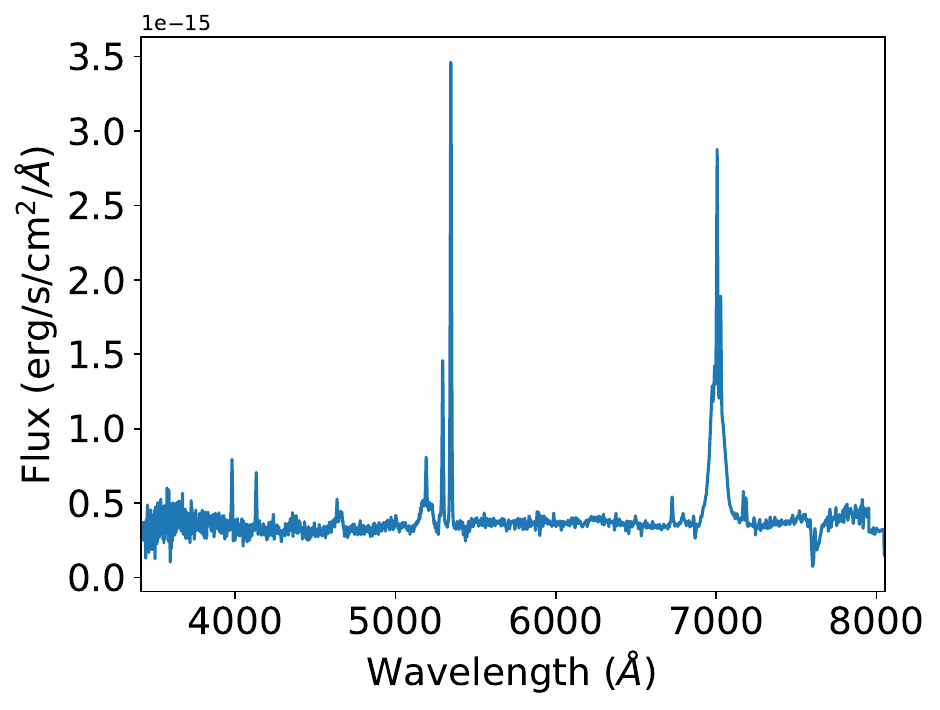}
\figsetgrpnote{Flux for the Lick, Palomar, and Keck spectra. See Table 3 for the observed spectral features and line measurements.}
\figsetgrpend

\figsetgrpstart
\figsetgrpnum{2.44}
\figsetgrptitle{Flux of 0650+2503}
\figsetplot{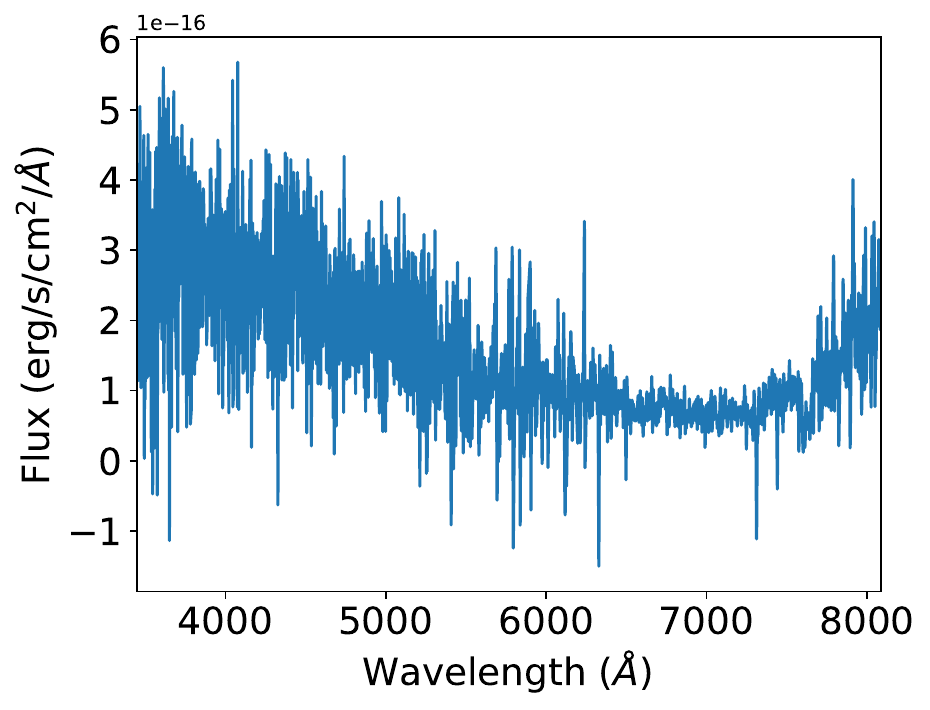}
\figsetgrpnote{Flux for the Lick, Palomar, and Keck spectra. See Table 3 for the observed spectral features and line measurements.}
\figsetgrpend

\figsetgrpstart
\figsetgrpnum{2.45}
\figsetgrptitle{Flux of 0829+1754}
\figsetplot{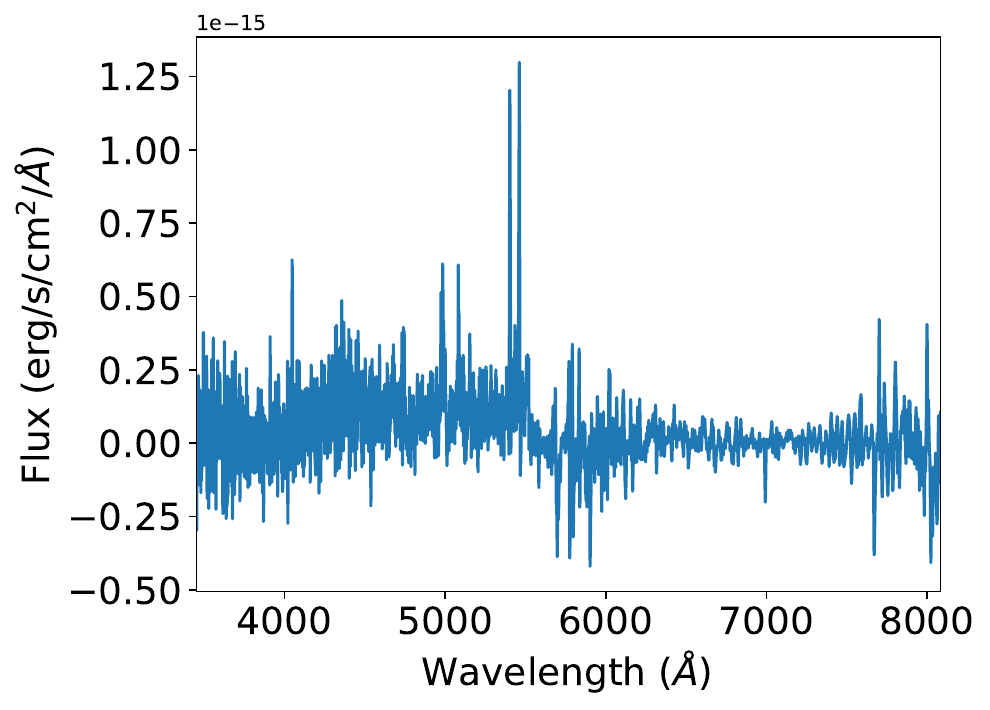}
\figsetgrpnote{Flux for the Lick, Palomar, and Keck spectra. See Table 3 for the observed spectral features and line measurements.}
\figsetgrpend

\figsetgrpstart
\figsetgrpnum{2.46}
\figsetgrptitle{Flux of 0829+2225}
\figsetplot{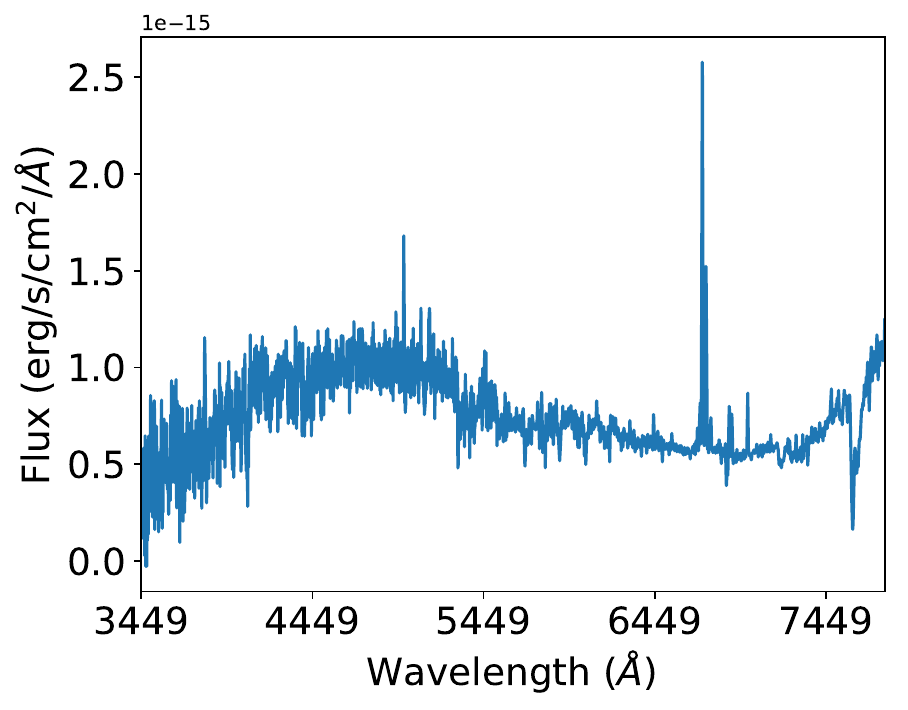}
\figsetgrpnote{Flux for the Lick, Palomar, and Keck spectra. See Table 3 for the observed spectral features and line measurements.}
\figsetgrpend

\figsetgrpstart
\figsetgrpnum{2.47}
\figsetgrptitle{Flux of 1124-0109}
\figsetplot{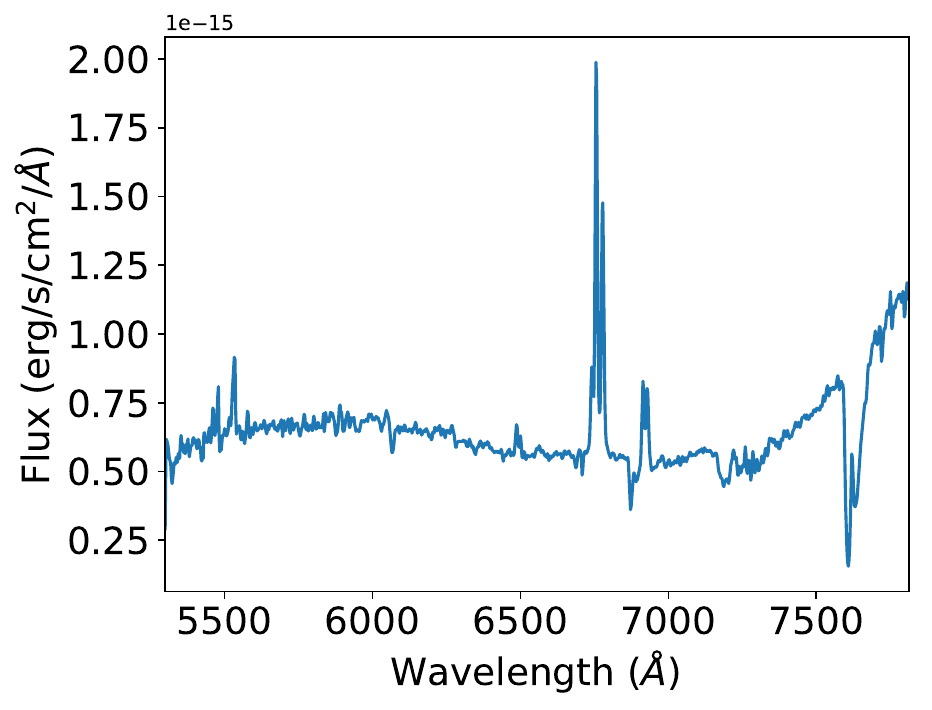}
\figsetgrpnote{Flux for the Lick, Palomar, and Keck spectra. See Table 3 for the observed spectral features and line measurements.}
\figsetgrpend

\figsetgrpstart
\figsetgrpnum{2.48}
\figsetgrptitle{Flux of 1129-0424}
\figsetplot{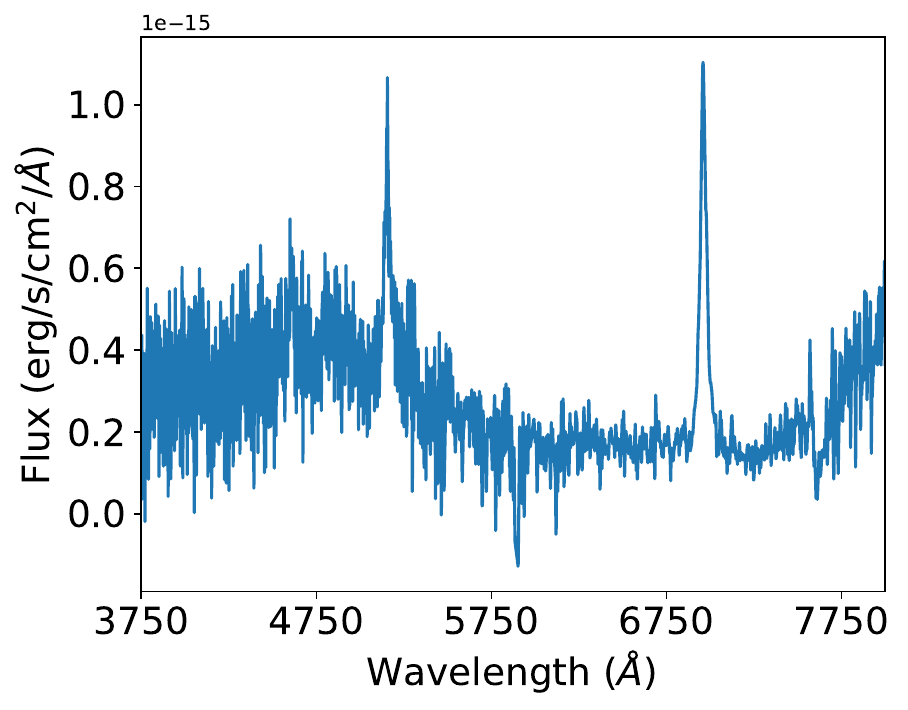}
\figsetgrpnote{Flux for the Lick, Palomar, and Keck spectra. See Table 3 for the observed spectral features and line measurements.}
\figsetgrpend

\figsetgrpstart
\figsetgrpnum{2.49}
\figsetgrptitle{Flux of 1203+0229}
\figsetplot{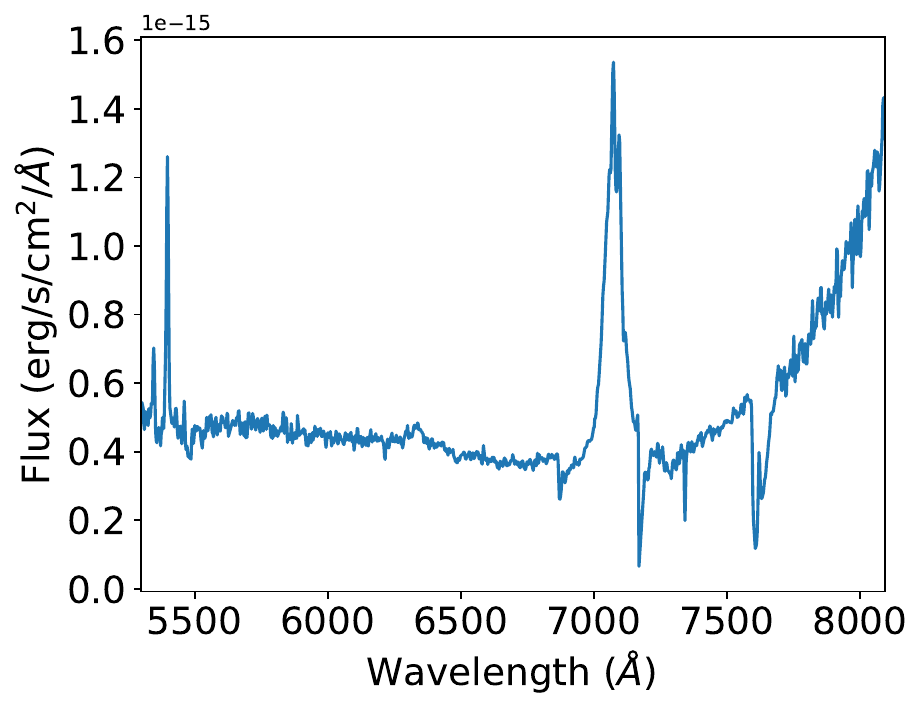}
\figsetgrpnote{Flux for the Lick, Palomar, and Keck spectra. See Table 3 for the observed spectral features and line measurements.}
\figsetgrpend

\figsetgrpstart
\figsetgrpnum{2.50}
\figsetgrptitle{Flux of 1205-0156}
\figsetplot{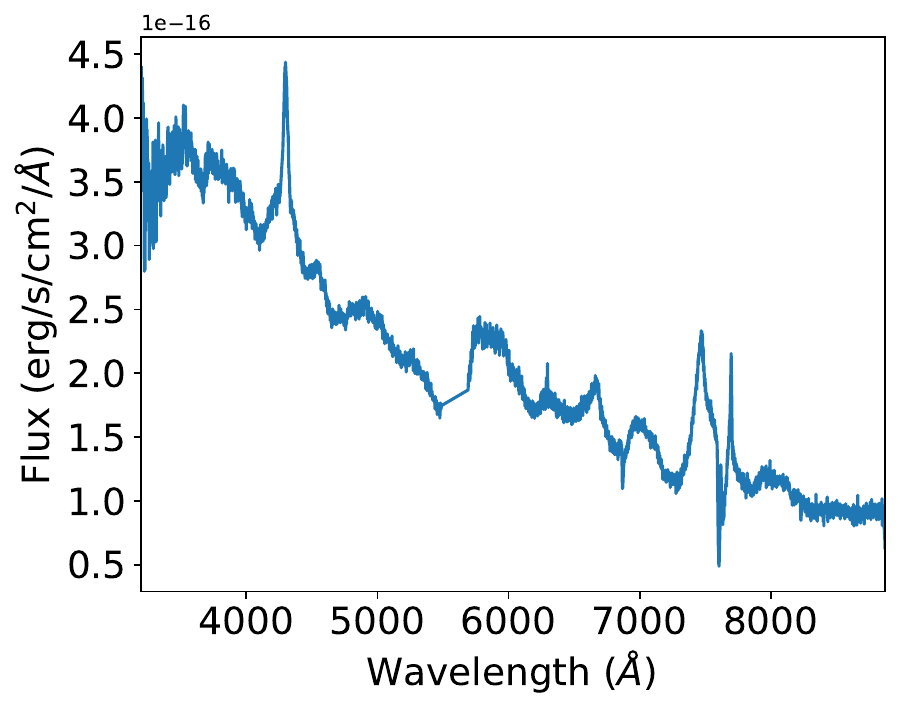}
\figsetgrpnote{Flux for the Lick, Palomar, and Keck spectra. See Table 3 for the observed spectral features and line measurements.}
\figsetgrpend

\figsetgrpstart
\figsetgrpnum{2.51}
\figsetgrptitle{Flux of 1207-0411}
\figsetplot{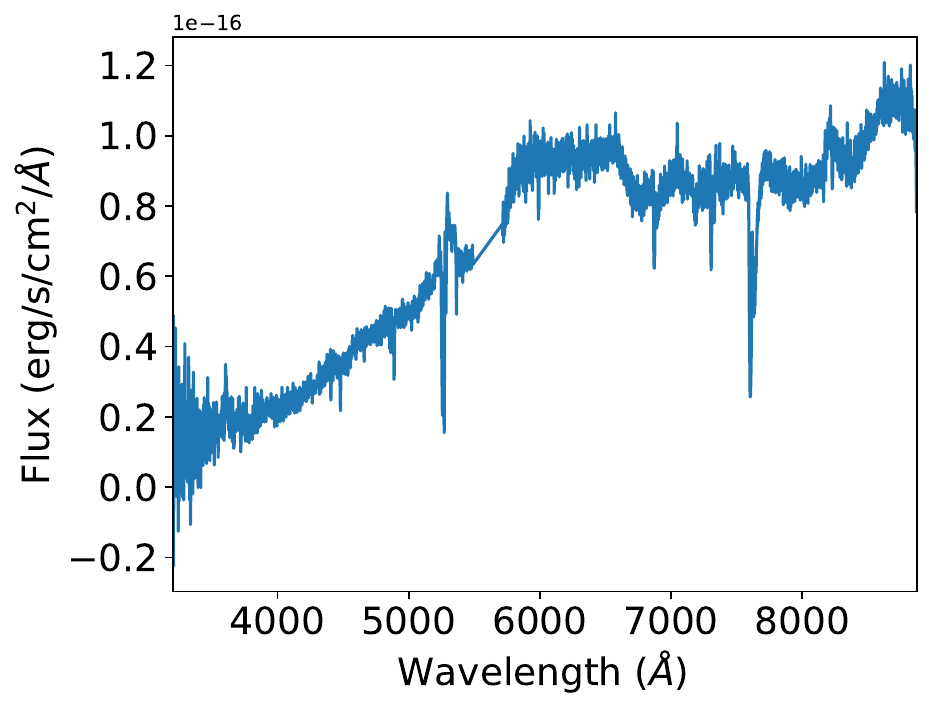}
\figsetgrpnote{Flux for the Lick, Palomar, and Keck spectra. See Table 3 for the observed spectral features and line measurements.}
\figsetgrpend

\figsetgrpstart
\figsetgrpnum{2.52}
\figsetgrptitle{Flux of 1209-0618}
\figsetplot{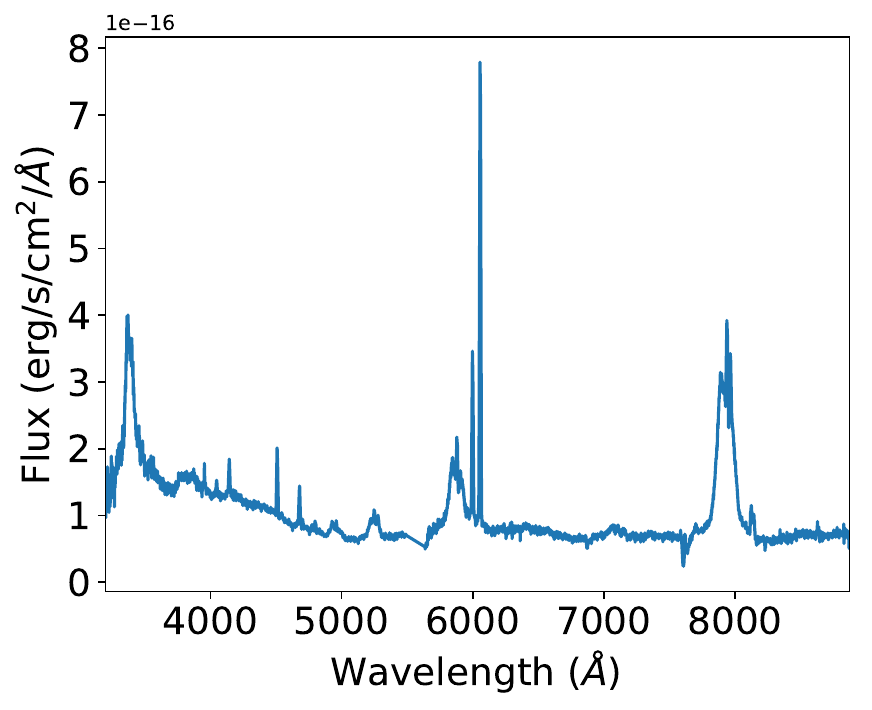}
\figsetgrpnote{Flux for the Lick, Palomar, and Keck spectra. See Table 3 for the observed spectral features and line measurements.}
\figsetgrpend

\figsetgrpstart
\figsetgrpnum{2.53}
\figsetgrptitle{Flux of 1212-0556}
\figsetplot{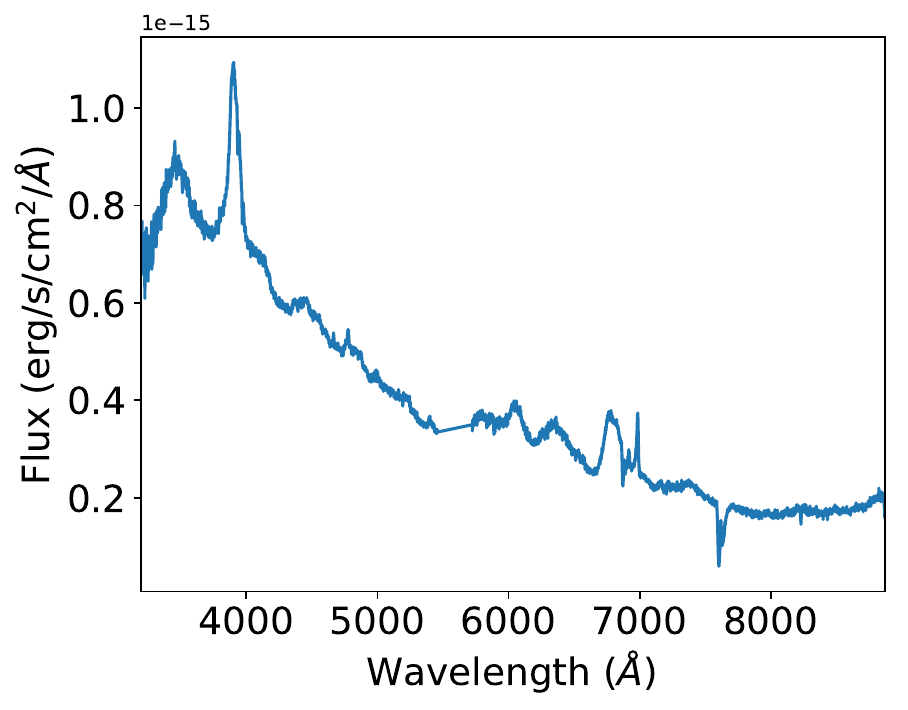}
\figsetgrpnote{Flux for the Lick, Palomar, and Keck spectra. See Table 3 for the observed spectral features and line measurements.}
\figsetgrpend

\figsetgrpstart
\figsetgrpnum{2.54}
\figsetgrptitle{Flux of 1220-0830}
\figsetplot{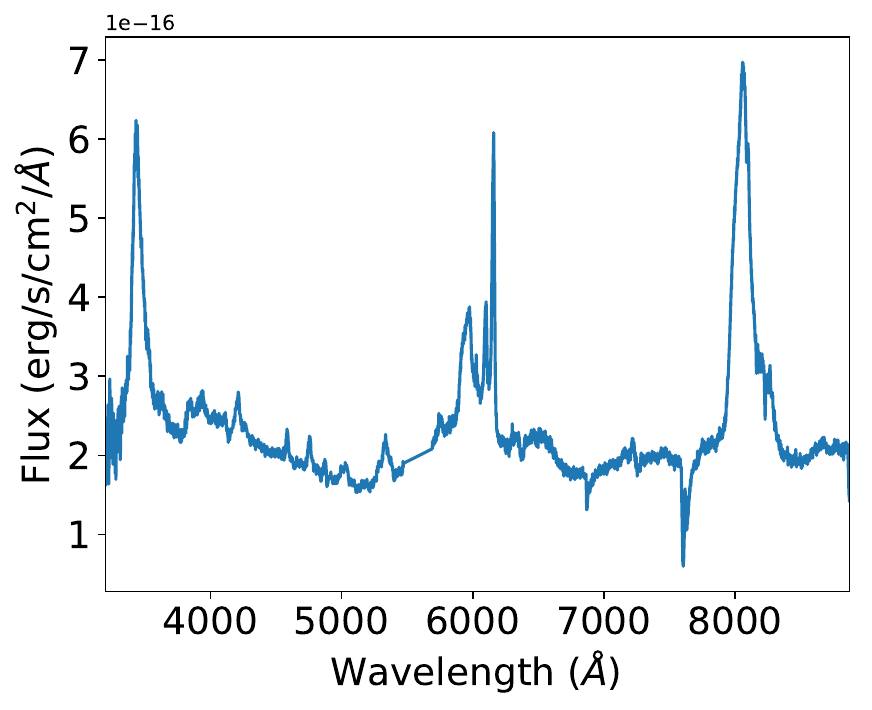}
\figsetgrpnote{Flux for the Lick, Palomar, and Keck spectra. See Table 3 for the observed spectral features and line measurements.}
\figsetgrpend

\figsetgrpstart
\figsetgrpnum{2.55}
\figsetgrptitle{Flux of 1228-0927}
\figsetplot{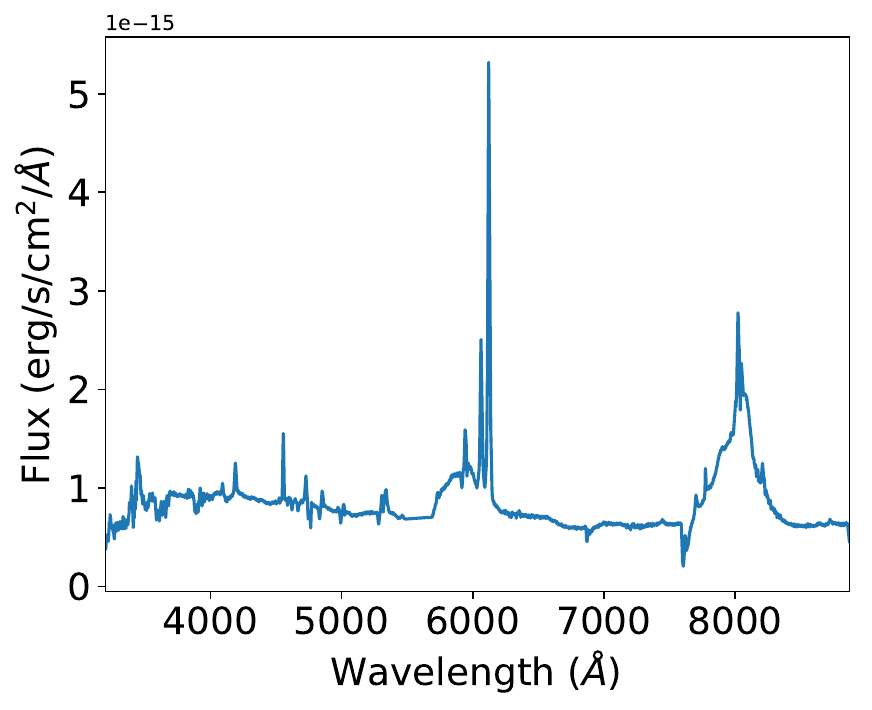}
\figsetgrpnote{Flux for the Lick, Palomar, and Keck spectra. See Table 3 for the observed spectral features and line measurements.}
\figsetgrpend

\figsetgrpstart
\figsetgrpnum{2.56}
\figsetgrptitle{Flux of 1240-0928}
\figsetplot{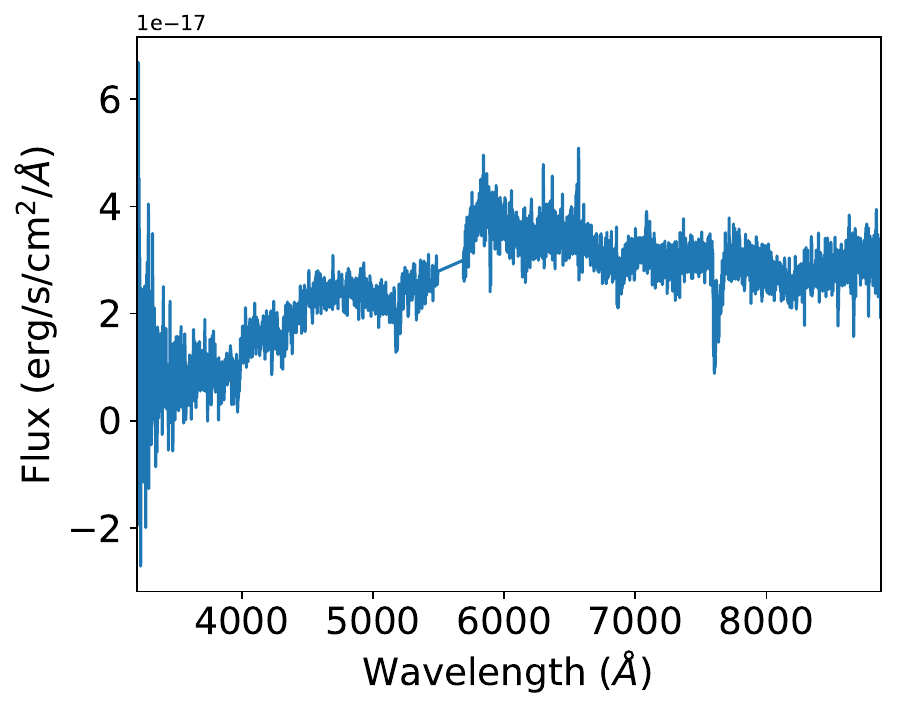}
\figsetgrpnote{Flux for the Lick, Palomar, and Keck spectra. See Table 3 for the observed spectral features and line measurements.}
\figsetgrpend

\figsetgrpstart
\figsetgrpnum{2.57}
\figsetgrptitle{Flux of 1248-0659}
\figsetplot{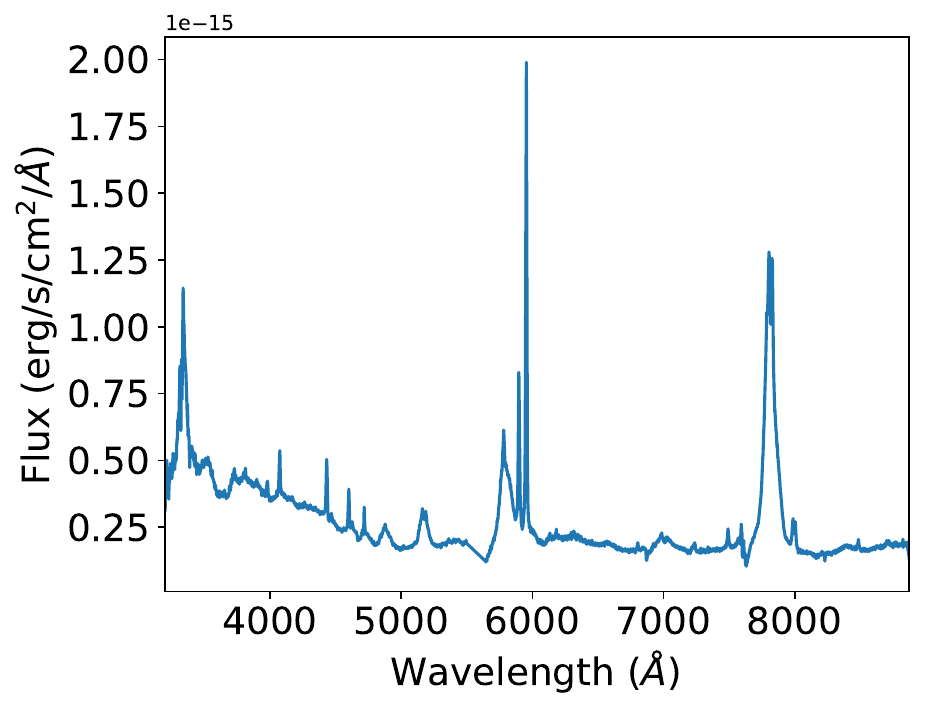}
\figsetgrpnote{Flux for the Lick, Palomar, and Keck spectra. See Table 3 for the observed spectral features and line measurements.}
\figsetgrpend

\figsetgrpstart
\figsetgrpnum{2.58}
\figsetgrptitle{Flux of 1252-0228}
\figsetplot{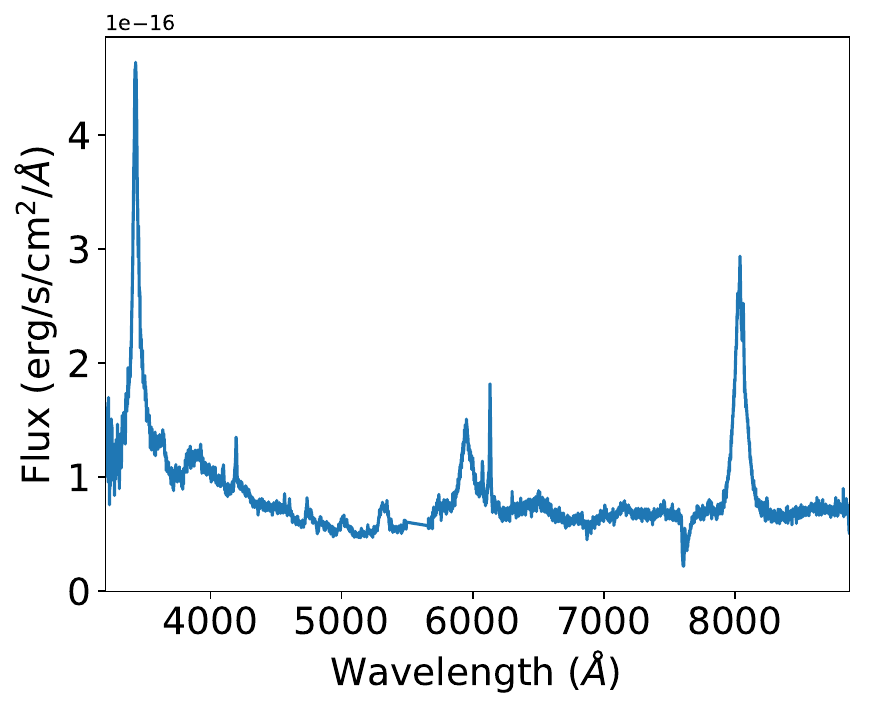}
\figsetgrpnote{Flux for the Lick, Palomar, and Keck spectra. See Table 3 for the observed spectral features and line measurements.}
\figsetgrpend

\figsetgrpstart
\figsetgrpnum{2.59}
\figsetgrptitle{Flux of 1253-0323}
\figsetplot{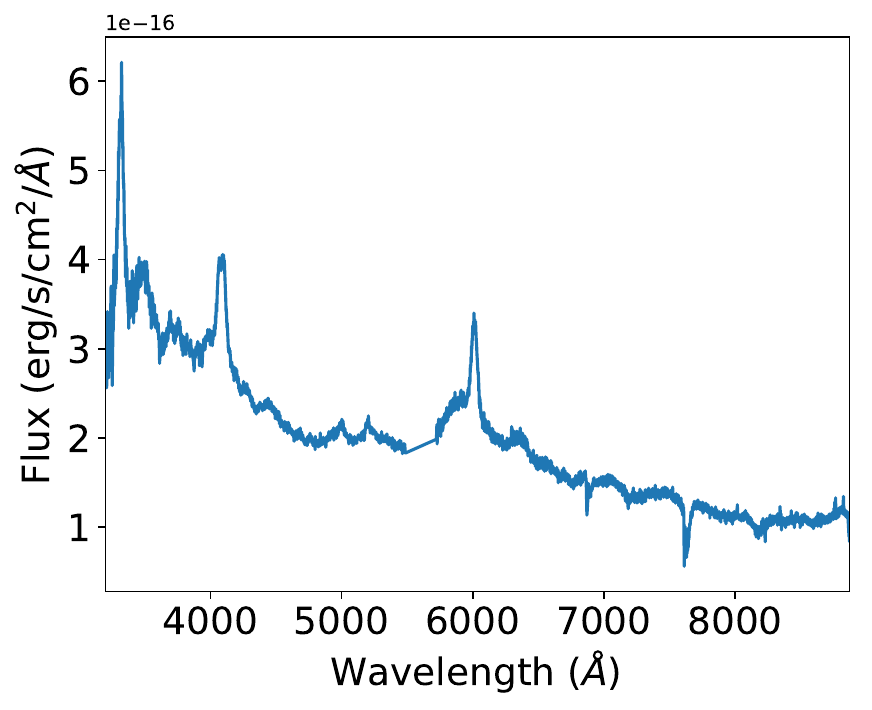}
\figsetgrpnote{Flux for the Lick, Palomar, and Keck spectra. See Table 3 for the observed spectral features and line measurements.}
\figsetgrpend

\figsetgrpstart
\figsetgrpnum{2.60}
\figsetgrptitle{Flux of 1602-1917}
\figsetplot{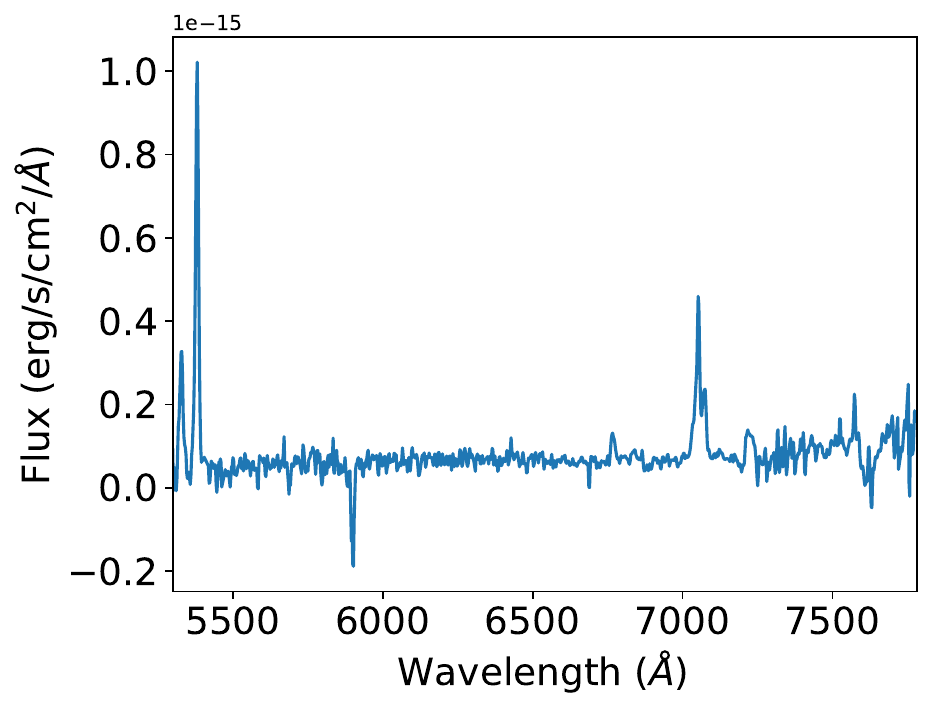}
\figsetgrpnote{Flux for the Lick, Palomar, and Keck spectra. See Table 3 for the observed spectral features and line measurements.}
\figsetgrpend

\figsetgrpstart
\figsetgrpnum{2.61}
\figsetgrptitle{Flux of 1603-2340}
\figsetplot{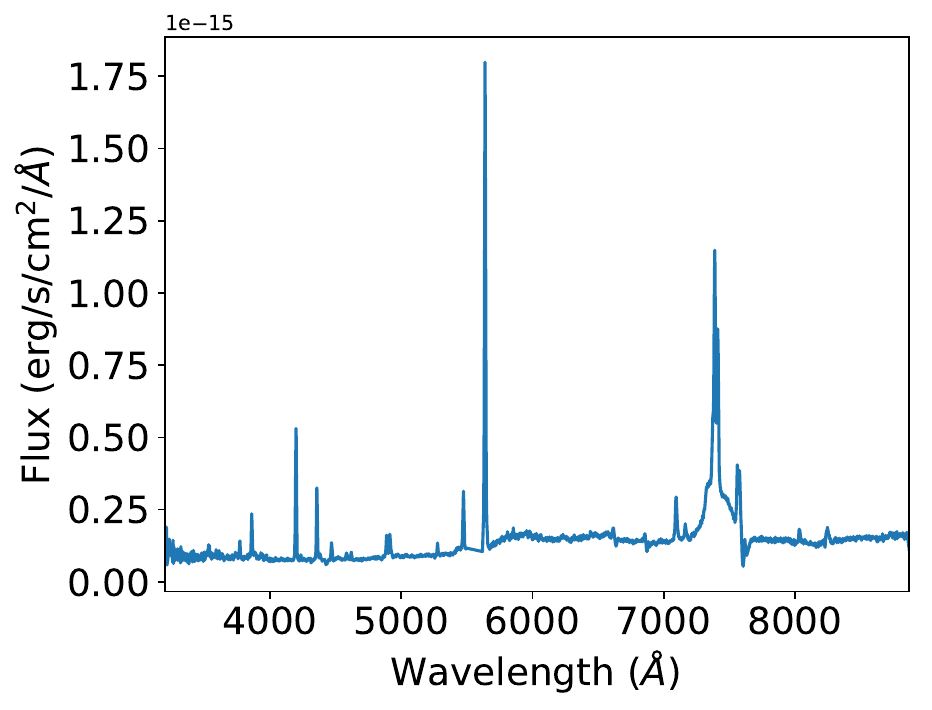}
\figsetgrpnote{Flux for the Lick, Palomar, and Keck spectra. See Table 3 for the observed spectral features and line measurements.}
\figsetgrpend

\figsetgrpstart
\figsetgrpnum{2.62}
\figsetgrptitle{Flux of 1608-2827}
\figsetplot{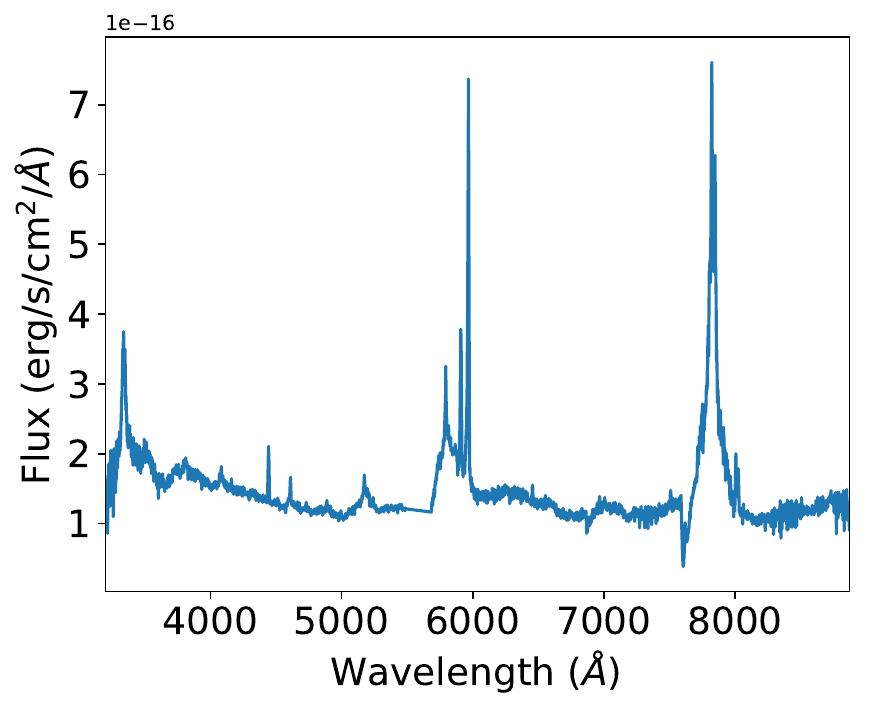}
\figsetgrpnote{Flux for the Lick, Palomar, and Keck spectra. See Table 3 for the observed spectral features and line measurements.}
\figsetgrpend

\figsetgrpstart
\figsetgrpnum{2.63}
\figsetgrptitle{Flux of 1610-2142}
\figsetplot{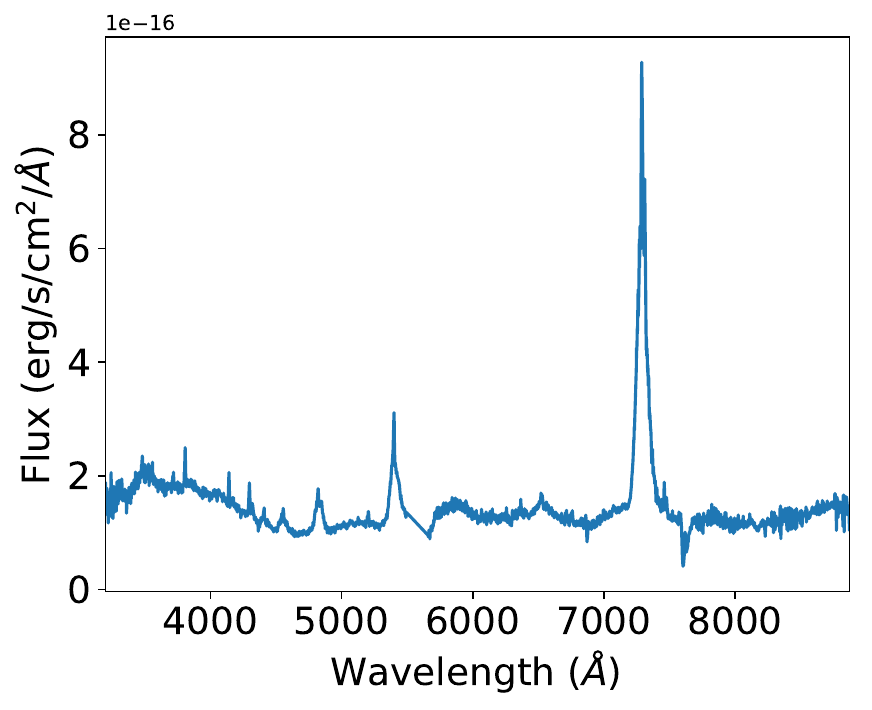}
\figsetgrpnote{Flux for the Lick, Palomar, and Keck spectra. See Table 3 for the observed spectral features and line measurements.}
\figsetgrpend

\figsetgrpstart
\figsetgrpnum{2.64}
\figsetgrptitle{Flux of 1610-1856}
\figsetplot{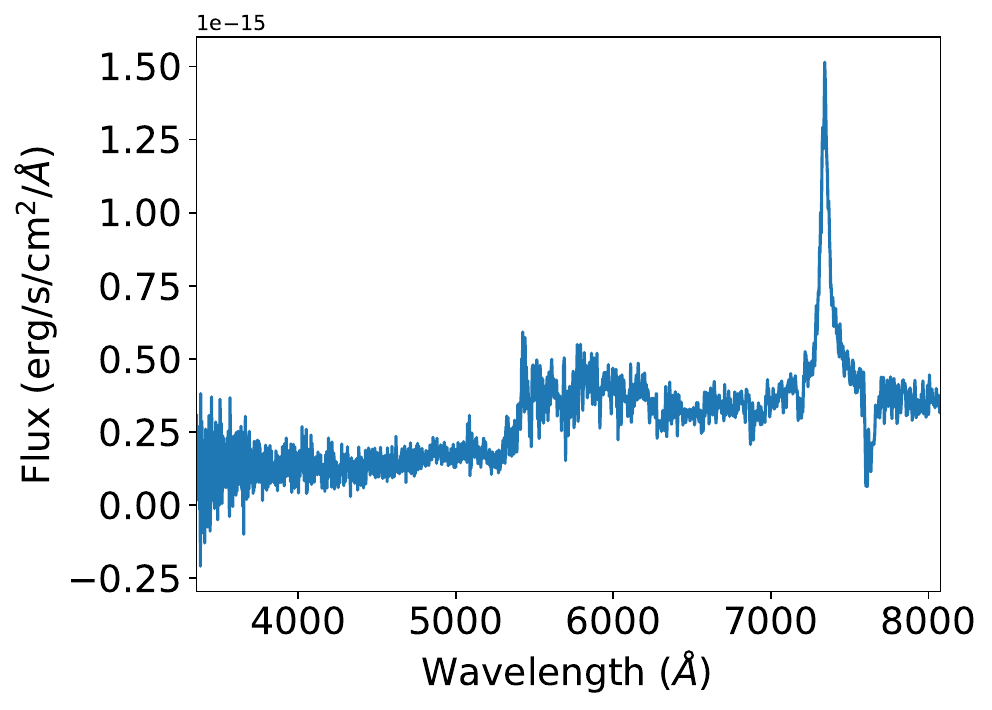}
\figsetgrpnote{Flux for the Lick, Palomar, and Keck spectra. See Table 3 for the observed spectral features and line measurements.}
\figsetgrpend

\figsetgrpstart
\figsetgrpnum{2.65}
\figsetgrptitle{Flux of 1611-1705}
\figsetplot{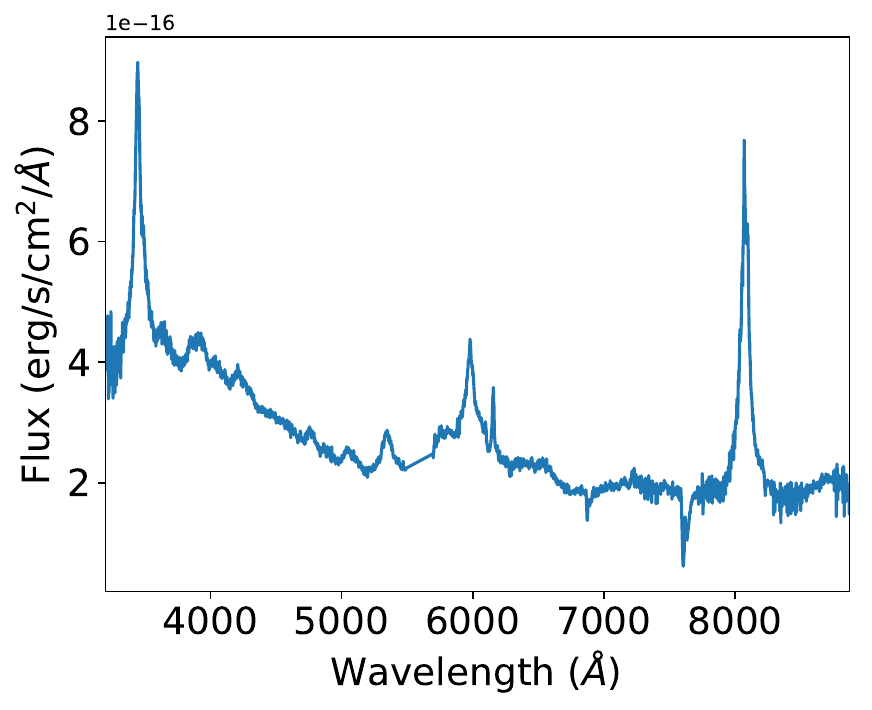}
\figsetgrpnote{Flux for the Lick, Palomar, and Keck spectra. See Table 3 for the observed spectral features and line measurements.}
\figsetgrpend

\figsetgrpstart
\figsetgrpnum{2.66}
\figsetgrptitle{Flux of 1620-1923}
\figsetplot{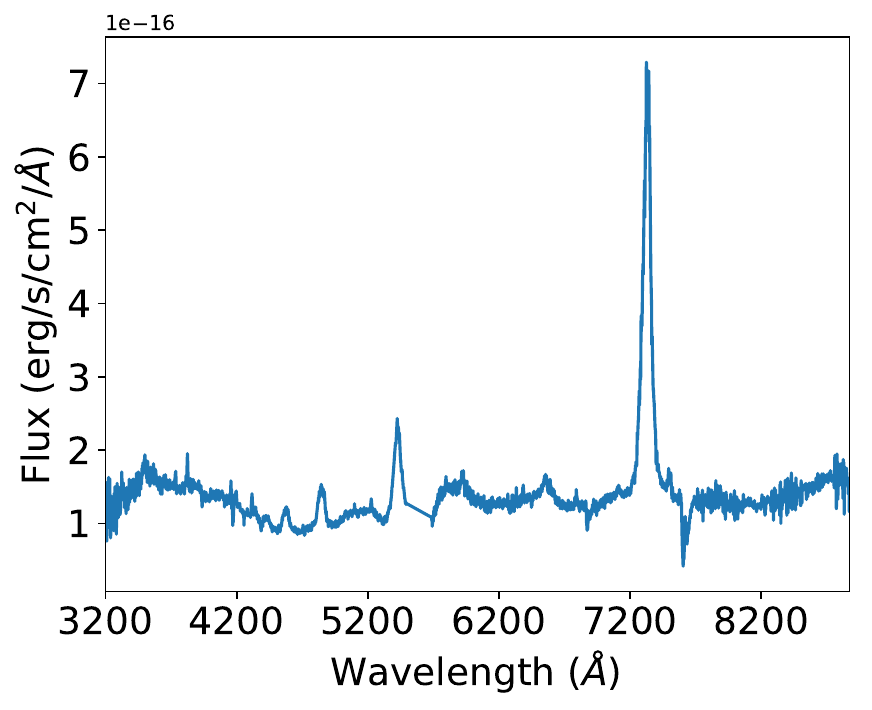}
\figsetgrpnote{Flux for the Lick, Palomar, and Keck spectra. See Table 3 for the observed spectral features and line measurements.}
\figsetgrpend

\figsetgrpstart
\figsetgrpnum{2.67}
\figsetgrptitle{Flux of 1630-2508}
\figsetplot{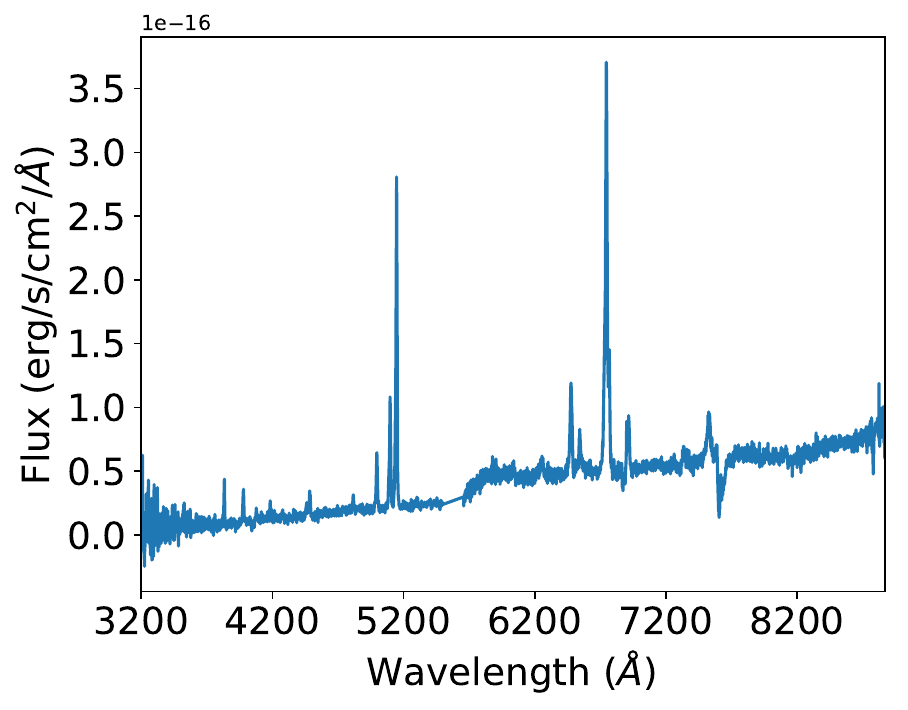}
\figsetgrpnote{Flux for the Lick, Palomar, and Keck spectra. See Table 3 for the observed spectral features and line measurements.}
\figsetgrpend

\figsetgrpstart
\figsetgrpnum{2.68}
\figsetgrptitle{Flux of 1631-1844}
\figsetplot{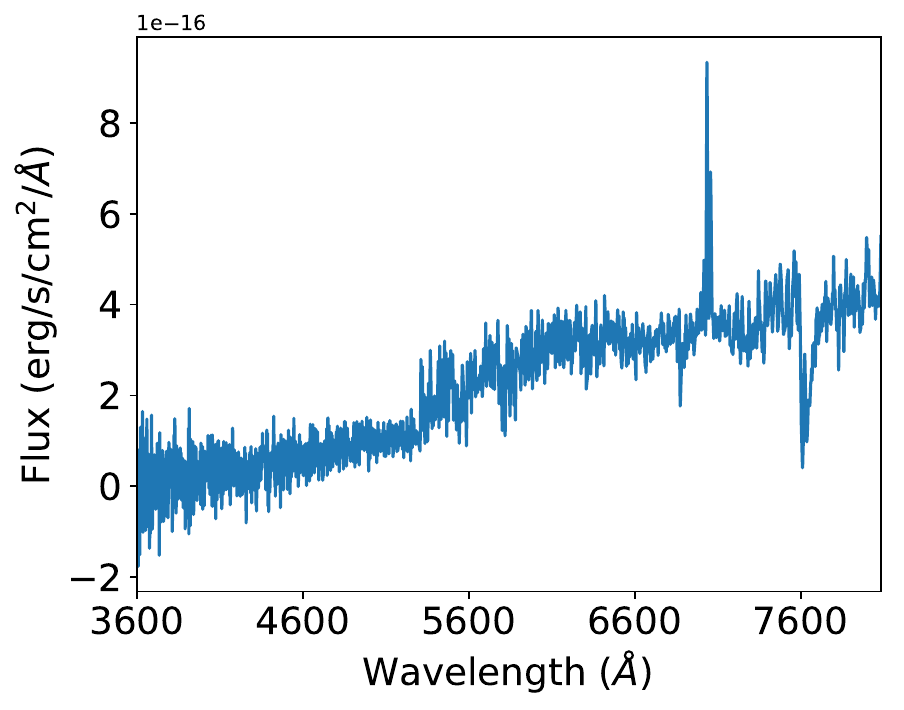}
\figsetgrpnote{Flux for the Lick, Palomar, and Keck spectra. See Table 3 for the observed spectral features and line measurements.}
\figsetgrpend

\figsetgrpstart
\figsetgrpnum{2.69}
\figsetgrptitle{Flux of 1641-2212}
\figsetplot{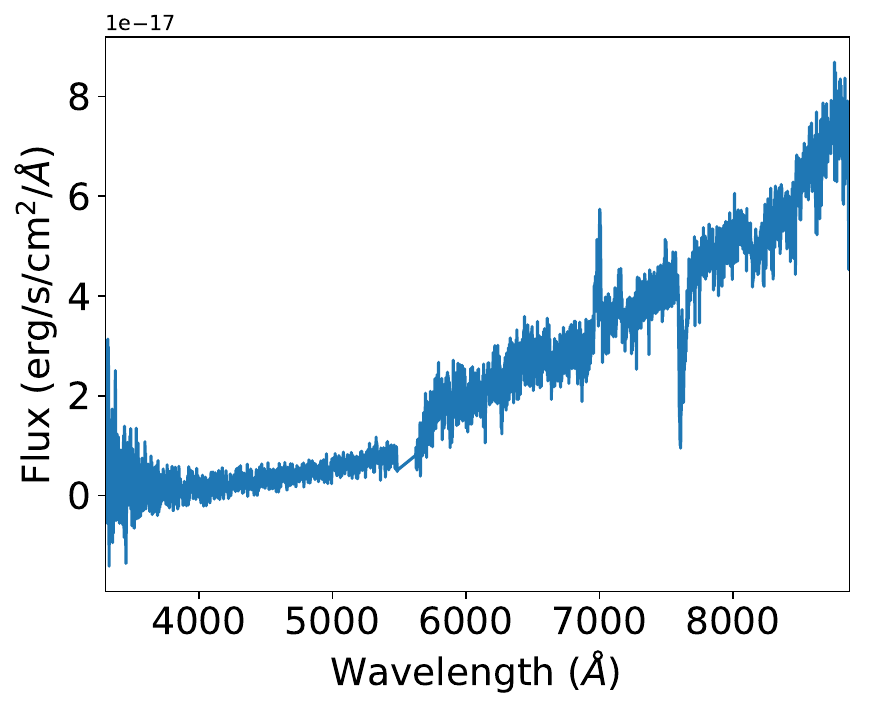}
\figsetgrpnote{Flux for the Lick, Palomar, and Keck spectra. See Table 3 for the observed spectral features and line measurements.}
\figsetgrpend

\figsetgrpstart
\figsetgrpnum{2.70}
\figsetgrptitle{Flux of 1644-1804}
\figsetplot{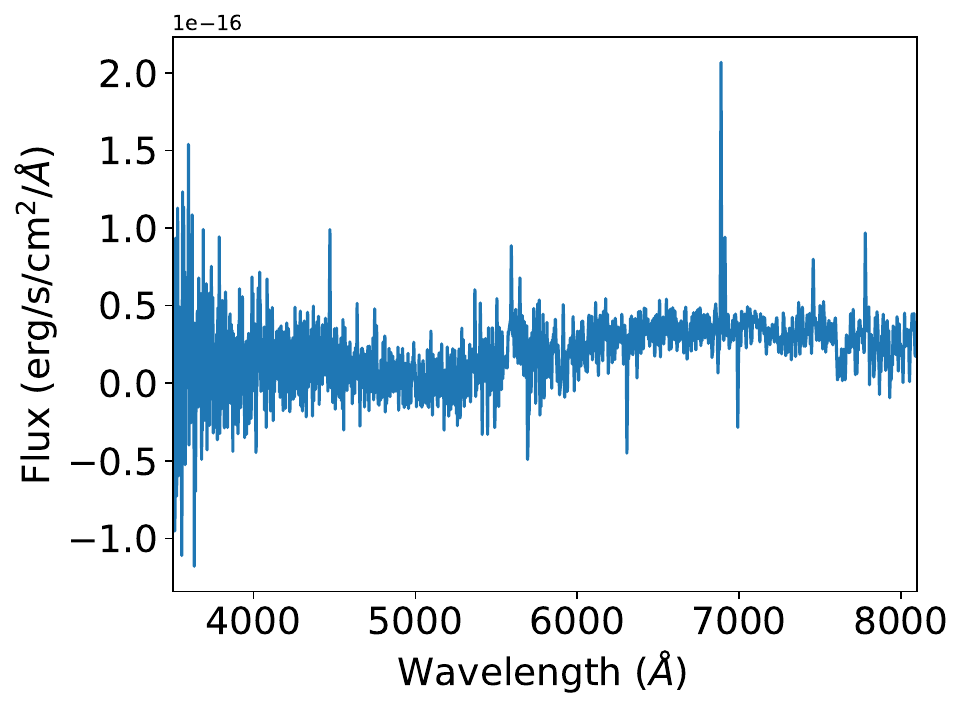}
\figsetgrpnote{Flux for the Lick, Palomar, and Keck spectra. See Table 3 for the observed spectral features and line measurements.}
\figsetgrpend

\figsetgrpstart
\figsetgrpnum{2.71}
\figsetgrptitle{Flux of 1652-2231}
\figsetplot{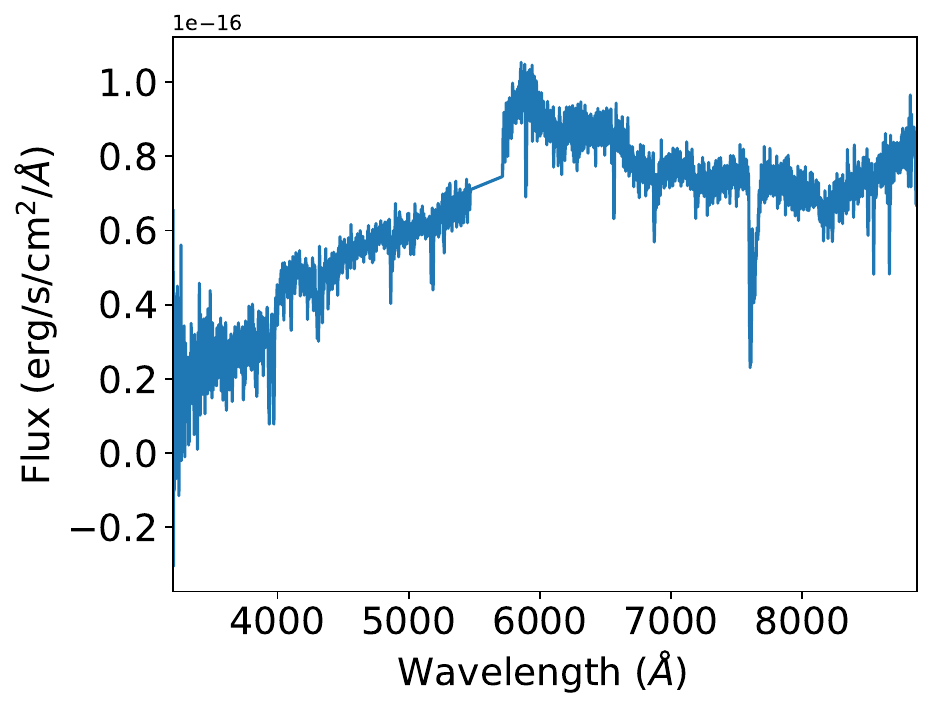}
\figsetgrpnote{Flux for the Lick, Palomar, and Keck spectra. See Table 3 for the observed spectral features and line measurements.}
\figsetgrpend

\figsetgrpstart
\figsetgrpnum{2.72}
\figsetgrptitle{Flux of 1824+4640}
\figsetplot{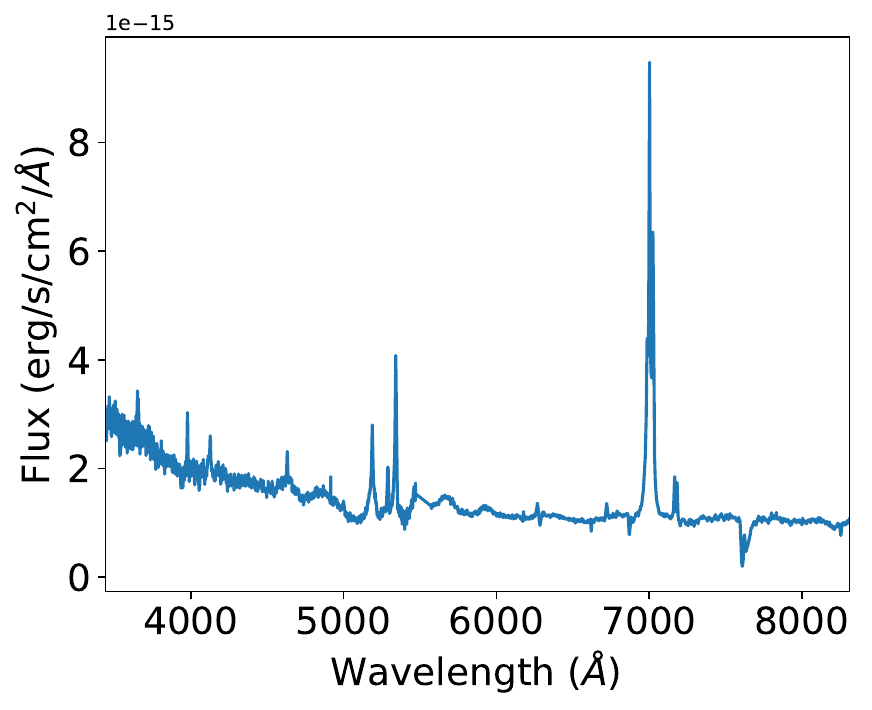}
\figsetgrpnote{Flux for the Lick, Palomar, and Keck spectra. See Table 3 for the observed spectral features and line measurements.}
\figsetgrpend

\figsetgrpstart
\figsetgrpnum{2.73}
\figsetgrptitle{Flux of 1840+4327}
\figsetplot{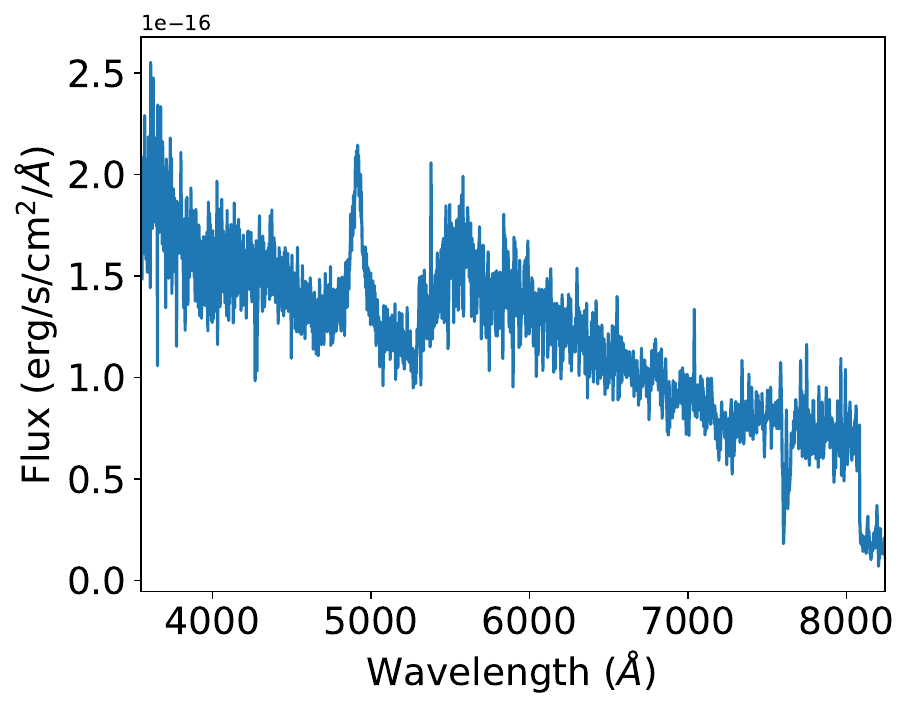}
\figsetgrpnote{Flux for the Lick, Palomar, and Keck spectra. See Table 3 for the observed spectral features and line measurements.}
\figsetgrpend

\figsetgrpstart
\figsetgrpnum{2.74}
\figsetgrptitle{Flux of 1841+4259}
\figsetplot{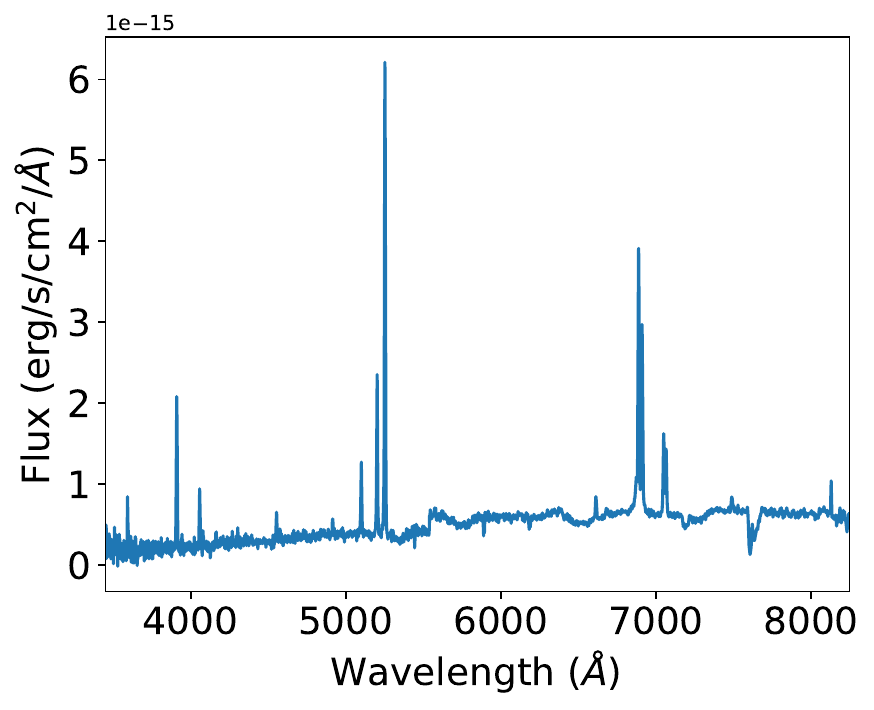}
\figsetgrpnote{Flux for the Lick, Palomar, and Keck spectra. See Table 3 for the observed spectral features and line measurements.}
\figsetgrpend

\figsetgrpstart
\figsetgrpnum{2.75}
\figsetgrptitle{Flux of 1841+4333}
\figsetplot{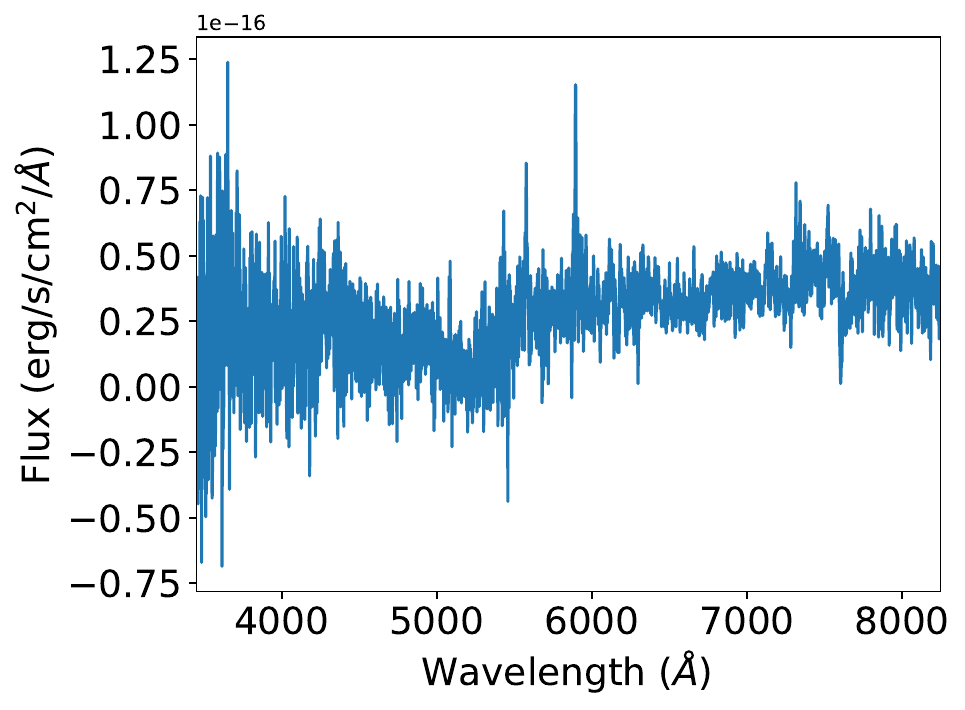}
\figsetgrpnote{Flux for the Lick, Palomar, and Keck spectra. See Table 3 for the observed spectral features and line measurements.}
\figsetgrpend

\figsetgrpstart
\figsetgrpnum{2.76}
\figsetgrptitle{Flux of 1842+4304}
\figsetplot{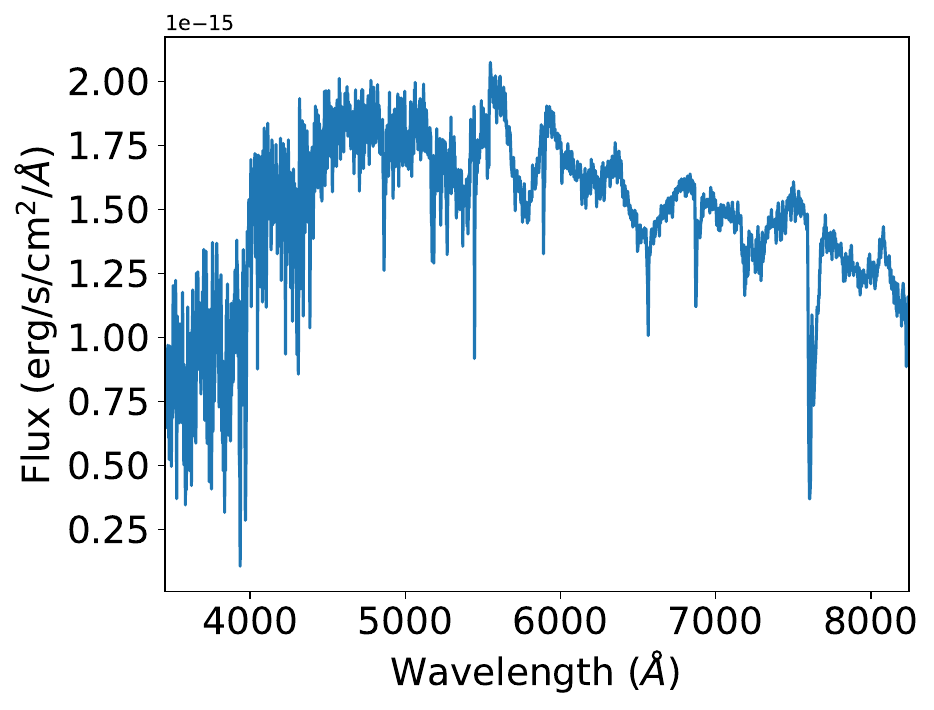}
\figsetgrpnote{Flux for the Lick, Palomar, and Keck spectra. See Table 3 for the observed spectral features and line measurements.}
\figsetgrpend

\figsetgrpstart
\figsetgrpnum{2.77}
\figsetgrptitle{Flux of 1843+4329}
\figsetplot{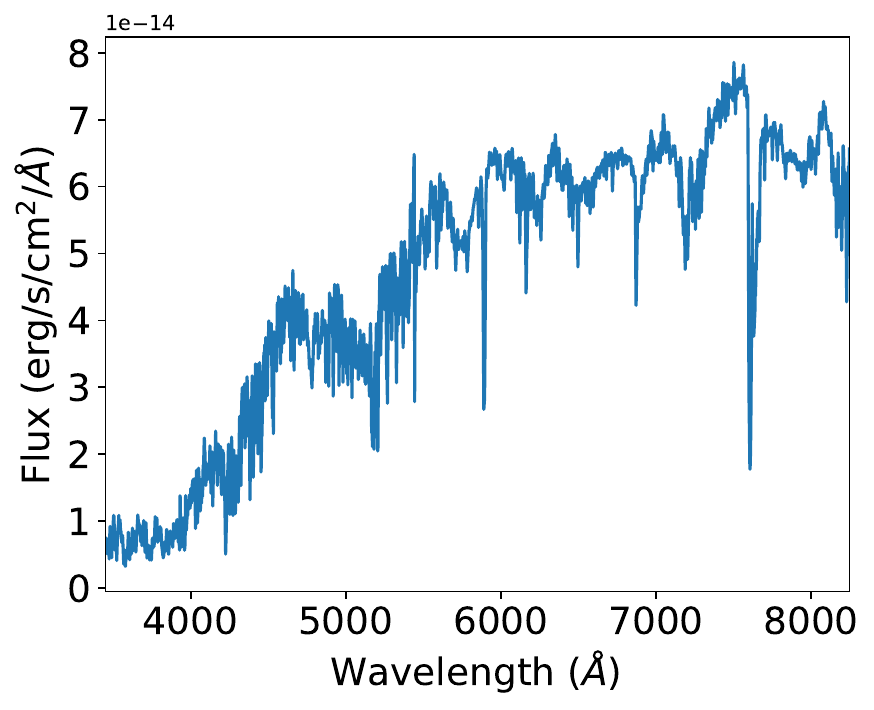}
\figsetgrpnote{Flux for the Lick, Palomar, and Keck spectra. See Table 3 for the observed spectral features and line measurements.}
\figsetgrpend

\figsetgrpstart
\figsetgrpnum{2.78}
\figsetgrptitle{Flux of 1845+4816}
\figsetplot{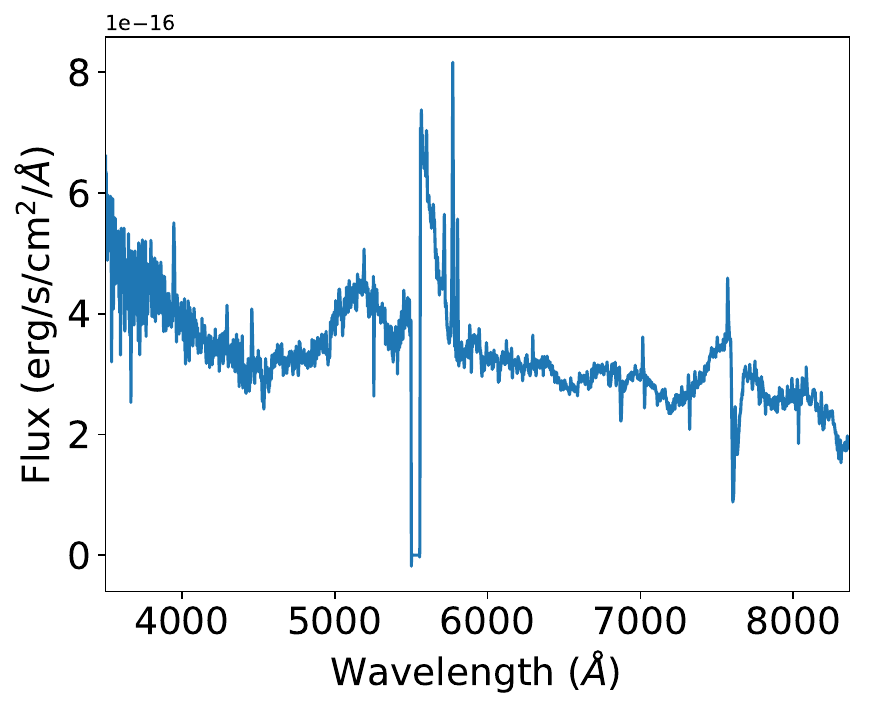}
\figsetgrpnote{Flux for the Lick, Palomar, and Keck spectra. See Table 3 for the observed spectral features and line measurements.}
\figsetgrpend

\figsetgrpstart
\figsetgrpnum{2.79}
\figsetgrptitle{Flux of 1847+4756}
\figsetplot{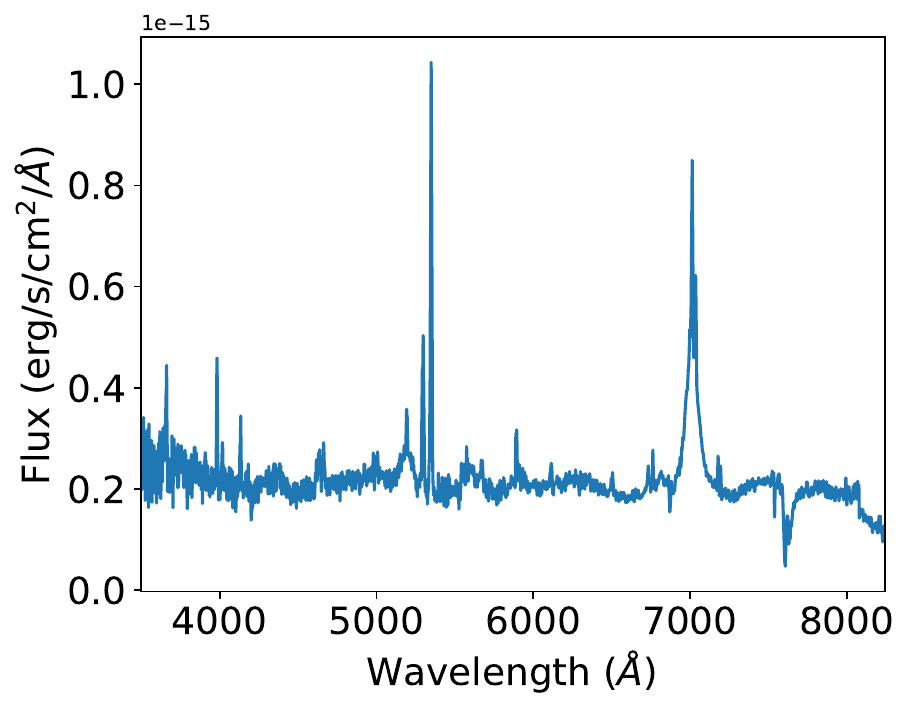}
\figsetgrpnote{Flux for the Lick, Palomar, and Keck spectra. See Table 3 for the observed spectral features and line measurements.}
\figsetgrpend

\figsetgrpstart
\figsetgrpnum{2.80}
\figsetgrptitle{Flux of 1848+4245}
\figsetplot{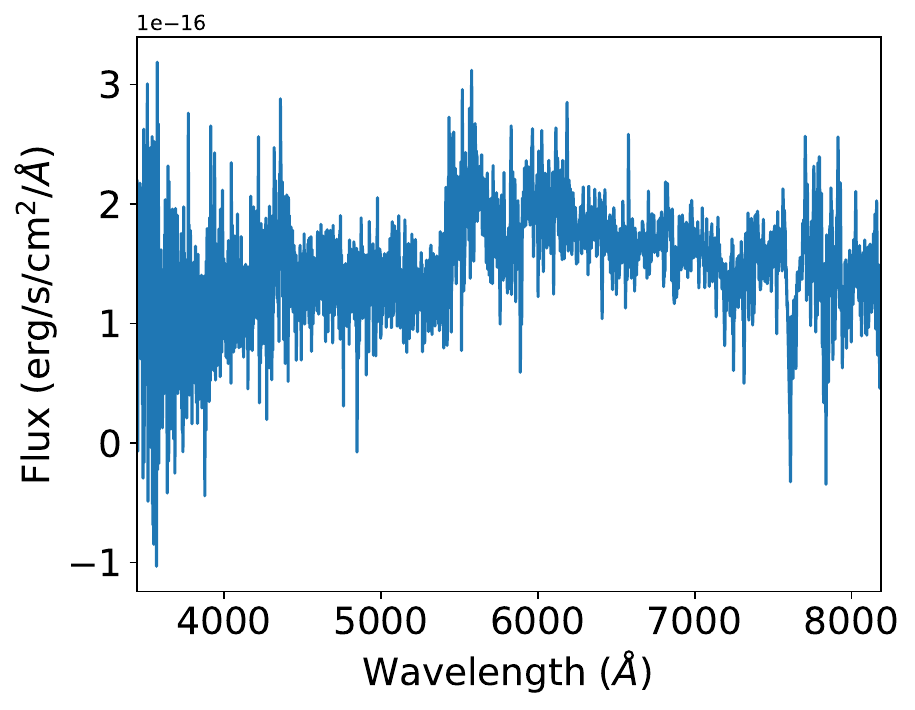}
\figsetgrpnote{Flux for the Lick, Palomar, and Keck spectra. See Table 3 for the observed spectral features and line measurements.}
\figsetgrpend

\figsetgrpstart
\figsetgrpnum{2.81}
\figsetgrptitle{Flux of 1848+4431}
\figsetplot{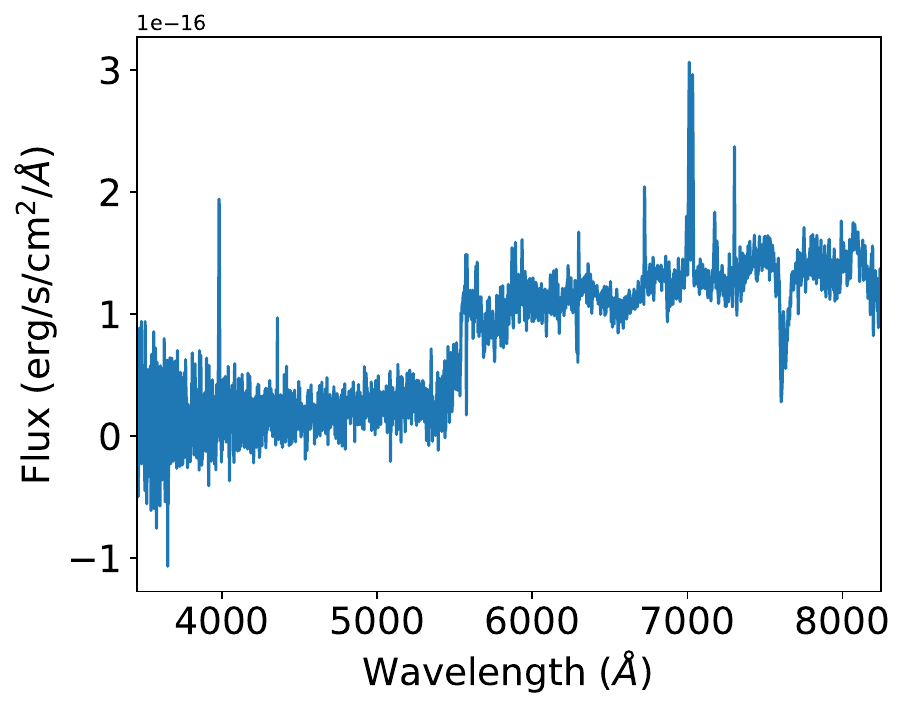}
\figsetgrpnote{Flux for the Lick, Palomar, and Keck spectra. See Table 3 for the observed spectral features and line measurements.}
\figsetgrpend

\figsetgrpstart
\figsetgrpnum{2.82}
\figsetgrptitle{Flux of 1850+4411}
\figsetplot{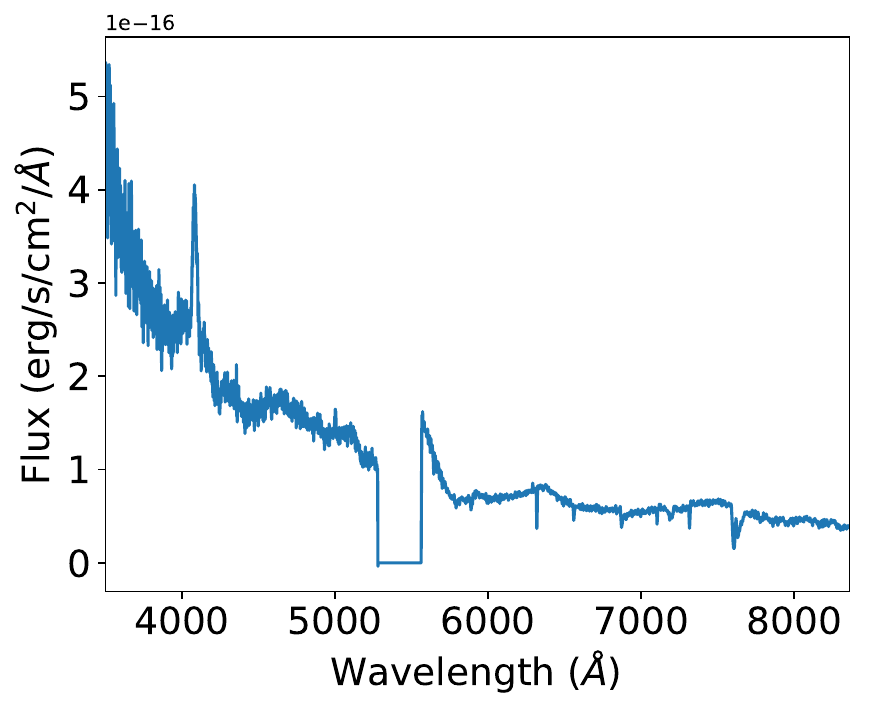}
\figsetgrpnote{Flux for the Lick, Palomar, and Keck spectra. See Table 3 for the observed spectral features and line measurements.}
\figsetgrpend

\figsetgrpstart
\figsetgrpnum{2.83}
\figsetgrptitle{Flux of 1853+4053}
\figsetplot{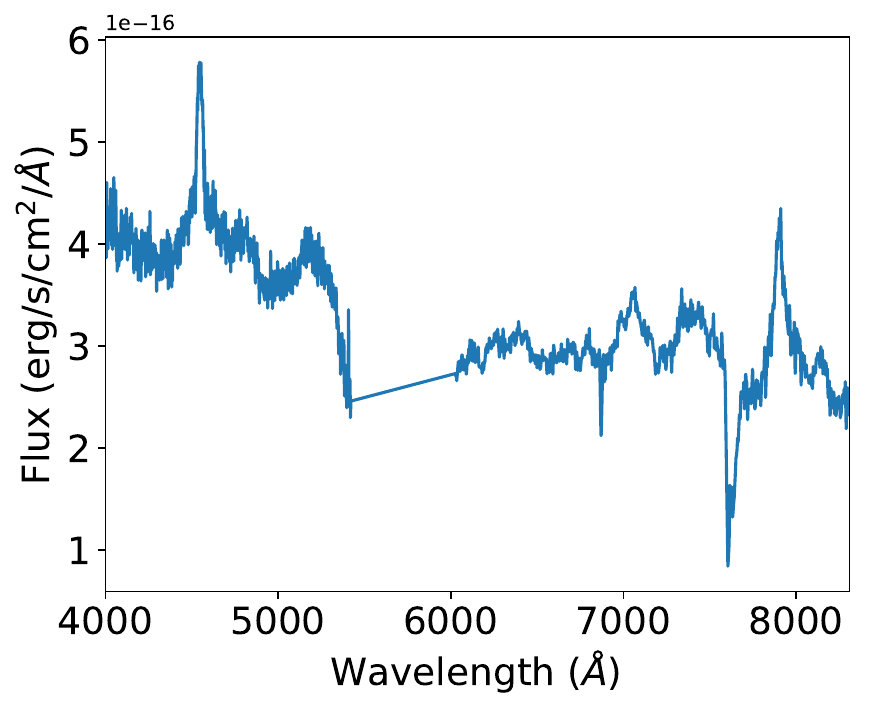}
\figsetgrpnote{Flux for the Lick, Palomar, and Keck spectra. See Table 3 for the observed spectral features and line measurements.}
\figsetgrpend

\figsetgrpstart
\figsetgrpnum{2.84}
\figsetgrptitle{Flux of 1855+4259}
\figsetplot{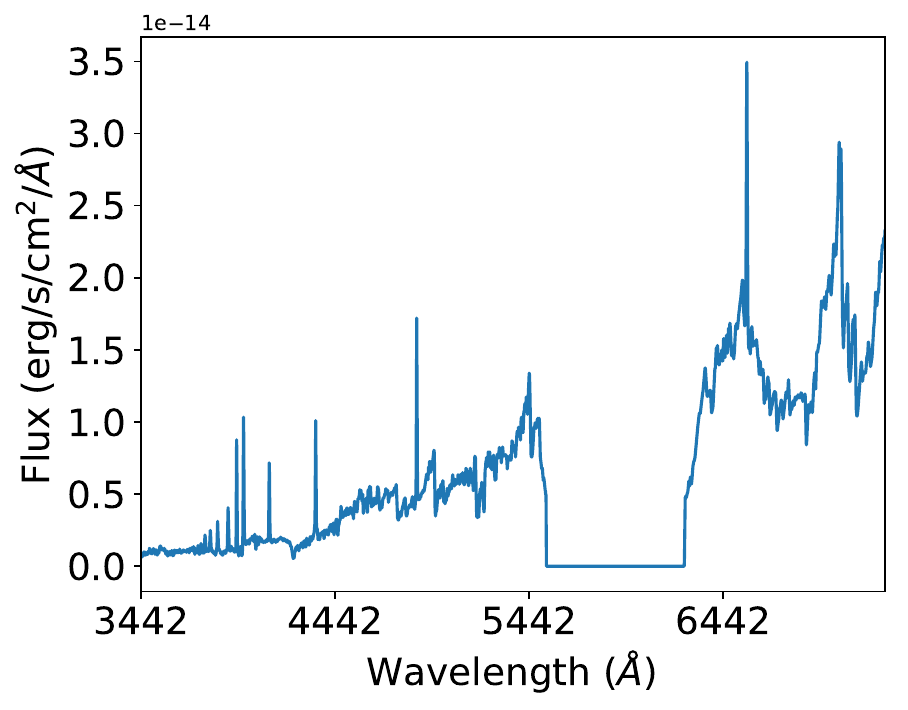}
\figsetgrpnote{Flux for the Lick, Palomar, and Keck spectra. See Table 3 for the observed spectral features and line measurements.}
\figsetgrpend

\figsetgrpstart
\figsetgrpnum{2.85}
\figsetgrptitle{Flux of 1856+4947}
\figsetplot{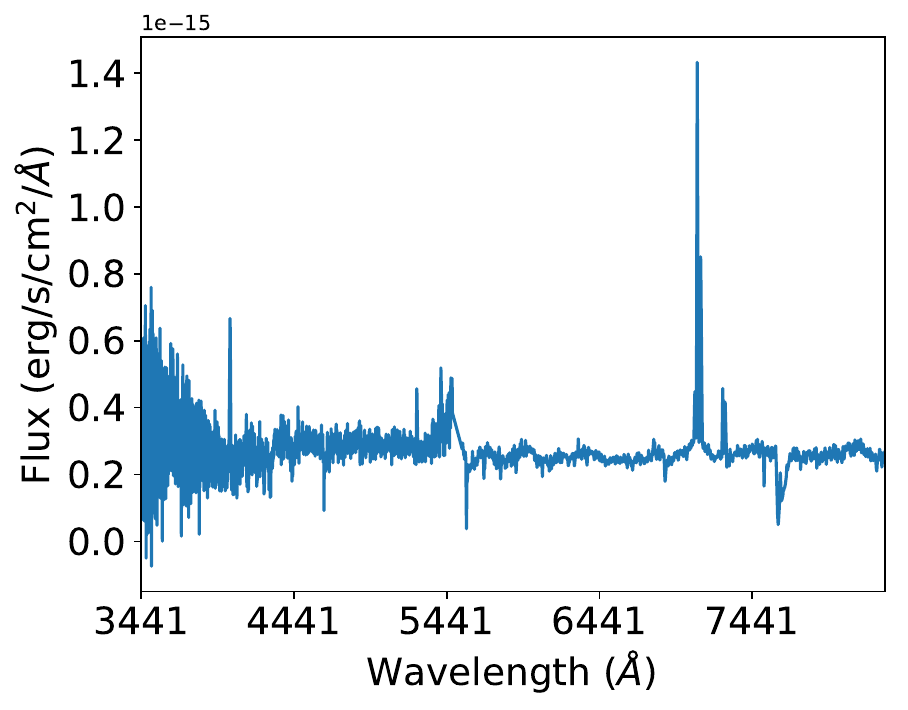}
\figsetgrpnote{Flux for the Lick, Palomar, and Keck spectra. See Table 3 for the observed spectral features and line measurements.}
\figsetgrpend

\figsetgrpstart
\figsetgrpnum{2.86}
\figsetgrptitle{Flux of 1858+4850}
\figsetplot{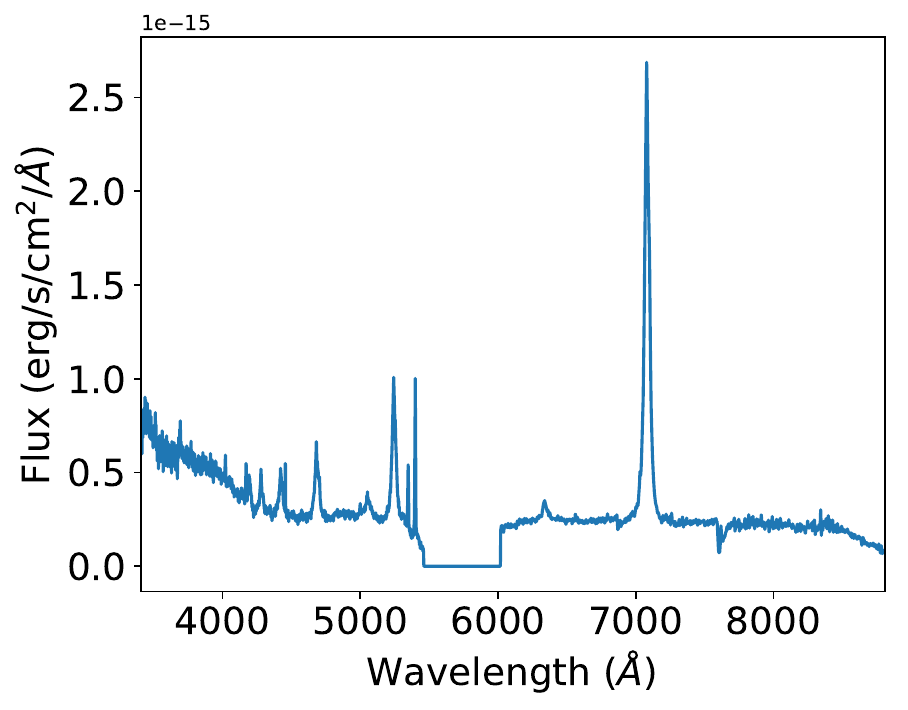}
\figsetgrpnote{Flux for the Lick, Palomar, and Keck spectra. See Table 3 for the observed spectral features and line measurements.}
\figsetgrpend

\figsetgrpstart
\figsetgrpnum{2.87}
\figsetgrptitle{Flux of 1858+4914}
\figsetplot{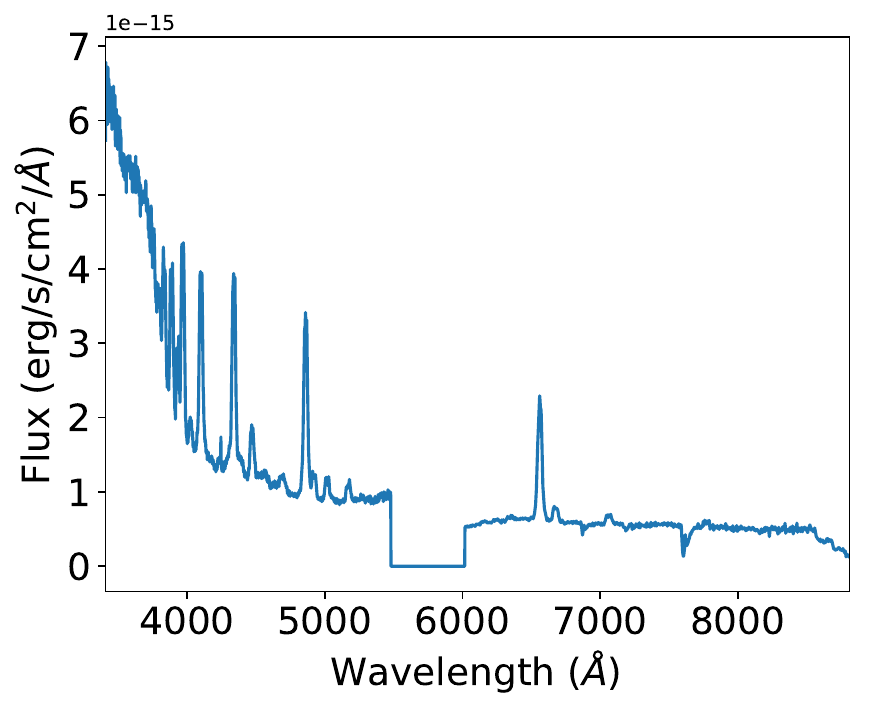}
\figsetgrpnote{Flux for the Lick, Palomar, and Keck spectra. See Table 3 for the observed spectral features and line measurements.}
\figsetgrpend

\figsetgrpstart
\figsetgrpnum{2.88}
\figsetgrptitle{Flux of 1904+3849}
\figsetplot{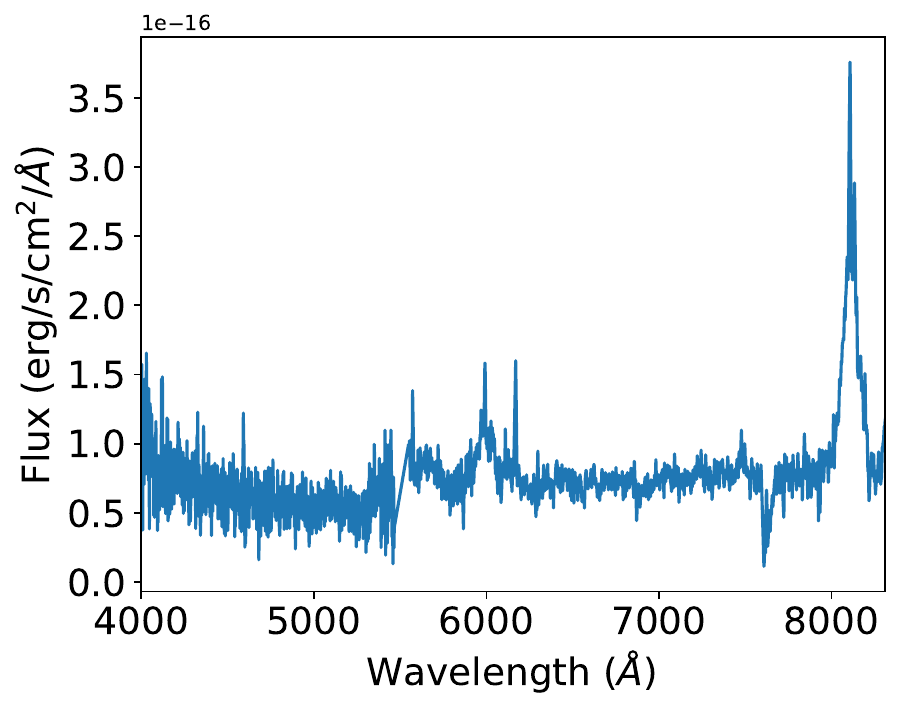}
\figsetgrpnote{Flux for the Lick, Palomar, and Keck spectra. See Table 3 for the observed spectral features and line measurements.}
\figsetgrpend

\figsetgrpstart
\figsetgrpnum{2.89}
\figsetgrptitle{Flux of 1904+3755}
\figsetplot{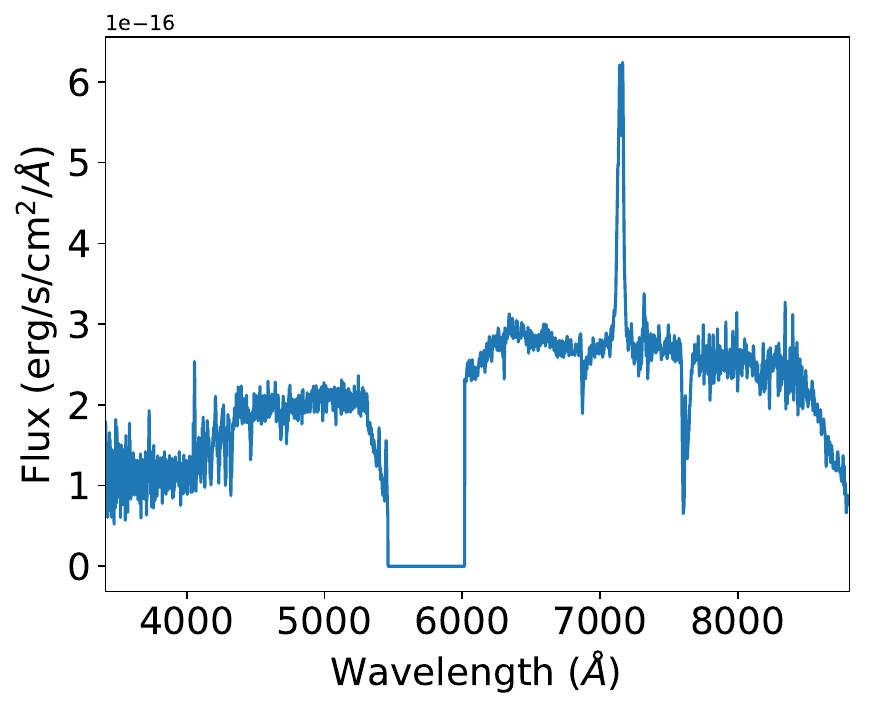}
\figsetgrpnote{Flux for the Lick, Palomar, and Keck spectra. See Table 3 for the observed spectral features and line measurements.}
\figsetgrpend

\figsetgrpstart
\figsetgrpnum{2.90}
\figsetgrptitle{Flux of 1904+4701}
\figsetplot{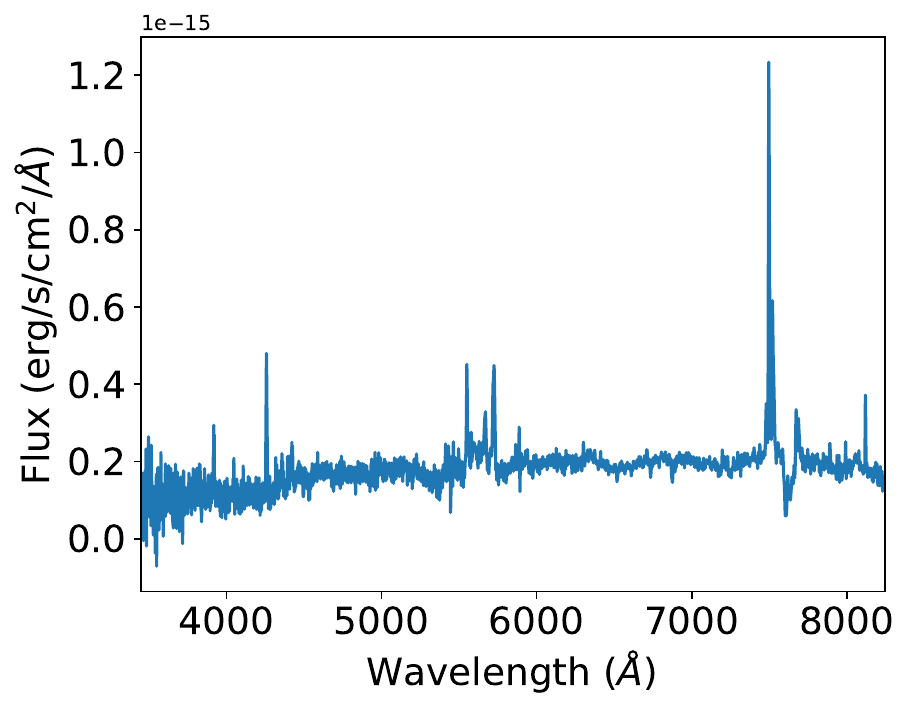}
\figsetgrpnote{Flux for the Lick, Palomar, and Keck spectra. See Table 3 for the observed spectral features and line measurements.}
\figsetgrpend

\figsetgrpstart
\figsetgrpnum{2.91}
\figsetgrptitle{Flux of 1908+3919}
\figsetplot{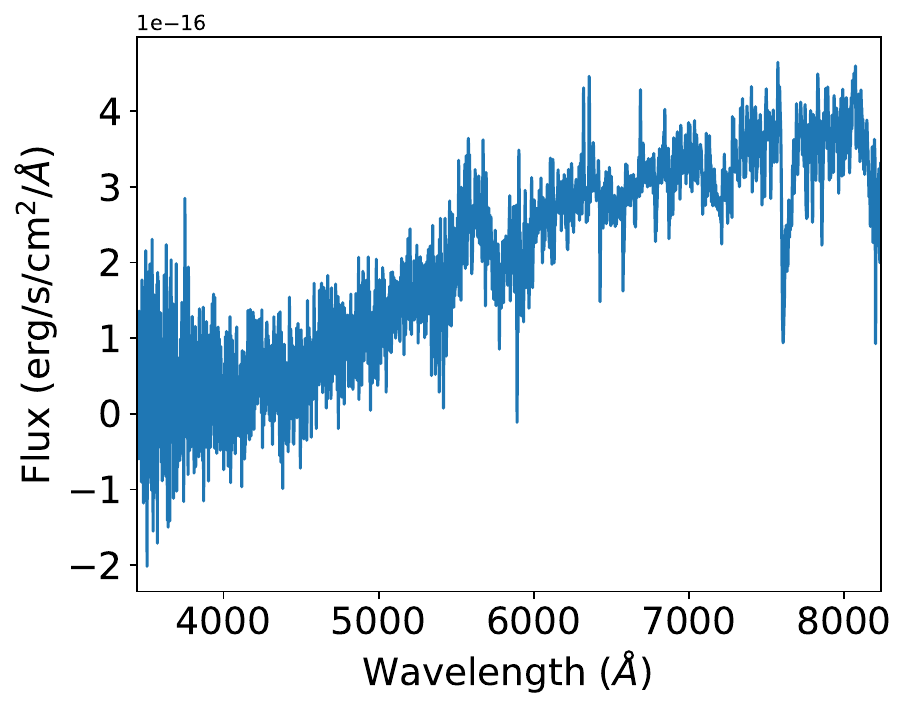}
\figsetgrpnote{Flux for the Lick, Palomar, and Keck spectra. See Table 3 for the observed spectral features and line measurements.}
\figsetgrpend

\figsetgrpstart
\figsetgrpnum{2.92}
\figsetgrptitle{Flux of 1909+4834}
\figsetplot{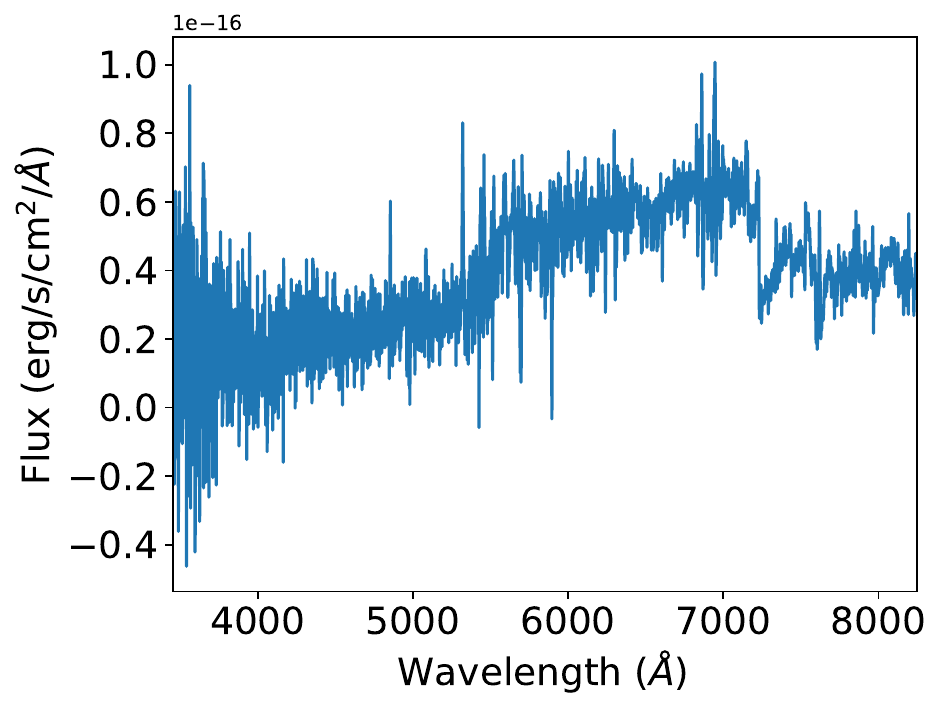}
\figsetgrpnote{Flux for the Lick, Palomar, and Keck spectra. See Table 3 for the observed spectral features and line measurements.}
\figsetgrpend

\figsetgrpstart
\figsetgrpnum{2.93}
\figsetgrptitle{Flux of 1910+3800}
\figsetplot{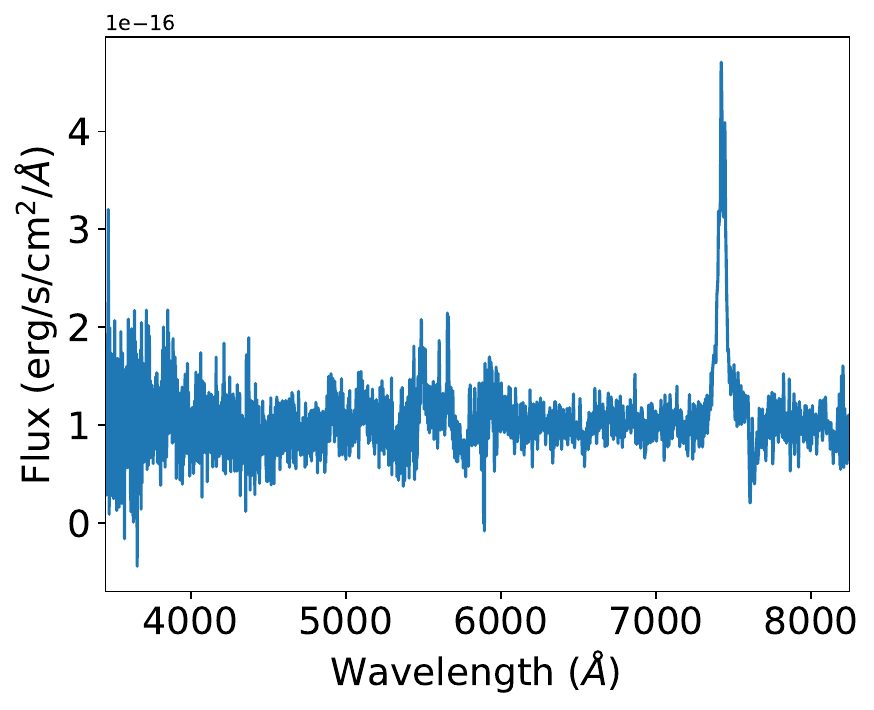}
\figsetgrpnote{Flux for the Lick, Palomar, and Keck spectra. See Table 3 for the observed spectral features and line measurements.}
\figsetgrpend

\figsetgrpstart
\figsetgrpnum{2.94}
\figsetgrptitle{Flux of 1911+4534}
\figsetplot{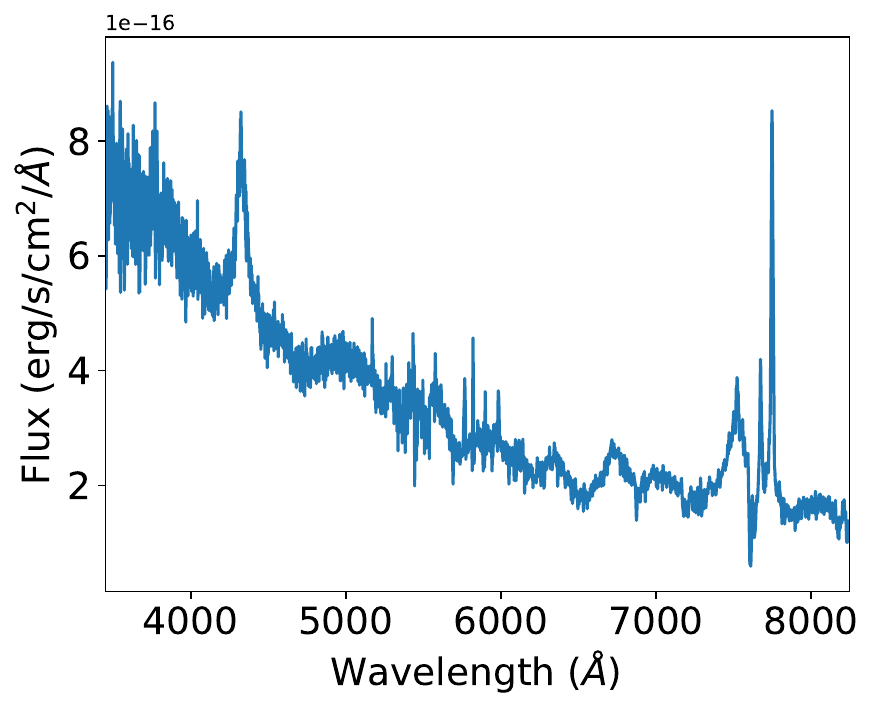}
\figsetgrpnote{Flux for the Lick, Palomar, and Keck spectra. See Table 3 for the observed spectral features and line measurements.}
\figsetgrpend

\figsetgrpstart
\figsetgrpnum{2.95}
\figsetgrptitle{Flux of 1911+5036}
\figsetplot{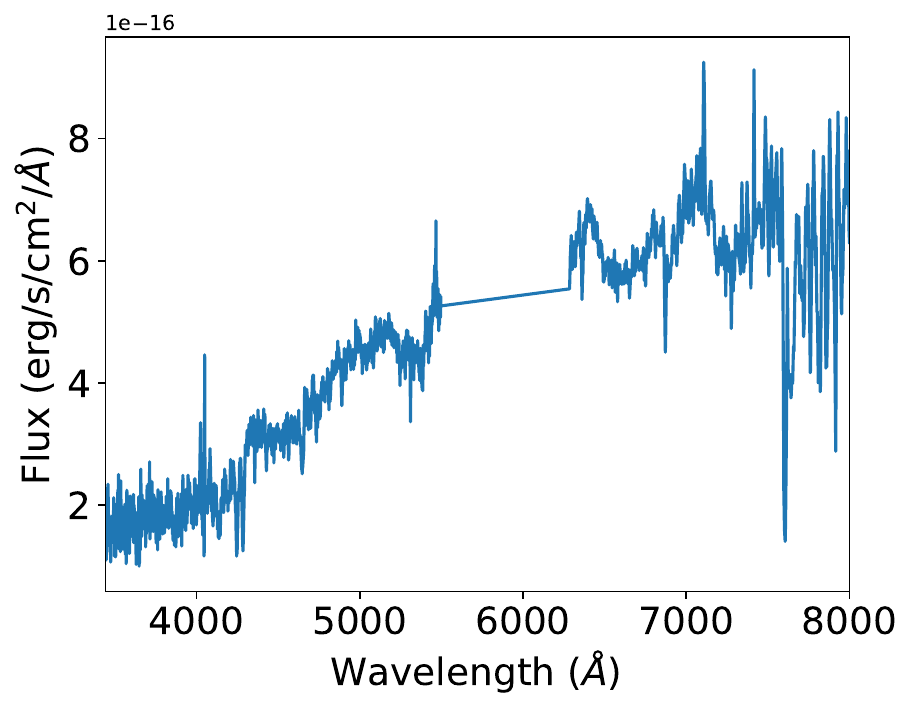}
\figsetgrpnote{Flux for the Lick, Palomar, and Keck spectra. See Table 3 for the observed spectral features and line measurements.}
\figsetgrpend

\figsetgrpstart
\figsetgrpnum{2.96}
\figsetgrptitle{Flux of 1911+5117}
\figsetplot{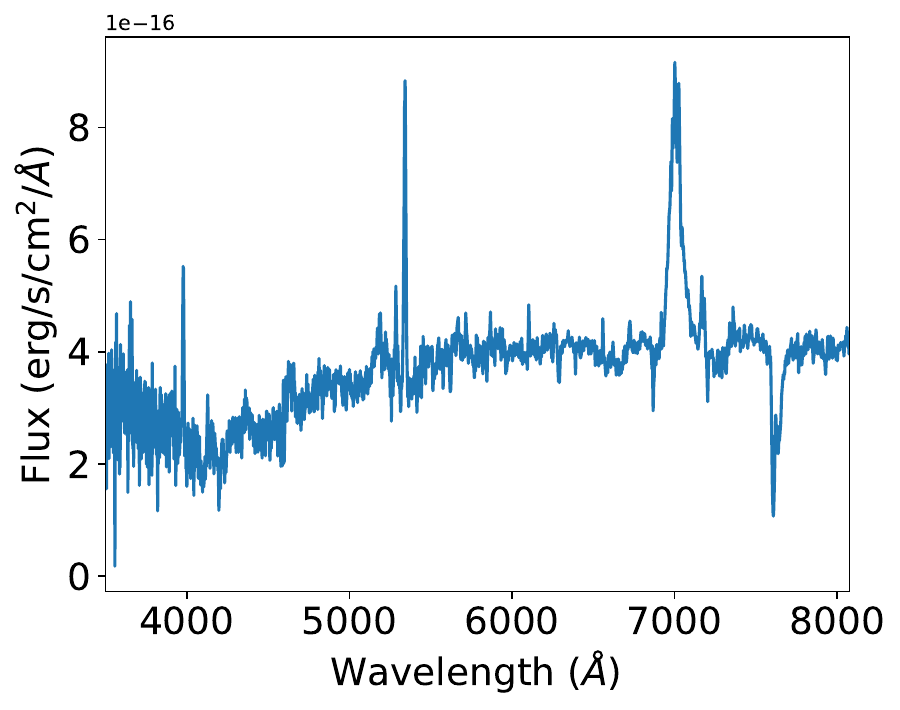}
\figsetgrpnote{Flux for the Lick, Palomar, and Keck spectra. See Table 3 for the observed spectral features and line measurements.}
\figsetgrpend

\figsetgrpstart
\figsetgrpnum{2.97}
\figsetgrptitle{Flux of 1913+3835}
\figsetplot{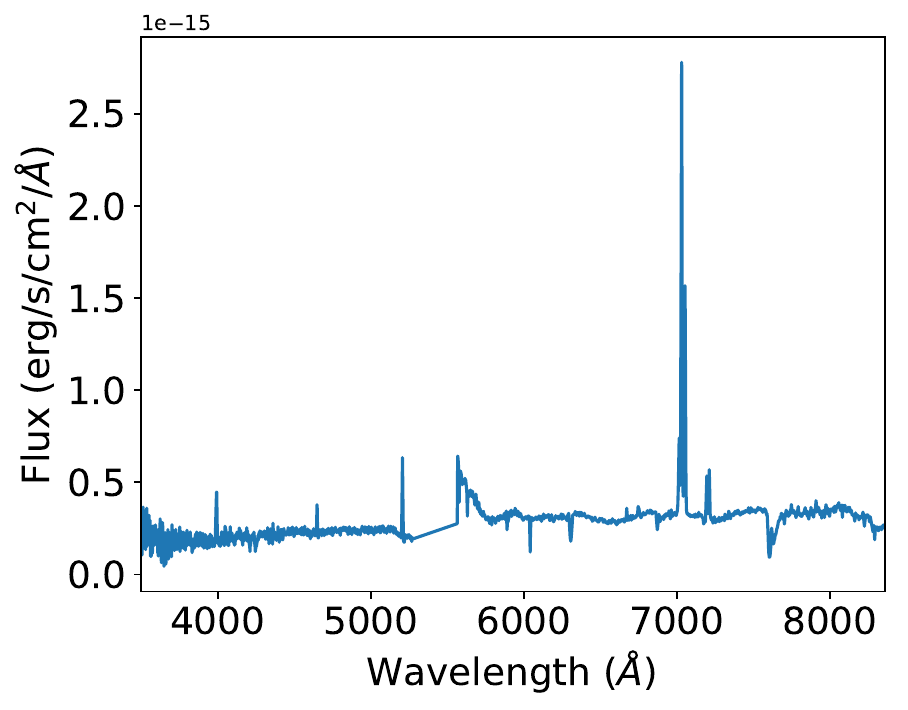}
\figsetgrpnote{Flux for the Lick, Palomar, and Keck spectra. See Table 3 for the observed spectral features and line measurements.}
\figsetgrpend

\figsetgrpstart
\figsetgrpnum{2.98}
\figsetgrptitle{Flux of 1913+4201}
\figsetplot{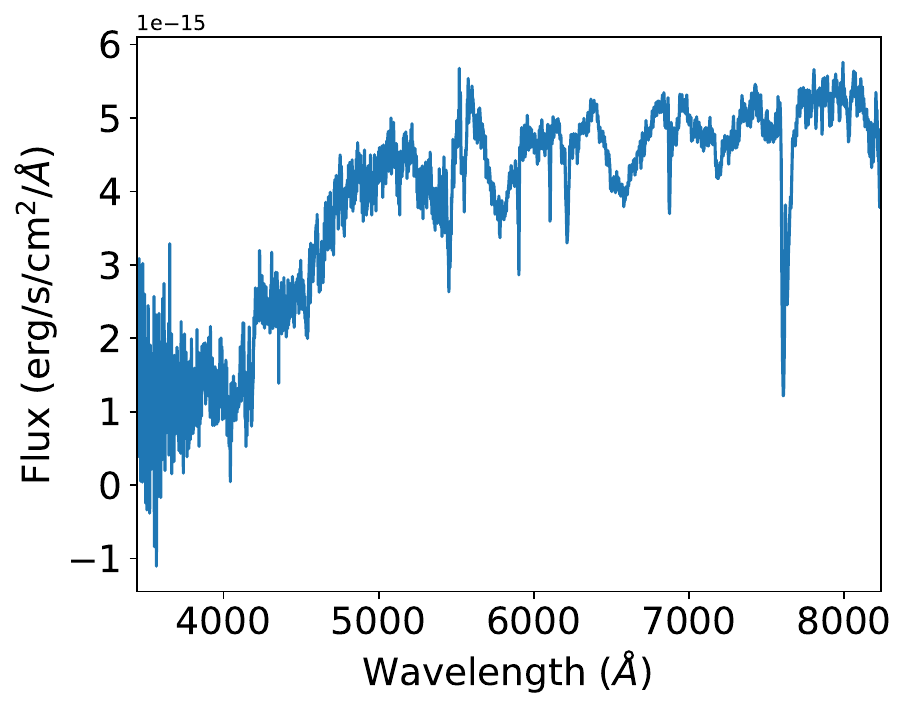}
\figsetgrpnote{Flux for the Lick, Palomar, and Keck spectra. See Table 3 for the observed spectral features and line measurements.}
\figsetgrpend

\figsetgrpstart
\figsetgrpnum{2.99}
\figsetgrptitle{Flux of 1914+4438}
\figsetplot{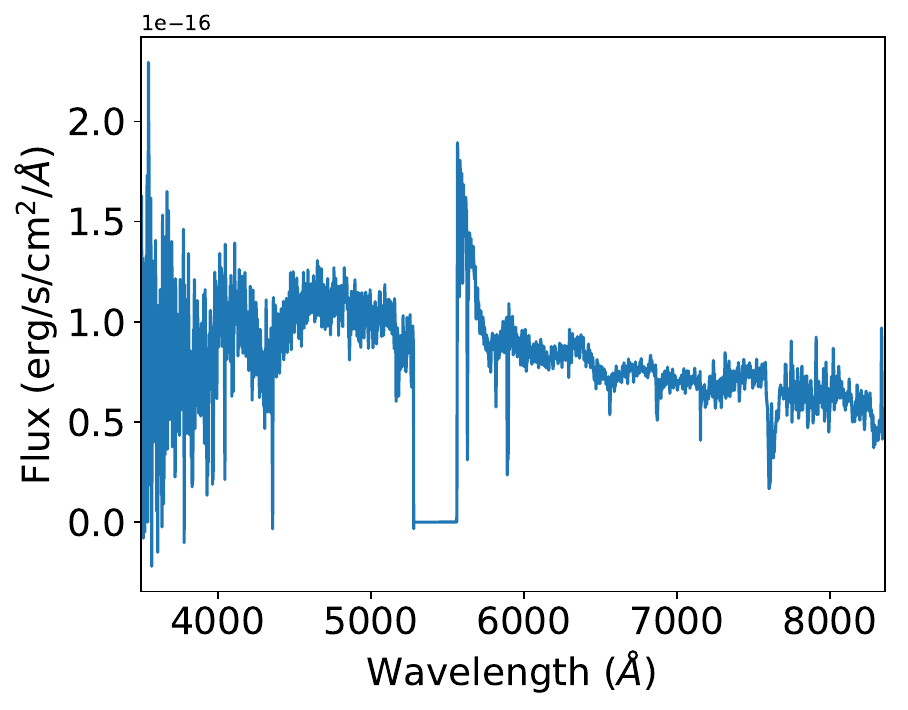}
\figsetgrpnote{Flux for the Lick, Palomar, and Keck spectra. See Table 3 for the observed spectral features and line measurements.}
\figsetgrpend

\figsetgrpstart
\figsetgrpnum{2.100}
\figsetgrptitle{Flux of 1914+4204}
\figsetplot{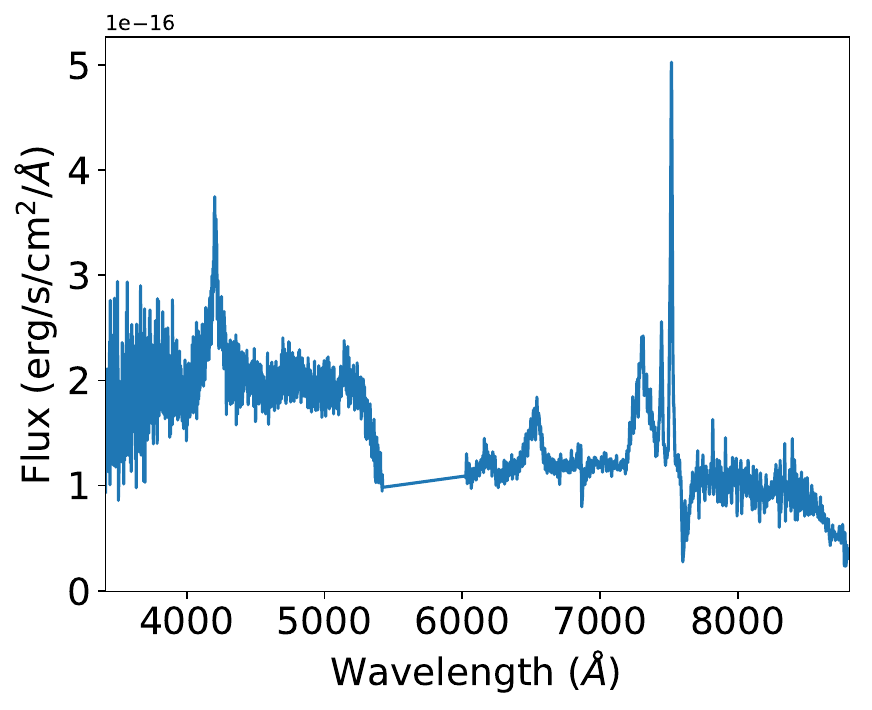}
\figsetgrpnote{Flux for the Lick, Palomar, and Keck spectra. See Table 3 for the observed spectral features and line measurements.}
\figsetgrpend

\figsetgrpstart
\figsetgrpnum{2.101}
\figsetgrptitle{Flux of 1915+4102}
\figsetplot{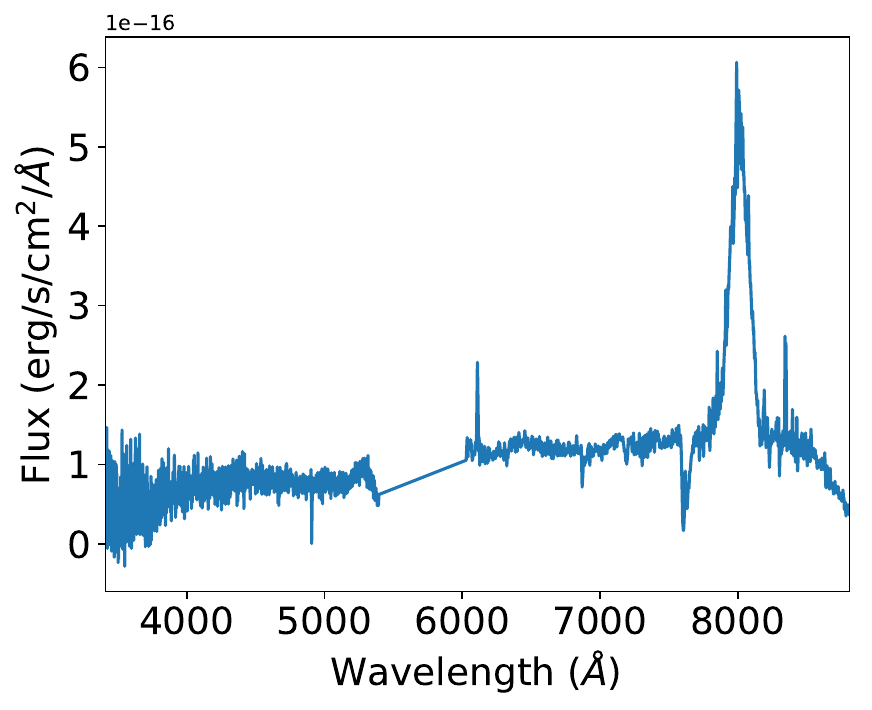}
\figsetgrpnote{Flux for the Lick, Palomar, and Keck spectra. See Table 3 for the observed spectral features and line measurements.}
\figsetgrpend

\figsetgrpstart
\figsetgrpnum{2.102}
\figsetgrptitle{Flux of 1916+3925}
\figsetplot{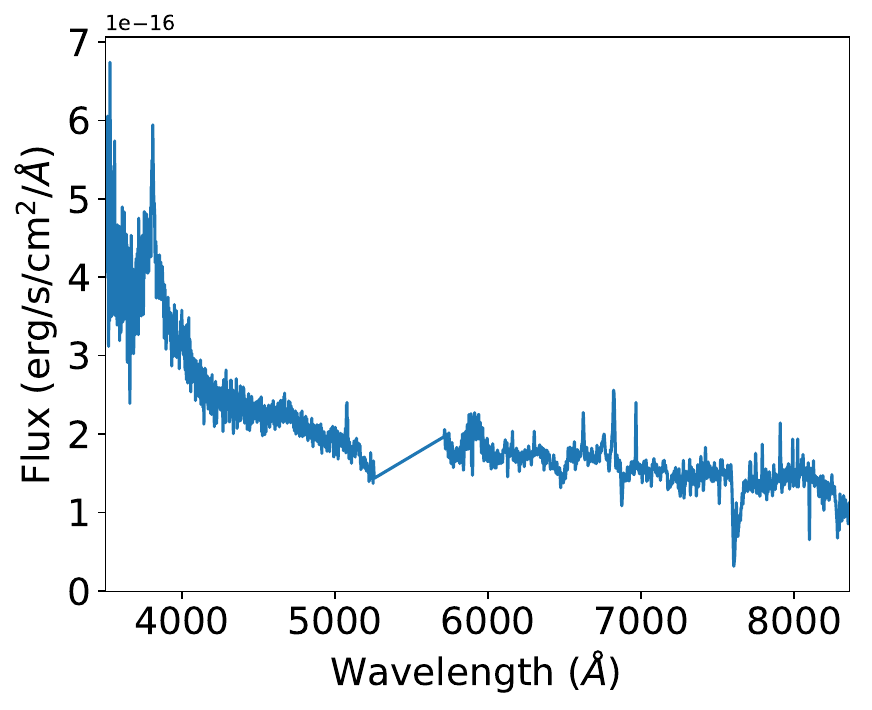}
\figsetgrpnote{Flux for the Lick, Palomar, and Keck spectra. See Table 3 for the observed spectral features and line measurements.}
\figsetgrpend

\figsetgrpstart
\figsetgrpnum{2.103}
\figsetgrptitle{Flux of 1916+4721}
\figsetplot{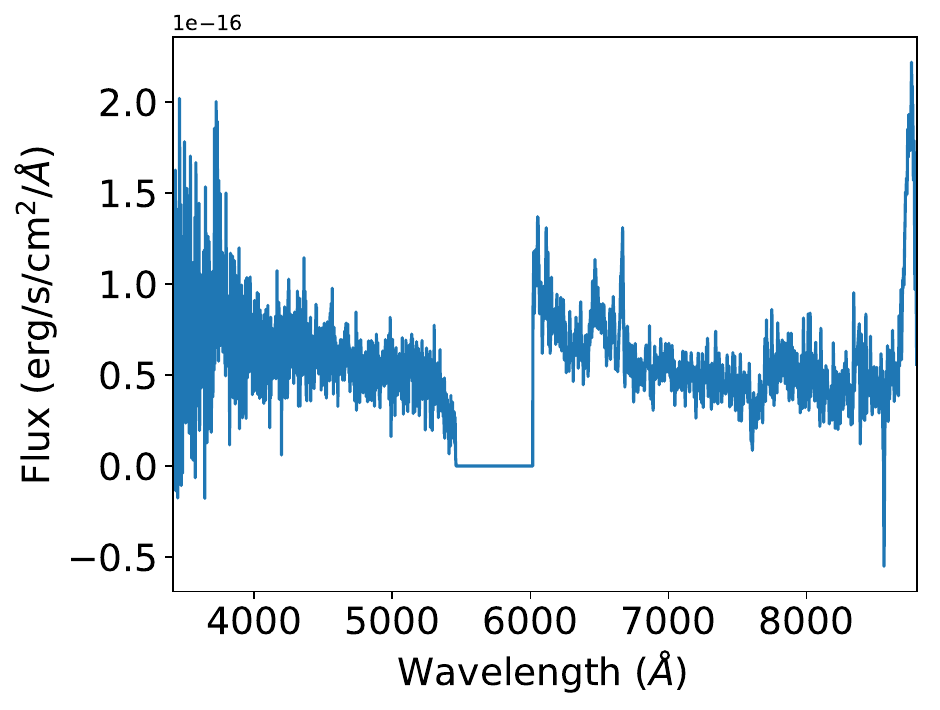}
\figsetgrpnote{Flux for the Lick, Palomar, and Keck spectra. See Table 3 for the observed spectral features and line measurements.}
\figsetgrpend

\figsetgrpstart
\figsetgrpnum{2.104}
\figsetgrptitle{Flux of 1916+5009}
\figsetplot{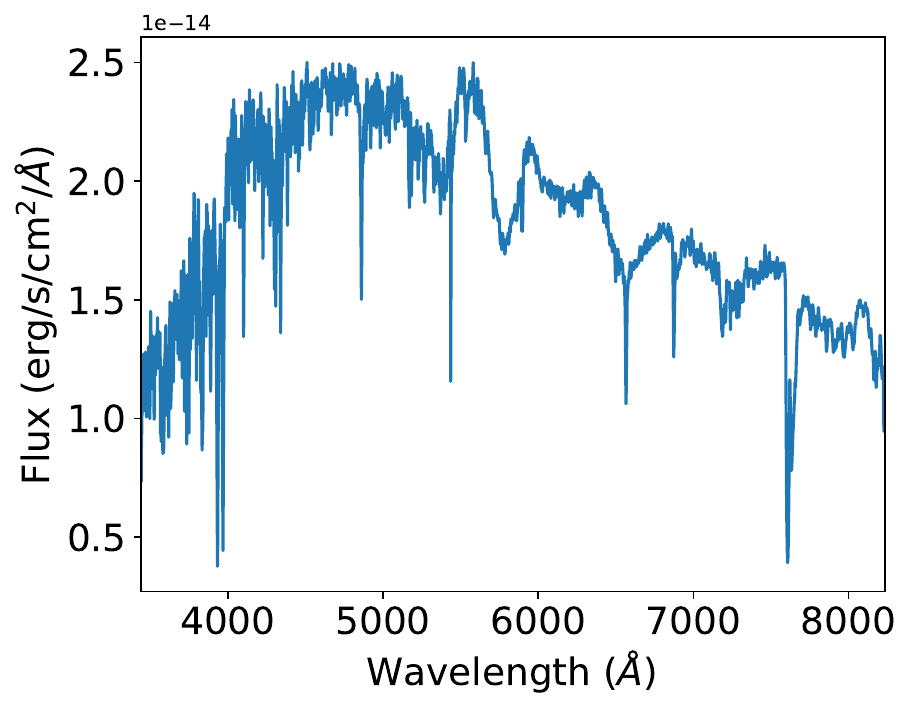}
\figsetgrpnote{Flux for the Lick, Palomar, and Keck spectra. See Table 3 for the observed spectral features and line measurements.}
\figsetgrpend

\figsetgrpstart
\figsetgrpnum{2.105}
\figsetgrptitle{Flux of 1917+4513}
\figsetplot{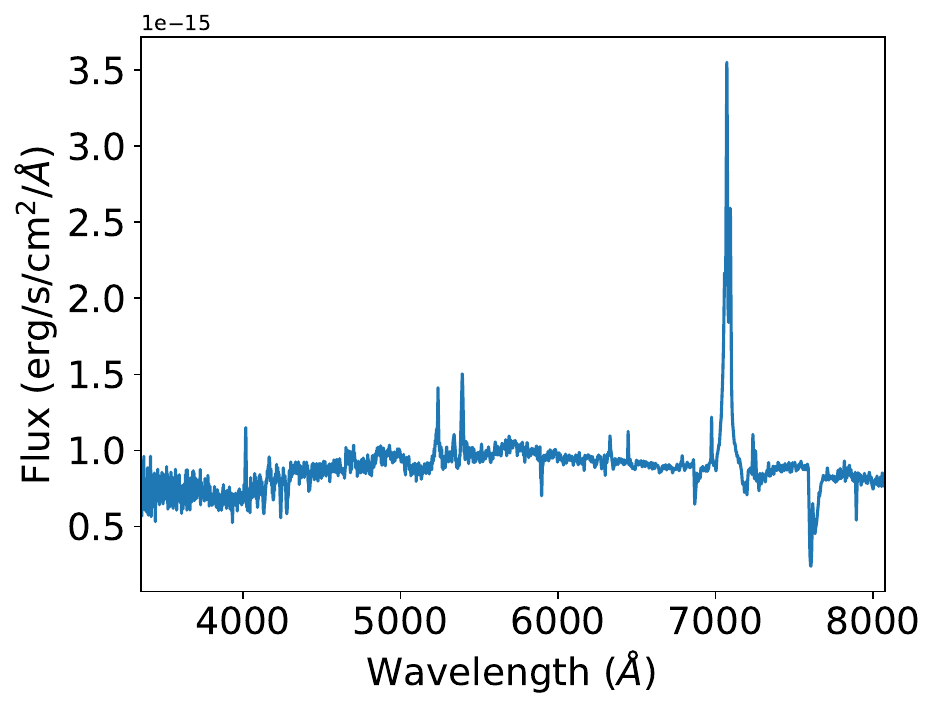}
\figsetgrpnote{Flux for the Lick, Palomar, and Keck spectra. See Table 3 for the observed spectral features and line measurements.}
\figsetgrpend

\figsetgrpstart
\figsetgrpnum{2.106}
\figsetgrptitle{Flux of 1918+4053}
\figsetplot{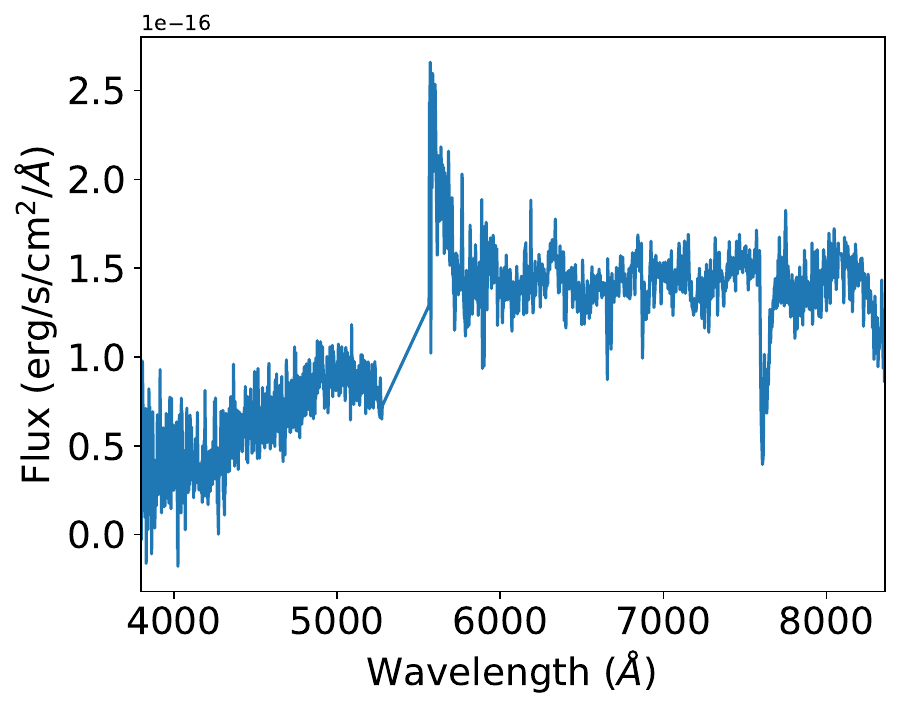}
\figsetgrpnote{Flux for the Lick, Palomar, and Keck spectra. See Table 3 for the observed spectral features and line measurements.}
\figsetgrpend

\figsetgrpstart
\figsetgrpnum{2.107}
\figsetgrptitle{Flux of 1918+4937}
\figsetplot{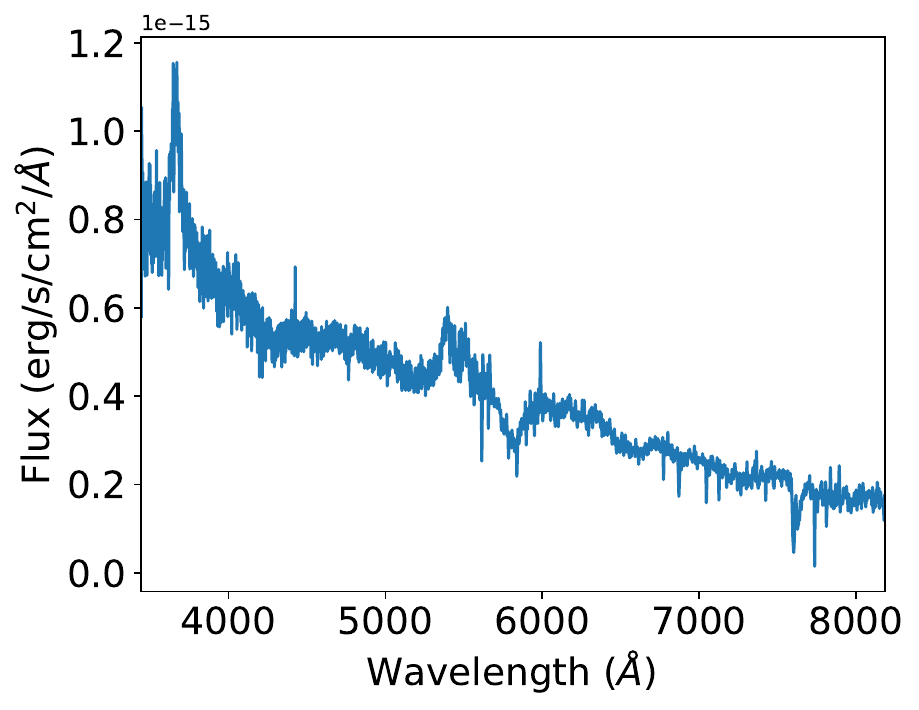}
\figsetgrpnote{Flux for the Lick, Palomar, and Keck spectra. See Table 3 for the observed spectral features and line measurements.}
\figsetgrpend

\figsetgrpstart
\figsetgrpnum{2.108}
\figsetgrptitle{Flux of 1918+5135}
\figsetplot{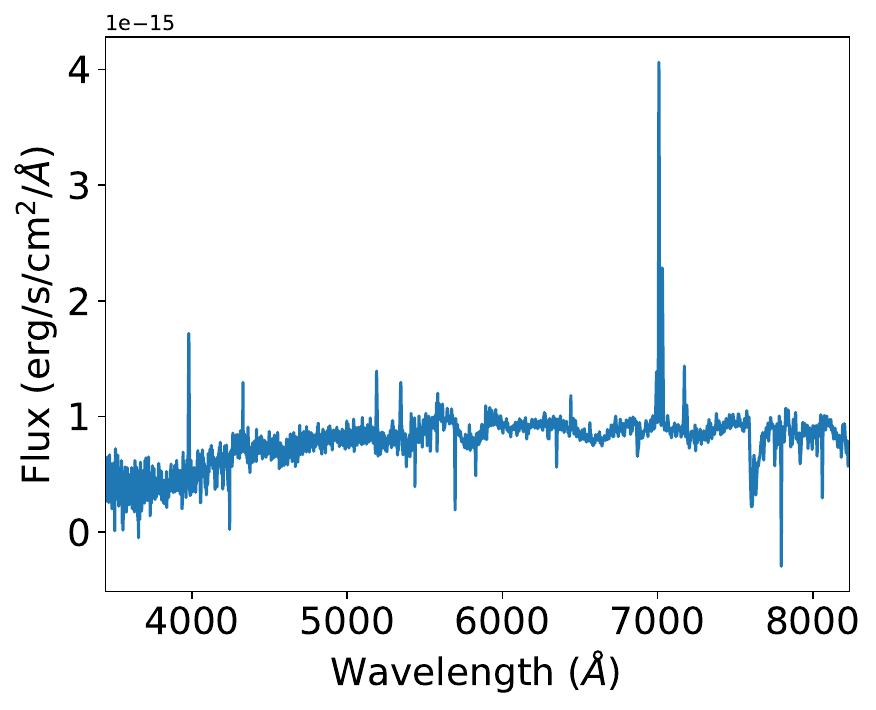}
\figsetgrpnote{Flux for the Lick, Palomar, and Keck spectra. See Table 3 for the observed spectral features and line measurements.}
\figsetgrpend

\figsetgrpstart
\figsetgrpnum{2.109}
\figsetgrptitle{Flux of 1919+4058}
\figsetplot{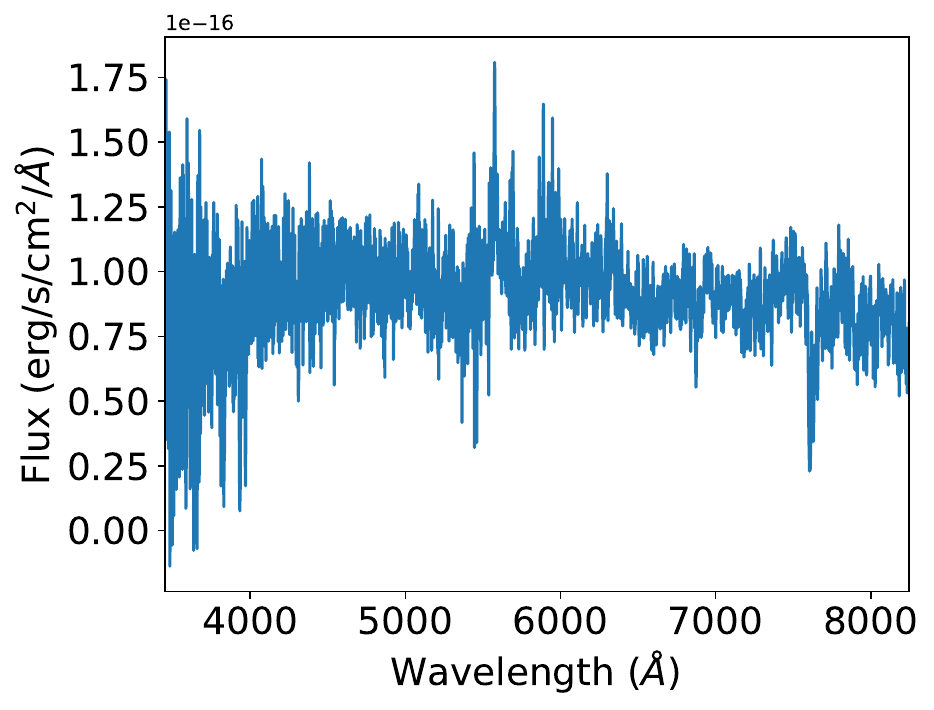}
\figsetgrpnote{Flux for the Lick, Palomar, and Keck spectra. See Table 3 for the observed spectral features and line measurements.}
\figsetgrpend

\figsetgrpstart
\figsetgrpnum{2.110}
\figsetgrptitle{Flux of 1919+4409}
\figsetplot{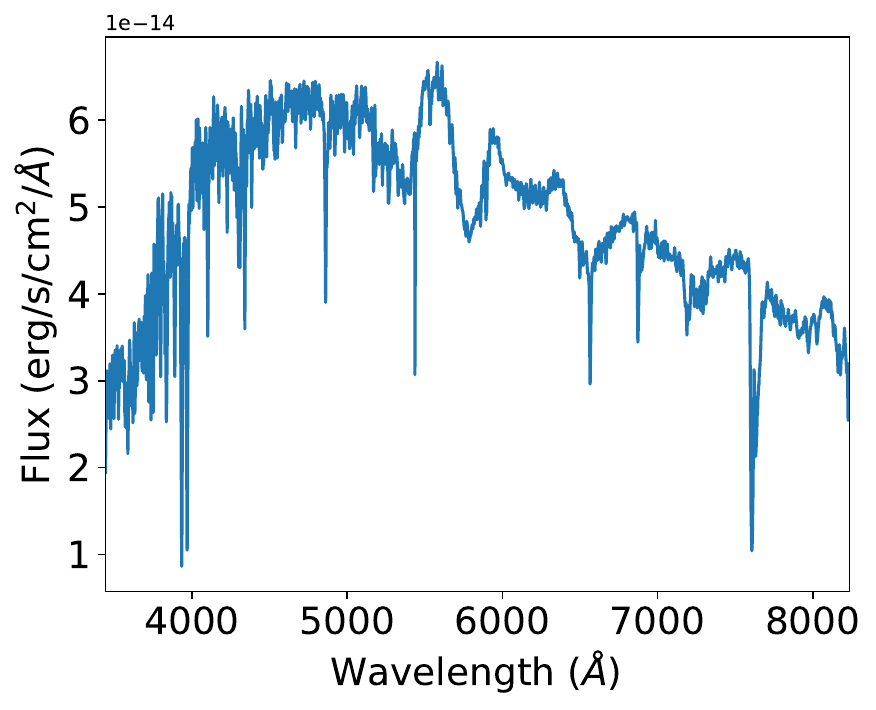}
\figsetgrpnote{Flux for the Lick, Palomar, and Keck spectra. See Table 3 for the observed spectral features and line measurements.}
\figsetgrpend

\figsetgrpstart
\figsetgrpnum{2.111}
\figsetgrptitle{Flux of 1919+5026}
\figsetplot{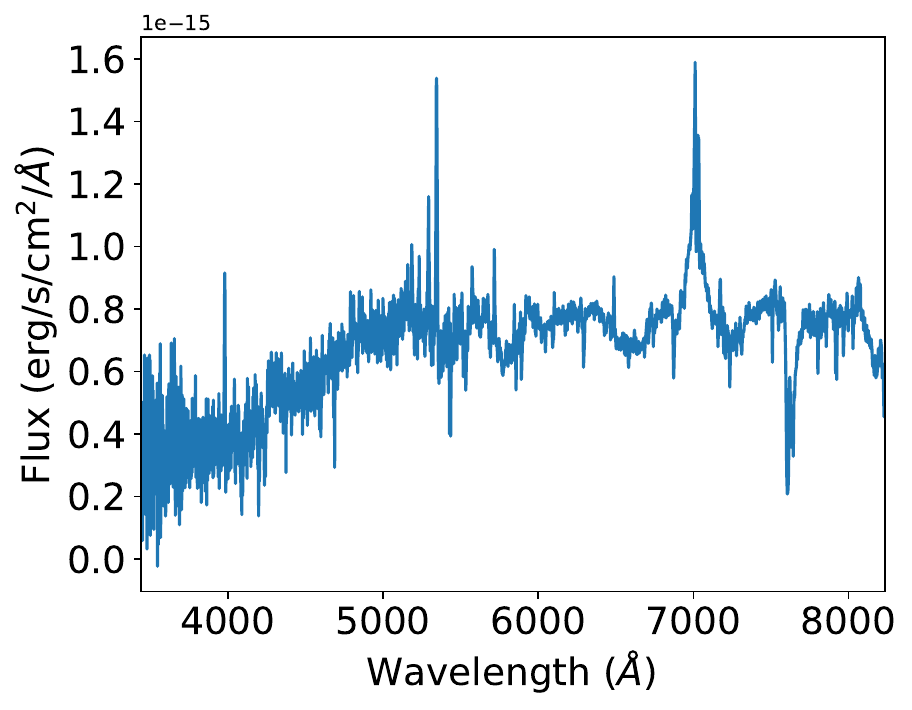}
\figsetgrpnote{Flux for the Lick, Palomar, and Keck spectra. See Table 3 for the observed spectral features and line measurements.}
\figsetgrpend

\figsetgrpstart
\figsetgrpnum{2.112}
\figsetgrptitle{Flux of 1920+3826}
\figsetplot{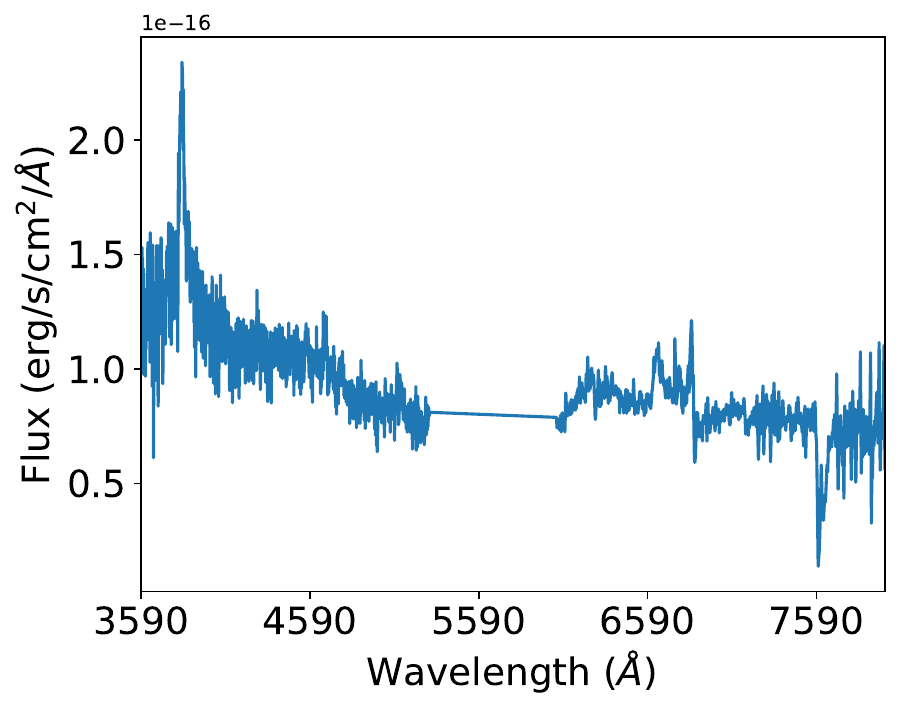}
\figsetgrpnote{Flux for the Lick, Palomar, and Keck spectra. See Table 3 for the observed spectral features and line measurements.}
\figsetgrpend

\figsetgrpstart
\figsetgrpnum{2.113}
\figsetgrptitle{Flux of 1920+4703}
\figsetplot{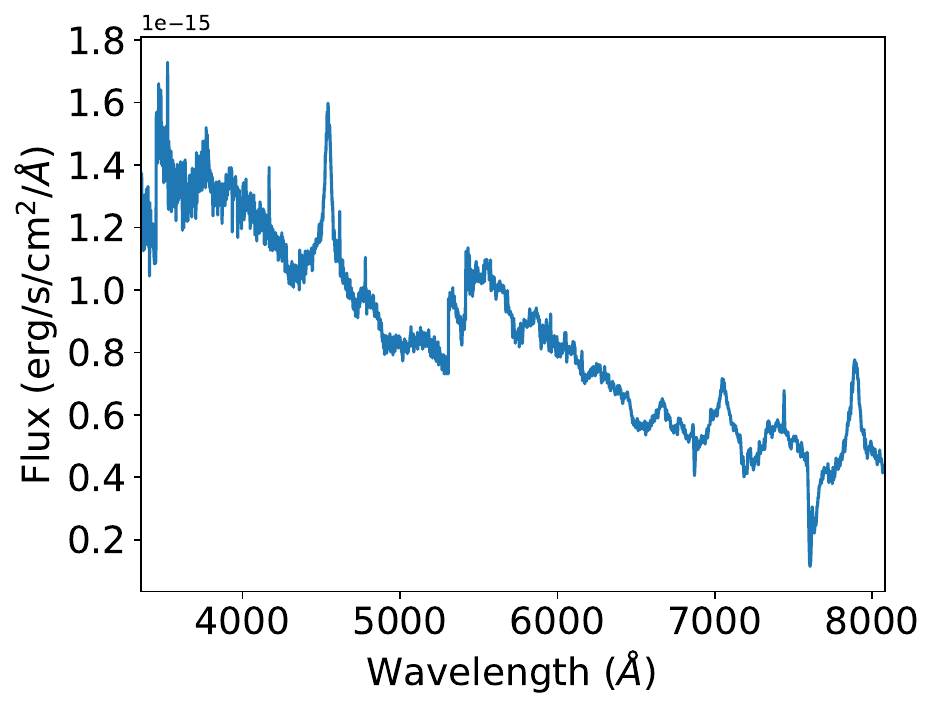}
\figsetgrpnote{Flux for the Lick, Palomar, and Keck spectra. See Table 3 for the observed spectral features and line measurements.}
\figsetgrpend

\figsetgrpstart
\figsetgrpnum{2.114}
\figsetgrptitle{Flux of 1922+4311}
\figsetplot{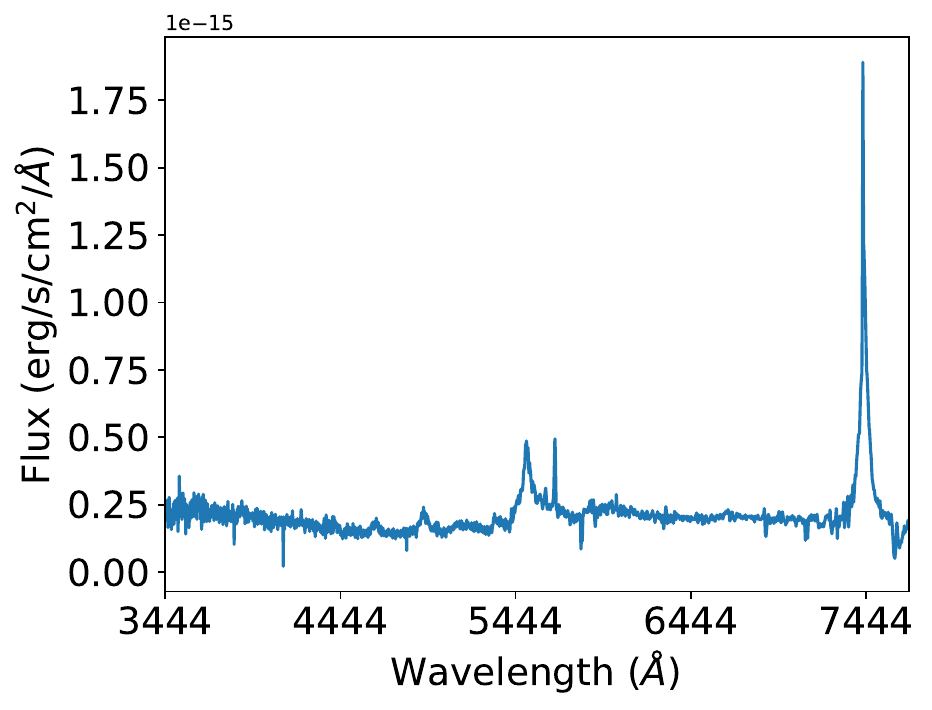}
\figsetgrpnote{Flux for the Lick, Palomar, and Keck spectra. See Table 3 for the observed spectral features and line measurements.}
\figsetgrpend

\figsetgrpstart
\figsetgrpnum{2.115}
\figsetgrptitle{Flux of 1922+4538}
\figsetplot{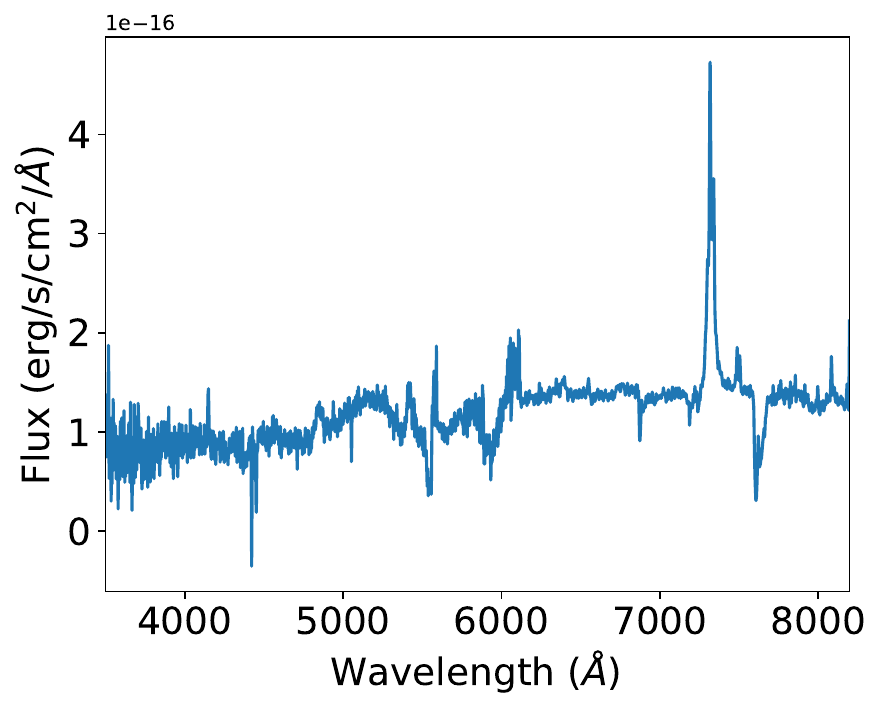}
\figsetgrpnote{Flux for the Lick, Palomar, and Keck spectra. See Table 3 for the observed spectral features and line measurements.}
\figsetgrpend

\figsetgrpstart
\figsetgrpnum{2.116}
\figsetgrptitle{Flux of 1923+4341}
\figsetplot{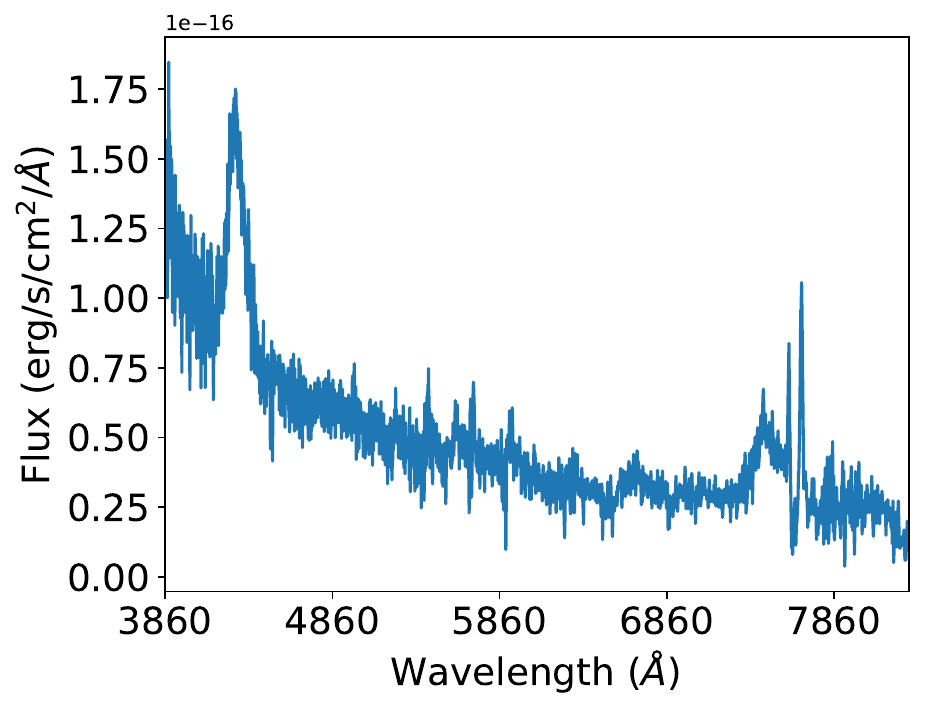}
\figsetgrpnote{Flux for the Lick, Palomar, and Keck spectra. See Table 3 for the observed spectral features and line measurements.}
\figsetgrpend

\figsetgrpstart
\figsetgrpnum{2.117}
\figsetgrptitle{Flux of 1923+4816}
\figsetplot{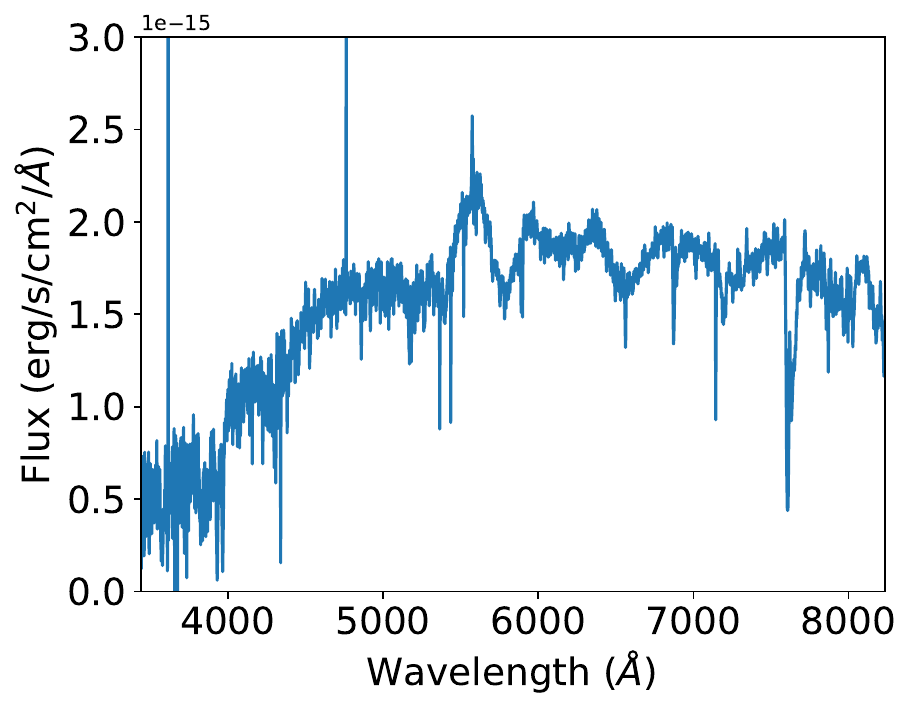}
\figsetgrpnote{Flux for the Lick, Palomar, and Keck spectra. See Table 3 for the observed spectral features and line measurements.}
\figsetgrpend

\figsetgrpstart
\figsetgrpnum{2.118}
\figsetgrptitle{Flux of 1923+4548}
\figsetplot{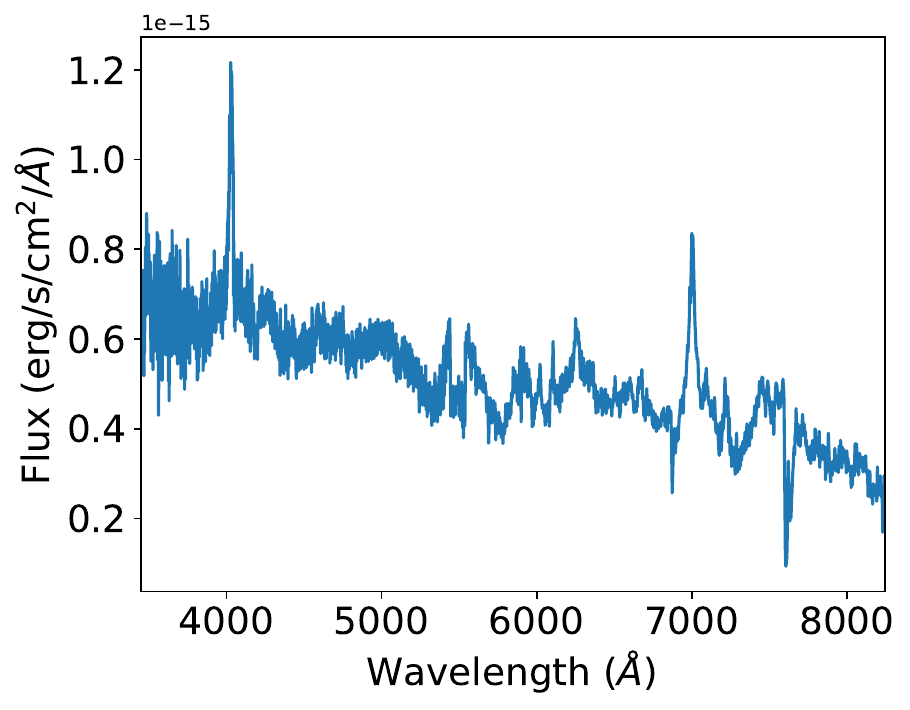}
\figsetgrpnote{Flux for the Lick, Palomar, and Keck spectra. See Table 3 for the observed spectral features and line measurements.}
\figsetgrpend

\figsetgrpstart
\figsetgrpnum{2.119}
\figsetgrptitle{Flux of 1923+4754}
\figsetplot{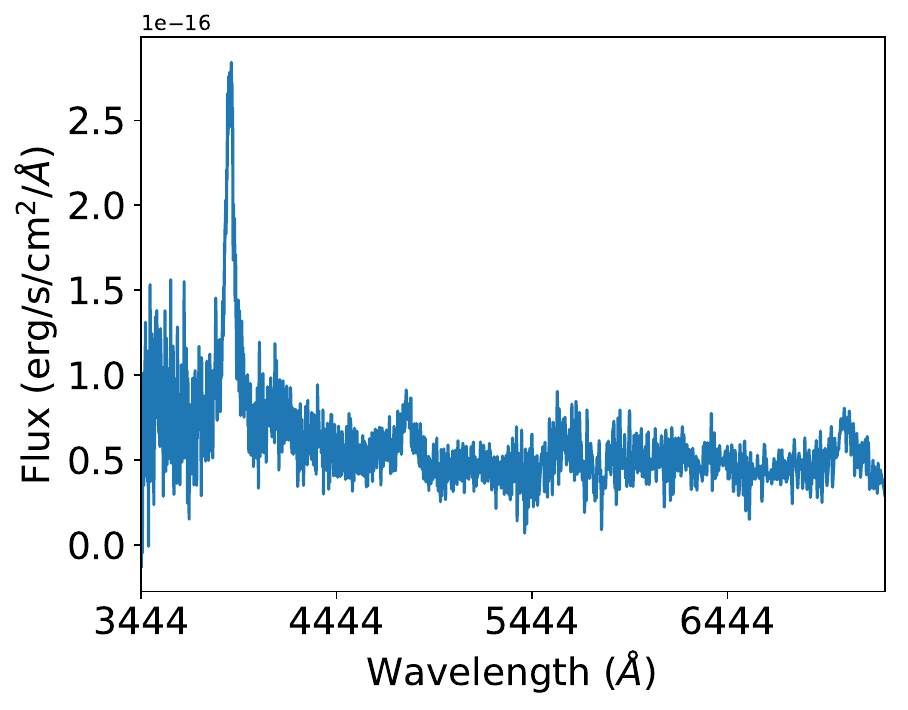}
\figsetgrpnote{Flux for the Lick, Palomar, and Keck spectra. See Table 3 for the observed spectral features and line measurements.}
\figsetgrpend

\figsetgrpstart
\figsetgrpnum{2.120}
\figsetgrptitle{Flux of 1924+4726}
\figsetplot{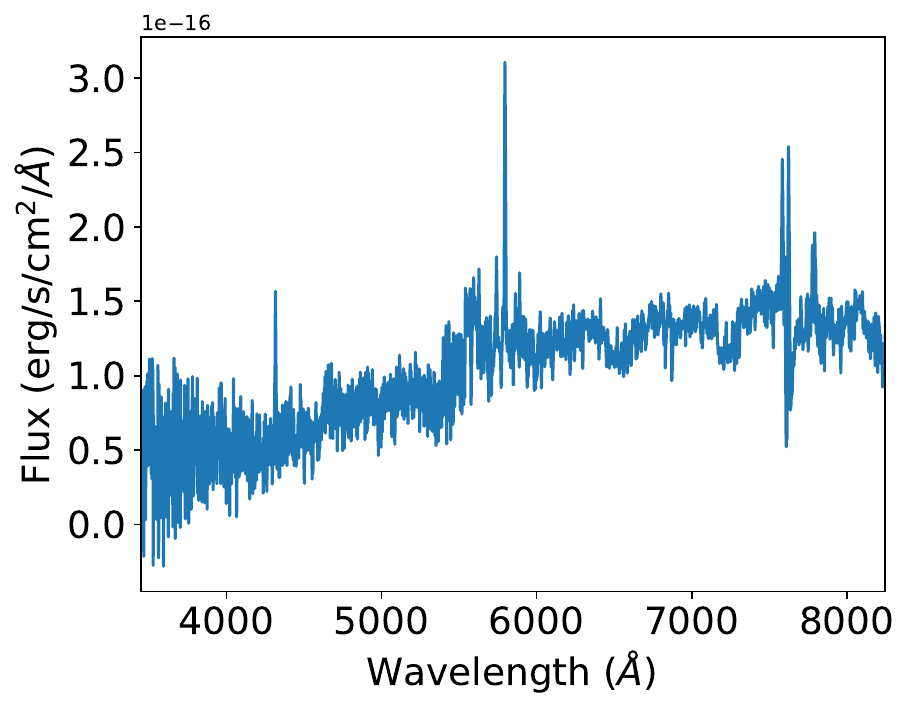}
\figsetgrpnote{Flux for the Lick, Palomar, and Keck spectra. See Table 3 for the observed spectral features and line measurements.}
\figsetgrpend

\figsetgrpstart
\figsetgrpnum{2.121}
\figsetgrptitle{Flux of 1925+5043}
\figsetplot{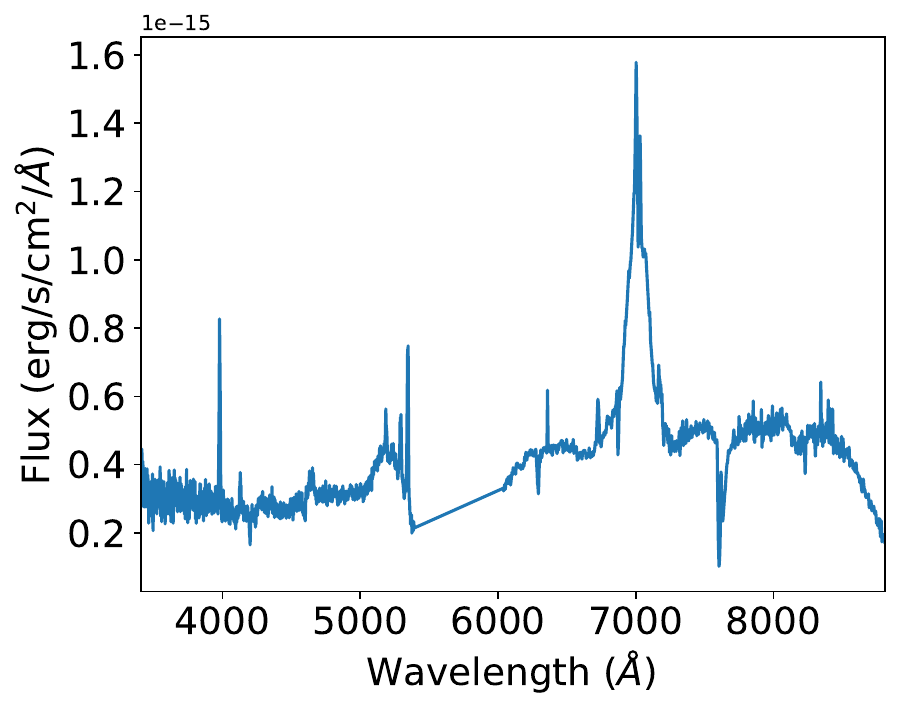}
\figsetgrpnote{Flux for the Lick, Palomar, and Keck spectra. See Table 3 for the observed spectral features and line measurements.}
\figsetgrpend

\figsetgrpstart
\figsetgrpnum{2.122}
\figsetgrptitle{Flux of 1926+4209}
\figsetplot{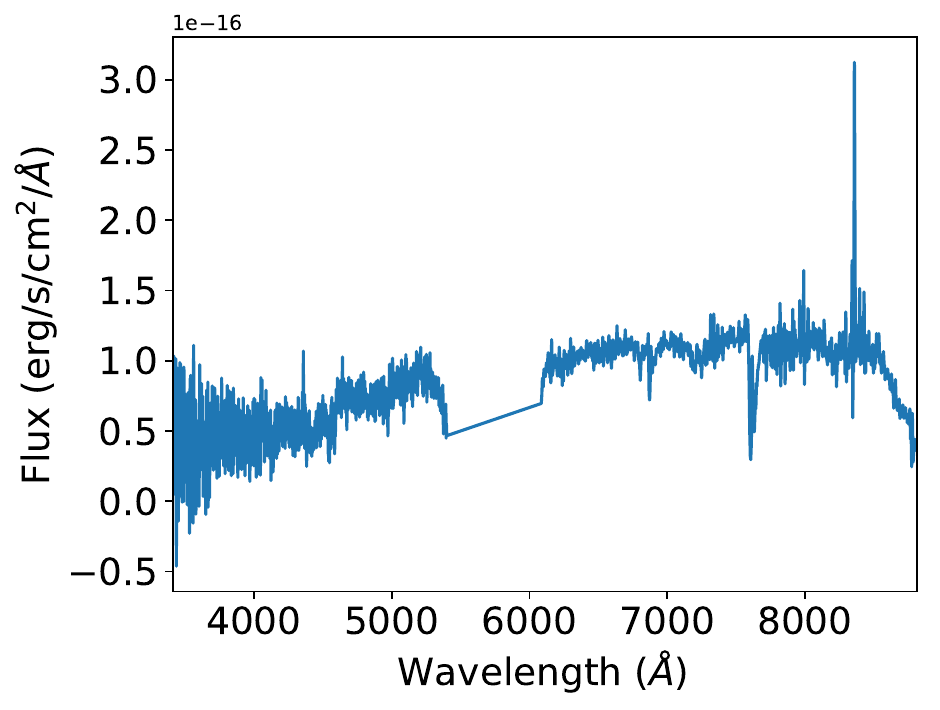}
\figsetgrpnote{Flux for the Lick, Palomar, and Keck spectra. See Table 3 for the observed spectral features and line measurements.}
\figsetgrpend

\figsetgrpstart
\figsetgrpnum{2.123}
\figsetgrptitle{Flux of 1926+5052}
\figsetplot{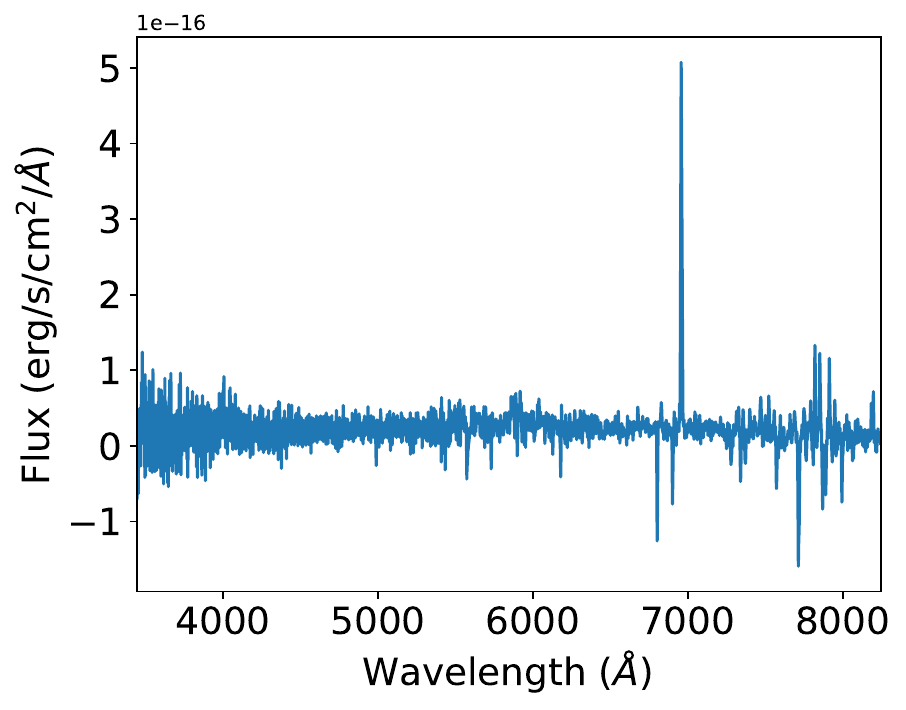}
\figsetgrpnote{Flux for the Lick, Palomar, and Keck spectra. See Table 3 for the observed spectral features and line measurements.}
\figsetgrpend

\figsetgrpstart
\figsetgrpnum{2.124}
\figsetgrptitle{Flux of 1926+4832}
\figsetplot{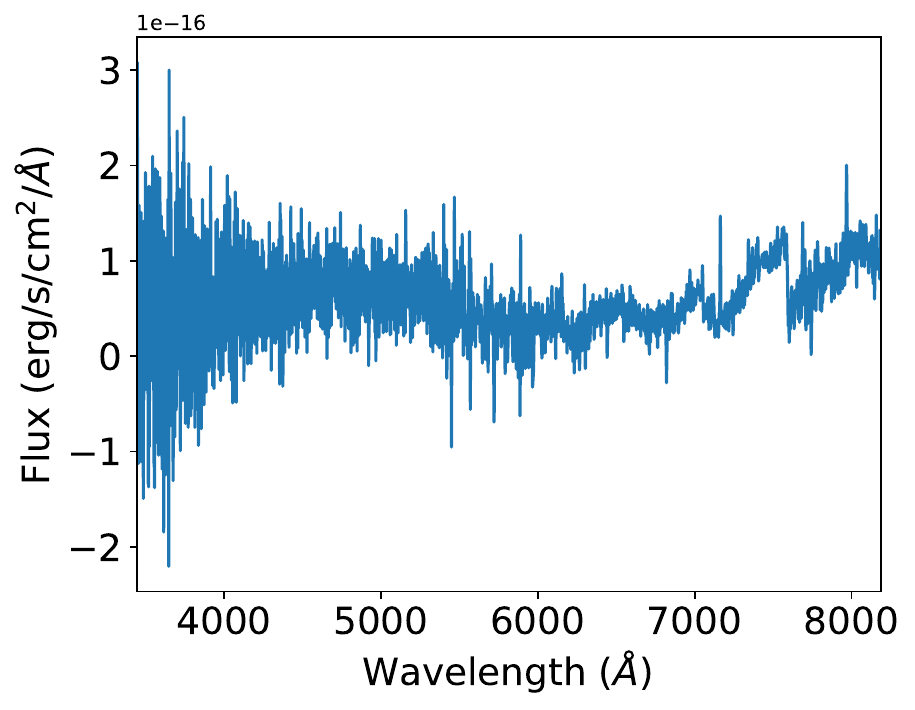}
\figsetgrpnote{Flux for the Lick, Palomar, and Keck spectra. See Table 3 for the observed spectral features and line measurements.}
\figsetgrpend

\figsetgrpstart
\figsetgrpnum{2.125}
\figsetgrptitle{Flux of 1926-1825}
\figsetplot{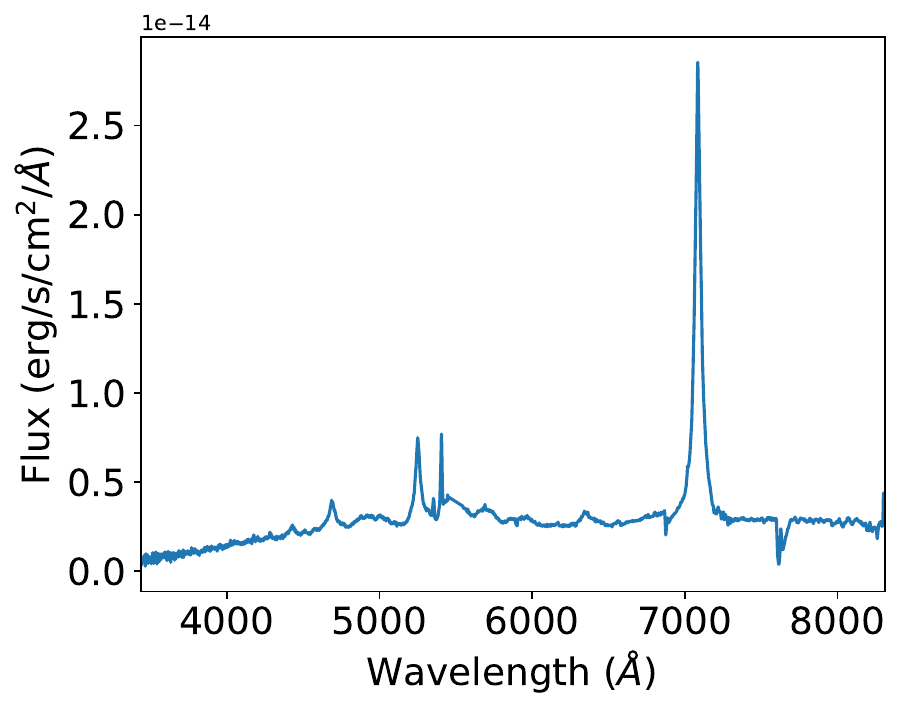}
\figsetgrpnote{Flux for the Lick, Palomar, and Keck spectra. See Table 3 for the observed spectral features and line measurements.}
\figsetgrpend

\figsetgrpstart
\figsetgrpnum{2.126}
\figsetgrptitle{Flux of 1929+4622}
\figsetplot{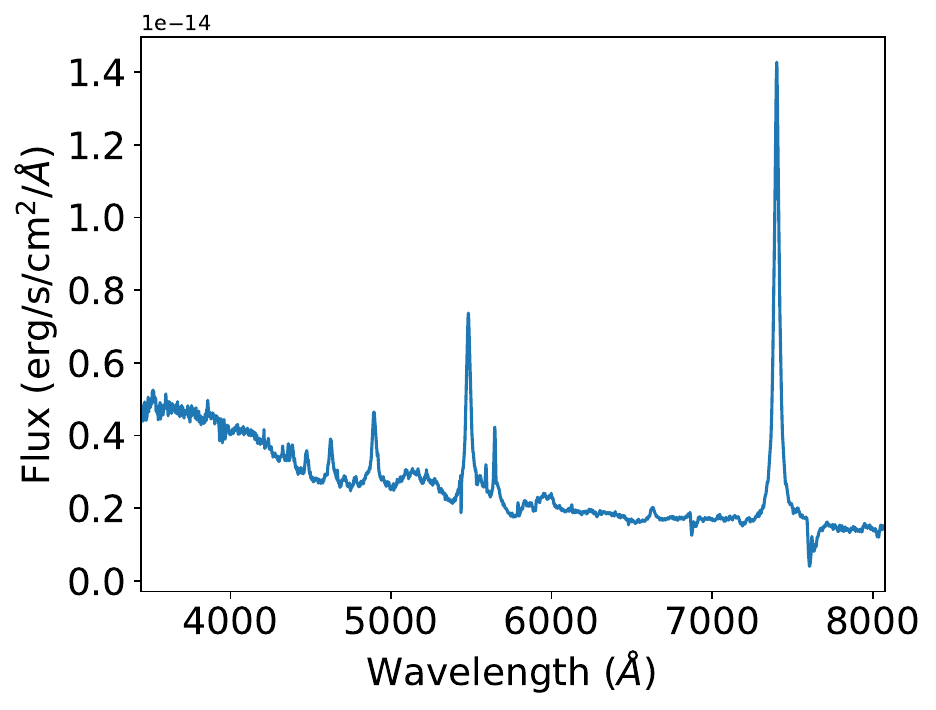}
\figsetgrpnote{Flux for the Lick, Palomar, and Keck spectra. See Table 3 for the observed spectral features and line measurements.}
\figsetgrpend

\figsetgrpstart
\figsetgrpnum{2.127}
\figsetgrptitle{Flux of 1930+4808}
\figsetplot{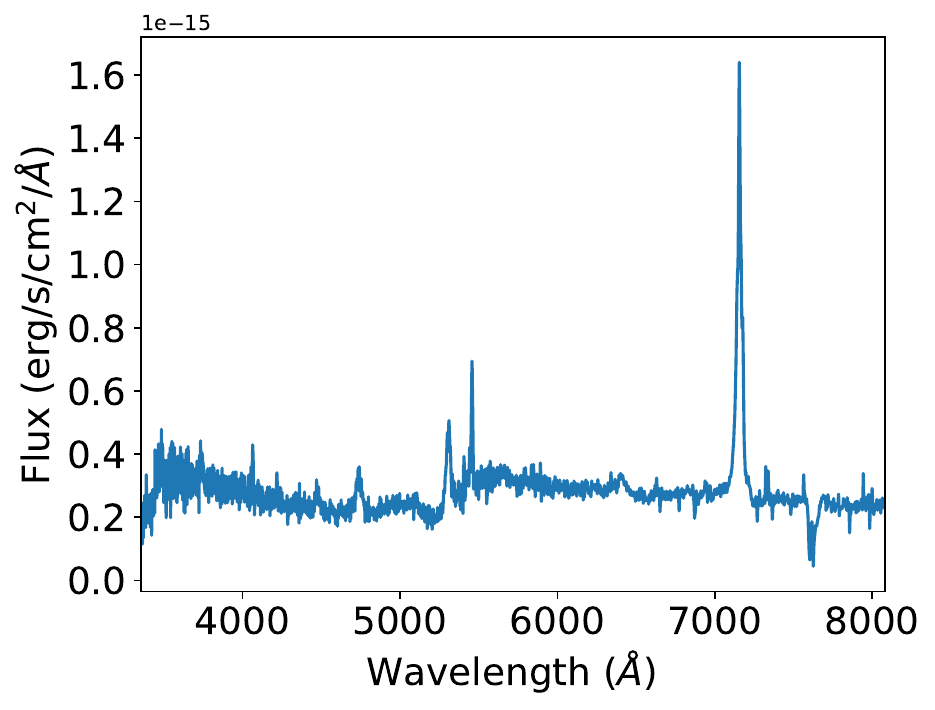}
\figsetgrpnote{Flux for the Lick, Palomar, and Keck spectra. See Table 3 for the observed spectral features and line measurements.}
\figsetgrpend

\figsetgrpstart
\figsetgrpnum{2.128}
\figsetgrptitle{Flux of 1930+3759}
\figsetplot{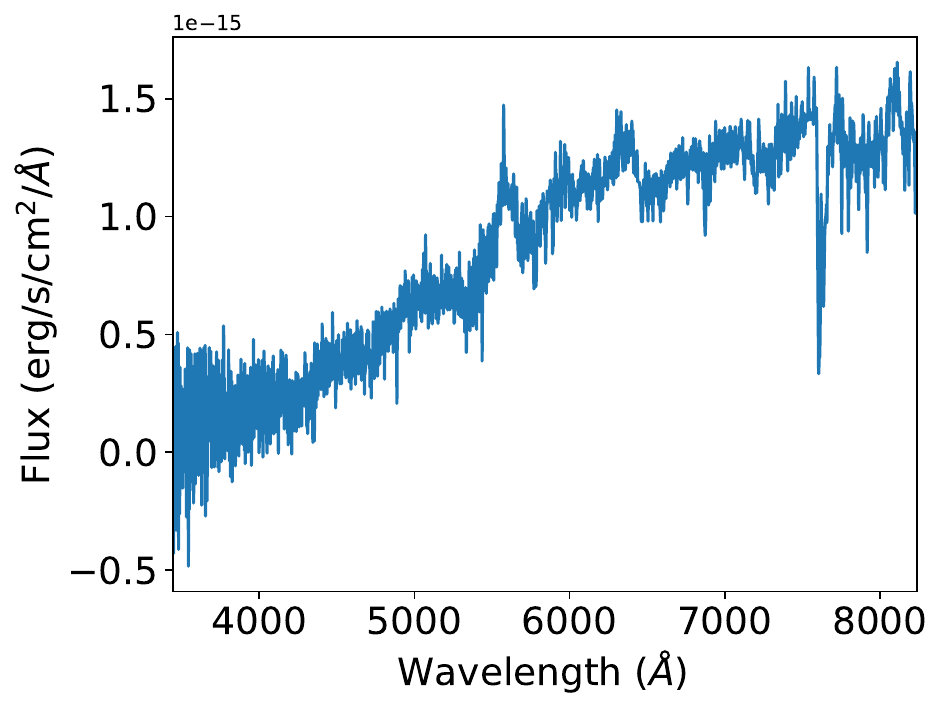}
\figsetgrpnote{Flux for the Lick, Palomar, and Keck spectra. See Table 3 for the observed spectral features and line measurements.}
\figsetgrpend

\figsetgrpstart
\figsetgrpnum{2.129}
\figsetgrptitle{Flux of 1931+4313}
\figsetplot{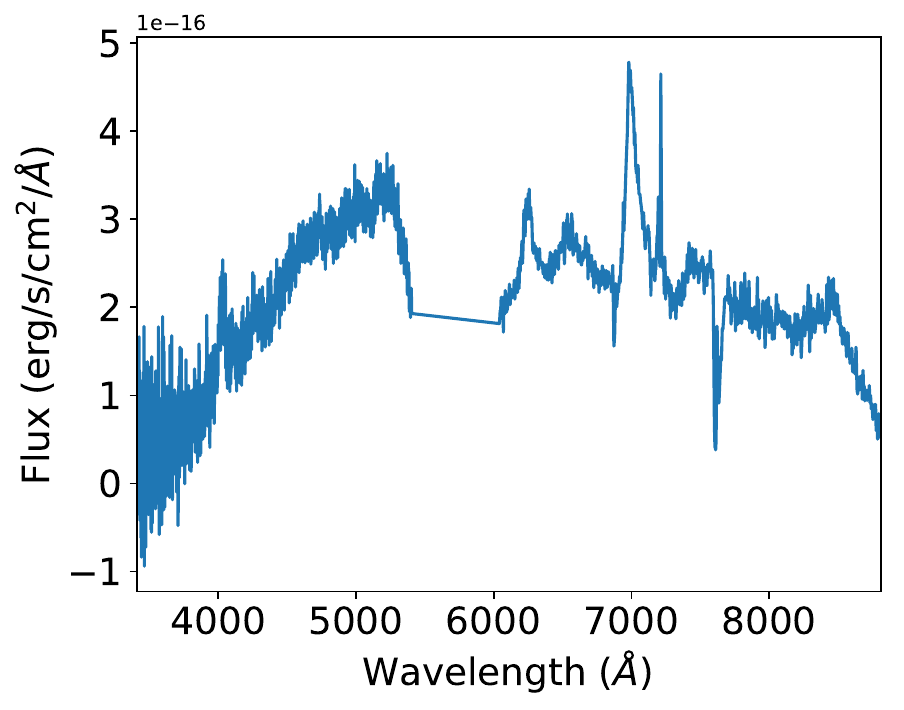}
\figsetgrpnote{Flux for the Lick, Palomar, and Keck spectra. See Table 3 for the observed spectral features and line measurements.}
\figsetgrpend

\figsetgrpstart
\figsetgrpnum{2.130}
\figsetgrptitle{Flux of 1931+4131}
\figsetplot{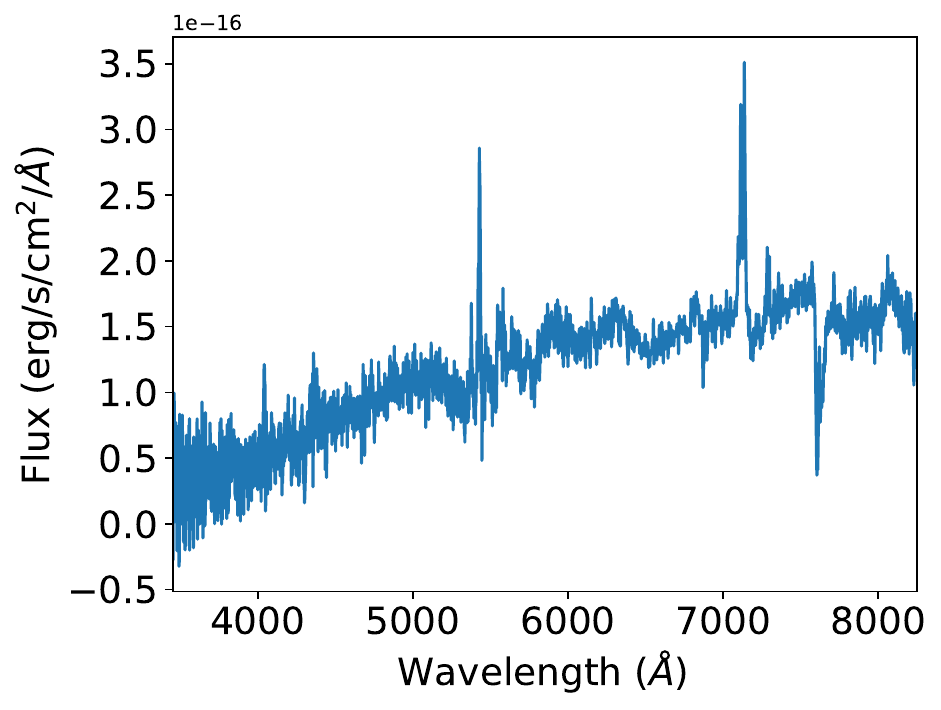}
\figsetgrpnote{Flux for the Lick, Palomar, and Keck spectra. See Table 3 for the observed spectral features and line measurements.}
\figsetgrpend

\figsetgrpstart
\figsetgrpnum{2.131}
\figsetgrptitle{Flux of 1931+3828}
\figsetplot{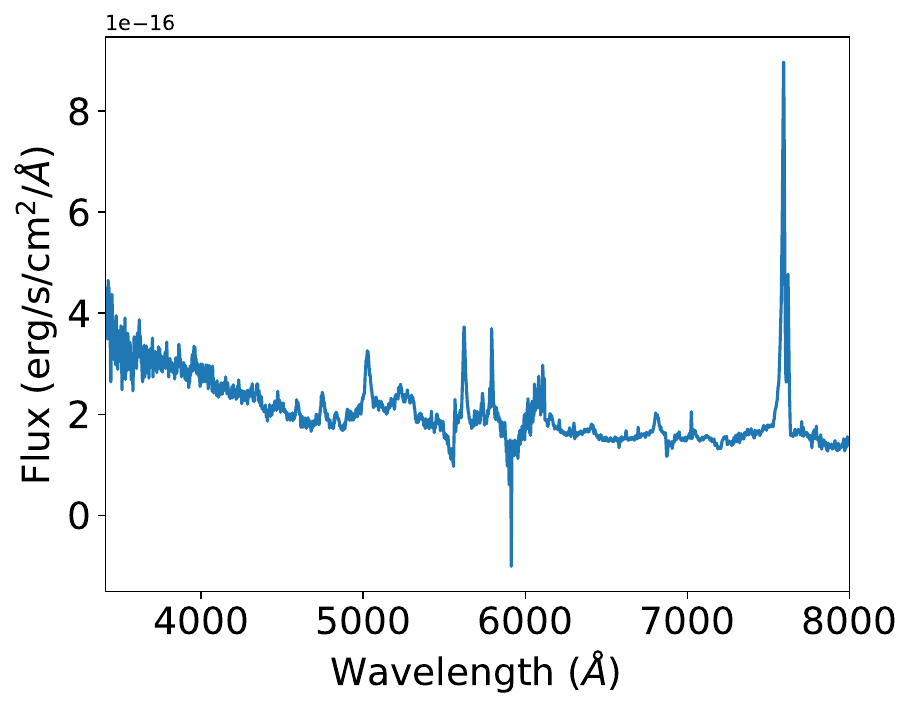}
\figsetgrpnote{Flux for the Lick, Palomar, and Keck spectra. See Table 3 for the observed spectral features and line measurements.}
\figsetgrpend

\figsetgrpstart
\figsetgrpnum{2.132}
\figsetgrptitle{Flux of 1931+4845}
\figsetplot{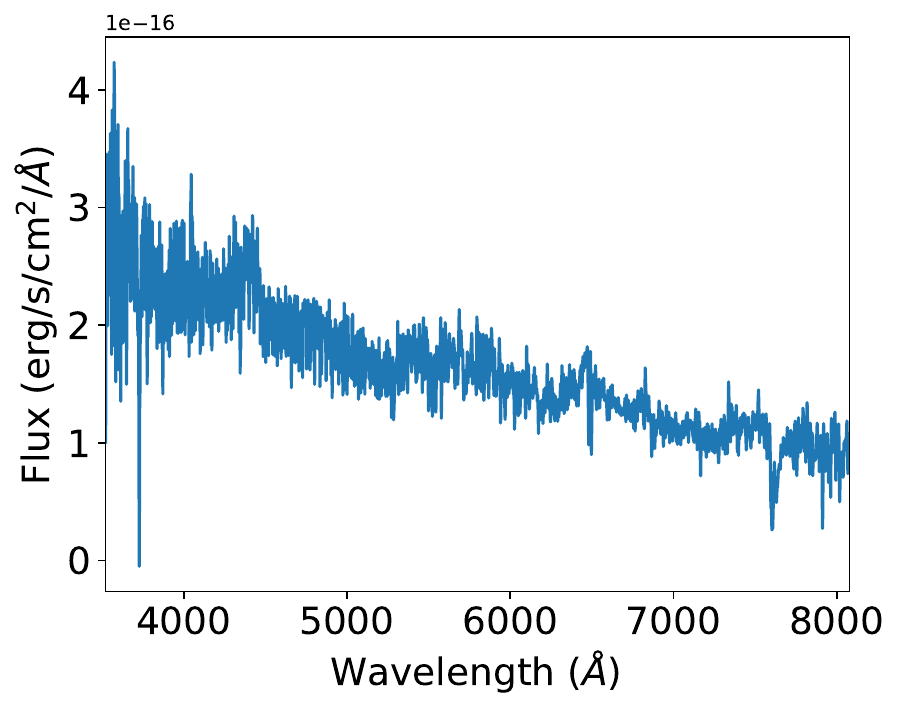}
\figsetgrpnote{Flux for the Lick, Palomar, and Keck spectra. See Table 3 for the observed spectral features and line measurements.}
\figsetgrpend

\figsetgrpstart
\figsetgrpnum{2.133}
\figsetgrptitle{Flux of 1932+4102}
\figsetplot{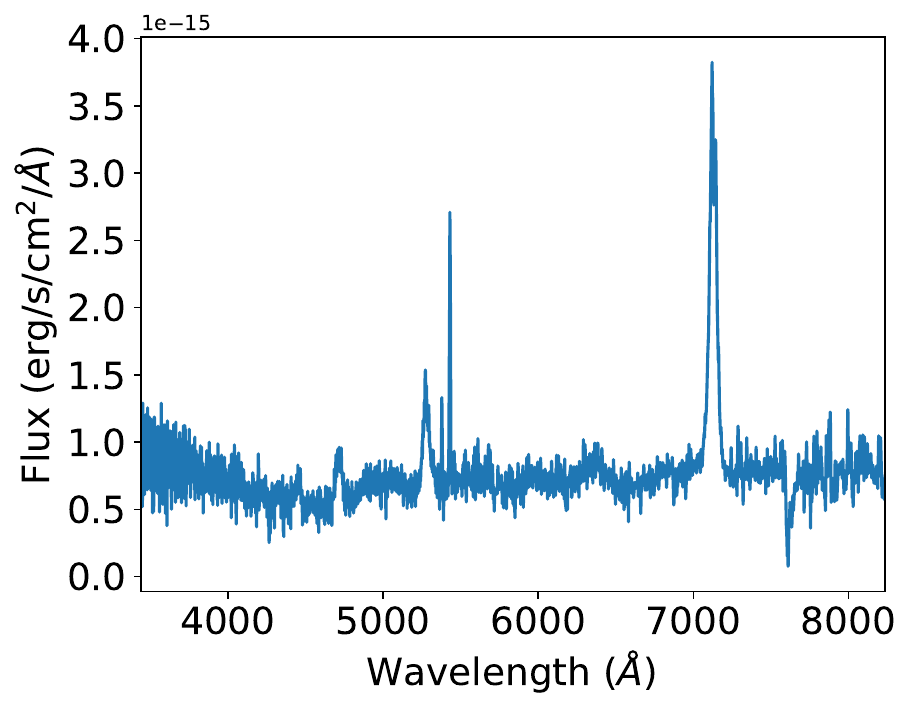}
\figsetgrpnote{Flux for the Lick, Palomar, and Keck spectra. See Table 3 for the observed spectral features and line measurements.}
\figsetgrpend

\figsetgrpstart
\figsetgrpnum{2.134}
\figsetgrptitle{Flux of 1933+4101}
\figsetplot{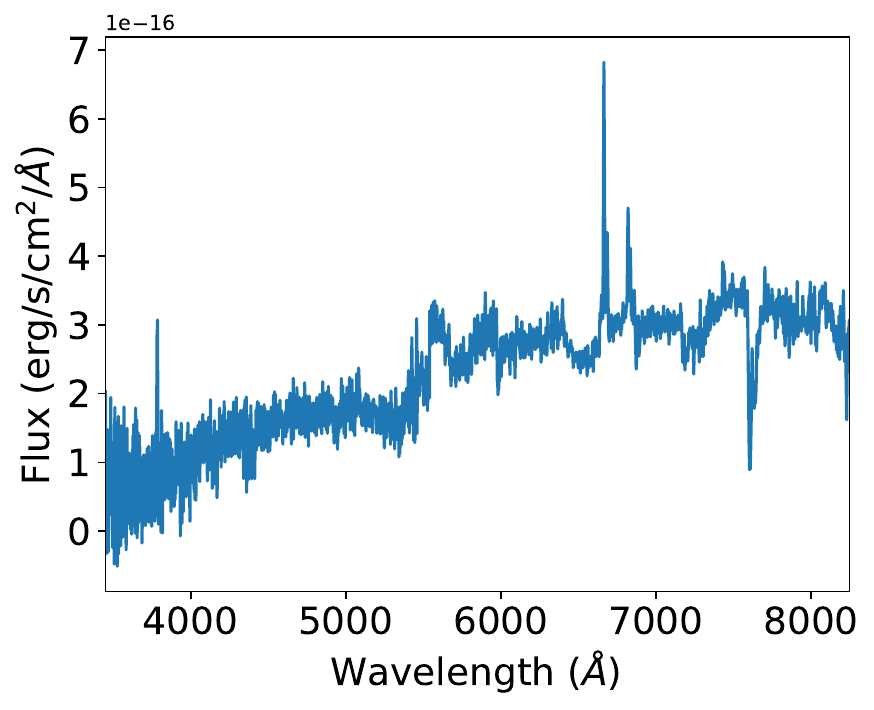}
\figsetgrpnote{Flux for the Lick, Palomar, and Keck spectra. See Table 3 for the observed spectral features and line measurements.}
\figsetgrpend

\figsetgrpstart
\figsetgrpnum{2.135}
\figsetgrptitle{Flux of 1936+4936}
\figsetplot{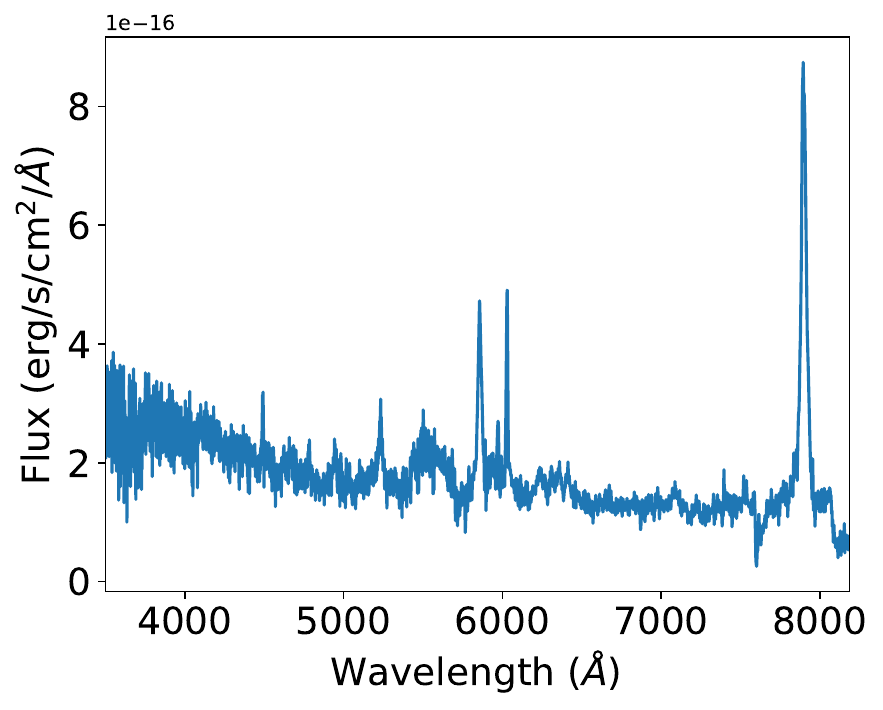}
\figsetgrpnote{Flux for the Lick, Palomar, and Keck spectra. See Table 3 for the observed spectral features and line measurements.}
\figsetgrpend

\figsetgrpstart
\figsetgrpnum{2.136}
\figsetgrptitle{Flux of 1937+4146}
\figsetplot{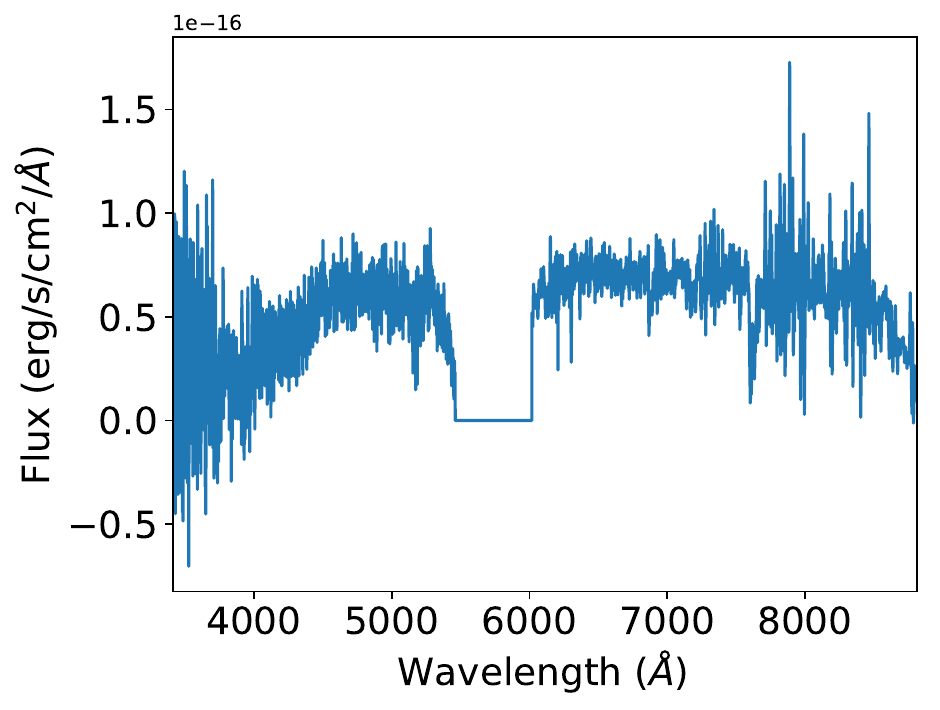}
\figsetgrpnote{Flux for the Lick, Palomar, and Keck spectra. See Table 3 for the observed spectral features and line measurements.}
\figsetgrpend

\figsetgrpstart
\figsetgrpnum{2.137}
\figsetgrptitle{Flux of 1937+4937}
\figsetplot{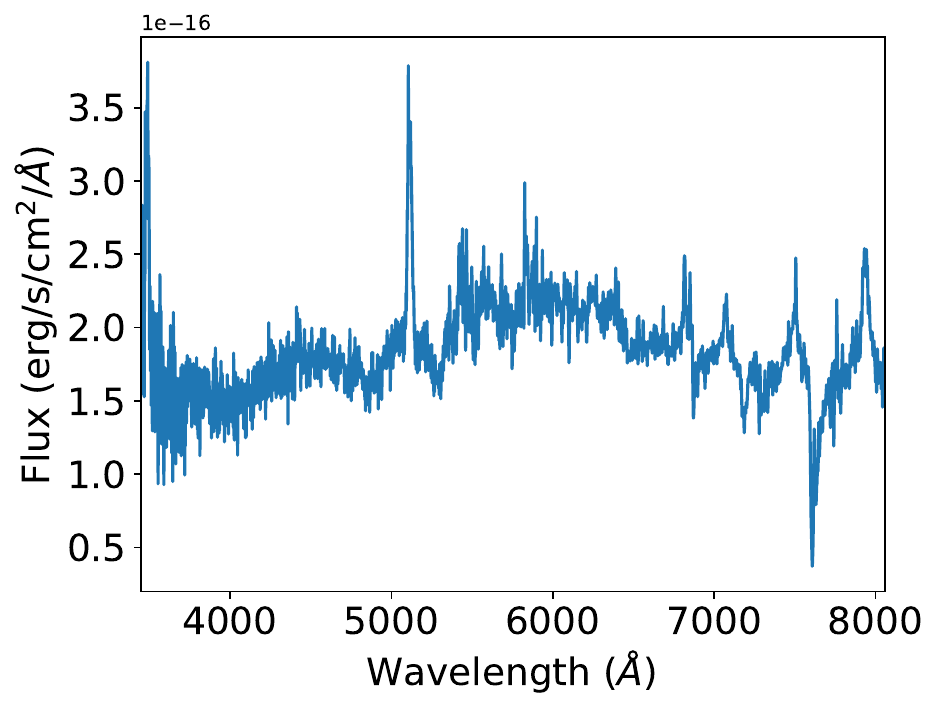}
\figsetgrpnote{Flux for the Lick, Palomar, and Keck spectra. See Table 3 for the observed spectral features and line measurements.}
\figsetgrpend

\figsetgrpstart
\figsetgrpnum{2.138}
\figsetgrptitle{Flux of 1939+4100}
\figsetplot{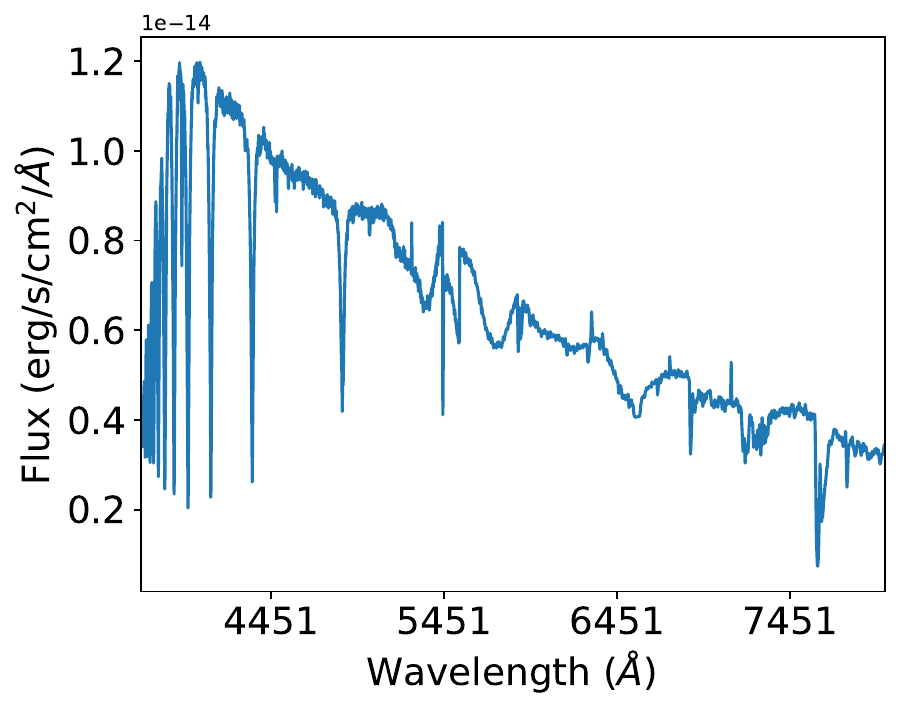}
\figsetgrpnote{Flux for the Lick, Palomar, and Keck spectra. See Table 3 for the observed spectral features and line measurements.}
\figsetgrpend

\figsetgrpstart
\figsetgrpnum{2.139}
\figsetgrptitle{Flux of 1939+4353}
\figsetplot{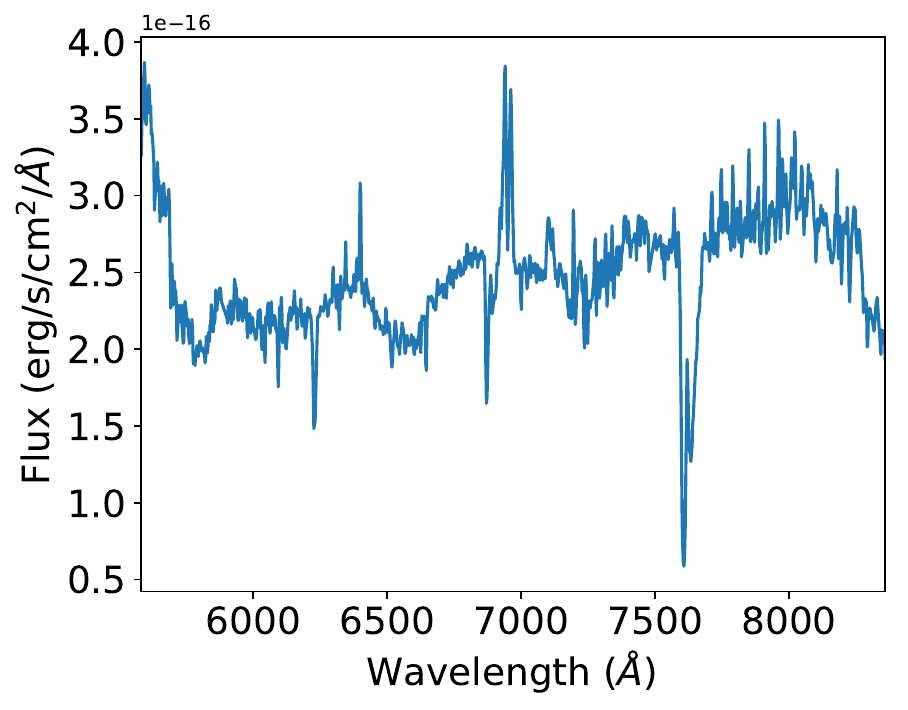}
\figsetgrpnote{Flux for the Lick, Palomar, and Keck spectra. See Table 3 for the observed spectral features and line measurements.}
\figsetgrpend

\figsetgrpstart
\figsetgrpnum{2.140}
\figsetgrptitle{Flux of 1942+4747}
\figsetplot{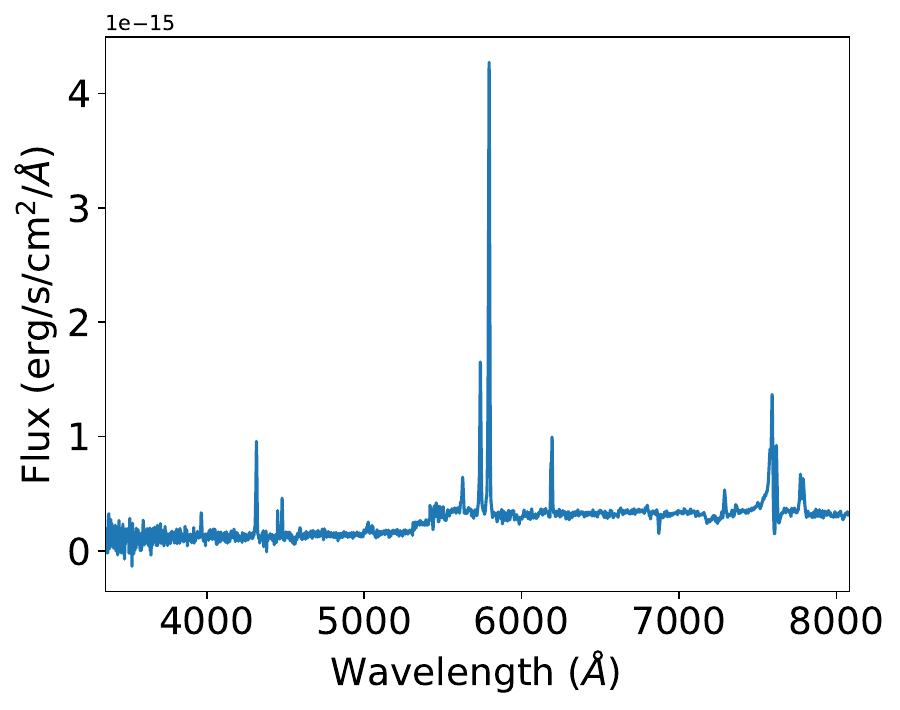}
\figsetgrpnote{Flux for the Lick, Palomar, and Keck spectra. See Table 3 for the observed spectral features and line measurements.}
\figsetgrpend

\figsetgrpstart
\figsetgrpnum{2.141}
\figsetgrptitle{Flux of 1943-1631}
\figsetplot{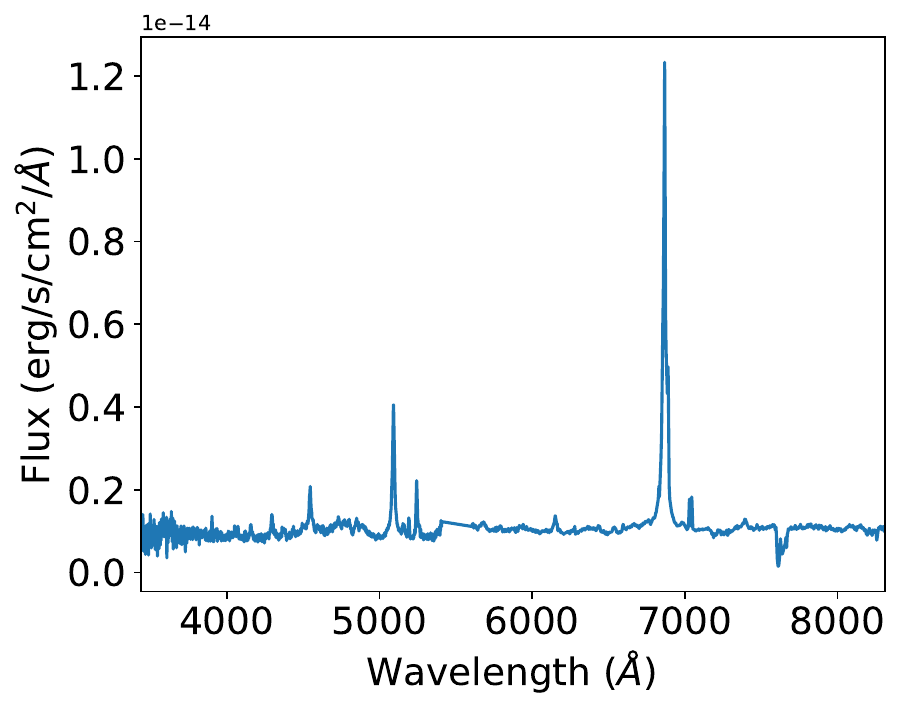}
\figsetgrpnote{Flux for the Lick, Palomar, and Keck spectra. See Table 3 for the observed spectral features and line measurements.}
\figsetgrpend

\figsetgrpstart
\figsetgrpnum{2.142}
\figsetgrptitle{Flux of 1943+5004}
\figsetplot{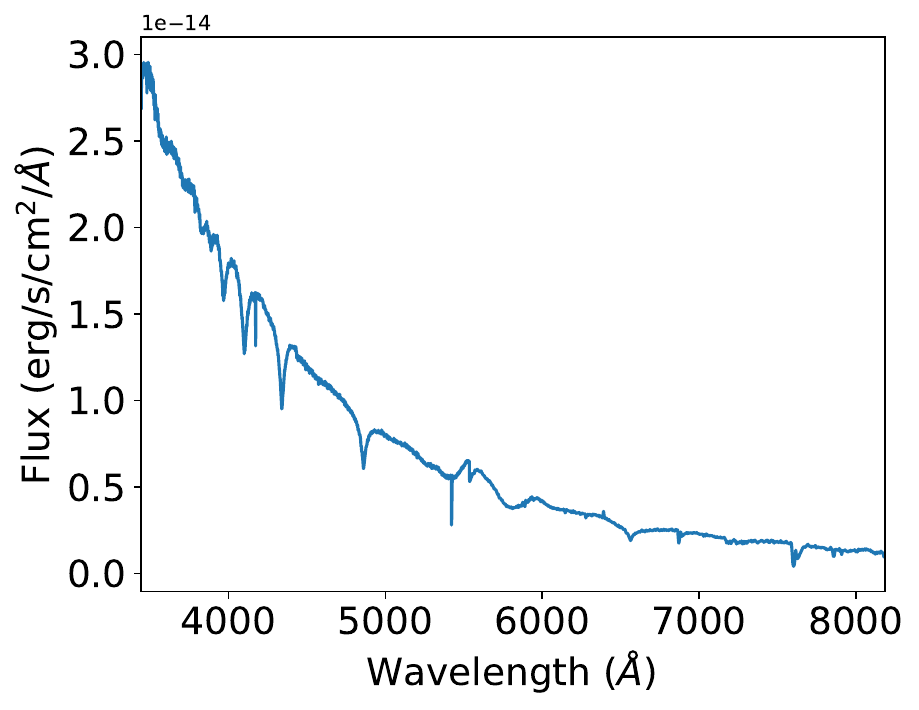}
\figsetgrpnote{Flux for the Lick, Palomar, and Keck spectra. See Table 3 for the observed spectral features and line measurements.}
\figsetgrpend

\figsetgrpstart
\figsetgrpnum{2.143}
\figsetgrptitle{Flux of 1943+4316}
\figsetplot{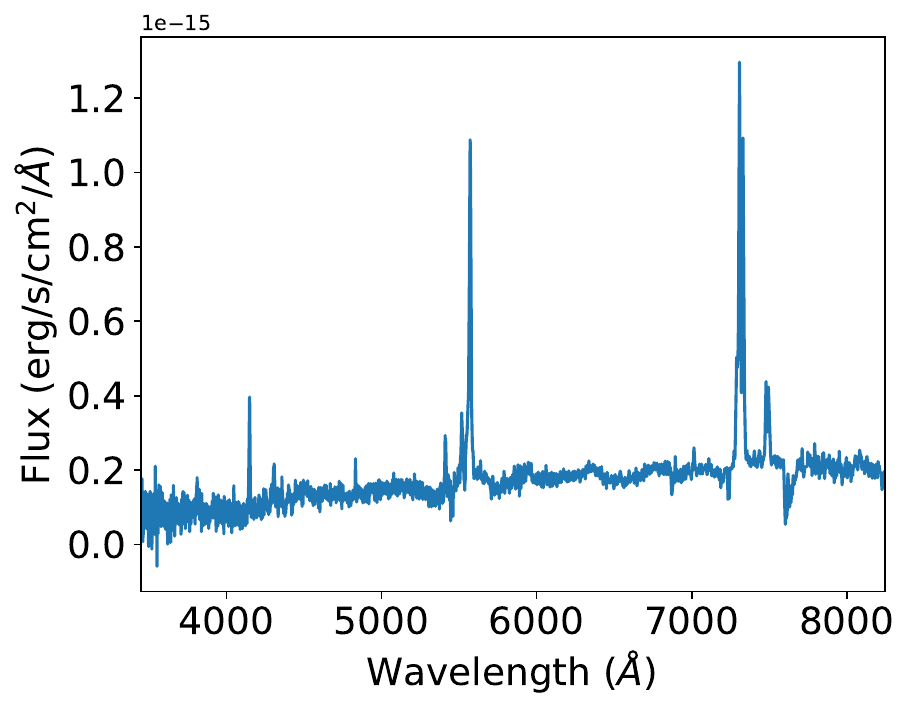}
\figsetgrpnote{Flux for the Lick, Palomar, and Keck spectra. See Table 3 for the observed spectral features and line measurements.}
\figsetgrpend

\figsetgrpstart
\figsetgrpnum{2.144}
\figsetgrptitle{Flux of 1944+4733}
\figsetplot{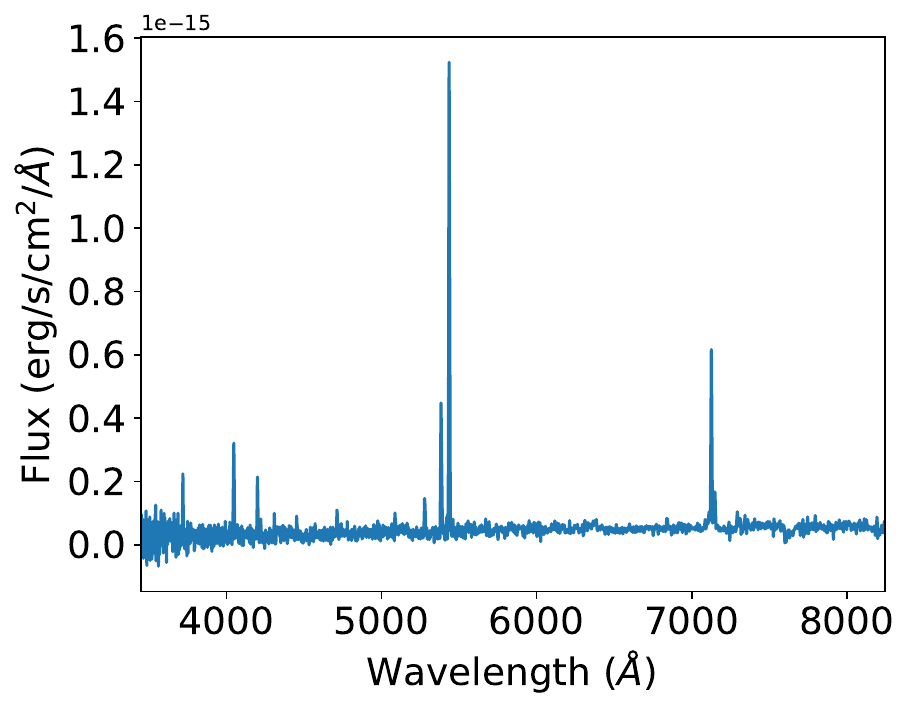}
\figsetgrpnote{Flux for the Lick, Palomar, and Keck spectra. See Table 3 for the observed spectral features and line measurements.}
\figsetgrpend

\figsetgrpstart
\figsetgrpnum{2.145}
\figsetgrptitle{Flux of 1946+4915}
\figsetplot{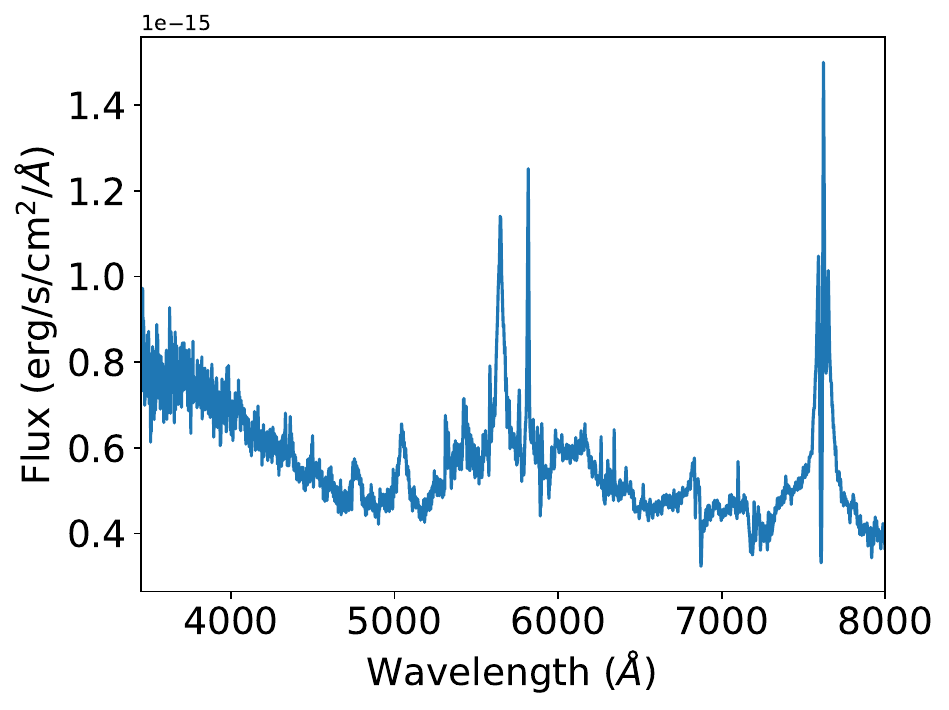}
\figsetgrpnote{Flux for the Lick, Palomar, and Keck spectra. See Table 3 for the observed spectral features and line measurements.}
\figsetgrpend

\figsetgrpstart
\figsetgrpnum{2.146}
\figsetgrptitle{Flux of 1946+4946}
\figsetplot{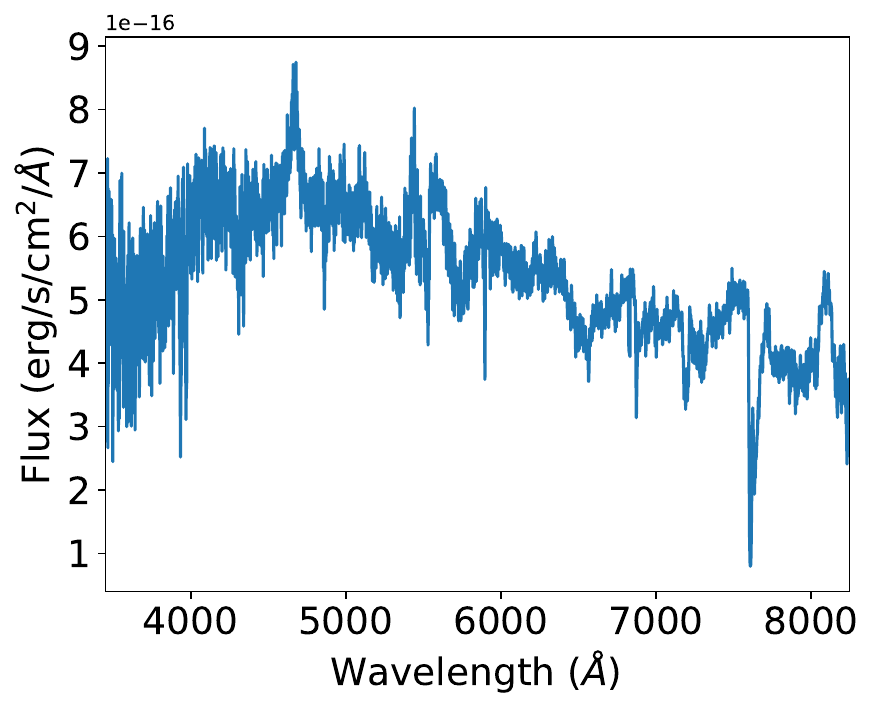}
\figsetgrpnote{Flux for the Lick, Palomar, and Keck spectra. See Table 3 for the observed spectral features and line measurements.}
\figsetgrpend

\figsetgrpstart
\figsetgrpnum{2.147}
\figsetgrptitle{Flux of 1947+4903}
\figsetplot{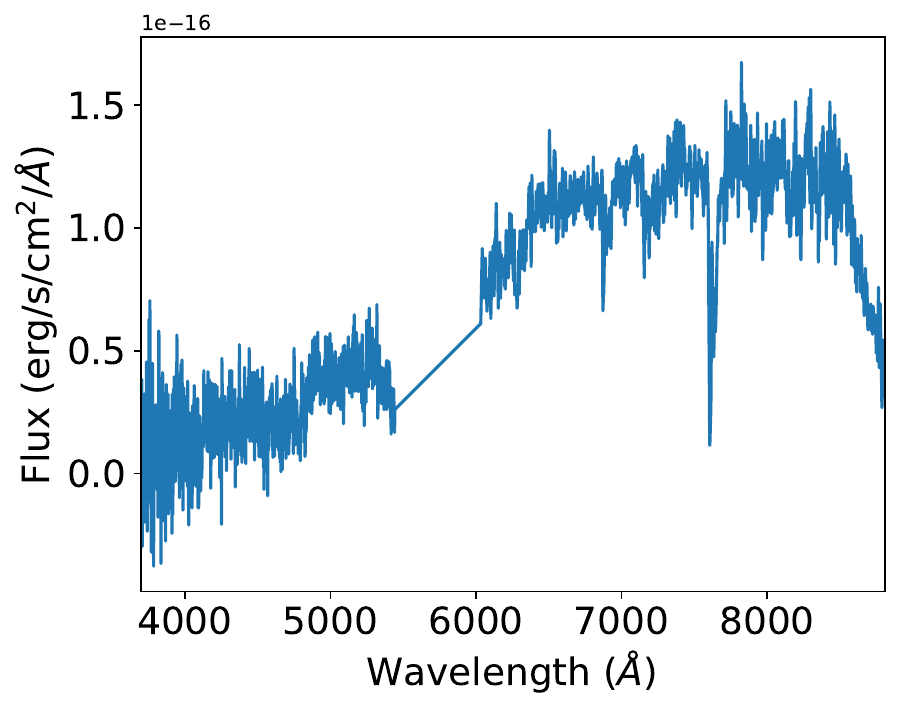}
\figsetgrpnote{Flux for the Lick, Palomar, and Keck spectra. See Table 3 for the observed spectral features and line measurements.}
\figsetgrpend

\figsetgrpstart
\figsetgrpnum{2.148}
\figsetgrptitle{Flux of 1947+4449}
\figsetplot{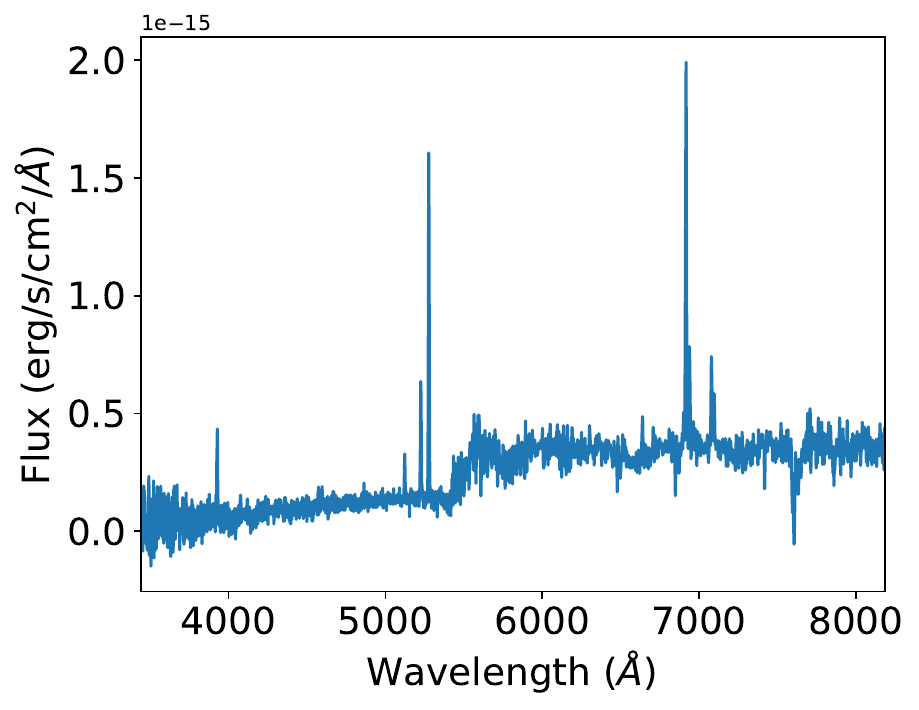}
\figsetgrpnote{Flux for the Lick, Palomar, and Keck spectra. See Table 3 for the observed spectral features and line measurements.}
\figsetgrpend

\figsetgrpstart
\figsetgrpnum{2.149}
\figsetgrptitle{Flux of 1949+4623}
\figsetplot{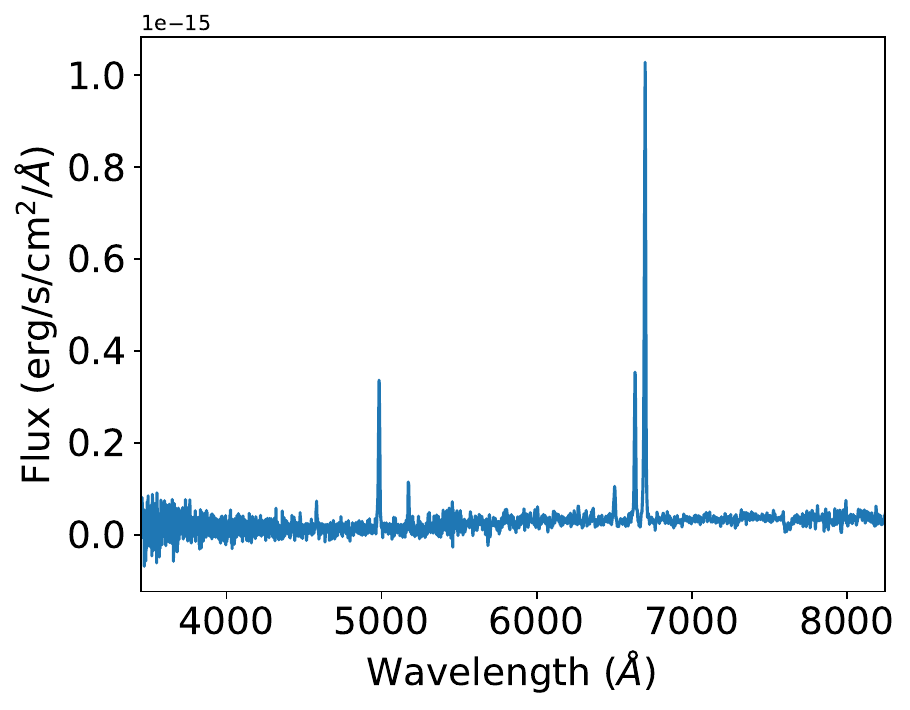}
\figsetgrpnote{Flux for the Lick, Palomar, and Keck spectra. See Table 3 for the observed spectral features and line measurements.}
\figsetgrpend

\figsetgrpstart
\figsetgrpnum{2.150}
\figsetgrptitle{Flux of 1950+4206}
\figsetplot{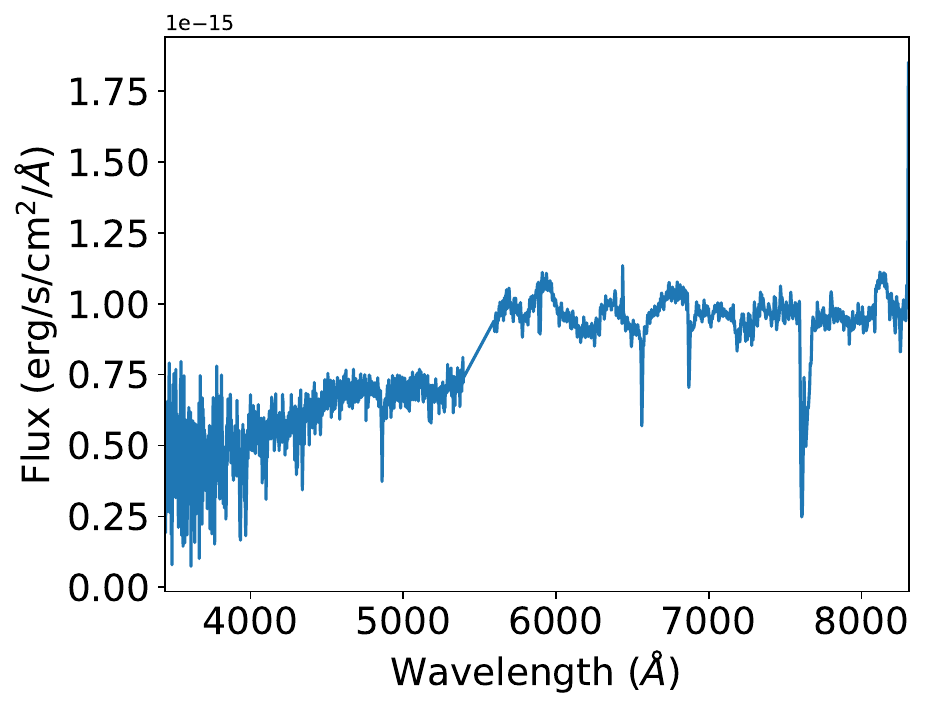}
\figsetgrpnote{Flux for the Lick, Palomar, and Keck spectra. See Table 3 for the observed spectral features and line measurements.}
\figsetgrpend

\figsetgrpstart
\figsetgrpnum{2.151}
\figsetgrptitle{Flux of 1950+4700}
\figsetplot{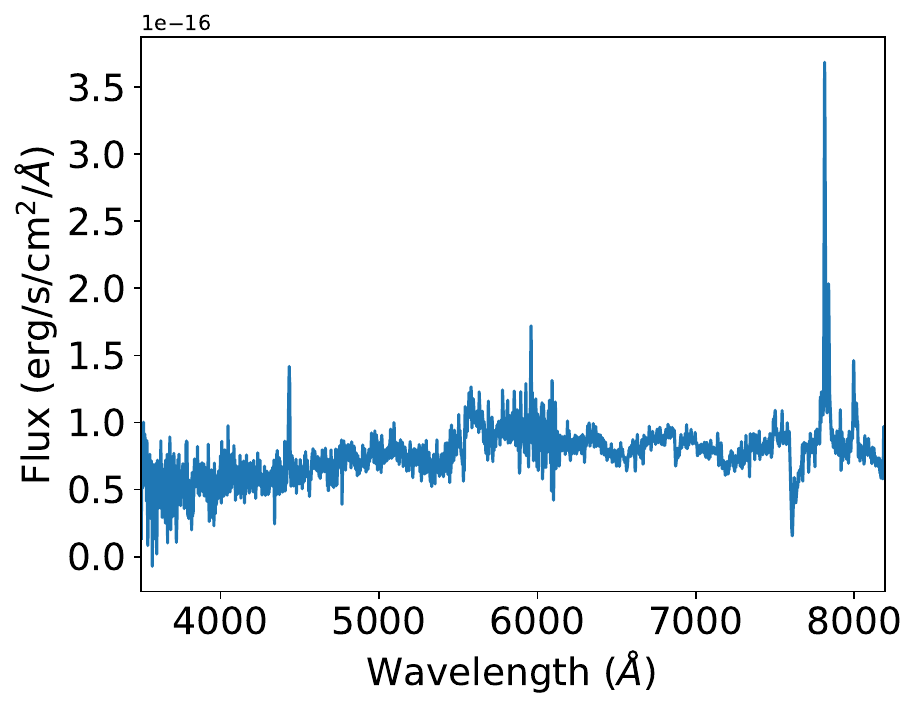}
\figsetgrpnote{Flux for the Lick, Palomar, and Keck spectra. See Table 3 for the observed spectral features and line measurements.}
\figsetgrpend

\figsetgrpstart
\figsetgrpnum{2.152}
\figsetgrptitle{Flux of 1951+4155}
\figsetplot{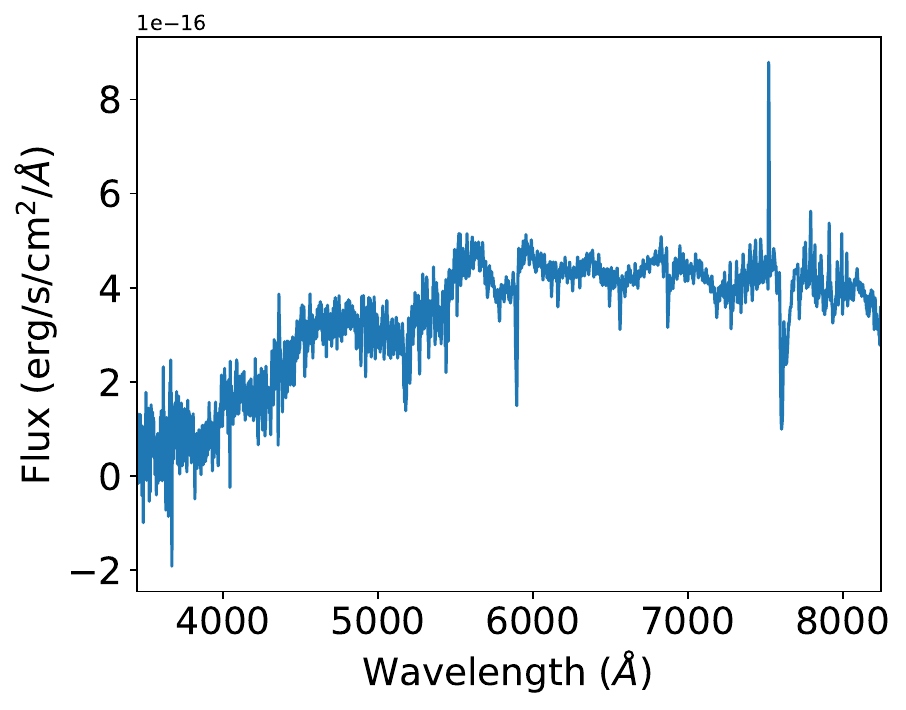}
\figsetgrpnote{Flux for the Lick, Palomar, and Keck spectra. See Table 3 for the observed spectral features and line measurements.}
\figsetgrpend

\figsetgrpstart
\figsetgrpnum{2.153}
\figsetgrptitle{Flux of 1951+3948}
\figsetplot{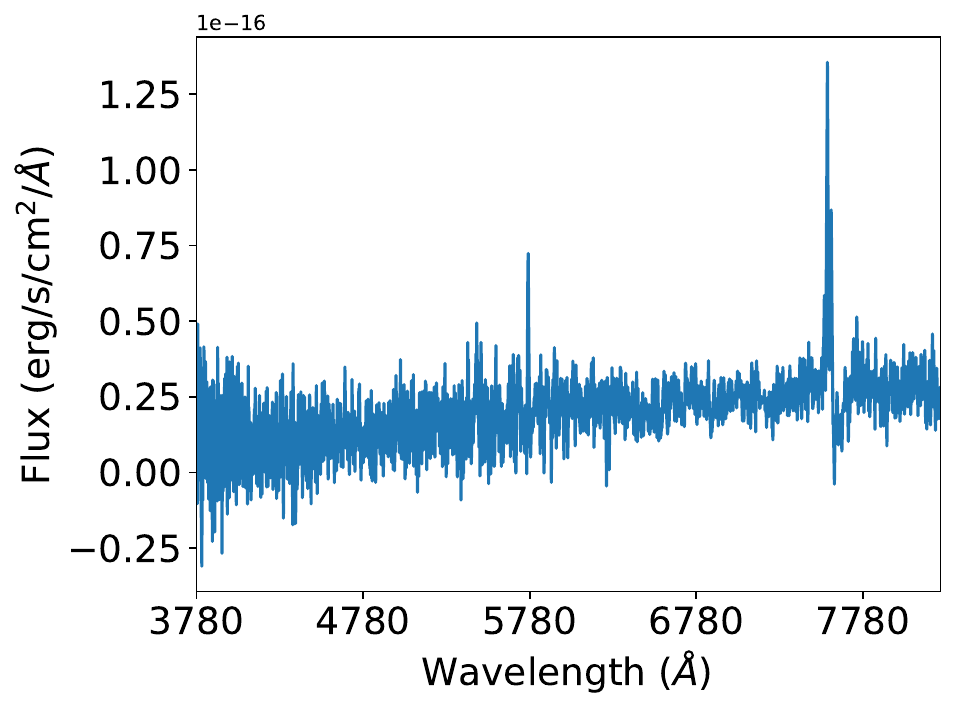}
\figsetgrpnote{Flux for the Lick, Palomar, and Keck spectra. See Table 3 for the observed spectral features and line measurements.}
\figsetgrpend

\figsetgrpstart
\figsetgrpnum{2.154}
\figsetgrptitle{Flux of 1953+4027}
\figsetplot{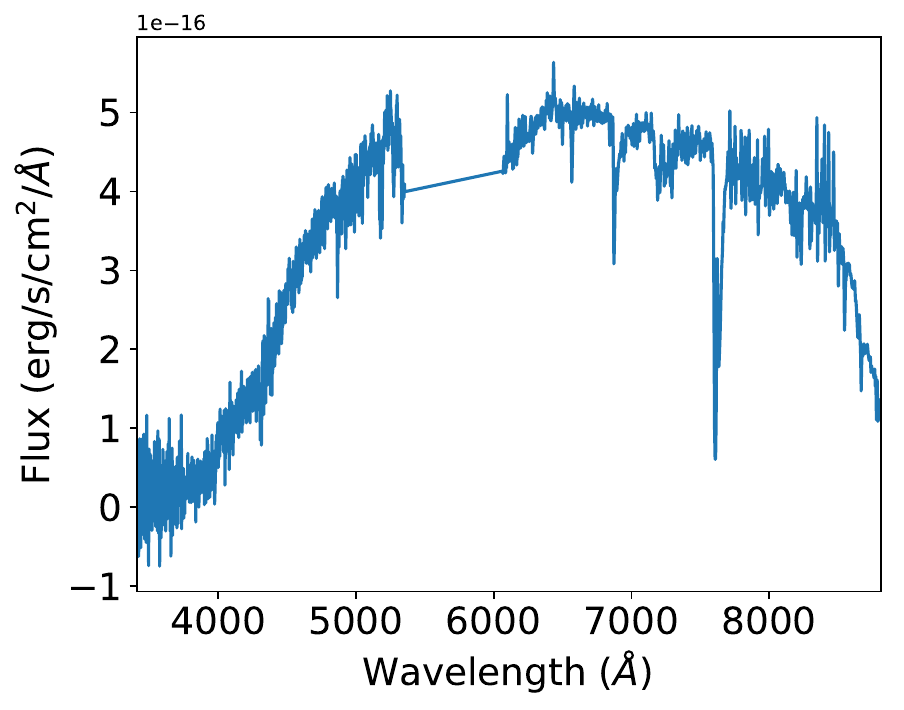}
\figsetgrpnote{Flux for the Lick, Palomar, and Keck spectra. See Table 3 for the observed spectral features and line measurements.}
\figsetgrpend

\figsetgrpstart
\figsetgrpnum{2.155}
\figsetgrptitle{Flux of 1954+4813}
\figsetplot{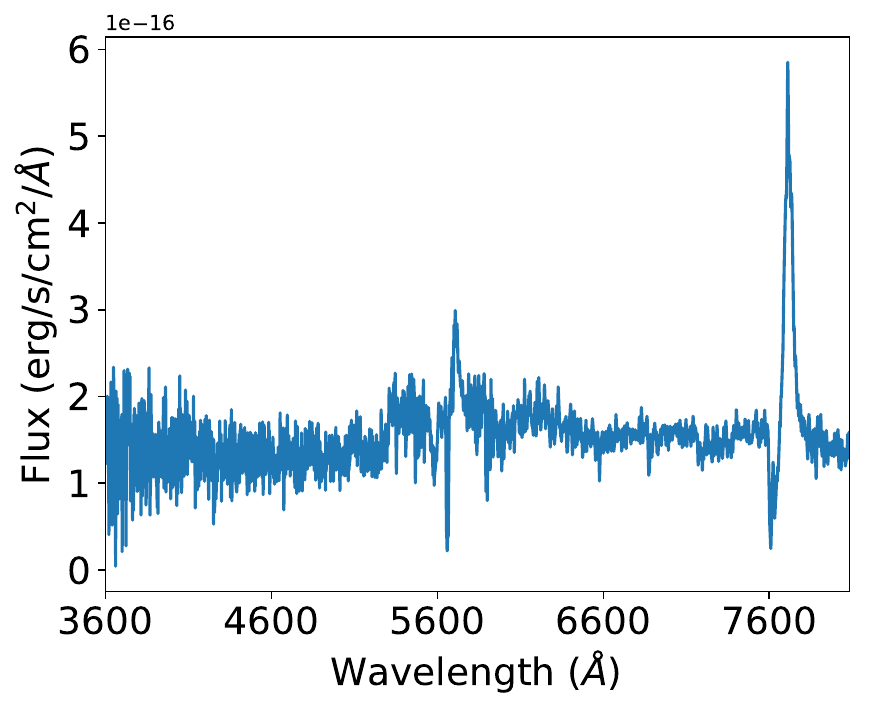}
\figsetgrpnote{Flux for the Lick, Palomar, and Keck spectra. See Table 3 for the observed spectral features and line measurements.}
\figsetgrpend

\figsetgrpstart
\figsetgrpnum{2.156}
\figsetgrptitle{Flux of 1958+4701}
\figsetplot{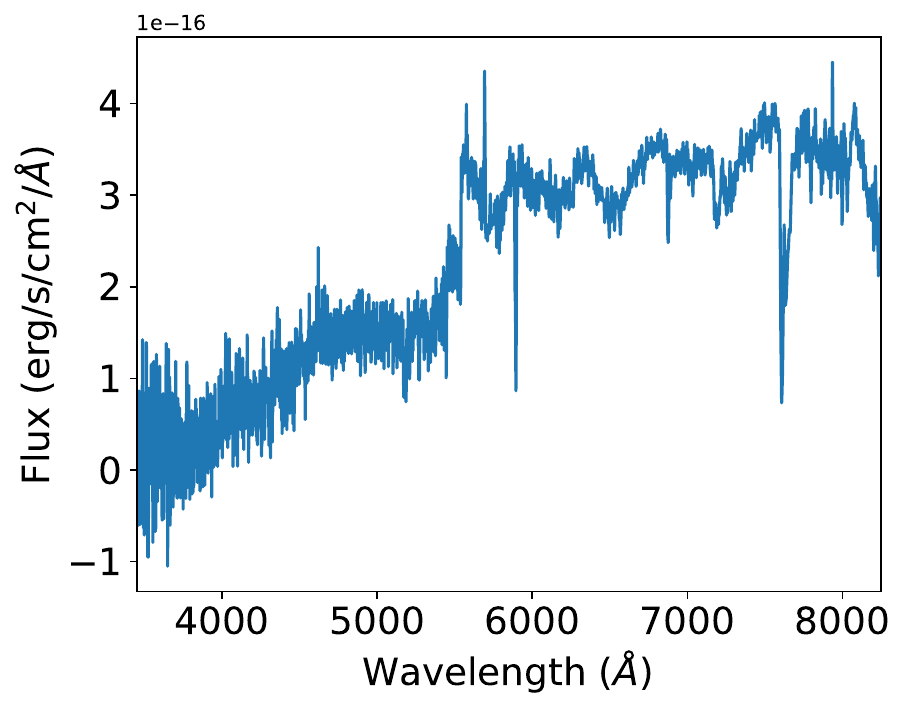}
\figsetgrpnote{Flux for the Lick, Palomar, and Keck spectra. See Table 3 for the observed spectral features and line measurements.}
\figsetgrpend

\figsetgrpstart
\figsetgrpnum{2.157}
\figsetgrptitle{Flux of 2158-1203}
\figsetplot{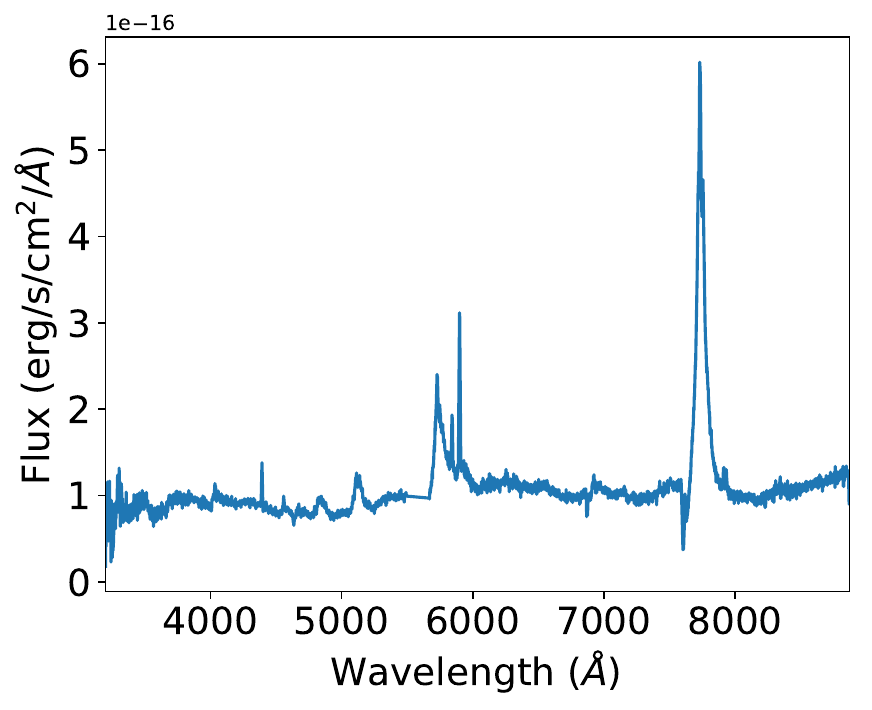}
\figsetgrpnote{Flux for the Lick, Palomar, and Keck spectra. See Table 3 for the observed spectral features and line measurements.}
\figsetgrpend

\figsetgrpstart
\figsetgrpnum{2.158}
\figsetgrptitle{Flux of 2202-0826}
\figsetplot{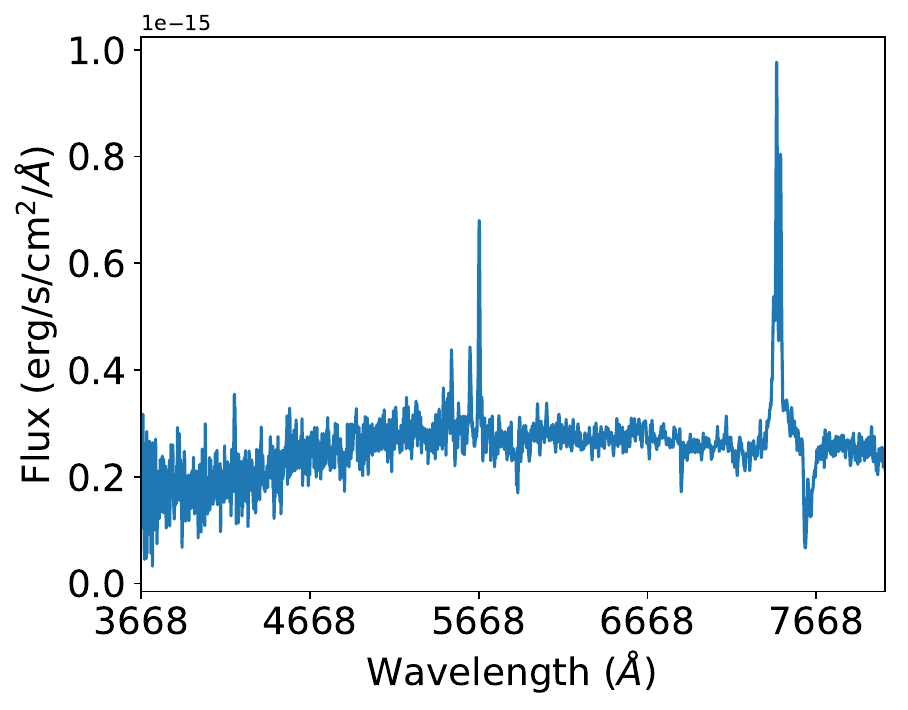}
\figsetgrpnote{Flux for the Lick, Palomar, and Keck spectra. See Table 3 for the observed spectral features and line measurements.}
\figsetgrpend

\figsetgrpstart
\figsetgrpnum{2.159}
\figsetgrptitle{Flux of 2206-0820}
\figsetplot{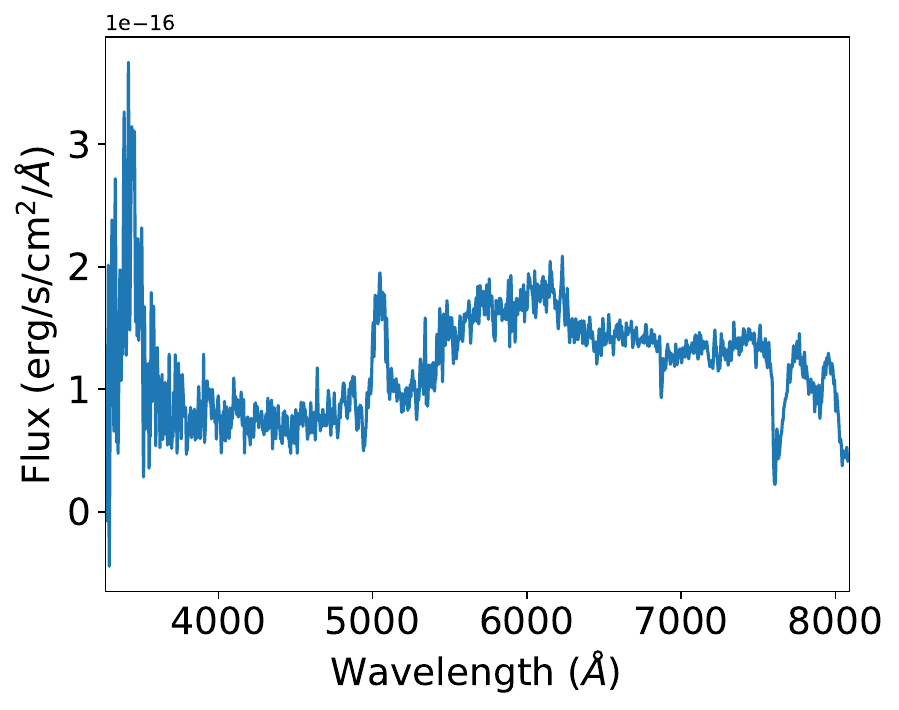}
\figsetgrpnote{Flux for the Lick, Palomar, and Keck spectra. See Table 3 for the observed spectral features and line measurements.}
\figsetgrpend

\figsetgrpstart
\figsetgrpnum{2.160}
\figsetgrptitle{Flux of 2212-0734}
\figsetplot{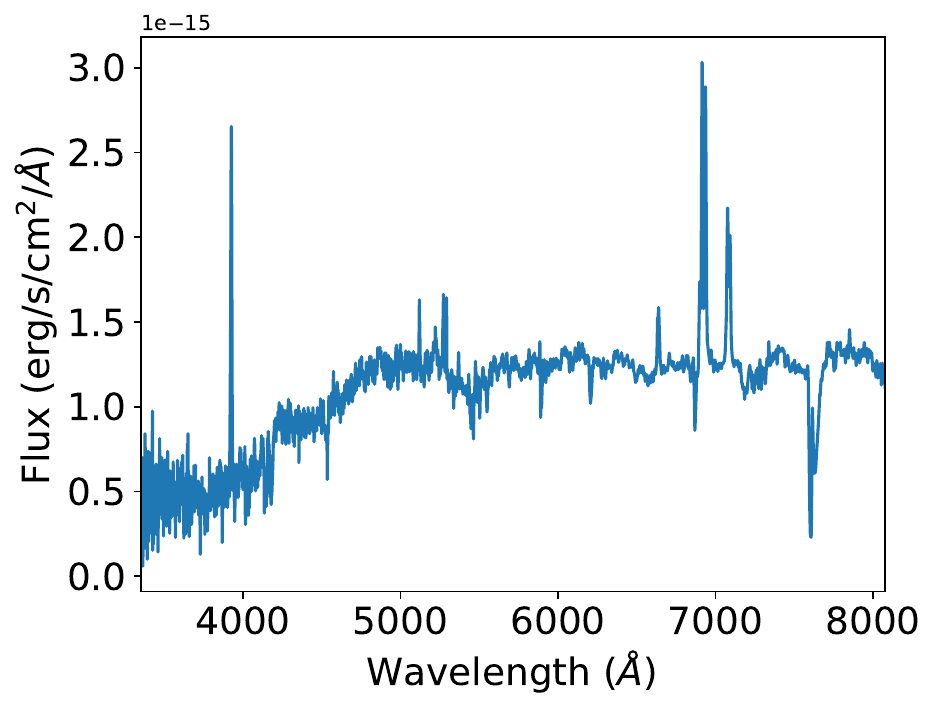}
\figsetgrpnote{Flux for the Lick, Palomar, and Keck spectra. See Table 3 for the observed spectral features and line measurements.}
\figsetgrpend

\figsetgrpstart
\figsetgrpnum{2.161}
\figsetgrptitle{Flux of 2213-1710}
\figsetplot{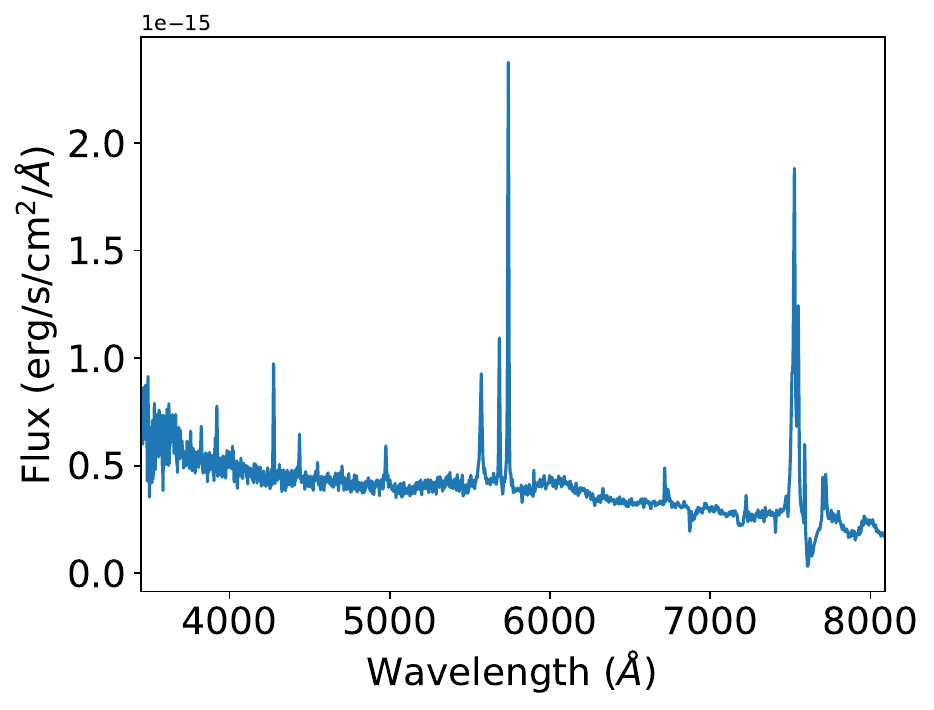}
\figsetgrpnote{Flux for the Lick, Palomar, and Keck spectra. See Table 3 for the observed spectral features and line measurements.}
\figsetgrpend

\figsetgrpstart
\figsetgrpnum{2.162}
\figsetgrptitle{Flux of 2213-1758}
\figsetplot{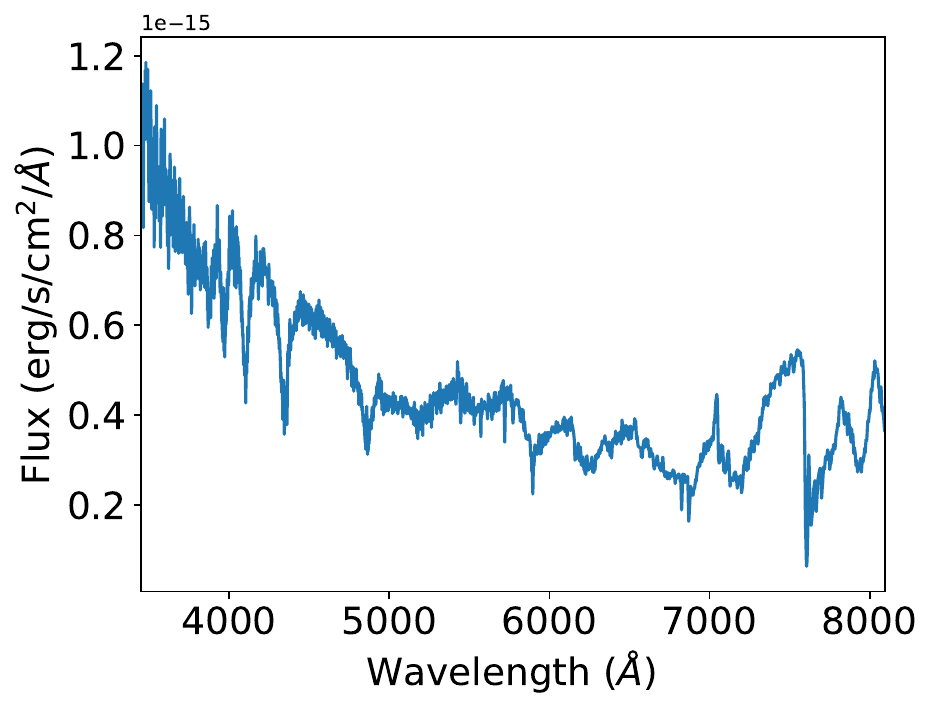}
\figsetgrpnote{Flux for the Lick, Palomar, and Keck spectra. See Table 3 for the observed spectral features and line measurements.}
\figsetgrpend

\figsetgrpstart
\figsetgrpnum{2.163}
\figsetgrptitle{Flux of 2214-0542}
\figsetplot{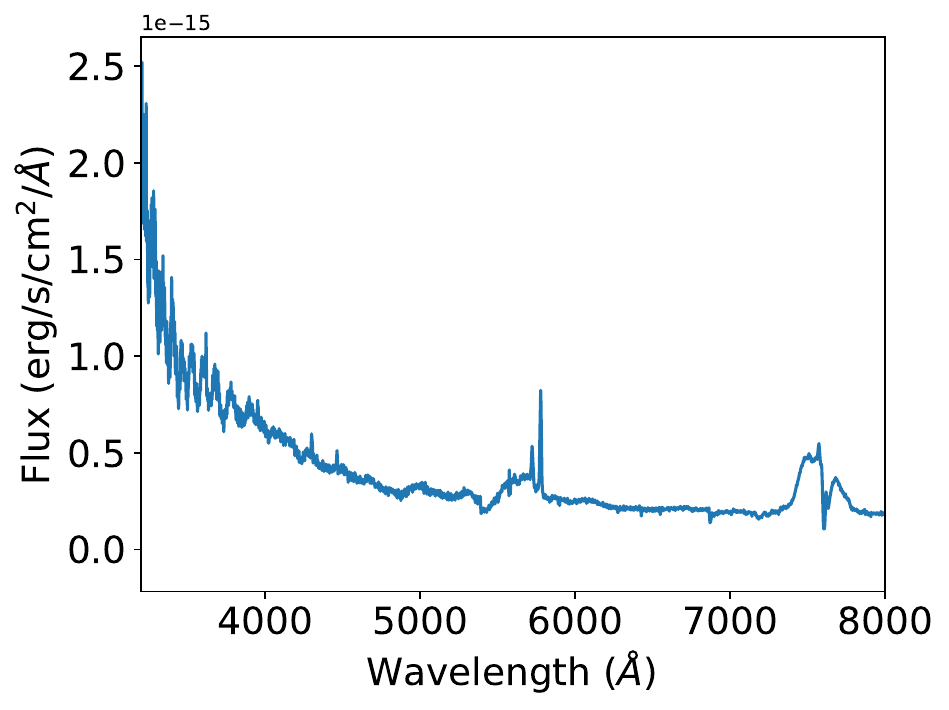}
\figsetgrpnote{Flux for the Lick, Palomar, and Keck spectra. See Table 3 for the observed spectral features and line measurements.}
\figsetgrpend

\figsetgrpstart
\figsetgrpnum{2.164}
\figsetgrptitle{Flux of 2214-1619}
\figsetplot{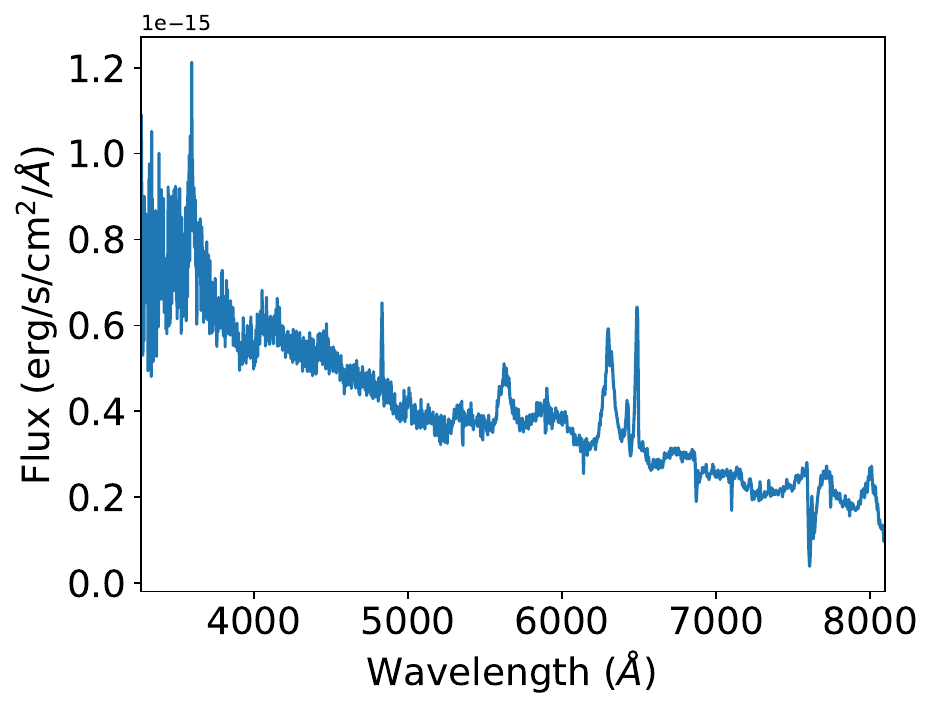}
\figsetgrpnote{Flux for the Lick, Palomar, and Keck spectra. See Table 3 for the observed spectral features and line measurements.}
\figsetgrpend

\figsetgrpstart
\figsetgrpnum{2.165}
\figsetgrptitle{Flux of 2215-1744}
\figsetplot{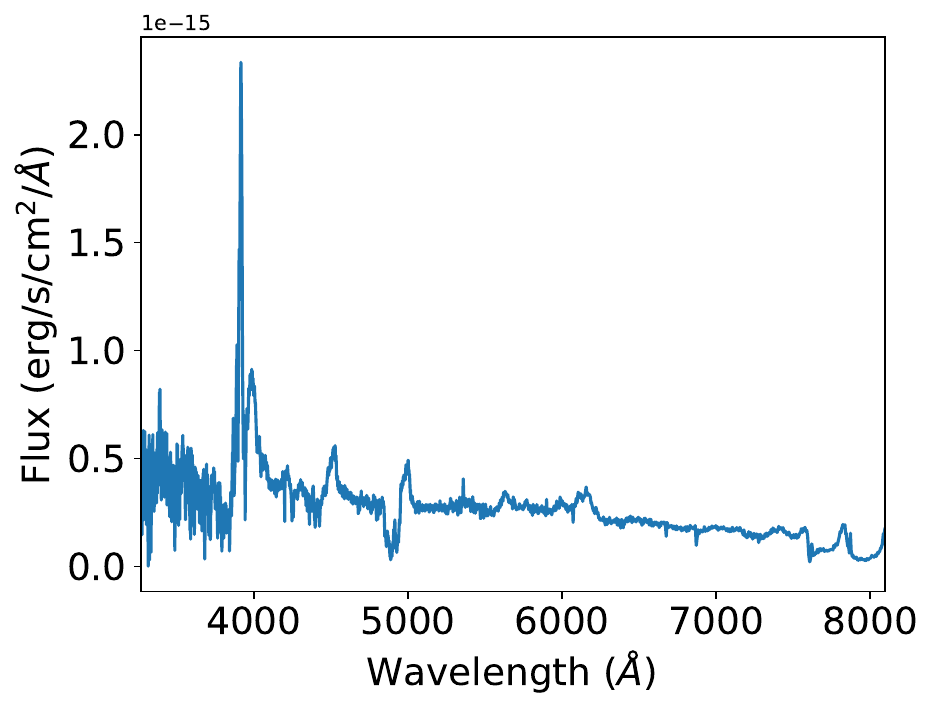}
\figsetgrpnote{Flux for the Lick, Palomar, and Keck spectra. See Table 3 for the observed spectral features and line measurements.}
\figsetgrpend

\figsetgrpstart
\figsetgrpnum{2.166}
\figsetgrptitle{Flux of 2216-1605}
\figsetplot{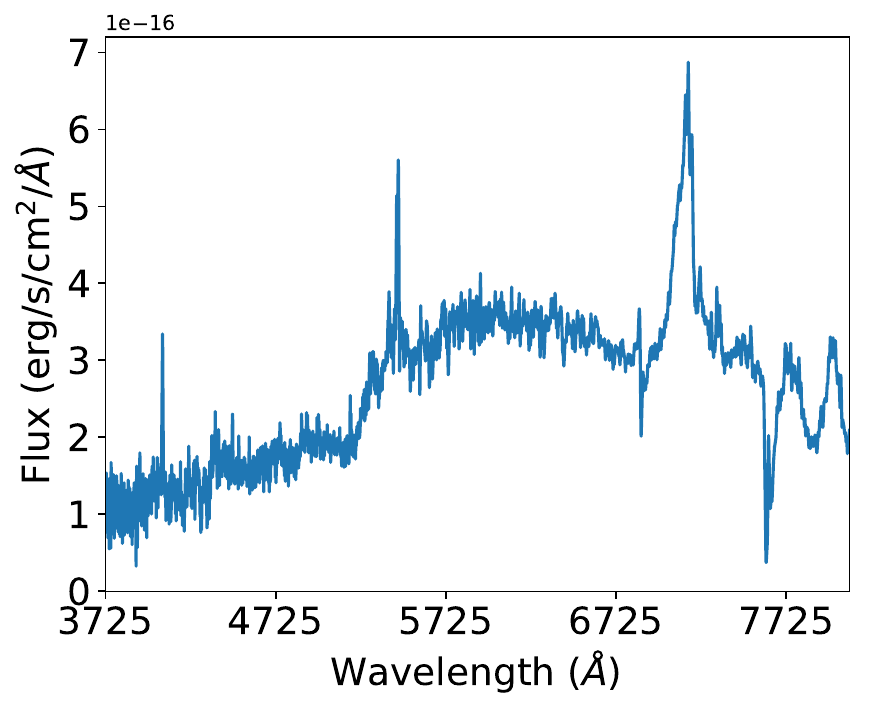}
\figsetgrpnote{Flux for the Lick, Palomar, and Keck spectra. See Table 3 for the observed spectral features and line measurements.}
\figsetgrpend

\figsetgrpstart
\figsetgrpnum{2.167}
\figsetgrptitle{Flux of 2216-0939}
\figsetplot{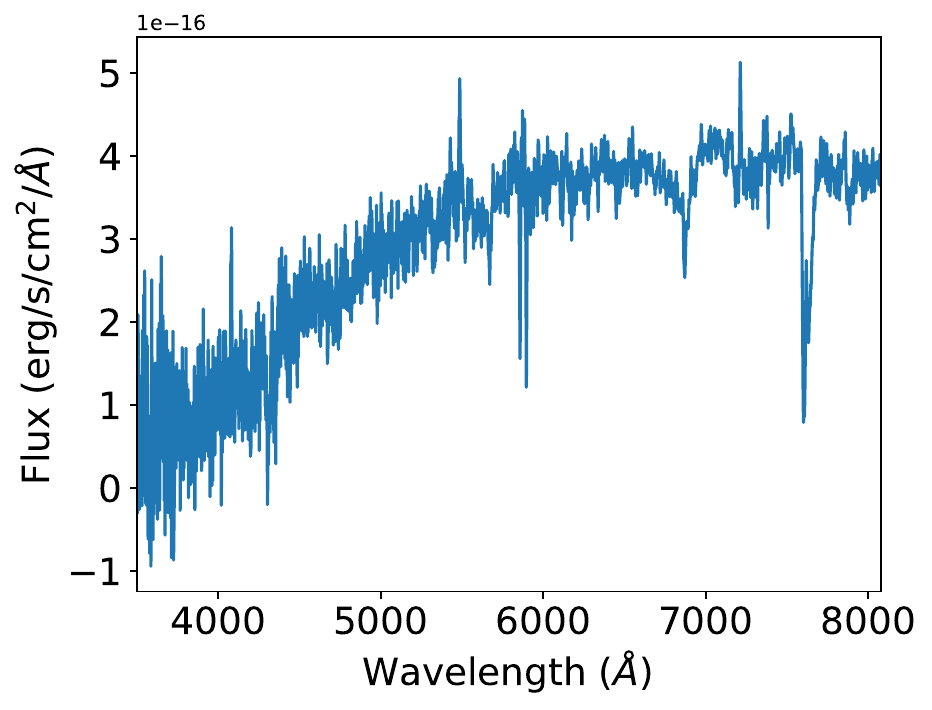}
\figsetgrpnote{Flux for the Lick, Palomar, and Keck spectra. See Table 3 for the observed spectral features and line measurements.}
\figsetgrpend

\figsetgrpstart
\figsetgrpnum{2.168}
\figsetgrptitle{Flux of 2217-0907}
\figsetplot{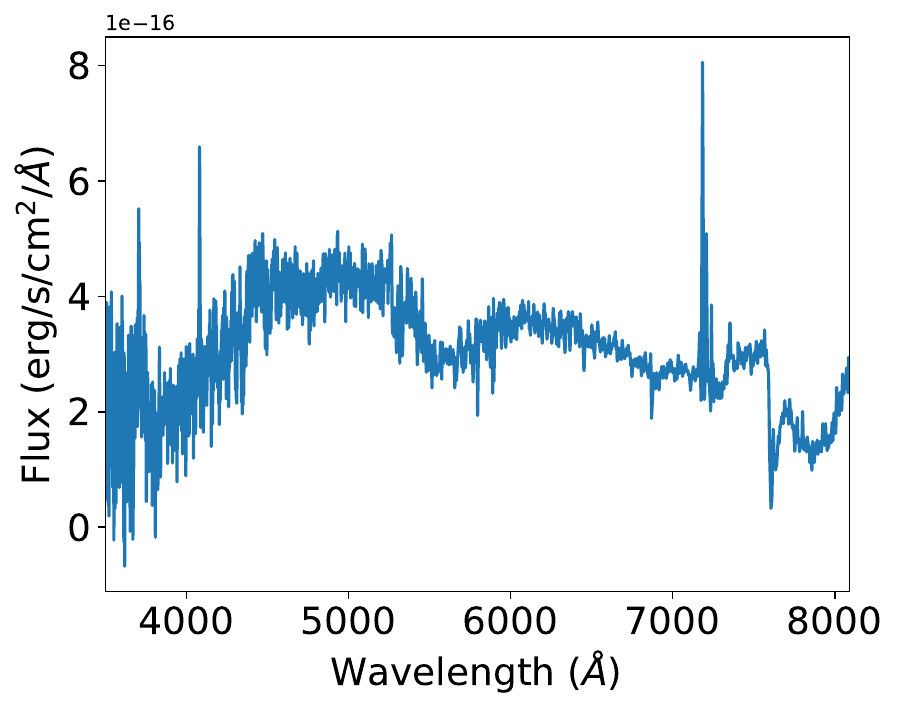}
\figsetgrpnote{Flux for the Lick, Palomar, and Keck spectra. See Table 3 for the observed spectral features and line measurements.}
\figsetgrpend

\figsetgrpstart
\figsetgrpnum{2.169}
\figsetgrptitle{Flux of 2217-1744}
\figsetplot{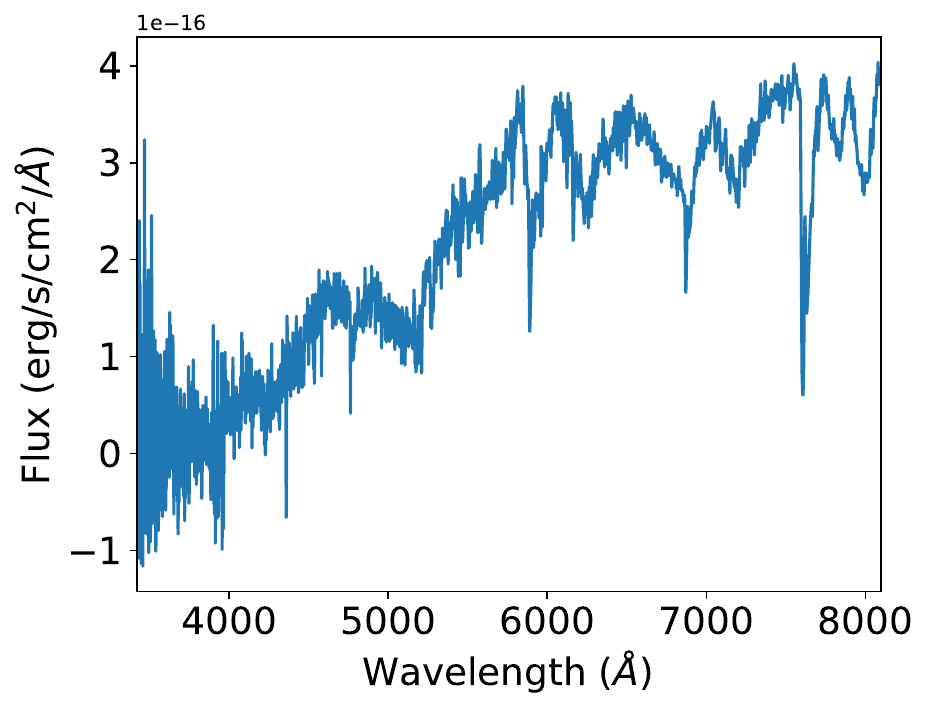}
\figsetgrpnote{Flux for the Lick, Palomar, and Keck spectra. See Table 3 for the observed spectral features and line measurements.}
\figsetgrpend

\figsetgrpstart
\figsetgrpnum{2.170}
\figsetgrptitle{Flux of 2225-0457}
\figsetplot{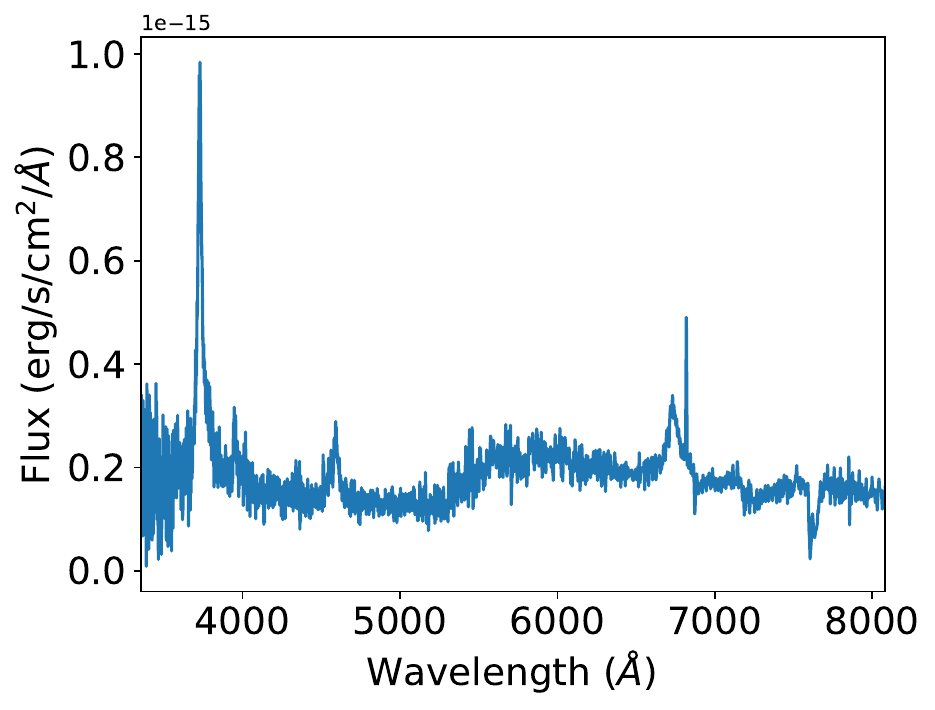}
\figsetgrpnote{Flux for the Lick, Palomar, and Keck spectra. See Table 3 for the observed spectral features and line measurements.}
\figsetgrpend

\figsetgrpstart
\figsetgrpnum{2.171}
\figsetgrptitle{Flux of 2226-0955}
\figsetplot{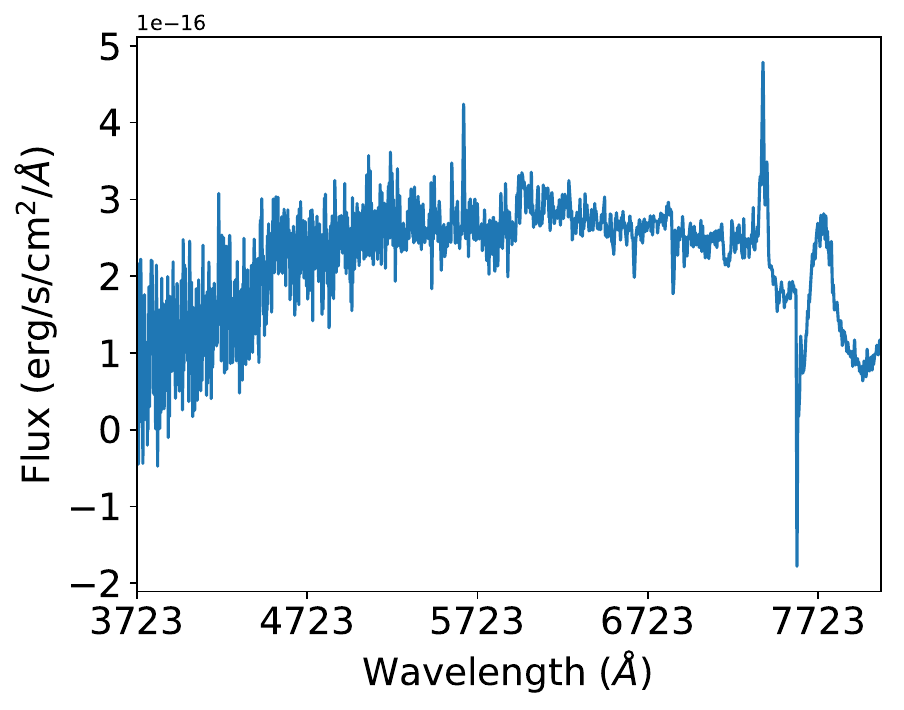}
\figsetgrpnote{Flux for the Lick, Palomar, and Keck spectra. See Table 3 for the observed spectral features and line measurements.}
\figsetgrpend

\figsetgrpstart
\figsetgrpnum{2.172}
\figsetgrptitle{Flux of 2230-0733}
\figsetplot{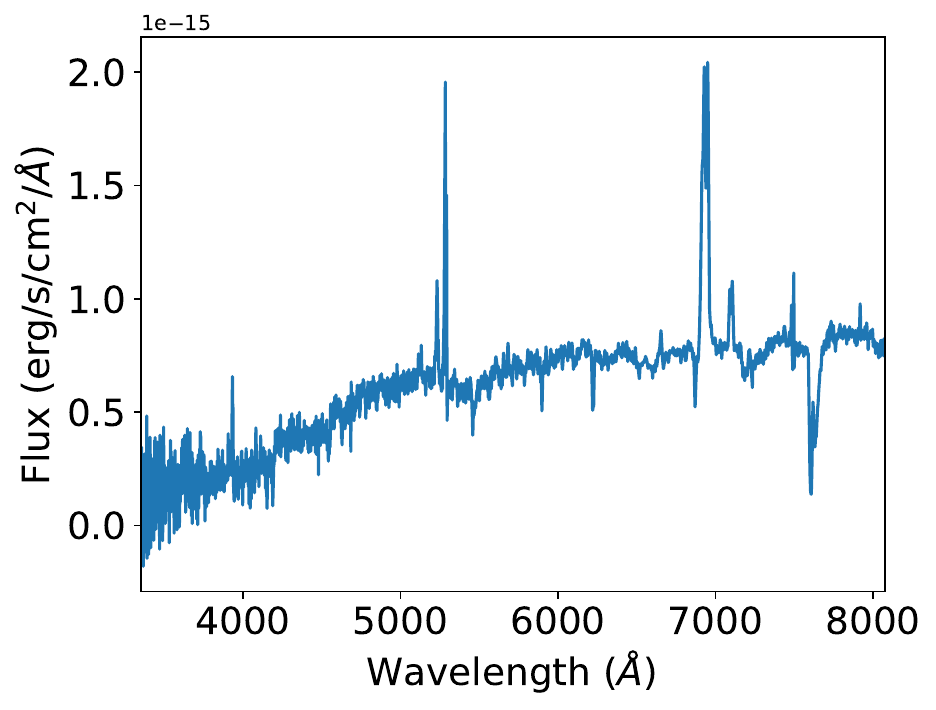}
\figsetgrpnote{Flux for the Lick, Palomar, and Keck spectra. See Table 3 for the observed spectral features and line measurements.}
\figsetgrpend

\figsetgrpstart
\figsetgrpnum{2.173}
\figsetgrptitle{Flux of 2235-0839}
\figsetplot{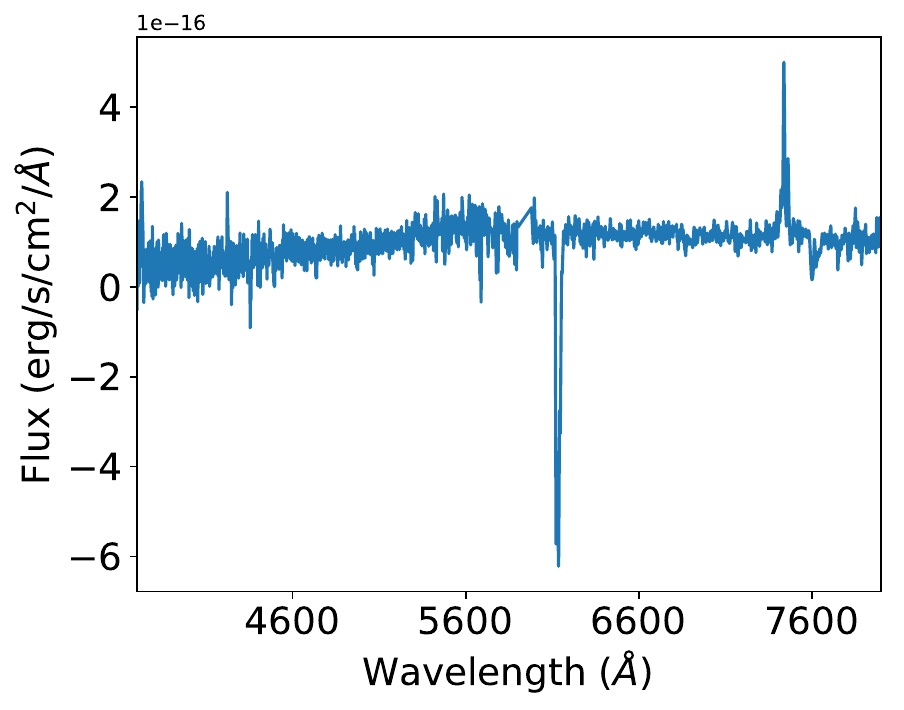}
\figsetgrpnote{Flux for the Lick, Palomar, and Keck spectra. See Table 3 for the observed spectral features and line measurements.}
\figsetgrpend

\figsetgrpstart
\figsetgrpnum{2.174}
\figsetgrptitle{Flux of 2238-0539}
\figsetplot{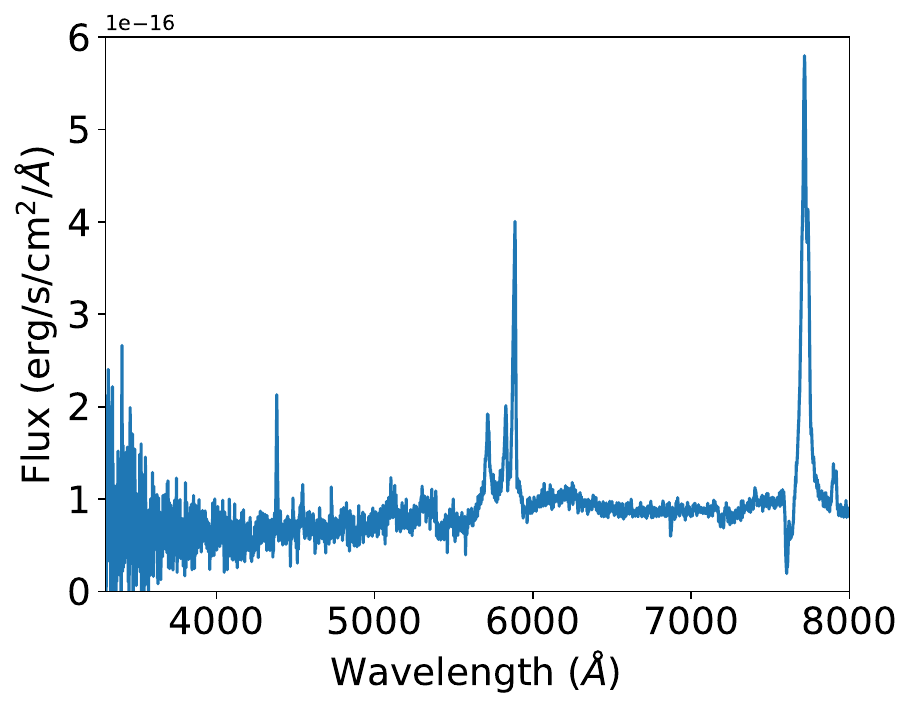}
\figsetgrpnote{Flux for the Lick, Palomar, and Keck spectra. See Table 3 for the observed spectral features and line measurements.}
\figsetgrpend

\figsetgrpstart
\figsetgrpnum{2.175}
\figsetgrptitle{Flux of 2251-1248}
\figsetplot{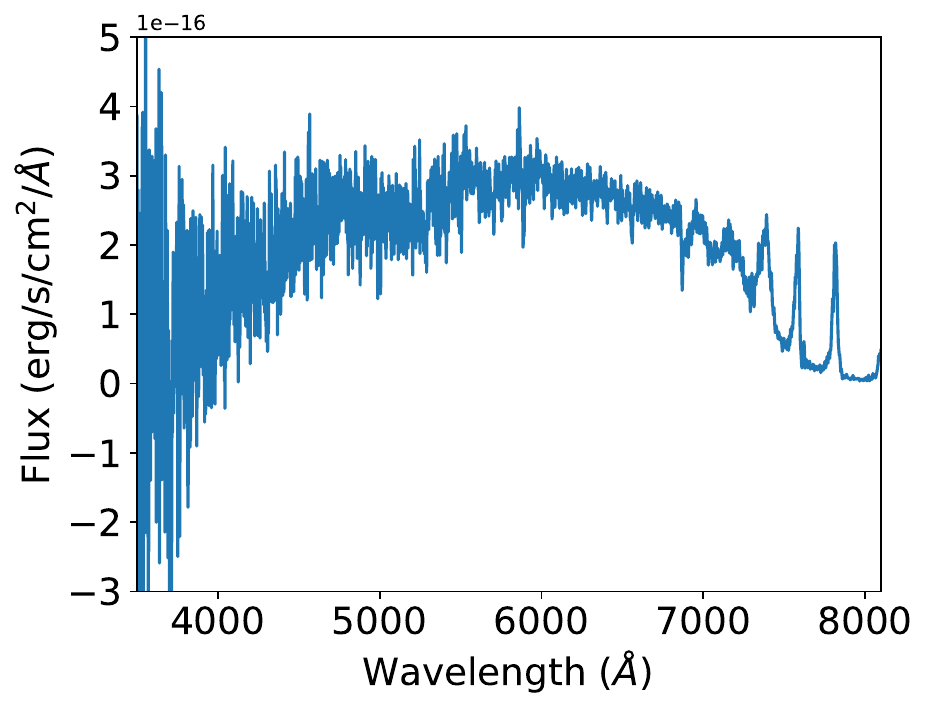}
\figsetgrpnote{Flux for the Lick, Palomar, and Keck spectra. See Table 3 for the observed spectral features and line measurements.}
\figsetgrpend

\figsetgrpstart
\figsetgrpnum{2.176}
\figsetgrptitle{Flux of 2253-0825}
\figsetplot{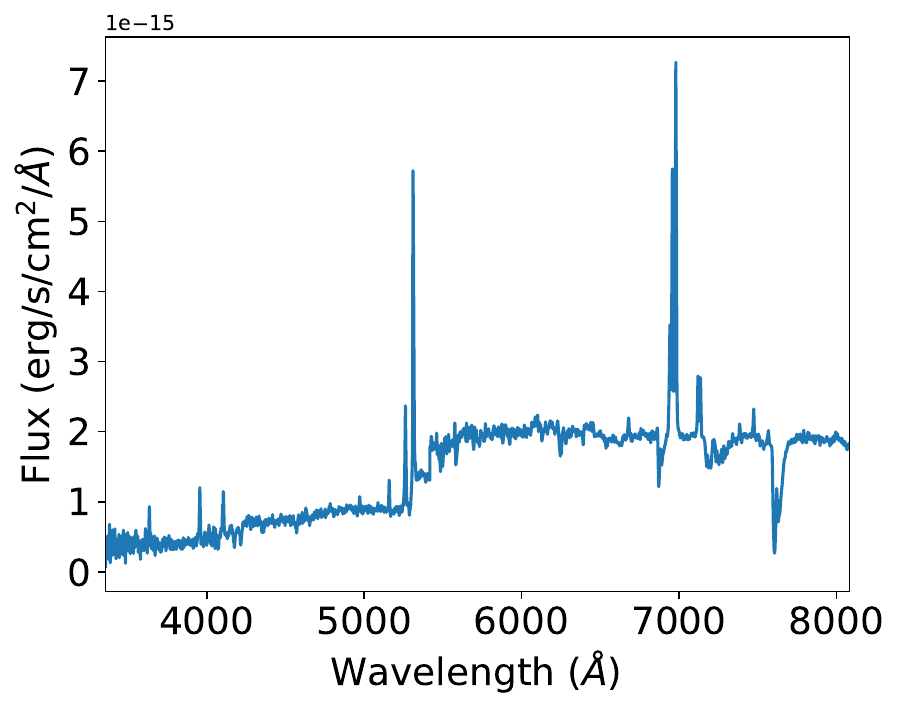}
\figsetgrpnote{Flux for the Lick, Palomar, and Keck spectra. See Table 3 for the observed spectral features and line measurements.}
\figsetgrpend

\figsetgrpstart
\figsetgrpnum{2.177}
\figsetgrptitle{Flux of 2301-0702}
\figsetplot{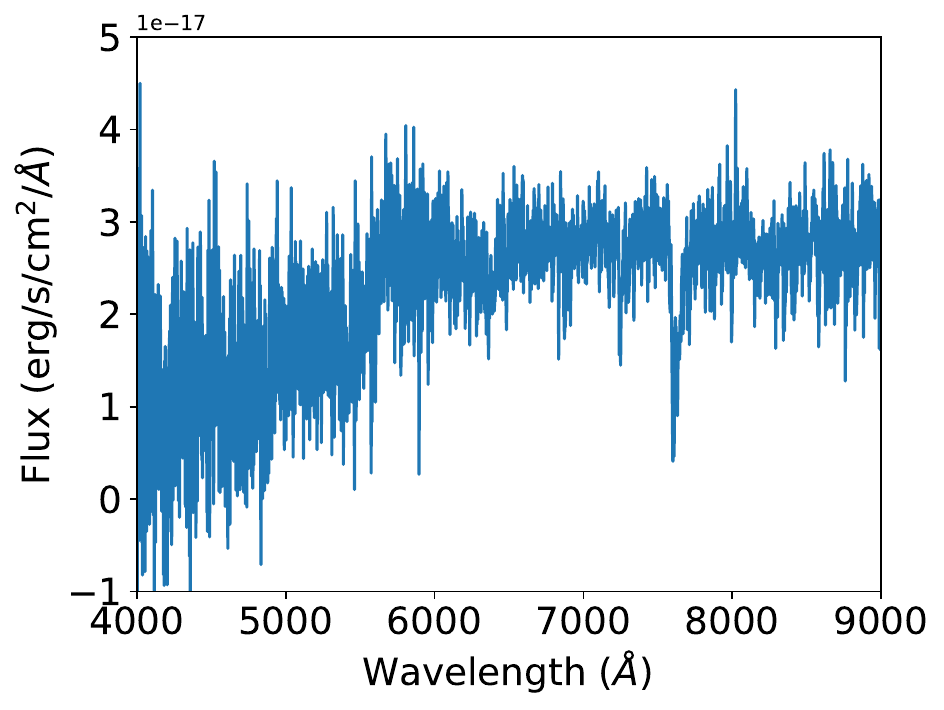}
\figsetgrpnote{Flux for the Lick, Palomar, and Keck spectra. See Table 3 for the observed spectral features and line measurements.}
\figsetgrpend

\figsetgrpstart
\figsetgrpnum{2.178}
\figsetgrptitle{Flux of 2303-0549}
\figsetplot{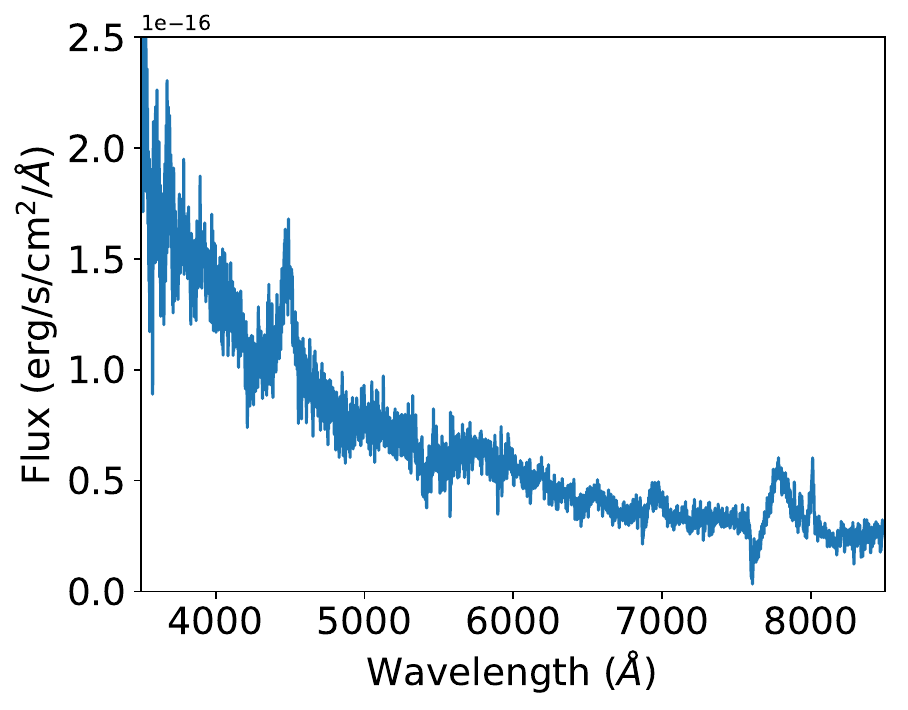}
\figsetgrpnote{Flux for the Lick, Palomar, and Keck spectra. See Table 3 for the observed spectral features and line measurements.}
\figsetgrpend

\figsetgrpstart
\figsetgrpnum{2.179}
\figsetgrptitle{Flux of 2304-0318}
\figsetplot{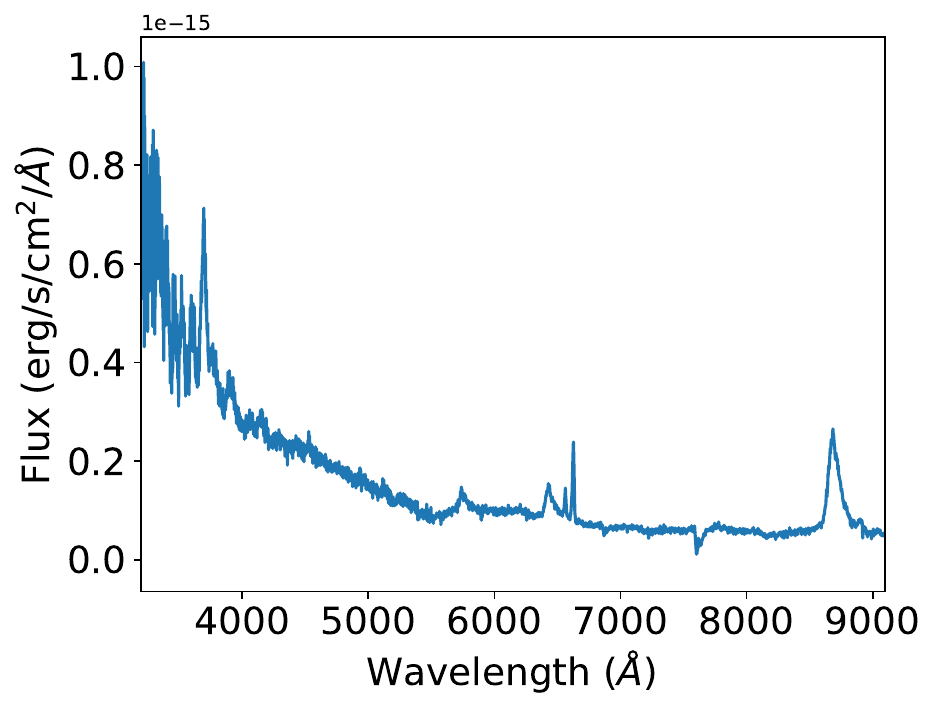}
\figsetgrpnote{Flux for the Lick, Palomar, and Keck spectra. See Table 3 for the observed spectral features and line measurements.}
\figsetgrpend

\figsetgrpstart
\figsetgrpnum{2.180}
\figsetgrptitle{Flux of 2306-0547}
\figsetplot{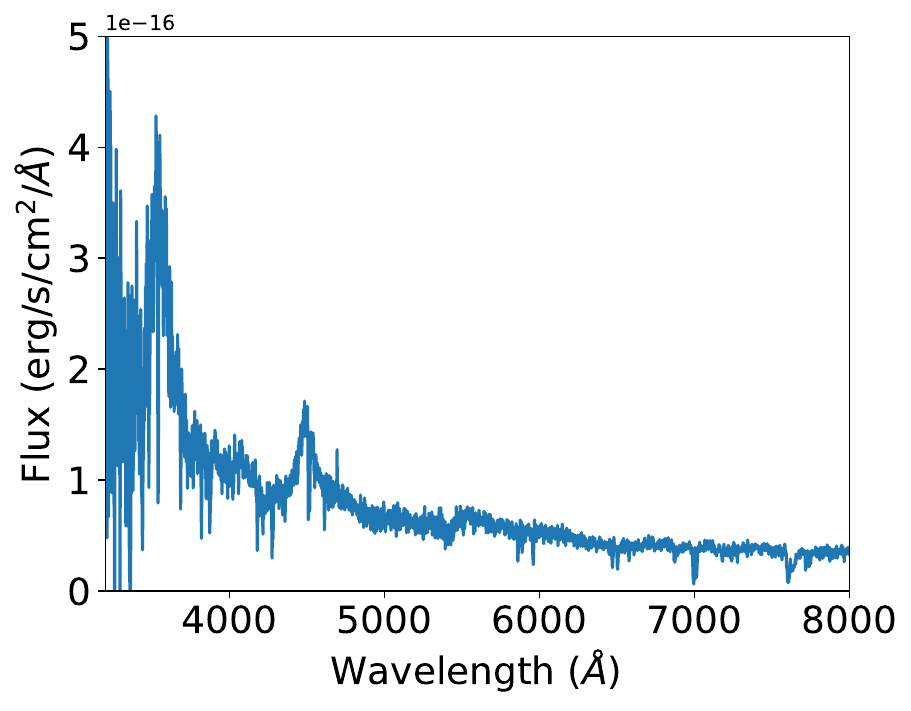}
\figsetgrpnote{Flux for the Lick, Palomar, and Keck spectra. See Table 3 for the observed spectral features and line measurements.}
\figsetgrpend

\figsetgrpstart
\figsetgrpnum{2.181}
\figsetgrptitle{Flux of 2307-0756}
\figsetplot{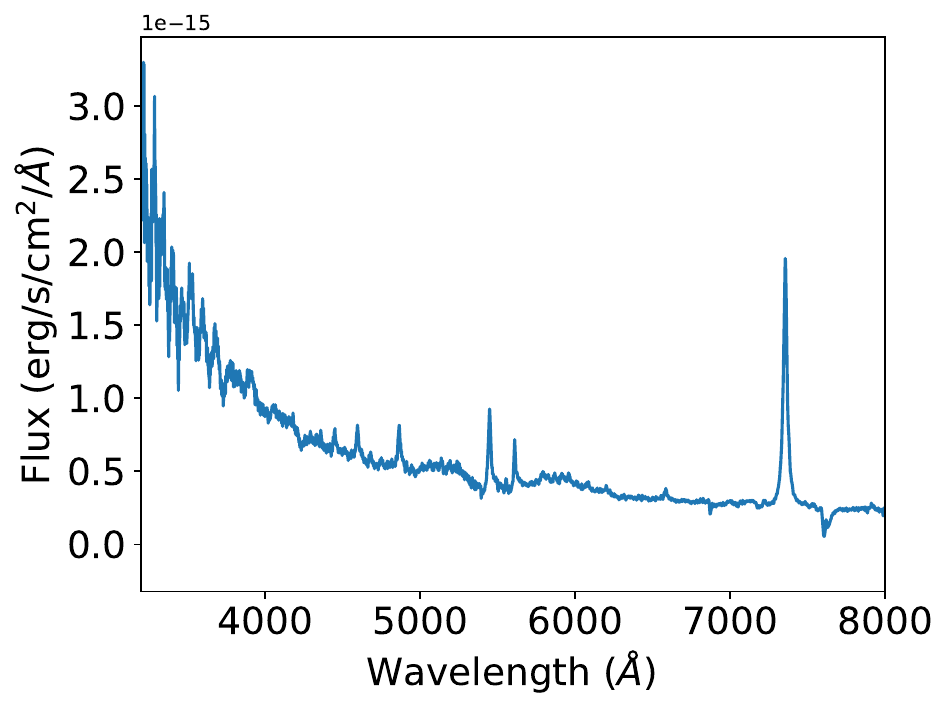}
\figsetgrpnote{Flux for the Lick, Palomar, and Keck spectra. See Table 3 for the observed spectral features and line measurements.}
\figsetgrpend

\figsetgrpstart
\figsetgrpnum{2.182}
\figsetgrptitle{Flux of 2311-0741}
\figsetplot{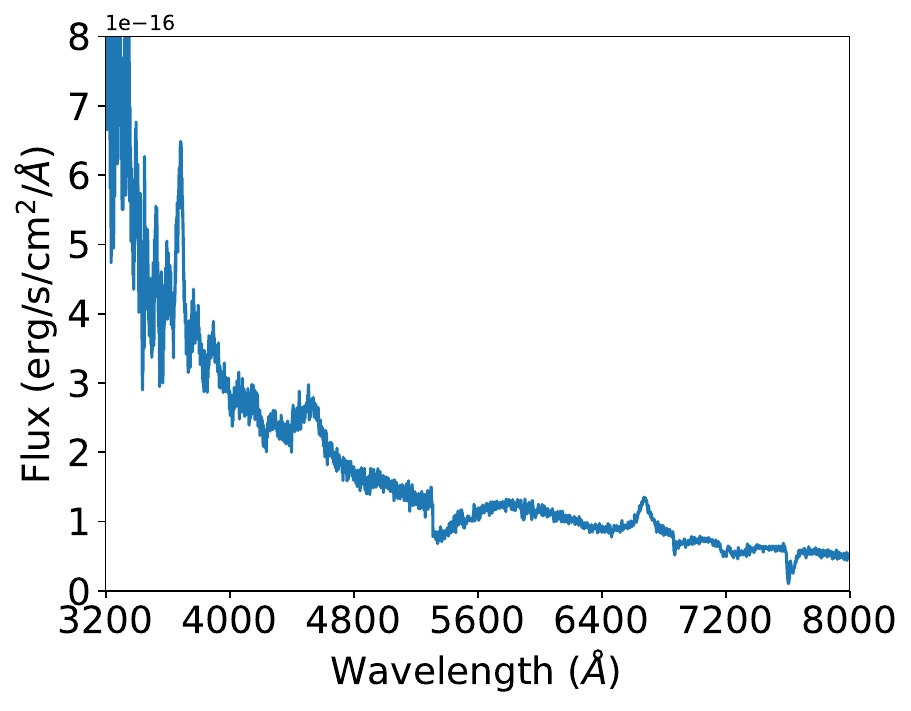}
\figsetgrpnote{Flux for the Lick, Palomar, and Keck spectra. See Table 3 for the observed spectral features and line measurements.}
\figsetgrpend

\figsetgrpstart
\figsetgrpnum{2.183}
\figsetgrptitle{Flux of 2319-0845}
\figsetplot{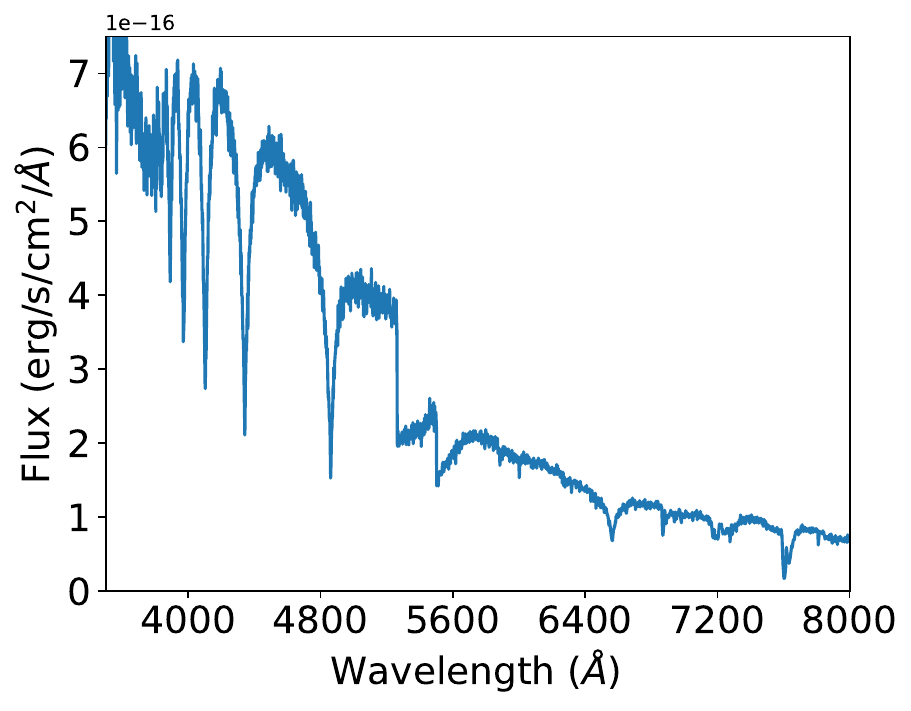}
\figsetgrpnote{Flux for the Lick, Palomar, and Keck spectra. See Table 3 for the observed spectral features and line measurements.}
\figsetgrpend

\figsetgrpstart
\figsetgrpnum{2.184}
\figsetgrptitle{Flux of 2333-0820}
\figsetplot{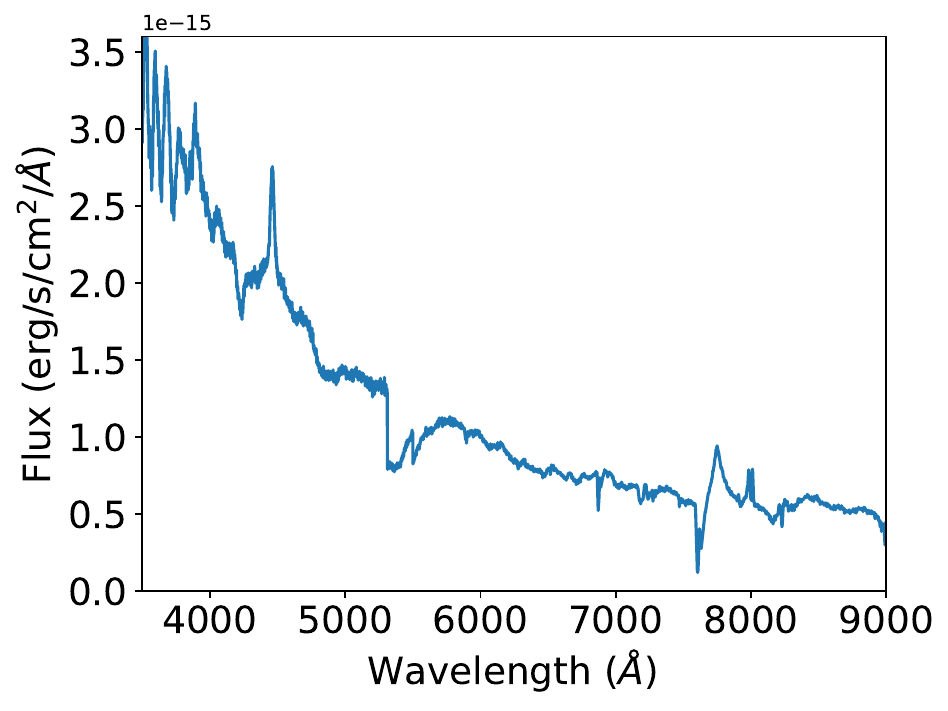}
\figsetgrpnote{Flux for the Lick, Palomar, and Keck spectra. See Table 3 for the observed spectral features and line measurements.}
\figsetgrpend

\figsetgrpstart
\figsetgrpnum{2.185}
\figsetgrptitle{Flux of 2348-0717}
\figsetplot{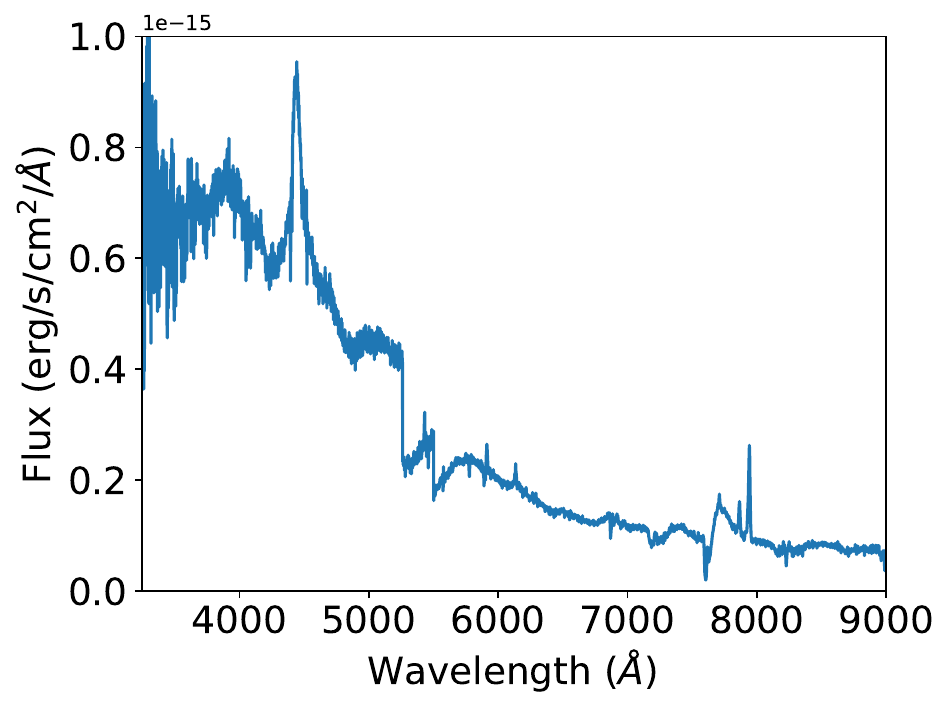}
\figsetgrpnote{Flux for the Lick, Palomar, and Keck spectra. See Table 3 for the observed spectral features and line measurements.}
\figsetgrpend

\figsetgrpstart
\figsetgrpnum{2.186}
\figsetgrptitle{Flux of 2356-0404}
\figsetplot{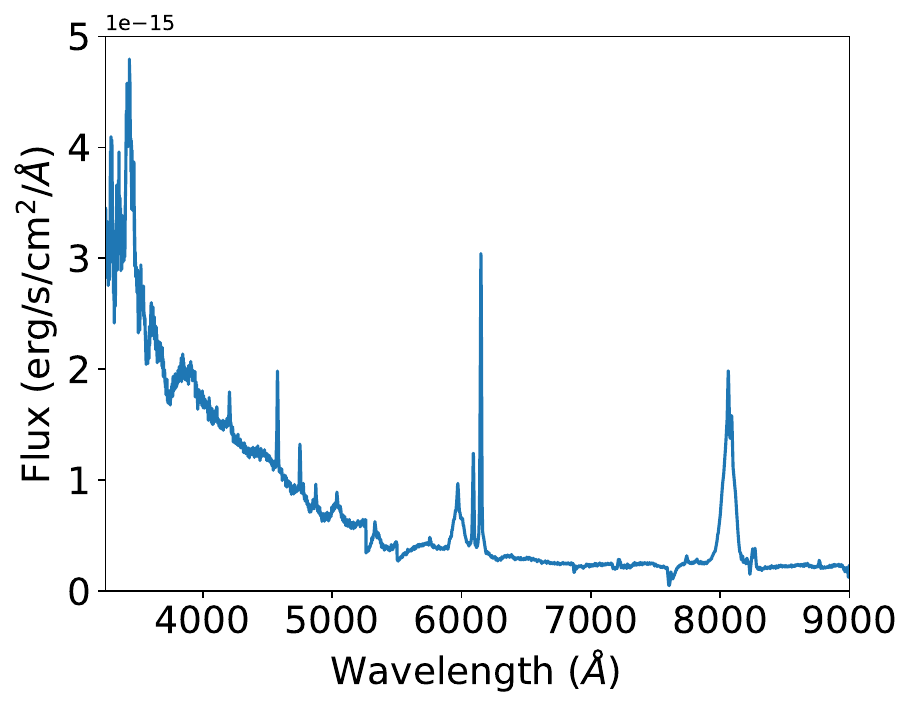}
\figsetgrpnote{Flux for the Lick, Palomar, and Keck spectra. See Table 3 for the observed spectral features and line measurements.}
\figsetgrpend

\figsetend

\section{Reduction \& Measurement}\label{sec:RM}
We reduced the two-dimensional slit spectra using long-slit spectroscopy routines in the Image Reduction and Analysis Facility (IRAF, \citet{IRAF}). First, we subtracted the bias level and flat-fielded the images using a uniformly illuminated dome ceiling to remove the sensitivity variation in the detector. Using the apall routine, we extracted the spectra from the two-dimensional images and computed an averaged sky subtraction by fitting the trace of the spectrum with a polynomial. 

\begin{figure}[hb]
\centering
	\includegraphics[width=20pc, height = 15pc]{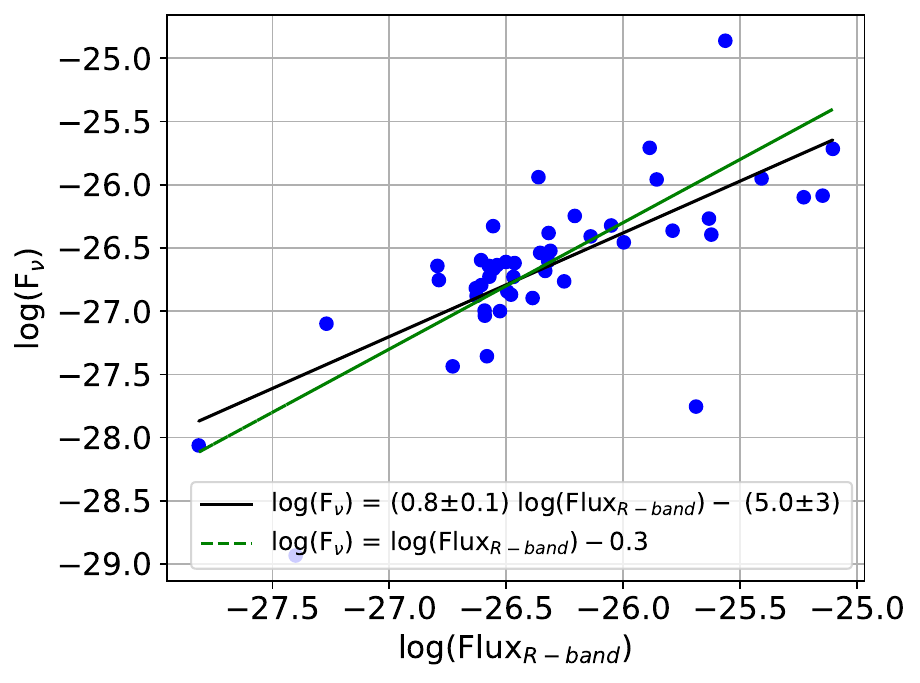}
\caption{Continuum flux. We plotted the continuum flux at 7000 \AA, $F_\nu$ (erg s$^{-1}$ cm$^{-2}$ Hz$^{-1}$), against their $r$-band flux for 47 objects. The black line is the best-fit line, while the green-dashed line is our estimate of the decrement $F_\nu/F_{r-band} = 1/2$. The RMSEs for both lines are 0.45.}
\label{fig:continuum}
\end{figure}


We obtained wavelength calibration using Ne arcs for the red spectra and a combination of Hg, He, Ar, and Cd arcs for the blue. Then, we performed a first-order spline fit with wavelength as a function of pixel numbers to typically 15-20 lamp emission lines. To verify the accuracy of the wavelength scale, we checked the wavelengths and positions of the sky lines. We then compared the observed wavelengths of the extracted sky lines to those in public databases\footnote{Mt. Hamilton:https://mthamilton.ucolick.org/ and Christian Buil's: http://www.astrosurf.com/buil/}. The uncertainties, $\delta z$, in Table \ref{tab:linetable} take into account the spread in the redshifts from the emission lines and the discrepancy in the applied wavelength scale. The redshift error could be as large as 0.003 for a discrepancy larger than 1-2 \AA. We attributed this error to the flexing of the telescope at different observing angles, compared with the zenith position where wavelength calibration arcs were obtained each afternoon.

For flux calibration, we used spectroscopy of the stars Feige 34, Feige 15, and BD+25 3941 to generate a flux scale, following the standard procedure in IRAF. To determine the accuracy of the flux scale, we checked 16 objects that we observed twice on different nights. The fluxes for different observing nights disagreed with each other, with the first night having, on average, 1.5 times more flux than the latter. Hence, we did not have confidence in the absolute flux; however, the relative flux between the blue and red sides was generally consistent. Nevertheless, 19 objects had mismatched fluxes based on the red/blue mismatch continuum at $\lambda \sim 5500 \angstrom $. We noted them in Table \ref{tab:linetable}. To analyze the luminosity, we compared the continuum flux at 7000 \AA, $F_\nu$ (erg s$^{-1}$ cm$^{-2}$ Hz$^{-1}$), with the flux in the $r$-band and fitted a linear line to the data as seen in Figure \ref{fig:continuum} for 47 objects. The best-fit line in black is Log($F_\nu$) = (0.8 $\pm$ 0.1)Log($F_{r-band}) - (5 \pm 3$). Although the line is slightly flatter than a linear relationship, we obtained an equally good fit by forcing the slope to unity (dashed green): Log($F_\nu$) = Log($F_{r-band}) - 0.3$. The spectroscopic continuum is half of the continuum from SDSS imaging. The root-mean-squared error (RMSE) for both lines is 0.4. 

\subsection{Measurement}
\begin{figure}[ht]
\begin{center}
\includegraphics[width=20pc, height = 15pc]{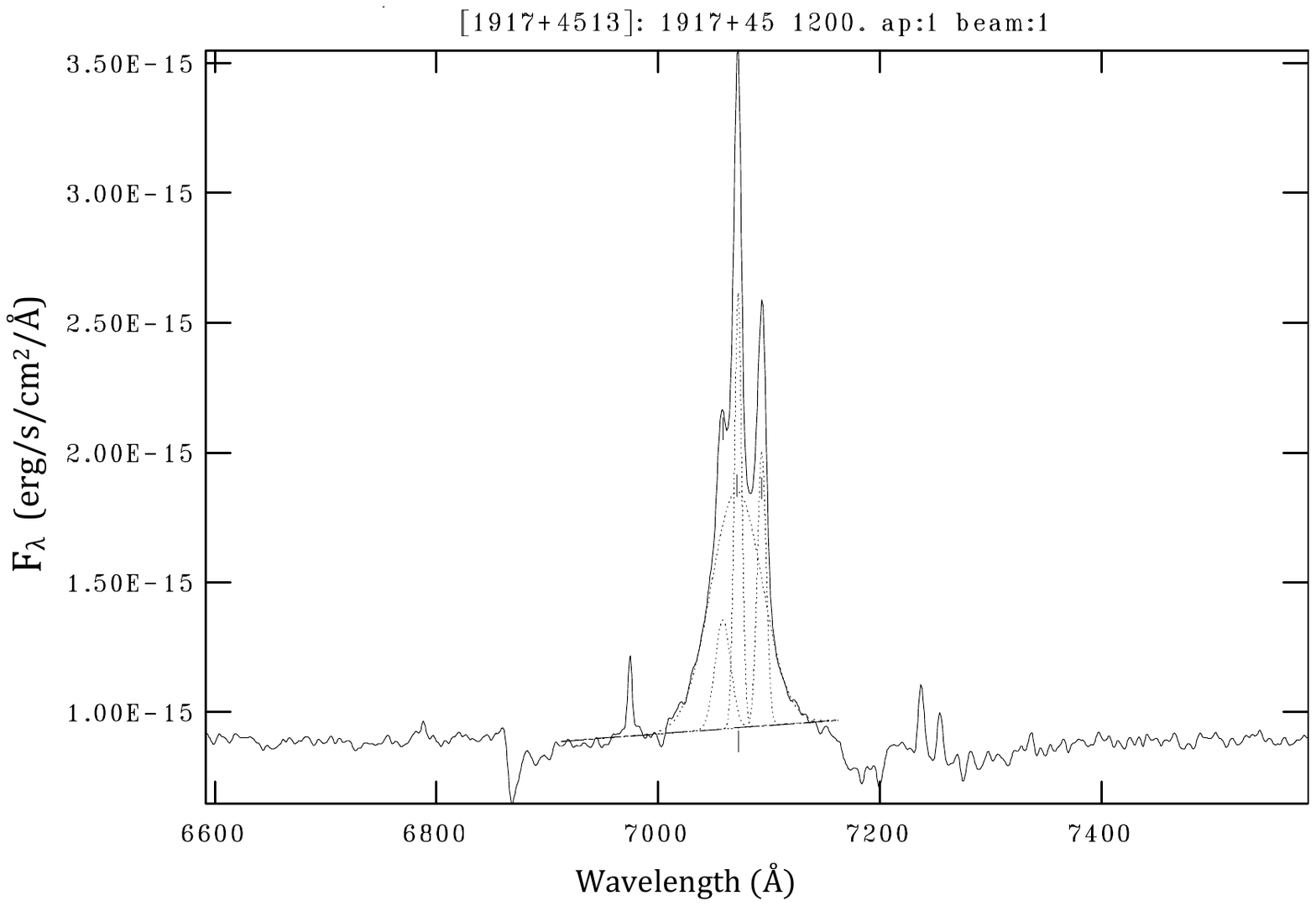}
\end{center}
\caption{Deblending of H$\alpha$ + [NII].The two [NII] 6548 $\angstrom$ and 6584 $\angstrom$ have FWHMs of 703 and 393 km s$^{-1}$, respectively. As the fit (dotted line) illustrates, \textit{splot} tried to fit narrow [NII]-6548; however, it mistook the wing of H$\alpha$ as part of [NII]. Hence, the fit overestimated the actual width. The broad and narrow components of H$\alpha$ have FWHMs of 2318 and 329 km s$^{-1}$, respectively.}	
\label{fig:HaD}
\end{figure}

\begin{figure}[hb]
\begin{center}
\includegraphics[width=20pc, height = 15pc]{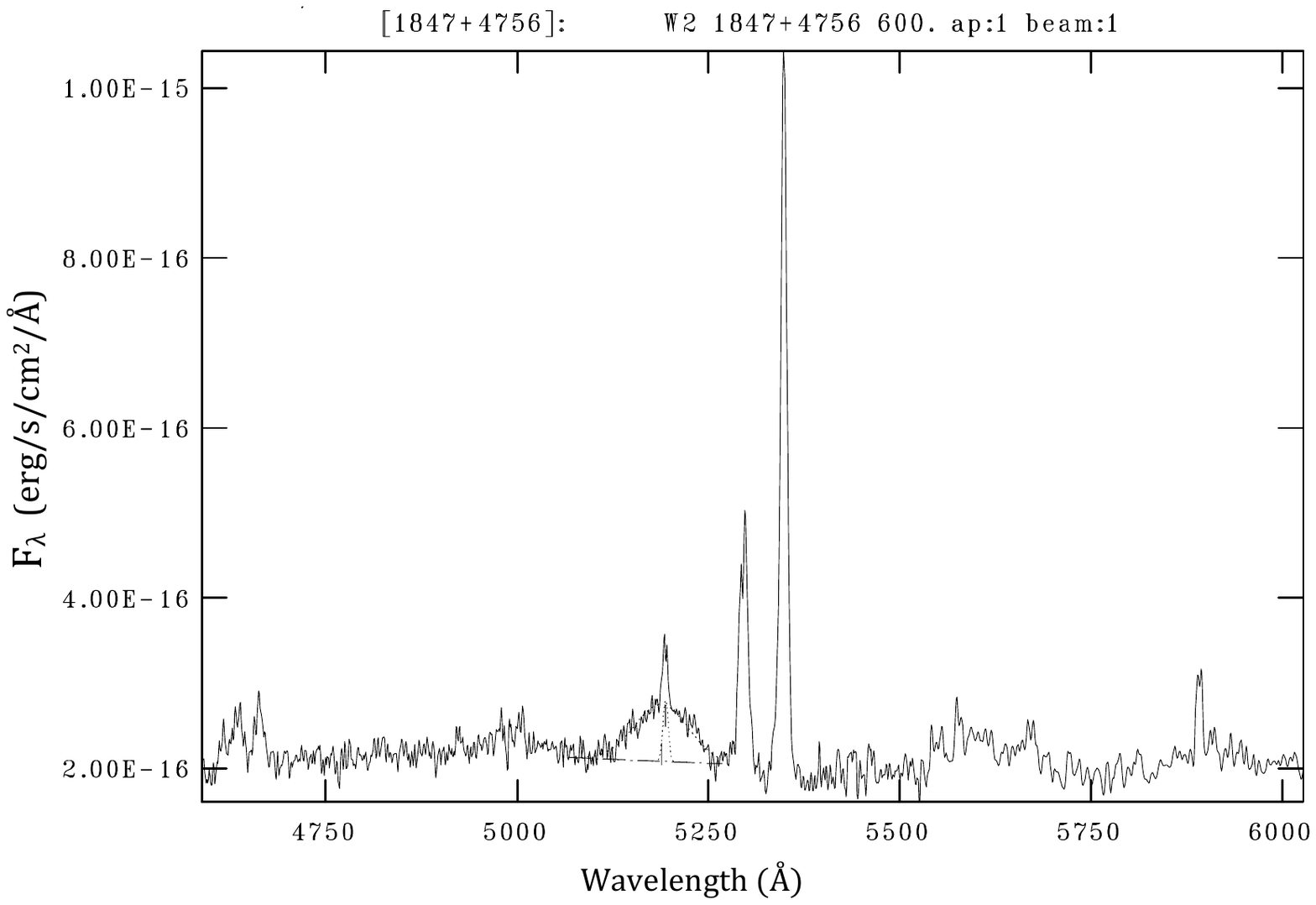}
\end{center}
\caption{Deblending of H$\beta$ in a Seyfert 1.5 galaxy. This candidate is a Seyfert 1.5, with comparable narrow and broad components of H$\beta$. The dotted curves are the fits. The FWHMs for H$\beta$ are 4800 (broad) and 470 (narrow) km s$^{-1}$. [OIII]-5007 has a FWHM of 570 km s$^{-1}$.}	
\label{fig:Sy1half}
\end{figure}

\begin{figure}
\begin{center}
\includegraphics[width=20pc, height = 15pc]{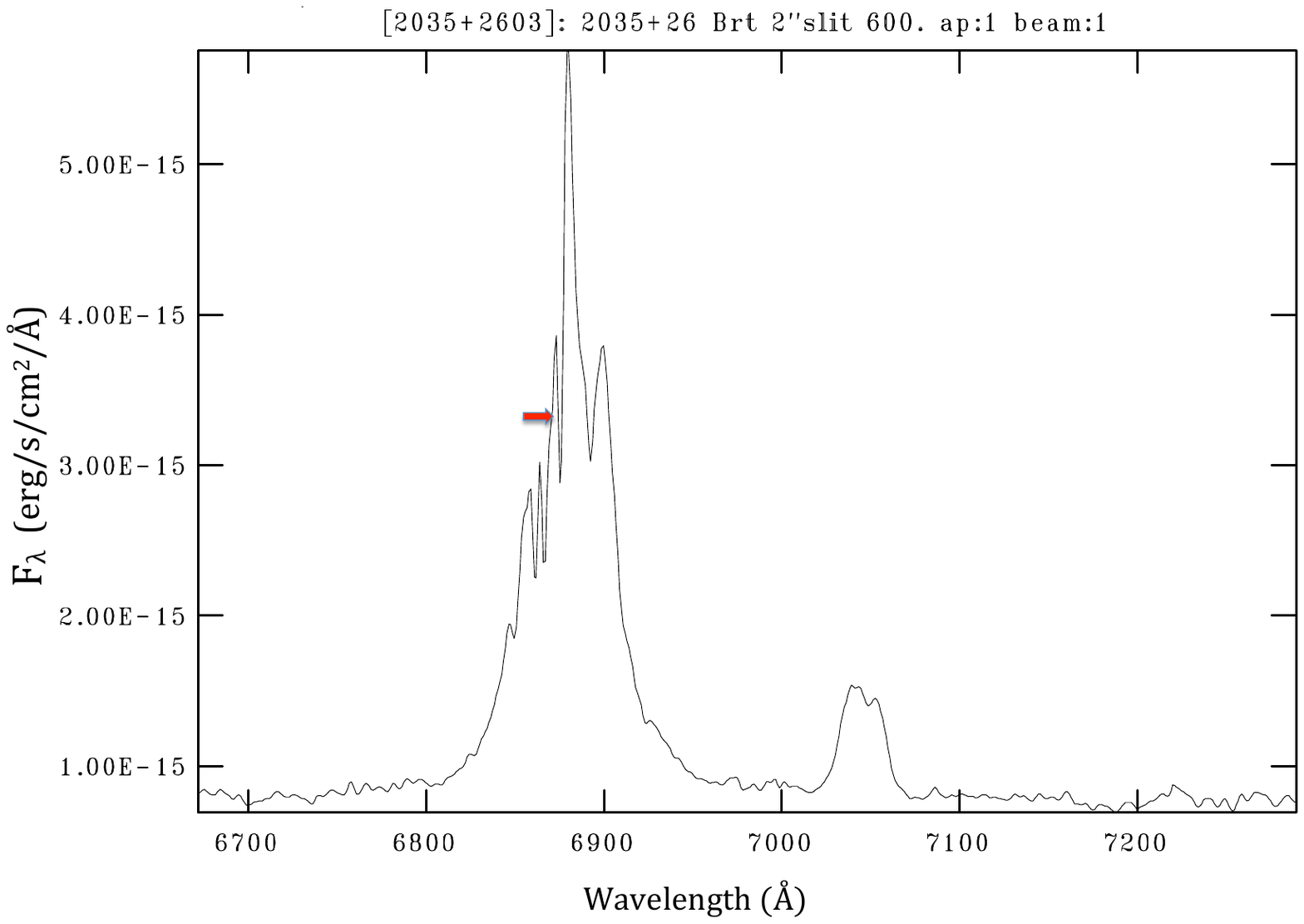}
\end{center}
\caption{An example where H$\alpha$ + [NII] could not be deblended. Due to $B$ band absorption at $\sim$ 6870 \AA, we could not confidently deblend H$\alpha$ from [NII]. We, therefore, estimated the width of H$\alpha$ by eye. However, the half-maximum falls at about 3.3 x 10$^{-16}$ erg s$^{-1}$cm$^{-2}$ \AA$^{-1}$, which is an ambiguous level due to [NII]-6548 absorption.}
\label{fig:HaNot}
\end{figure}

We measured the emission lines using the IRAF routine $splot$, which calculated the line fluxes, equivalent width, and the FWHM. Table \ref{tab:linetable} details these measurements. The most commonly measured lines were H$\beta$, [OIII], H$\alpha$, [NII], and occasionally CIII\big] and MgII. For single-profile fits of lines such as CIII\big], MgII, H$\beta$, and [OIII], we fitted with a single Gaussian. For blended lines such as H$\alpha$ + [NII] and H$\beta$ with broad wings but narrow peaks, we deblended the line complex. Figures \ref{fig:HaD} and \ref{fig:Sy1half} present examples of deblended emission lines. In the few instances that we could not deblend the lines (Figure \ref{fig:HaNot}), we estimated the width at the half-point by eye and noted them in Table \ref{tab:linetable}.

\begin{figure}
\begin{center}
\includegraphics[width=20pc, height = 15pc]{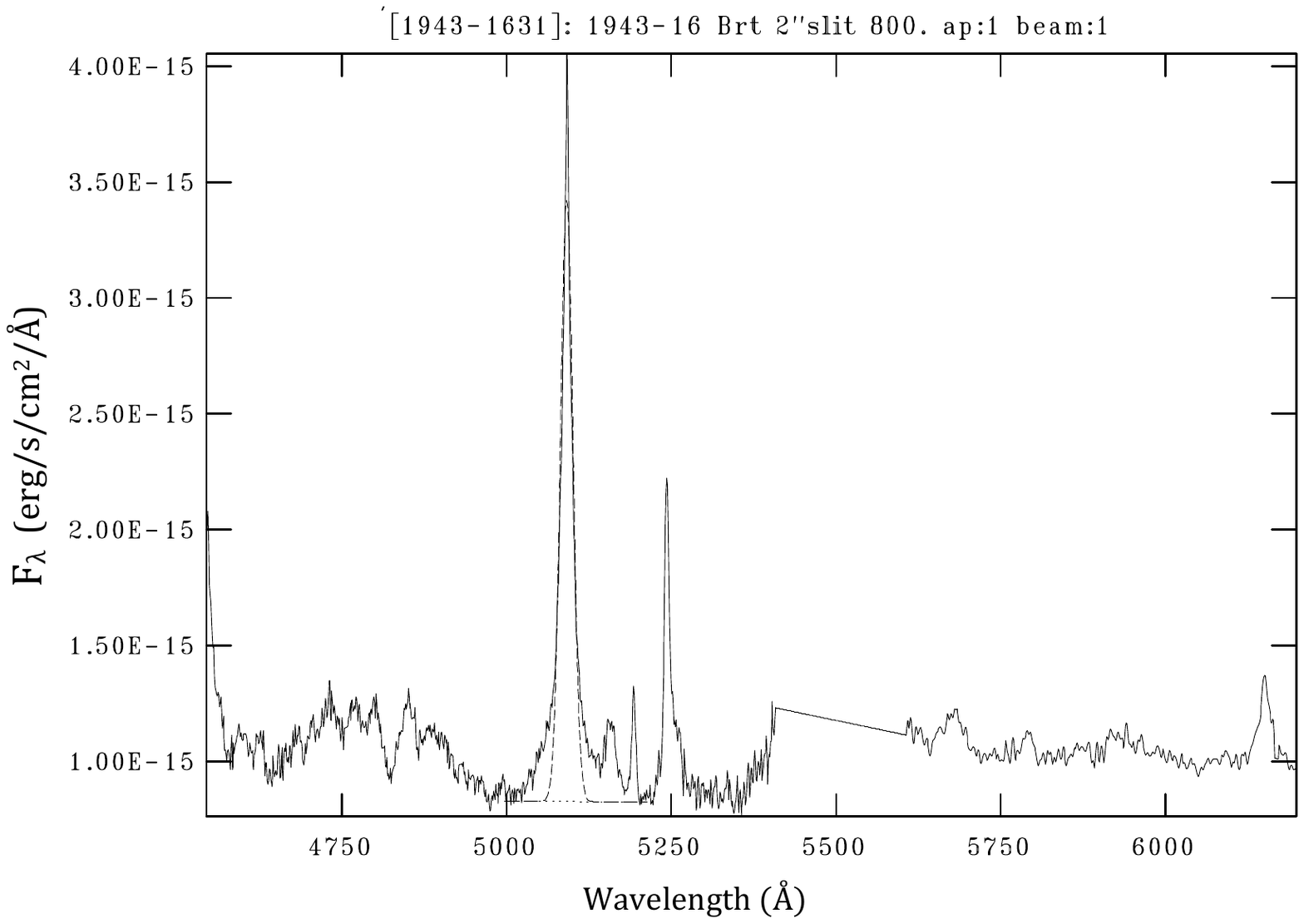}
\end{center}
\caption{A spectrum of a narrow line Type 1 AGN galaxy. H$\beta$ has an FWHM of 1300 km s$^{-1}$ ($\Delta v < 2000$ km s$^{-1}$) and a much broader base than that of [OIII]-5007. Hence, we classified such cases as``narrow-line" Type I AGN. [OIII]-5007's FWHM is 670 km s$^{-1}$.}	
\label{fig:NSy1}
\end{figure}

We converted the total integrated fluxes into luminosity \citep{Wrightcal}. To estimate the error in the fluxes, we tested six cases where we had difficulty separating the continuum level from the broad line wings. For these cases, we picked broad lines such as CIII\big] in 0058+0841, CIV and CIII\big] in 0104+0516, MgII in 0340+1450, and CIV and CIII\big] in 0428+2104, since the uncertainty tends to be larger for these lines. From these six cases, we established that the fluxes have at worst 20 - 27\% uncertainty. This propagated into a 0.2-0.3 logarithmic error for luminosity. Since these cases were on the extreme end, the error determined from them conservatively reflected the accuracy of the integrated fluxes for all the objects. 

To measure the uncertainty for the FWHM, we used two different wavelength centers: ones from \textit{splot} ($\lambda_1$'s) and ones where we judge that the ``true" centers should be ($\lambda_2$'s) (usually the peaks). This error was most prominent in high-z and broad-line objects with sharp peaks and wide wings because sometimes Gaussians could not perfectly fit them. With $\lambda_2$ as the center, we looked at seven cases (the aforementioned six lines and an additional MgII in 0104+0516) and manually estimated the FWHM by eye. Finally, we compared the measurements between the two different FWHMs, which typically indicated a 25\% uncertainty. As mentioned above, these lines had significant uncertain redshifts and FWHMs. We flagged any lines that significantly exceeded this error in Table \ref{tab:linetable}, which could be due to a bad deblend or interference from the atmospheric A-band absorption at $z \sim 0.15 - 0.17$. 
\section{Testing the Reliability of the Kepler AGN Candidate Sample}
\label{sec:sample_reliability}
\begin{table*}
    \centering
    \caption{Observed AGN candidates in Kepler Prime and K2 fields. The sample was selected prior to the Kepler campaigns using the procedure outlined in Section \ref{sec:sample_sel}. This table highlights two main selection techniques: \cite{EM12} (EM12) and \cite{Stern} (S12). In the EM12 column, "W2R" refers to candidates drawn from EM12 WISE/2MASS/ROSAT sample, while "W2" indicates selection based only on WISE/2MASS. The S12 column marks candidates that satisfy the WISE color criterion $W1 - W2 \geq 0.8$ with a ``y". For any Kepler observed candidates, we include their Kepler ID and associated campaign(s). The full table is available online.} \label{tab:logsheet}
    \begin{tabular}{lllllllll}
        \hline
        Name & RA & DEC & Camp & Kepler ID & EM12 & S12 & Type & Obs \\
        \hline
        0039+0322&00 39 48.8&03 22 18.6&8&220334557&&y&Ty 1&Palomar\\
        0042+0730&00 42 25.6&07 30 08.5&8&220532935&&y&Ty 1&Palomar\\
        0047+0831&00 47 27.2&08 31 15.5&8&220576589&&y&Ty 1&Palomar\\
        0052+0526&00 52 08.2&05 26 50.6&8&220437798&&y&Ty 1&Palomar\\
        0058+0841&00 58 36.6&08 41 09.0&8&220584265&&y&Ty 1&Palomar\\
        0058+1145&00 58 44.0&11 45 25.5&8&220710118&&y&Ty 1&Palomar\\
        0104+0516&01 04 38.1&05 16 14.1&8&220428701&&y&Ty 1&Palomar\\
        0109+0724&01 09 40.2&07 24 46.4&8&220528946&&y&Ty 1&Palomar\\
        0128+0505&01 28 26.7&05 05 58.6&&&&&Ty 1&Palomar\\
        0336+1547&03 36 54.4&15 47 45.5&4&210523272&W2R&y&Ty 1&Lick\\
        0337+2335&03 37 01.4&23 35 54.0&4&211039300&W2R&y&Ty 1&Lick\\
        \hline
    \end{tabular}
\end{table*}


Our group proposed for the bulk of AGN observed by Kepler. As discussed in Section~\ref{sec:sample_sel}, this included both known AGN and ``AGN candidates'' that were highly likely to be AGN, but could not be verified before the Kepler observations occurred. The main goal of the spectroscopic observations reported in this paper is to test the reliability of this sample of AGN candidates that Kepler observed: how many are actual AGN, and how many turned out not to be AGN?

For the purposes of this paper, an AGN is defined to be a quasar, Type 1 or 2 Seyfert galaxy, as indicated by ground-based optical spectroscopy. Stating the null hypothesis to be that the target is {\it not} an AGN, a non-spectroscopic test will make a Type I error when an object designated as an AGN Candidate turns out, in fact, not to be an AGN. The reliability of such an identification is then the probability that an object called an AGN Candidate is, in fact, an AGN, that is, that the null hypothesis is rejected when in fact it is false, or (1 - the Type I error rate). In this section, we first address the reliability of the two main techniques used to find AGN candidates, and then extend that test to the full sample.

\subsection{EM12 Reliability}
\begin{table*}[th!]
    \centering
    \caption{AGN candidate sample reliability data.
    Column 1: Sample name.
    Column 2: Total number of candidates in each sample.
    Column 3: Number of (and percentage of total) candidates confirmed as Type 1 AGN.
    Column 4: Same as Column 3 for Type 2 AGN.
    Column 5: Same as Column 3 for all types of AGN. This is simply the sum of Columns 3 and 4.
    Column 6: Same as Column 3 for targets for which spectroscopy indicated they are not AGN.
    Column 7: The fraction of AGN that are Type 2 AGN. This is simply the ratio of Column 4 divided by Column 5.
    } \label{tab:AGN_accuracy}
    \begin{tabular}{lcccccc}
        \hline
        Sample    & Total & Type 1    & Type 2 &  Total AGN & Other     & Type 2 fraction \\
        \hline
        EM12 W2R     &  49 &  44 (90\%) &  3  (6\%) & 47 (96\%) &    2  (4\%) & 6\% \\
        EM12 W2 only &  39 &  20 (51\%) & 14 (36\%) &  34 (87\%) &  5 (13\%) & 41\% \\
        \cite{Stern} & 118 &  87 (74\%) & 19 (16\%) & 106 (90\%) & 12 (10\%) & 18\% \\
        Full         & 186 & 105 (56\%) & 35 (19\%) & 140 (75\%) & 46 (25\%) & 25\% \\
        \hline
    \end{tabular}
\end{table*}

The first sample of AGN candidates we test in this paper is from EM12.
That paper produced two samples of relatively bright ($ J \gtrsim 16 $) AGN candidates: the smaller and more reliable ``W2R'' (WISE/2MASS/ROSAT) sample and the larger and less reliable ``W2'' sample that utilized only WISE and 2MASS data.

The W2R survey utilized infrared data from the WISE and 2MASS all-sky surveys and the ROSAT all-sky X-ray survey to produce a list of 4316 high-likelihood AGN candidates with an estimated reliability of $\sim$95\%. In short, to be identified as an AGN candidate, a target had to show AGN-like infrared colors and associated X-ray emission to pass this test. We used this list to identify AGN candidates within 7 degrees of the boresight center of each planned Kepler prime and K2 pointing; that is, potential AGN that were likely to be observed by Kepler. The final list of observed targets was smaller, due to the loss of two Kepler detectors and gaps between chips, which made some of these targets unobservable.

Forty-nine AGN candidates from the W2R list made this cut and then were spectroscopically observed from the ground as described earlier. (Those targets are listed with ``W2R'' in Column~EM12 of Table~ \ref{tab:logsheet}.) We spectroscopically identified 44 of these as Type~1 AGN, 3 as Type~2 AGN, and 2 as ``other'' (e.g., stars or normal galaxies). That is, a total of 47/49 = 96\% were identified as AGN. Further, of the AGN, only 2/47=6\% were Type 2 AGN, which was not surprising since Type~2 AGN were generally not strong soft X-ray sources. Both of these results were consistent with the findings of EM12, which claimed a $\gtrsim$95\% reliability and only $\sim$5\% Type 2 AGN.

Similarly, the ``W2'' (WISE/2MASS) sample utilized the WISE and 2MASS data, producing a much larger list of 55,050 moderate-likelihood AGN candidates, for which the target only had to show AGN-like infrared colors to be identified as an AGN candidate. (That is, X-ray data were not used for that sample.) However, the reliability of this larger sample was much worse, and not able to be strongly constrained. By the definitions in EM12, all W2R sources are also W2 sources. An additional 39 sources that met all the infrared criteria listed above were included in the Kepler list and follow-up spectroscopy. Note, however, that not all W2 targets were included because their lower assumed reliability made them less persuasive Kepler observation candidates. Of the 39 W2-only targets for which we obtained spectra, 20 were found to be Type~1 AGN, 14 were Type~2 AGN, and 5 were not identified as AGN. That is, 34/39=87\% of the W2 candidates turned out to be AGN. It is interesting that a much larger 14/34=41\% of the AGN were Type~2 AGN, indicating that using infrared colors alone yields a much larger fraction of Type~2 AGN. Again, these results are consistent with the (less well-determined) findings of EM12.

\subsection{Stern et al. Reliability}
\begin{figure}
\begin{center}
	\includegraphics[width=18pc, height = 12pc]{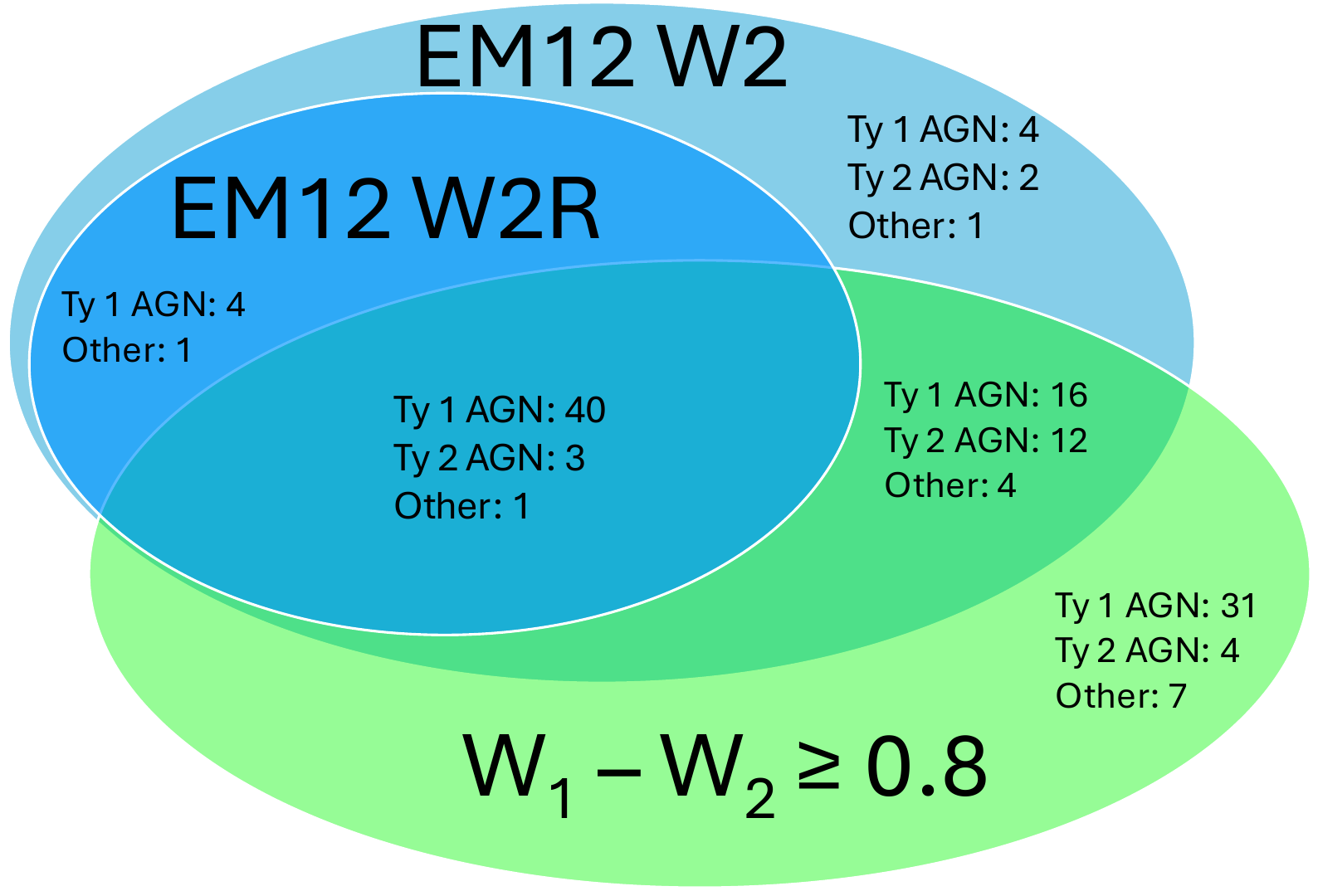}
\end{center}
\caption{Venn diagram of the different AGN selection methods. Our Kepler AGN sample is selected through several methods, as mentioned in Section \ref{sec:sample_sel} and \ref{sec:sample_reliability}. This Venn diagram shows the distribution of the candidates for the EM12 and \cite{Stern} methods: EM12 W2R, EM12 W2, and the WISE color cut. The sample is further divided into Type 1/2 AGN and Other non-AGN, such as stars and normal galaxies.}
\label{fig:VennDiagram}
\end{figure}

The other main source of AGN candidates in this program was derived from \cite{Stern}.
That paper did not provide a list of AGN candidates, but instead a simple infrared test derived from the WISE data: galaxies with $ W1-W2 \geq 0.8 $ were considered to be good AGN candidates. A total of 118 targets in this program satisfied this AGN candidate criterion and are denoted with a ``y'' in Column~S12 of Table~\ref{tab:logsheet}. Of these, our spectroscopy identified 87 as Type~1 AGN, 19 as Type~2 AGN, and 12 as non-AGN. That is, the reliability was 106/118=90\%, and 19/106=18\% of the AGN were Type~2 AGN. This success rate is slightly lower than the 95\% rate reported by \cite{Stern}, but \cite{Hviding2022} found similar reliability. Using a sample of over 400,000 SDSS galaxies, \cite{Hviding2022} observed an 89\% accuracy for identifying AGN and composite galaxies using the same WISE color cut. Between the EM12 and \cite{Stern} methods, a total of 130 AGN candidates were included, due to an overlap of 76 sources that were identified as AGN candidates by both methods. Figure \ref{fig:VennDiagram} shows the distribution of the candidates for these samples.


\subsection{Kepler/K2 AGN sample Reliability}

An additional 56 candidates were identified by a number of other methods. However, none of the subsample sizes were large enough to test reliably. In total, 186 AGN candidates had spectra measured, with 105 being identified as Type~1 AGN, 35 as Type~2 AGN, and 46 as non-AGN. Table \ref{tab:AGN_accuracy} provides a summary of these results. For the sample of Kepler AGN as a whole, 140/186=75\% were AGN, and 35/140=25\% of the AGN were Type~2 AGN. Thus, it is clear that the main AGN candidate identification techniques, EM12 and \cite{Stern}, yielded significantly higher reliability than the other methods did.

\section{Examining The Properties of The AGN Sample}\label{sec:Ana_properties}
\begin{splitdeluxetable*}{lhhlllllllllBllllllllllll}
\tablecaption{Line Measurements for the AGN candidates. ``b" means broad components (FWHM $>$ 1000 km s$^{-1}$), and ``..." means not measured. $L$ is the luminosity (erg s$^{-1}$) in the log scale.}\label{tab:linetable}
\tablewidth{0pt}
\tablehead{
\colhead{Name} & \nocolhead{z} & \nocolhead{Spectral Feature} & \colhead{$\rm log(L_{bCIV})$} & \colhead{$ \rm EW_{bCIV}$} & \colhead{$\rm FWHM_{bCIV}$} & \colhead{$\rm log(L_{bCIII\big]})$} & \colhead{$\rm EW_{bCIII\big]}$} & \colhead{$ \rm FWHM_{bCIII\big]}$} & \colhead{$\rm log(L_{bMgII})$} & \colhead{$ \rm EW_{bMgII}$} & \colhead{$ \rm FWHM_{bMgII}$} & \colhead{$ \rm log(L_{bH\beta})$} & \colhead{$ \rm EW_{bH\beta}$} & \colhead{$ \rm FWHM_{bH\beta}$} & \colhead{$ \rm log(L_{H\beta})$} & \colhead{$ \rm EW_{H\beta}$} & \colhead{$ \rm FWHM_{H\beta}$} & \colhead{$ \rm log(L_{bH\alpha}$)} & \colhead{$\rm EW_{bH\alpha}$} & \colhead{$\rm FWHM_{bH\alpha}$} & \colhead{$ \rm log(L_{H\alpha})$} & \colhead{$ \rm EW_{H\alpha}$} & \colhead{$ \rm FWHM_{H\alpha}$} \\ 
\colhead{} & \nocolhead{}& \nocolhead{} & \colhead{} &\colhead{(\AA)} & \colhead{(km s$^{-1}$)} & \colhead{} & \colhead{(\AA)} & \colhead{(km s$^{-1}$)}& \colhead{} & \colhead{(\AA)} & \colhead{(km s$^{-1}$)} & \colhead{} & \colhead{(\AA)} & \colhead{(km s$^{-1}$)} & \colhead{} & \colhead{(\AA)} & \colhead{(km s$^{-1}$)} & \colhead{} & \colhead{(\AA)} & \colhead{(km s$^{-1}$)}& \colhead{} & \colhead{(\AA)} & \colhead{(km s$^{-1}$)} 
}
\startdata
0042+0730&$1.3880 \pm 0.0004$&bCIV, bCIII\big], bMgII, OII&44.6&130.4&3630&43.8&39.9&4400&43.3&29.9&2020&...&...&...&...&...&...&...&...&...&...&...&...\\
0047+0831&$0.570 \pm 0.002 $&bMgII, [OII], bH$\delta$, bH$\gamma$, [OIII], [NII], bH$\alpha$&...&...&...&...&...&...&43.2&39.9&3120&...&...&...&...&...&...&42.8&227.5&2020&...&...&...\\
0058+1145&$1.260 \pm 0.001$&bCIV, bCIII\big], bMgII, [OII]&44.7&48.7&6590&44.1&22.8&5490&44.0&56.4&5040&...&...&...&...&...&...&...&...&...&...&...&...\\
0104+0516&$1.544 \pm 0.003$&bCIV, bCIII\big], bMgII&44.7&63.4&7140&44.2&32.7&6530&43.9&53.7&3930&...&...&...&...&...&...&...&...&...&...&...&...\\
0128+0505&$2.013 \pm 0.005$&bLy$\alpha$, bCIV, bCIII\big], bMgII&44.6&93.4&6660&44.5&129.8&10800&44.2&108.0&6840&...&...&...&...&...&...&...&...&...&...&...&...\\
0510+1630&$0.0180 \pm 0.0007$&[OII], bH$\gamma$, bH$\beta$, [OIII], bH$\alpha$, [NII] &...&...&...&...&...&...&...&...&...&40.5&96.9&3690&39.8&20.1&350&41.6&579.2&3280&40.6&72.3&360\\
\enddata
\end{splitdeluxetable*}

This section outlines the classification of Type 1 AGN into various subcategories (Section \ref{subsec:class}) and examines the properties of the AGN sample through the BPT diagram (Section \ref{subsec:BPT}) and their spectral energy distributions (Section \ref{subsec:SEDs}). Table \ref{tab:linetable} provides photometric and line measurements of all data. Details on the measurement uncertainties appear in Section \ref{sec:RM}. 

\subsection{Type 1 AGN Sub-Classification} \label{subsec:class}
Among the 105 Type 1 AGN, 16 spectra fall under the Seyfert 1.5 category due to the broad and narrow components of H$\beta$ of comparable strength as seen in Figure \ref{fig:Sy1half}. Another 11 spectra are Narrow Line Type 1 AGN, characterized by a slight wing at the base of a narrow H$\beta$, exemplified in Figure \ref{fig:NSy1}. We established any objects that have a $\Delta v (FWHM)_{H\beta} < 2000$ km s$^{-1}$ and a broader H$\beta$'s base compared to [OIII] to be narrow Type 1 AGN. One AGN exhibits a slight wing in H$\alpha$ thus, it falls under the Seyfert 1.9 category. Two candidates, 0339+1306 and 1930+3759, display a featureless continuum with a good signal-to-noise ratio, suggesting they are likely BL Lac objects.

\subsection{BPT: Line Ratio Diagram} \label{subsec:BPT}
\begin{figure}
\begin{center}
\includegraphics[width=20pc, height = 15pc]{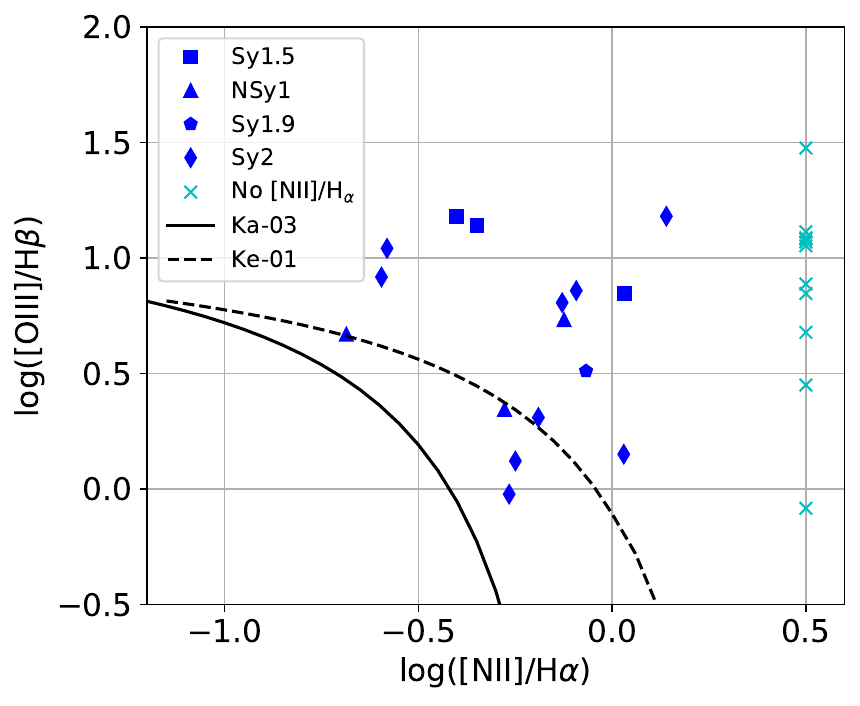}
\end{center}
\caption{BPT diagram of log([OIII]/H$\beta$) vs log([NII]/H$\alpha$). This graph displays 16 galaxies with narrow components in H$\beta$, [OIII]-5007, H$\alpha$, and [NII]-6584. The triangle indicates a narrow Type 1 AGN, in which we tried to fit the narrow components. 4 out of these 16 objects (25\%) are composite galaxies (between Ka03 and Ke01 lines) with major narrow line emissions from HII regions. The rest of the galaxies above the Ke01 line (75\%) are in the AGN region. Another 11 objects, marked by the crosses, are either at high redshifts where we lost H$\alpha$+[NII] past the red end of the spectra or we could not get a good fit of H$\alpha$+[NII].}	
\label{fig:bpt}
\end{figure}

In the BPT diagram (Figure \ref{fig:bpt}), we plotted 16 galaxies with narrow-line flux ratios against the theoretical boundaries between AGN and normal star-forming galaxies as defined:
\begin{equation}
\label{eq:ka03}
	\log(\rm[OIII]/H\beta) > \frac{0.61}{\log([NII]/H\alpha) - 0.05} + 1.3
\end{equation}
\begin{equation}
\label{eq:ke01}
	\log(\rm[OIII]/H\beta) > \frac{0.61}{\log([NII]/H\alpha) - 0.47} + 1.19
\end{equation}
where (\ref{eq:ka03}) and (\ref{eq:ke01}) are from \citet{ka} and \cite{ke01}, respectively. Objects above the Ke01 line are in the AGN regime, while the ones that fell in between the Ke01 and Ka03 are ``composite" galaxies. These galaxies have both star-forming regions and AGN \citep{ke06}. Four out of 16 (25\%) appear to be ``composite" galaxies with significant narrow line emission from HII regions. Anything below the Ka03 line is are star-forming galaxy. In the plot, we have narrow Type 1 AGN galaxies (triangle markers), for which we tried to fit the narrow component of H$\beta$. However, the broad region at the base still slightly influenced the fits. For objects at high redshifts where we could not observe H$\alpha$+[NII] or those Seyfert 1.5 galaxies where we could not extract a good fit for the narrow component of H$\alpha$ and [NII], we measured their [OIII] to H$\beta$ ratios and included these galaxies on the right-hand side of Figure \ref{fig:bpt}, all at an arbitrary log([NII]/H$\alpha$) value of 0.5. These spectra were consistent with AGN-dominated spectra.

\subsection{Type 1 AGN Spectral Energy Distributions (SEDs)}
\label{subsec:SEDs}
\begin{figure}
\begin{center}
	\includegraphics[width=20pc, height = 15pc]{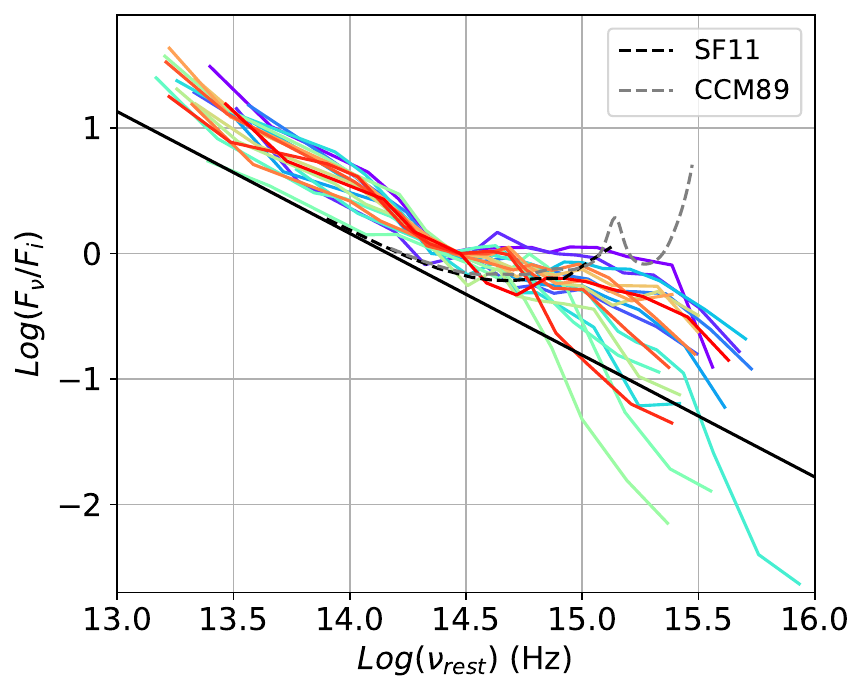}
\end{center}
\caption{SEDs for Type 1 AGN with WISE/2MASS \citep{WISE, 2MASS}, GALEX (\dataset[10.17909/25nx-tp05]{http://dx.doi.org/10.17909/25nx-tp05}), and SDSS photometry. 19 Type 1 AGN have photometry from WISE/2MASS, GALEX and SDSS. SDSS photometry is from DR14 Skyserver database. The flux, $F_\nu$ (erg s$^{-1}$ cm$^{-2}$ Hz$^{-1}$) for W1 - W4, J, H, Ks, has been normalized by the corresponding $F_i$ and plotted as a function of the shifted rest frequency. The black line is -0.97log($\nu_{rest}$) + 13.74, a typical standard slope of a quasar spectrum averaged over a wide wavelength baseline. We then de-redden this black line power law using \cite{SF11}'s extinction coefficients (black dashed line) and \cite{CCM89} model (grey dashed line) for $R_V = 3.1$.}
\label{fig:SEDWGS}
\end{figure}

\begin{figure}
\begin{center}
	\includegraphics[width=20pc, height = 15pc]{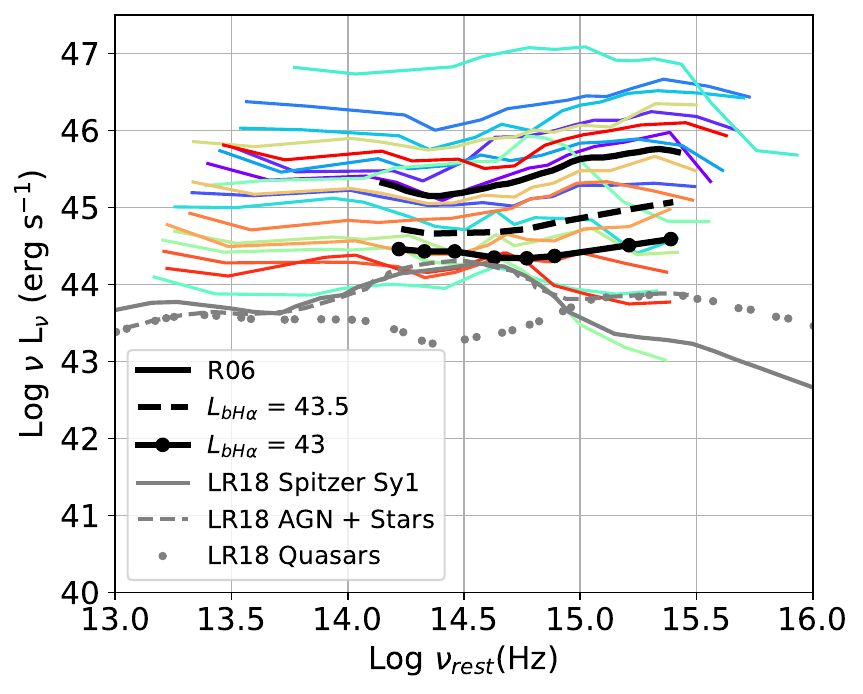}
\end{center}
\caption{SED of Kepler AGNs against different templates. The black lines were taken from \cite{SL12Fig11} Figure 11. These lines are the mean SEDs with different L$_{bH\alpha}$ bins, with the R06 line being the net AGN SED. The grey line is the composite SED of Spitzer AGNs, while the dashed grey line is the AGN with stellar contribution template from \cite{LR18}. The grey dots are the template for normal quasars that \cite{LR18} has taken from \cite{Xu}. In the mid-IR (log $\nu_{rest} \leq 14$), the plot is independent of luminosity, while there is a characteristic blue bump in the UV for very luminous quasars.}
\label{fig:SEDSL12LR18}
\end{figure}

We present the multi-band spectral energy distributions (SEDs) of Type 1 AGN, excluding Seyfert 1.5 and Narrow Line Type 1, that extend from 22 $\micron$ to the ultraviolet (UV) in Figure \ref{fig:SEDWGS} and \ref{fig:SEDSL12LR18}. Given the larger sample of Type 1 AGN and their characteristically stronger flux and variability, we focus our studies on these instead of the smaller population of Type 2 AGN. The SED in Figure \ref{fig:SEDWGS} plots the normalized flux photometry ($F_\nu/ F_i$) from WISE/2MASS, GALEX, and SDSS as a function of the rest frequency. $F_i$ (erg s$^{-1}$ cm$^{-2}$ Hz$^{-1}$) is the flux of the $i^{th}$ candidate at a particular band. The normalization follows the subsequent scheme: J band for z $\leq$ 0.25, H band for 0.25 $ < $ z $ \leq $ 0.6, K band for 0.6 $ < $ z $ \leq $ 1.4, and W1 band for z $ > $1.4. 19 spectra have photometry from WISE/2MASS, GALEX, and SDSS. The SED demonstrates a power-law relationship with a ``bump" at around 3 $\micron$ or log($\nu_{\rm rest}$) = 14 and another one starting around 14.75 in Figure \ref{fig:SEDWGS}. The latter one is attributed to Balmer continuum emission and optically thick thermal emission \citep{MS82}. The strengths of these ``bumps" vary widely, with nearly half of the AGN lacking a detectable bump. 

From our measurements of the flux ratios of the broad and narrow components of H$\beta$ and H$\alpha$, we determined the Balmer decrements of H$\alpha$/H$\beta$ to be 4 (broad) to 4.5 (narrow). These decrements are sensitive to the amount of internal dust reddening inside each AGN. However, even if the reddening of the broad lines and the continuum were known to be equal, the reddening correction cannot be determined without knowing the intrinsic (unreddened) Balmer decrement of the broad line regions (BLR). Unfortunately, several effects, especially collisions, can produce BLR H$\alpha$/H$\beta$ ratios considerably higher than the conventional Case B's value of $\sim$ 3.1 \citep{Lacy1982, Gaskell1984}. As an exercise, let us naively assume that the intrinsic ratios are describable with Case B's, in which case the observed Balmer decrements would imply AGN reddenings E(B-V) of 0.2--0.3 magnitude. We then compare this large reddening correction with the observed SED in Figure \ref{fig:SEDWGS}. In this figure, we de-redden a typical power law for a quasar using \cite{SF11}'s extinction coefficients and \cite{CCM89}'s analytical model (at the courtesy of \cite{Gordon2024} dust extinction package). From this comparison, we can see that the redder minority of these AGN could indeed have a continuum reddening of 0.3 mag. However, the larger majority of them cannot have such large continuum reddenings. As the diagram illustrates, a typical continuum power law, when dereddened for E(B-V), would yield an unreasonably blue intrinsic AGN. That would lead us to the implausible conclusion that many of these AGN are intrinsically bluer than even the bluest known quasars, which have constant $F_\nu$ in the visual to UV bands \citep{Malkan1983, Malkan1984, Malkan1988, Malkan1991}. We conclude that although our selection methods can find significantly reddened quasars, they nonetheless mostly discover quasars with low continuum extinctions.

Traditionally, the big blue bump due to the excess in the UV served as a main way to characterize AGN. Many works, such as \citet{SL12Fig11} and \citet{LR18}, have made detailed advances to study the SEDs and their dependence on luminosity. In Figure \ref{fig:SEDSL12LR18}, we consider the luminosity distribution of SED for Type 1 AGN and compare it with the mean SEDs in \citet{SL12Fig11} and the different AGN templates from \citet{LR18}. To the left of the graph (log $\nu_{\rm rest} \leq 14$), the SED is independent of luminosity, while it peaks in the UV. This characteristic blue bump becomes weaker in less luminous AGN. The black lines were mean SEDs from \citet{SL12Fig11} Figure 11, binning at different L$_{bH\alpha}$ with the R06 line being the net AGN SED. The more luminous quasars exhibit the same pattern as these black lines though at different luminosity levels. We further compare our data set against a few templates from \cite{LR18} as a comparison. The Spitzer (solid grey) and the AGN + Stars (dashed grey) templates describe the pattern behavior of the less luminous objects, while the quasar templates (dotted grey) were much too biased. This result implies the presence of star formation in some of the Type 1 AGN. These comparisons with previous literature suggest that we will not find a whole different population of AGN than those identified through optical spectroscopy and the spectral features such as the big blue bump in AGN SEDs.
\section{Conclusion}\label{sec:conclusion}	
This paper presents the reduced 125 new Kast spectra from Lick, 58 DBSP spectra from Palomar, and 3 LRIS spectra from Keck, mostly in the Kepler prime field and K2 fields from 0-5, 7, 8, 10, 12, and 13. Out of the 186 spectra, we identified 105 candidates as Type 1 and 35 as Type 2 AGN. After we confirmed the new AGN, we checked the reliability of our selection methods. Starting with \cite{EM12} IR and X-ray sample, we identified 47/49 as AGN, giving a reliability of 96\%, consistent with the established likelihood in the EM12 paper. For the WISE color cut from \cite{Stern}, we found a 90\% reliability, slightly below the suggested rate from the paper but similar to what \cite{Hviding2022} has found. The next step is to check that the sample of confirmed Kepler AGN exhibits properties similar to those previously studied in the literature. First, we plotted a BPT line ratio diagram of the narrow components. Four out of 16 narrow-line galaxies turn out to be composite galaxies with prominent narrow-line emissions from HII regions, and 12 are in the AGN region. Then, we constructed the Spectral Energy Distributions (SEDs) of Type 1 AGN from 22 $\micron$ to the UV. The SEDs showed the usual excess emission ``bumps" around 3 $\micron$. However, the strength of these bumps ranged widely, being undetectable in almost half the objects. The overall line and continuum properties of the newly discovered AGN do not differ markedly from those of previous samples (e.g. \citet{Malkan2017}). We then compared the AGN sample with a few works in previous literature and determined that the sample yielded the same AGN population as those that would have been found using other methods based on optical spectroscopy and color selection.
\appendix
\section{Analysis of emission line properties and luminosities in the Kepler AGN Candidates}
\label{sec:AGNemissionlineproperties}
Emission lines in AGN are critical for understanding the properties of both broad line regions (e.g. via the Balmer lines) and narrow-line regions (e.g. via the forbidden lines like [NII], [OIII], and [OII]). Numerous studies, including \citet{Osterbrock}, \cite{SL12Fig11} \cite{SL12Fig3}, and \cite{SL13}, have systematically explored AGN emission lines using their optical spectroscopic information. We present the analysis of AGN candidates' emission line properties, focusing on line-width ratios, fluxes, and luminosities. We compare our findings with existing literature to assess their consistency with established trends. This approach allows us to determine whether the AGN sample exhibits consistent properties with those identified through optical spectroscopy.

First, we analyzed full-width-half-maximum (FWHM) ratios for key emission lines such as H$\beta$ to H$\alpha$, using both narrow and broad-line Seyferts. The H$\beta$ to H$\alpha$ ratio follows the relation: 
\begin{equation}
    \rm FWHM_{H\beta} = (1.16 \pm 0.11) FWHM_{H\alpha} + (251 \pm 250)\, km s^{-1},
\end{equation}
which suggests that H$\beta$ profiles tends to be bigger than that of H$\alpha$, consistent with \citet{Osterbrock}'s result of $(1.16 \pm 0.05)$. 

To further explore AGN broad and narrow line regions, we examined the flux ratios of the Balmer lines. The correlation between logarithmic luminosities (erg s$^{-1}$) of the broad (b) and narrow (n) components of H$\beta$ and H$\alpha$ yielded consistent linear trends, which resulted in Balmer decrements of H$\alpha$/H$\beta$ = 4 (broad) to 4.5 (narrow).

\begin{figure}
\begin{center}
	\includegraphics[width=19pc, height = 15pc]{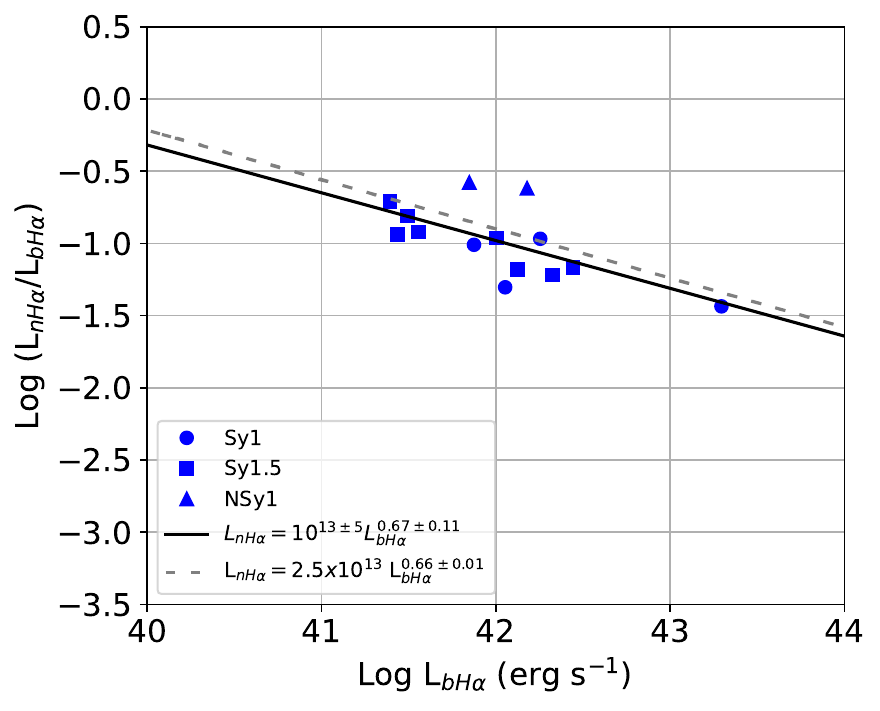}
\end{center}
\caption{The luminosity distribution of narrow H$\alpha$ to broad H$\alpha$ vs. L$_{bH\alpha}$. The black solid line is the best fit of the data. The grey dashed line is the best-fit power law of the T1 sample from \cite{SL12Fig3} Figure 3. The relationship between L$_{nH\alpha}$ and L$_{bH\alpha}$ in our data is consistent with that of \cite{SL12Fig3}, and the NLR/BLR ratio is increasing with decreasing luminosity.}
\label{fig:LnHa_LbHa}
\end{figure}

\begin{figure}[b]
\begin{center}
	\includegraphics[width=20pc, height = 15pc]{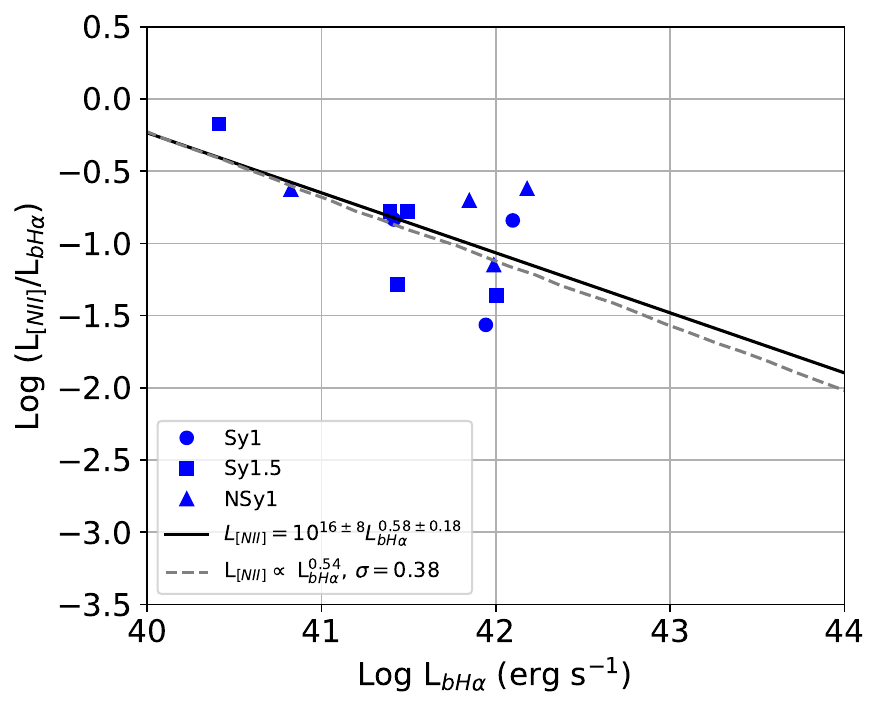}
\end{center}
\caption{The luminosity ratio of [NII] to broad H$\alpha$ vs. L$_{bH\alpha}$. The black solid line best fits the data, while the grey dashed line is from \cite{SL13} Figure 1. The same tendency for the narrow components to dominate at low luminosity appears here as in the previous two plots.}
\label{fig:LNii_LbHa}
\end{figure}

Our analysis extended to the luminosity ratio of narrow components of H$\alpha$ (Figure \ref{fig:LnHa_LbHa}) and [NII]-6584 (Figure \ref{fig:LNii_LbHa}) to broad H$\alpha$. Using L$_{bH\alpha}$ as the determining characteristic, we established relationships with the aforementioned lines through the best-fits power law. In Figure \ref{fig:LnHa_LbHa}, we got L$_{nH\alpha}$ $\propto$ L$_{bH\alpha}^{0.67 \pm 0.11}$, which is consistent with \cite{SL12Fig3}'s result of L$_{nH\alpha}$ $\propto$ L$_{bH\alpha}^{0.66 \pm 0.01}$. Moreover, these Seyfert 1.5 candidates are also within \cite{SL12Fig3}'s defined range of intermediate-type AGN. The narrow-to-broad ratio is increasing with decreasing luminosity, which suggests that narrow H$\alpha$ becomes more prominent at low luminosity. [NII]-6584's results also follow this same tendency as seen in Figure \ref{fig:LNii_LbHa}, giving a proportionality of L$_{[NII]}$ $\propto$ L$_{bH\alpha}^{0.58 \pm 0.18}$. Looking at the relationship between these three narrow lines with broad H$\alpha$, the narrow components of AGN tend to dominate at lower luminosity.

\begin{figure}
\begin{center}
\includegraphics[width=20pc, height = 15pc]{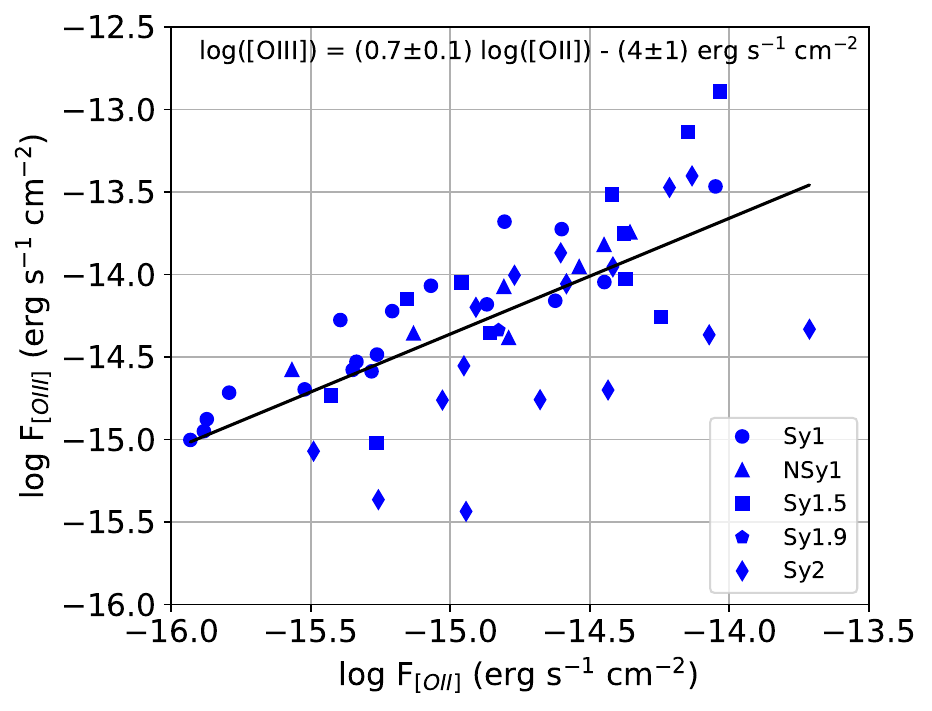}
\end{center}
\caption{Flux measurements of [OIII] vs [OII] in logarithm. The result yields a correlation of log [OIII] (erg s$^{-1}$ cm$^{-2}$) = (0.7 $\pm$ 0.1) log [OII] - (4 $\pm$ 1),
indicating that [OII] rose much faster in flux than [OIII].}	
\label{fig:oxygen}
\end{figure}

Finally, we explored the relationship between the fluxes of [OIII] and [OII] (Figure \ref{fig:oxygen}). The result yields a correlation of 
\begin{equation}
	\rm \log [OIII] (erg\,s^{-1} cm^{-2}) = (0.7 \pm 0.1) \log [OII] - (4 \pm 1),
\end{equation}
indicating that [OII] flux rose much faster than [OIII]. From this analysis, we determined that AGN selected via the IR colors yield the same population and properties as those identified through optical spectroscopy.

\bibliography{main}{}
\bibliographystyle{aasjournal}

\end{document}